\documentclass[oneside,12pt]{CUEDthesisPSnPDF}

\pdfoutput=1

\ifpdf
	\title{\LARGE\textsc{Miniaturized Microwave Devices and Antennas for Wearable, Implantable and Wireless Applications}}
  \author{Muhammad Ali Babar Abbasi}
  \collegeordept{{Department of Electrical Engineering}}
  \university{\href{http://www.frederick.ac.cy}{Frederick University, Nicosia, Cyprus}}
  \crest{\includegraphics[width=14mm]{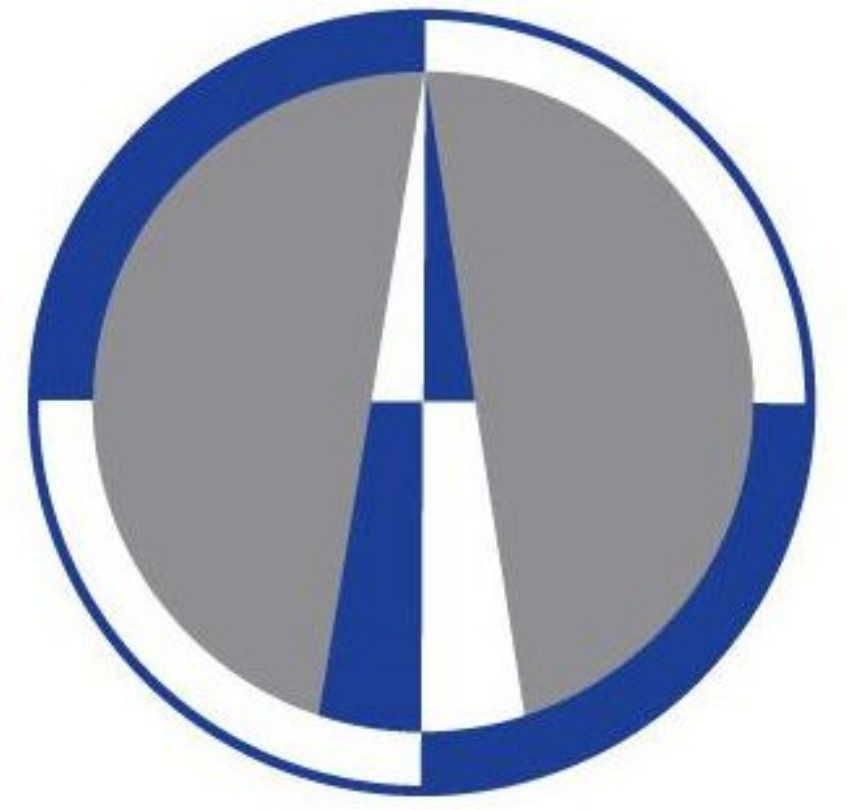}}
\else
  \author{Muhammad Ali Babar Abbasi}
  \collegeordept{Department of Electrical Engineering}
  \university{Frederick University, Nicosia, Cyprus}
  \crest{\includegraphics[bb = 0 0 292 336, width=30mm]{Frederick_logo}}
\fi

\degree{PhD -- Electrical Engineering}
\degreedate{2017}

\usepackage{watermark}
\usepackage{subcaption}
\usepackage{graphicx}
\usepackage{arabtex}
\usepackage{utf8}
\usepackage{float}
\usepackage{perpage}

\usepackage{amsmath}

\begin{document}




\maketitle

\setcounter{secnumdepth}{3}
\setcounter{tocdepth}{3}

\frontmatter 
\pagenumbering{roman}


\begin{dedication} 
	Dedicated to my parents. 
	
	%
	%

\end{dedication}





\begin{acknowledgements} 
	
	\setcode{utf8}
	
	Praise be to Allah \<سبحانَه و تعالى>, the Most Gracious, the Most Merciful, for blessing me with the strength to complete this work. I pray that the growth I achieved would be used in his cause. I am infinitely grateful to my parents for passing down the values and constant encouragement throughout my educational career. I thank the Frederick University for giving me the opportunity to pursue my doctoral studies. I am also thankful to the Erasmus Mundus INTACT doctoral level scholarship program for funding my research. I am thankful to the School of Electrical Engineering and Computer Sciences, National University of Sciences and Technology (NUST), Islamabad, Pakistan in grooming my research aptitude. Specifically, I am grateful to Dr. Munir Ahmed Tarar for advising my thesis. I am thankful to COMSATS Institute of Information and Technology, Islamabad, Pakistan for providing me strong undergraduate education and to Dr. Syed Muzahir Abbas for his guidance. 
	
	I would like to extend my profound gratitude to my mentor and principal advisor Dr. Symeon Nikolaou for his constant guidance and encouragement that I received from him. He trusted in me, always held me to a high standard, and encouraged me to work hard. In addition to being my advisor, he is an amazing person and I have been thoroughly enriched working with him. It has been an honor for me to work under his supervision. I am also thankful to Dr. Marco A. Antoniades for his active participation in my research endeavors throughout my tenure at Frederick University. He always encouraged and supported me. He extended his help by allowing me to be a part of his research laboratory at the University of Cyprus (UCY). I would like to thank Dr. Photos Vrionides for his constant support and kind words. I would like to thank Professor Antonije R. Djordjevic and Dr. Marija M. Nikolic for their help and guidance while my scientific mission in the University of Belgrade, Serbia. Also, thanks to Tim and Harish from Cambridge University for contributing in public CUED Latex material. 
	
	I would like to thank my friends who made my stay in Cyprus the most memorable, always supported me and encouraged me, tolerated my sarcasm, and accepted me with my workaholic nature. Thanks to Waqar (the guide), Salman (the information), Naveed (doctor sab), Taqua (the kindness), Christina (the humbleness), Mostafa (the inspiration), Rafay (the discipline), (sweet) Abdul, Agn\'e (the amazing), Yousaf (the curious), Haroon (the help), Zunaib (the calm), Aqeela (the taste) and Zainab (the fun). I could always count on the support and encouragement from Badar bhai, Haroon bhai and family, also from Saleem, Rizwan, Ilyas, Hamza, Sahar, Ambreen and Hassan, even remotely. I also enjoyed the company of Dr. Hassaan, Maria Api and kids, Sir Zeeshan, Sir Ajmal, Dr. Adnan, Dr. Umer, Ms. Rania, Birendra (late), Diogo, Daniela, Kinga, Andreas(s), Hanne, Constantinos, Ouan, Mohamed, Kata, Simran, George, Marilena(s), Angelena, Krisztina, Veronica, Dora, Chloe, Mainul Haque, James, Krish, Ahmed, Winnie, Talha, Parash, Melanie and Sergio. The time spent with my colleagues, working as a team, deep discussions, long walks, adventures, travelling within Cyrus and abroad, road trips, and most of all the sleepless nights working in the lab and time spent in "PhD House" will always be a valuable part of my memory.

\end{acknowledgements}


\begin{abstracts}    

	This thesis presents a number of microwave devices and antennas that maintain high operational efficiency and are compact in size at the same time. One goal of this thesis is to address several miniaturization challenges of antennas and microwave components by using the theoretical principles of metamaterials, metasurfaces coupling resonators and stacked radiators, in combination with the elementary antenna and transmission line theory. While innovating novel solutions, standards and specifications of next generation wireless and bio-medical applications were considered to ensure advancement in the respective scientific fields. A compact reconfigurable phase-shifter based on negative-refractive-index transmission-line (NRI-TL) materialist unit cells is presented. The corresponding work focuses on implementation of two inter-switchable NRI-TL metamaterial unit cells collocated within the same board area in a reconfigurable manner. In addition to phase shifter, a practical implementation of a miniaturized microwave crossover using NRI-TL metamaterials is demonstrated. The metamaterial theory has signify that by loading a conventional transmission line with lumped LC elements, a large phase shift per-unit-length can be achieved. Using this feature of NRI-TL metamaterials, the miniaturization of a crossover has been achieved by replacing the conventional microstrip transmission lines in a three-section branch-line coupler crossover with NRI-TL metamaterials lines, representing a compact NRI-TL crossover with an area 47 times smaller than the area of the conventional counterpart. By moving from 1D metamaterials to 2D metasurfaces, an array of Electromagnetic Band Gap (EBG) structures is introduce. The corresponding study discussed the use of an EBG array structure as a director for efficient radiation of a wearable antenna. A planar monopole printed on a partially grounded substrate is first loaded with a \textit{M}${\times}$\textit{N} EBG structure to achieve a maximum directivity along the \textit{z}-axis when the antenna is placed in the \textit{x-y} plane. A successful attempt of miniaturizing such structure to a 2${\times}$1 element array excited by a monopole to form an efficient antenna is presented. The reflection phase of a single EBG unit cell has been studied and exploited towards efficient radiation of planar monopole. The shape of the EBG unit cell and the gap between the ground and the EBG layer are adjusted so that the antenna operates at 2.45 GHz. The proposed antenna retained its impedance matching when placed directly upon a living human subject with an impedance bandwidth of 5\%, while it exhibits a measured gain of 6.88 dBi. A novel equivalent array model is presented to qualitatively explain the radiation mechanism of the antenna. In the next study, a fork shaped wearable sensor antenna is designed by utilizing the available ground space of the EBG backed monopole. The antenna is designed to efficiently radiate towards the human body while keeping the detuning and impedance mismatching affects to a minimum, which are reported to be inevitable due to body loading. For this study, a high resolution numerical human body phantom was used and Finite Difference Time-domain (FDTD) solver was used as a simulation environment. Challenges and possible solutions for through-body communication has been investigated. Next, a fully parametrized solution for an implantable antenna is proposed using metallic coated stacked substrate layers. The intended solution makes it possible to use the proposed antenna structure as an implant at several different areas of the human body with potentially distinct electrical properties with consequently different detuning effect. Particle swarm algorithm has been implemented to optimize the antenna performance while operating inside distinct human tissue surroundings. Within dead tissue (quasi \textit{in-vivo}) operation of the implantable antenna, on-body operation of the wearable sensor antenna, and an efficient communication of EBG backed monopole with an off-body transceiver depicts a practical solution for body-centric wireless communication system. The entire study answered several research questions of applied electromagnetic in the field of bio-medicine. Next, miniaturization and implementation of a UWB antenna along with an analytical model to predict the resonance is presented. Lastly, several miniaturized rectifiers designed specifically for efficient wireless power transfer are proposed, experimentally verified, and discussed.

\end{abstracts}

\tableofcontents
\listoffigures
\printnomenclature  
\addcontentsline{toc}{chapter}{Nomenclature}

\mainmatter 

\chapter{Introduction}
\ifpdf
\graphicspath{{Introduction/IntroductionFigs/PNG/}{Introduction/IntroductionFigs/PDF/}{Introduction/IntroductionFigs/}}
\else
\graphicspath{{Introduction/IntroductionFigs/EPS/}{Introduction/IntroductionFigs/}}
\fi

\section{Motivation}

Wireless devices are an essential part of our everyday life these days. Advancements in the wireless technology seem to further support the necessity of the wireless devices in such a way that we are moving from person-to-person wireless communication towards person-to-device and device-to-device communication. With this trend, a future can be envisioned in which "everything" is connected, especially in a high density urban environment. This leads to a huge number of wireless nodes, trying to establish an uninterrupted connectivity with each other, either directly, or with an aid of a relaying node. To support such connectivity, applied electromagnetic hold the grave importance since all the wireless technologies relies primarily on the physics of electromagnetic waves. Improving the overall efficiency of wireless communication system by studying and proposing improved radiating antennas and microwave devices to support an enormous demand of connectivity is the main motivation behind this work. Possible applications of these devices should cover a wide spectrum of prospective next generation wireless technologies including 5G, health care, and wireless power transfer.

\section{Objectives}

The major objectives of this work are: (1) designing, miniaturization and realization of microwave devices/transmission lines; (2) the design of compact high gain wearable antenna; (3) to develop an efficient implantable antenna; (4) to develop a wearable sensor node for the communication with off-body transceiver and an implantable antenna; (5) the development of techniques to reduce the decoupling and impedance miss-matching of a wearable antenna; (6) to miniaturize the UWB antenna and (7) to design compact rectifier for low power input.

\section{Organization of the Dissertation }

As stated before, in this thesis, a number of compact microwave devices and antennas are discussed and compared to current state-of-the-art. In second chapter, elementary theory of metamaterials and metasurfaces is introduced. A compact reconfigurable phase shifter solution inspired by NRI-TL metamaterial unit cell is discussed. Same theory is implemented to miniaturize the size of a branch-line coupler microwave crossover.  In third chapter, a planar monopole backed with a 2$\boldsymbol{\times}$1 array of Electromagnetic Band Gap (EBG) structures is introduced, operating as a wearable antenna. In detail discussion on miniaturization procedure along with the use of an EBG array structure  as  a  director  for  efficient  resonance  of  a  printed  monopole antenna is presented. A novel radiation mechanism model to quantify the antenna's power patterns has been explained. Antenna has further been tested for human body loading and structural deformation.  Chapter 4 discusses an electrically small implantable antenna designed using stacked microstrip layer technology. The antenna performance has been tested in simplistic and high resolution numerical phantom and practical implementation has been verified by measurements. In Chapter 5, wearable sensor antenna is introduce to study the challenges of body centric wireless communication. Through-body communication has been verified between wearable sensor antenna and an implantable antenna. Chapter 6 presents miniaturization techniques for UWB antenna and analytical model to predict the antenna resonances using proposed geometry. Finally, a number of rectifiers for efficient wireless power transfer are presented and discussed at the end of the thesis in Appendix A. The list of publications that are either already in press or in the process of review are enlisted in Appendix B.



\chapter{Metamaterial Based Electrically Small Microwave Devices}
\ifpdf
\graphicspath{{Chapter1/Chapter1Figs/PNG/}{Chapter1/Chapter1Figs/PDF/}{Chapter1/Chapter1Figs/}}
\else
\graphicspath{{Chapter1/Chapter1Figs/EPS/}{Chapter1/Chapter1Figs/}}
\fi

This chapter presents an introduction and miniaturization applications of negative-refractive-index transmission-line (NRI-TL) metamaterial. A compact reconfigurable phase-shifter based on NRI-TL metamaterial unit cells is presented and discussed. Two inter-switchable NRI-TL metamaterial unit cells are collocated on the same board area, and can be reconfigured based on the biasing polarity of embedded PIN diodes to provide two discrete phase states. The PIN diodes are located on the shunt branches of the metamaterial line in order to reduce losses in the direct signal path. Design limitations in terms of return loss and insertion loss are discussed in relation to the two phase states. A proof-of-concept module is designed for the phase advance of 55.5${}^{\circ}$ and 128.5${}^{\circ}$ in two reconfigurable states of the module when the insertion loss for the two states is ${-}$1.43 dB and ${{-}}$0.89 dB respectively. Simulation and measurements are compared to test the validity of designed phase shifter. In the end a practical implementation of a miniaturized microwave crossover using NRI-TL metamaterials is presented.

\section{Introduction to Metamaterials}
The Metamaterials can be generally defined as a class of an artificial material, or a wave propagation media, exhibiting extraordinary electromagnetic properties that cannot be found in natural ones \cite{engheta2006metamaterials}. The name is based on a Greek letter $\mu\varepsilon\tau\alpha$ (meta) that means "after" or "beyond" It means that the these material do not follow the natural principles and can be terms as "$\mu\varepsilon\tau\alpha$" physical in nature. Metamaterials has a number of exciting applications that has drawn the attention from the scientific community of multiple disciplines. One interesting fact about Metamaterials is that it draw the physics and engineering community close together based on their superior priorities shocking for the physicists, while incredible for the engineers. There is nothing new in metamaterial theory in terms of fundamental science and the whole theory can be understood based on the classical electromagnetic theory. The theoretical concept of the existence of backward-waves based on the signs of permittivity was first developed in 1904 \cite{lamb1904group} and since then, the artificially arranged complex materials have been the subject to scientific advancement since then.

A key feature of these artificial materials is the negative refraction. When an Electromagnetic wave passes through these materials, the electric vector $\vec{E}$, the magnetic vector $\vec{H}$ and the wave vector $\vec{k}$ do not follow the "right hand rule" for the electromagnetic wave propagation. In other words, the wave propagation is in --$\vec{k}$ direction. Due to this property, occasionally "left handed materials (LHM)" is the title given to these materials in literature. In other literature, the property of the materials is not focused, but the wave propagation is, so the name given is "backward-wave materials", which is also the term used in this thesis. The two parameters used to characterize the electric and magnetic properties of materials are the permittivity ($\epsilon$) and the permeability ($\mu$), when electromagnetic fields are studied to interact the material. The measure of the changes in the medium to absorb the electric energy is described as ($\epsilon$) and it relates to the terms $\vec{E}$ known as electric field strength and $\vec{D}$, known as electric displacement as a result of $\vec{E}$. The term "relative dielectric constant" or ($\epsilon_r$) used normally in literature is actually the ratio of permittivity of the material to that of free space. Where the free space permittivity denoted by ($\epsilon_0$) is $8.85\times10^{-12} F/m$. On the other hand, the Permeability is a constant of proportionality that exists between 
magnetic induction and magnetic field intensity while the free space permeability is denoted by $\mu_0$ and is approximately $4\pi\times10^{-7} H/m$. It has been discussed previously that the for the normal materials, the wave propagation's electric and magnetic properties are defined by the permittivity and permeability of the material. Both these terms are positive in nature. However, for the Metamaterials, both these terms tend to be negative, while the electromagnetic wave propagation is still possible because the product of these terms still gives a positive. The constitutive material parameters in terms of refractive index can be written as: 
\begin{equation}\label{I1}
{\eta} =\pm\sqrt{\epsilon_r\mu_r} 
\end{equation}

So the squared $\eta$ is not affected by the signs of the $\epsilon_r$ and $\mu_r$. 

\subsection{Electromagnetic wave propagation in metamaterials}

Let us see the Maxwell's equation reduced to a wave equation and the general equation of a plane wave is: 

\begin{equation}\label{I2}
\vec{E} =\vec{E_0}e^{-j\vec{k}\hat{r}} 
\end{equation}

\begin{equation}
\vec{H}={\vec{H}_0}e^{-j\vec{k}\hat{r}}
\label{I3}
\end{equation}

Where $\vec{E_0}$ and $\vec{H_0}$ are the vectors in the direction of propagation of the wave $\vec{k}$ while $\vec{r}$ is the position vector for the observation point. The expressions can be written as: 

\begin{equation}
\vec{k}={k_x\hat{x} + k_y\hat{y} + k_z\hat{z}}
\end{equation}

\begin{equation}
k=\sqrt{k_x^2 + k_y^2 + k_z^2}
\end{equation}

\begin{equation}
\vec{r} = x.\hat{x} + y.\hat{y} + z.\hat{z}
\end{equation}

We can solve the Maxwell's equations by transforming the system into second order partial differential equation. Writing the equations in time harmonic form yields:

\begin{equation}
\nabla \times \vec{E} = -j\omega\vec{B} 
\end{equation}

\begin{equation}
\nabla \times \vec{H} = j\omega\vec{D} + \vec{J} 
\end{equation}

We also know that:

\begin{equation}\label{i4}
\nabla\cdot{\vec{B}} = \nabla\cdot{\mu\vec{H}} = 0
\end{equation}

\begin{equation}\label{i5}
\nabla\cdot{\vec{D}} = \nabla\cdot{\epsilon\vec{E}} = \rho_v
\end{equation}

Equations \ref{i4} and \ref{i5} can be rewritten in the form: 

\begin{equation}\label{i7}
\nabla \times \vec{E} = -j\omega\mu\vec{H}
\end{equation}

\begin{equation}
\nabla \times \vec{H} = j\omega\epsilon\vec{E} + \vec{J} 
\end{equation}

In the medium having free charges that allow the current to flow ($\vec{J}=\sigma\vec{E}$), re-writing the Ampere's Law will give: 

\begin{equation}\label{i6}
\nabla \times \vec{H} = j\omega\epsilon\vec{E} + \sigma\vec{E} = j\omega[\epsilon+\frac{\sigma}{j\omega}]\vec{E} 
\end{equation}

Where the term $[\epsilon+\frac{\sigma}{j\omega}]$, written as $[\epsilon-j\frac{\sigma}{\omega}]$ in the phaser domain shows that the effective complex permittivity can be represented by the a lumped expression of the conductivity and the permittivity. So this expression can be represented as: 

\begin{equation}
\epsilon_{eff} = [\epsilon-j\frac{\sigma}{\omega}] = \epsilon_0[\epsilon_r-j\frac{\sigma}{\omega\epsilon_0}] = \epsilon_0\epsilon_{ref}
\end{equation}

And the equation \ref{i6} becomes: 

\begin{equation}\label{i8}
\nabla \times \vec{H} = j\omega\epsilon_{eff}\vec{E}
\end{equation}
While the $\epsilon_{eff}$ is complex. For the field vectors, we know that:

\begin{equation}
\vec{E} \perp \vec{H} \perp \vec{k}
\end{equation}

So, 
\begin{equation}
\vec{E} = {\eta}\vec{H}\times\vec{k}\ 
\end{equation}

\begin{equation}
\vec{H} = \frac{1}{\eta}\vec{k}\times\vec{E}
\end{equation}

For the plane waves with expressions $\vec{E} =\vec{E_0}e^{-j\vec{k}\hat{r} + j\omega t} $ and $\vec{H} =\vec{H_0}e^{-j\vec{k}\hat{r} + j\omega t} $, the equations \ref{i7} and \ref{i8} will eventually reduce to: 

\begin{equation}
\vec{k}\times \vec{E} = 
\begin{cases}
\omega\mu\vec{H}, & \text{if}\ \mu > 0 \\
-\omega|\mu|\vec{H}, & \text{if}\ \mu < 0
\end{cases}
\end{equation}

\begin{equation}
\vec{k}\times \vec{H} = 
\begin{cases}
-\omega\epsilon\vec{E}, & \text{if}\ \mu > 0 \\
\omega|\epsilon|\vec{E}, & \text{if}\ \mu < 0
\end{cases}
\end{equation}

Note that in case of both aforementioned cases i-e backward-wave and forward wave propagation, the time averaged flux of energy is determined by: 

\begin{equation}
\vec{S} = \frac{1}{2}\vec{E}\times\vec{H}^*
\end{equation}

Where $\vec{S}$ shows the actual propagation of energy. For the case of metamaterial, the backward-wave has $\vec{S}$ and $\vec{K}$ in reverse directions. In the proceeding discussions in the thesis chapters, it will be shown that the most extraordinary electromagnetic and transmission line based properties of metamaterial revolves around this backward-wave propagation.

\subsection{Metamaterial Applications}

Artificial material engineering has enabled the design and development of compact, cost-effective and high-performance solutions at microwave, terahertz and optical frequencies. The electromagnetic behavior of metamaterial structures is modified intentionally by adjusting either their electrical or geometrical properties to formulate non-linear materials, semiconductor structures, liquid crystals, microfluids and so on \cite{oliveri2015reconfigurable}. Within the same framework, the increasing demand for tunable microwave systems has further advanced research towards reconfigurability within metamaterials. At microwave frequencies, recent developments in negative-refractive-index transmission line (NRI-TL) metamaterials \cite{eleftheriades2002planar,antoniades2003compact} have enabled the design and realization of electrically small reconfigurable metamaterial lines. The reconfigurability is achieved by loading a host transmission line with lumped capacitors and inductors. The practical use of such NRI-TL metamaterials still poses a number of technological and theoretical challenges in terms of operational bandwidth, losses, realization complexity, tuning speed and tolerance sensitivity. Over and above the aforementioned challenges, the inclusion of the tuning mechanism (biasing lines and control mechanism etc.) must not affect the desired electromagnetic behavior of the reconfigurable system. Some prominent techniques to achieve reconfigurability include microfluidic channels \cite{choi2015novel}, CMOS-based active inductors \cite{abdalla2007printed}, BST thick-film substrates \cite{duran2011electrically}, GaAs \cite{kim2005linear,kim2007compact} and Schottky varactor diodes \cite{damm2005compact}, ferroelectric varactors \cite{kuylenstierna2006composite}, MEMS capacitors \cite{perruisseau2009low}, and metal-insulator-metal MIM capacitors \cite{michishita2011tunable}. 

\section{Reconfigurable NRI-TL Metamaterial Phase Shifter}

In this work, a reconfigurable NRI-TL metamaterial phase shifter is presented that uses PIN diodes to change the insertion phase through the device. Two phase states have been achieved by reciprocating the polarity of the biasing voltage. Using the same physical microstrip transmission line to host more than one NRI-TL metamaterial unit cells, leads to a degradation in the performance of the phase shifter when switching from one state to another. These issues are thoroughly discussed, leading to the major contribution of this work, which is the application of a multivariable optimization algorithm that allows the device to exhibit low insertion loss and good impedance matching for both phase states. A detailed analysis relating to design limitations and the matching topology used to match both phase states in a common design is presented in the following sections.

\begin{figure}[t]
	\centering
	\begin{subfigure}{0.4\textwidth}
		\includegraphics[width=1\textwidth]{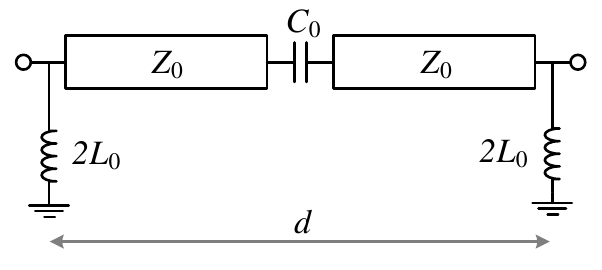}
		\caption{}
	\end{subfigure}
	\begin{subfigure}{0.4\textwidth}
		\includegraphics[width=1\textwidth]{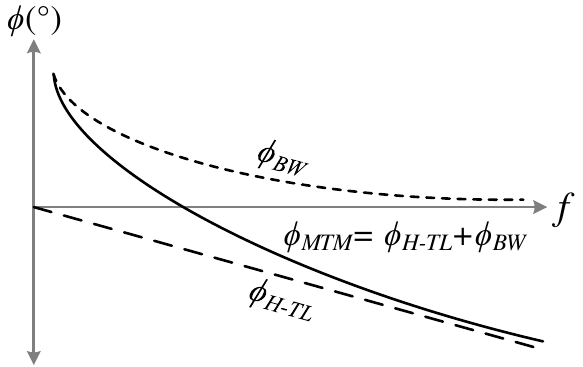}
		\caption{}
	\end{subfigure}
	\begin{subfigure}{0.4\textwidth}
		\includegraphics[width=1\textwidth]{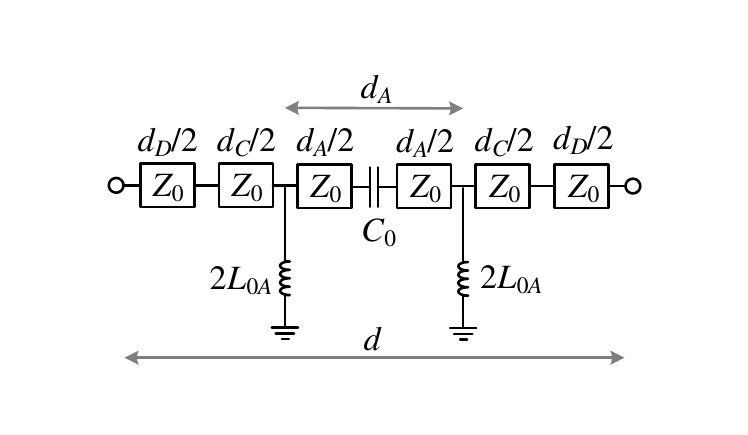}
		\caption{}
	\end{subfigure}
	\begin{subfigure}{0.4\textwidth}
		\includegraphics[width=1\textwidth]{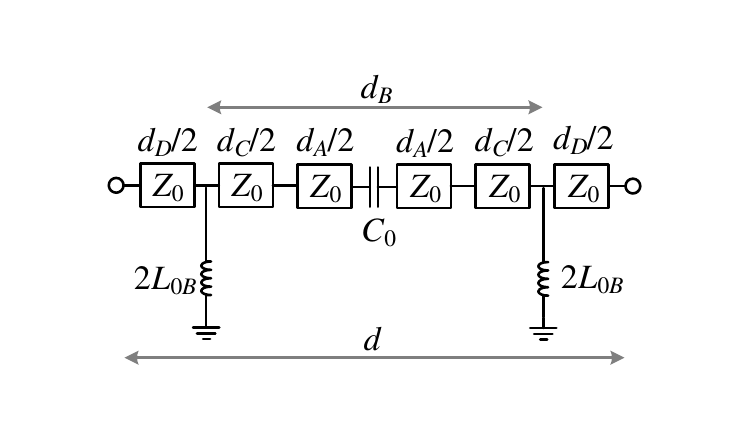}
		\caption{}
	\end{subfigure}
	\caption{Fig.1 (a) NRI-TL metamaterial $\Pi$ unit cell, (b) phase response of a metamaterial line as a composite response of the transmission line (TL) and backward-wave (BW) line responses, and metamaterial line architecture for (c) phase \textit{State A}, and (d) phase \textit{State B} (\textit{d${}_{D}$} represent extra length of transmission line outside the unit cells).}
	\label{mtmf1}
\end{figure}

NRI-TL metamaterials are synthesized by periodically loading a conventional microwave transmission line with lumped-element series capacitors and shunt inductors. Figure \ref{mtmf1}(a) shows the equivalent circuit of a symmetric $\Pi$-shaped NRI-TL metamaterial unit cell \cite{antoniades2016transmission}. Here, a host transmission line with characteristic impedance \textit{Z}${}_{0}$, inductance per unit length \textit{L}, capacitance per unit length \textit{C}, and length \textit{d} has been loaded with a lumped-element series capacitor \textit{C}${}_{0}$ and two shunt inductors \textit{2L}${}_{0}$. The conventional host transmission line realizes a negative phase response, \textit{$\phi$${}_{H-TL}$}, while the series capacitive and shunt inductive loading forms a backward-wave (BW) line with a positive phase response,\textit{ $\phi$${}_{BW}$}, as shown in Figure \ref{mtmf1}(b). Under effective medium conditions, the phase shift per NRI-TL metamaterial unit cell is given by \cite{antoniades2003compact}: 

\begin{eqnarray}\label{mtm1}
\phi _{MTM} =\phi _{H-TL} +\phi _{BW} 
\end{eqnarray}

\begin{equation}\label{mtm2}
\phi _{MTM} =-\omega \sqrt{LC} d+\frac{1}{\omega \sqrt{L_{0} C_{0} } } 
\end{equation}

This is valid under the impedance matching condition below, which stipulates that both the lines must have the same characteristic impedance.

\begin{equation}\label{impedance}
Z_{0} =\sqrt{\frac{L}{C} } =\sqrt{\frac{L_{0} }{C_{0} } } 
\end{equation}

In this work, a microstrip transmission line with an impedance of \textit{Z}${}_{0}$ \textit{= }50 $\Omega$ and a fixed length of \textit{d} = 35 mm is used to host two collocated and reconfigurable NRI-TL metamaterial phase-shifting lines that realize two discrete phase states, \textit{State A} and \textit{State B}, given by \textit{$\phi$${}_{MTM\_A}$} and \textit{$\phi$${}_{MTM\_B}$}, respectively. The architecture of the two individual NRI-TL metamaterial lines is shown in Figure \ref{mtm1}(c) and (d). Both lines have the same series components, and differ only in the locations and values of the shunt loading inductors, \textit{L}${}_{0}$\textit{${}_{A}$} and \textit{L}${}_{0}$\textit{${}_{B}$}${}_{. }$By switching between the two inductor values, a fixed phase change can be effected, according to (1b). A change in the value of \textit{L}${}_{0}$ without changing \textit{C}${}_{0}$, however, will lead to a mismatch along the line, since the impedance matching condition of (\ref{impedance}) will no longer apply. This is mitigated by the fact that the two metamaterial unit cells used in Figure \ref{mtmf1}(c) and (d) also differ in size, given by \textit{d${}_{A}$} and \textit{d${}_{B}$} respectively, providing an additional degree of freedom that is used to provide good impedance matching in both states. 


\begin{figure}[htb]
	\centering
	\includegraphics[width=1\textwidth]{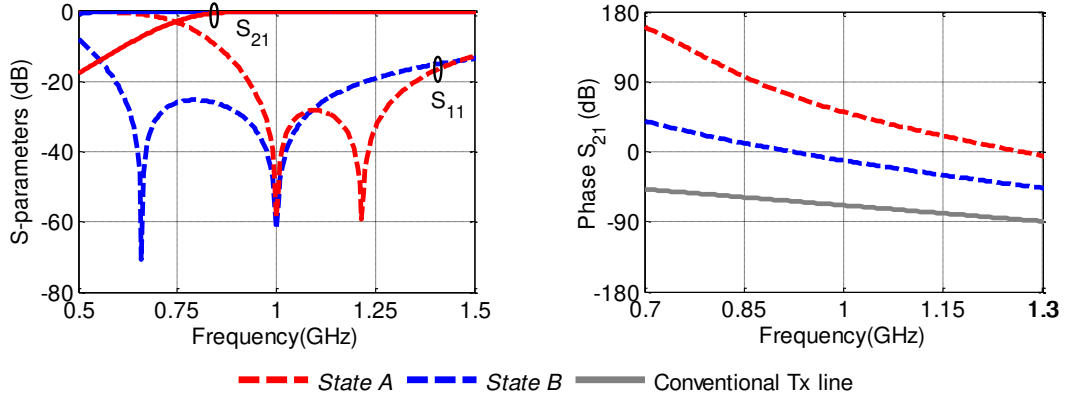}
	\caption{Simulated responses for magnitude of S-parameters (left), and phase of S${}_{21}$ (right) of the two-state NRI-TL metamaterial lines of Figure \ref{mtmf1} (c) and (d), compared to the same length conventional transmission line. }
	\label{mtmf2}
\end{figure}

\begin{figure}[htb]
	\centering
	\includegraphics[width=1\textwidth]{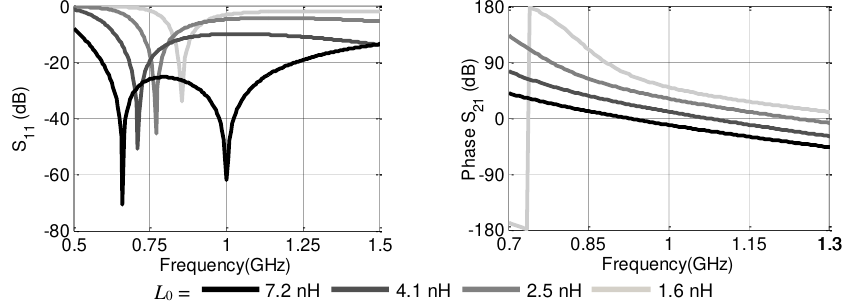}
	\caption{Simulated responses for magnitude of S-parameters (left), and phase of \textit{S$_{21}$} (right) of the NRI-TL metamaterial line as the value of the shunt loading inductor L0 is varied from 8.0 nH in State A with a phase of \textit{$\phi$${}_{MTM\_A}$} = $\mathrm{-}$15${}^\circ$ to 1.5 nH in State B with a phase of \textit{$\phi$${}_{MTM\_B}$} = +60${}^\circ$. The length of the unit cell size, d, and the loading capacitance \textit{C$_0$} remain the same for \textit{$\phi$${}_{H-TL}$} = $\mathrm{-}$70${}^\circ$, \textit{f$_0$} = 1 GHz, and \textit{C$_0$} = 3.2 pF.}
	\label{mtmf3}
\end{figure}

Using \ref{mtm1} and \ref{mtm2} and the design procedure outlined in \cite{antoniades2016transmission}, p. 17 with \textit{$\phi$${}_{H-TL}$} = ${-}$70${}^\circ$ at a design frequency of \textit{f${}_{0}$} = 1 GHz, the loading element values required to achieve a phase of \textit{$\phi$${}_{MTM\_A}$} = ${-}$15${}^\circ$${}^{ }$in \textit{State A} are \textit{C}${}_{0}$ = 3.2 pF, and \textit{L}${}_{0}$ = 8.0 nH, while to achieve a phase of \textit{$\phi$${}_{MTM\_B}$} = +60${}^\circ$${}^{ }$in \textit{State B} the values are \textit{C}${}_{0}$ = 1.4 pF, and \textit{L}${}_{0}$ = 3.5 nH. The transmission and reflection characteristics of the two-state metamaterial line are presented in Figure \ref{mtmf2} when the NRI-TL metamaterial unit cell presented in Figure \ref{mtmf1}(a) is used. It can be observed from Figures \ref{mtmf2}(a) and (b) that in both states the metamaterial line exhibits excellent transmission and reflection characteristics around 1 GHz, while the two desired phase states are achieved. 

\subsection{Design Process}

Combining two NRI-TL metamaterial unit lines representing two distinct phase states poses two major challenges. Firstly, when the same transmission line is used to host two NRI-TL unit cells with reconfigurability only in the shunt branch, a single series loading capacitor \textit{C}${}_{0}$ must be used for both phase states, which results in impedance matching degradation. To further elaborate on this problem, consider an NRI-TL metamaterial unit cell that incurs a phase of \textit{$\phi$${}_{MTM\_A}$} = ${-}$15${}^\circ$ at 1 GHz in \textit{State A} that is required to change its phase to \textit{$\phi$${}_{MTM\_B}$} = +60${}^\circ$ at 1 GHz in \textit{State B}. In order to increase the phase incurred to achieve a higher phase in \textit{State B} using the same unit cell size and series capacitive loading, the shunt loading inductance \textit{L}${}_{0}$ must be decreased according to (1b), since \textit{$\phi$${}_{H-TL}$} and \textit{C}${}_{0}$ will remain constant. The gradual decrease in the value of \textit{L}${}_{0}$ that leads to a higher incurred phase can be seen clearly in Figure \ref{mtmf3}, together with the commensurate degradation of the impedance matching at 1 GHz. This is due to the mismatch between the characteristic impedance of the transmission line and the backward-wave line, which makes the impedance matching condition of equation \ref{impedance} invalid. The second major challenge in the implementation of the two-state phase shifter is that a reconfigurable mechanism is required that does not disturb the matching and phase response of the metamaterial line while switching between the two phase states. 

\begin{figure}[htb]
	\centering
	\includegraphics[width=1\textwidth]{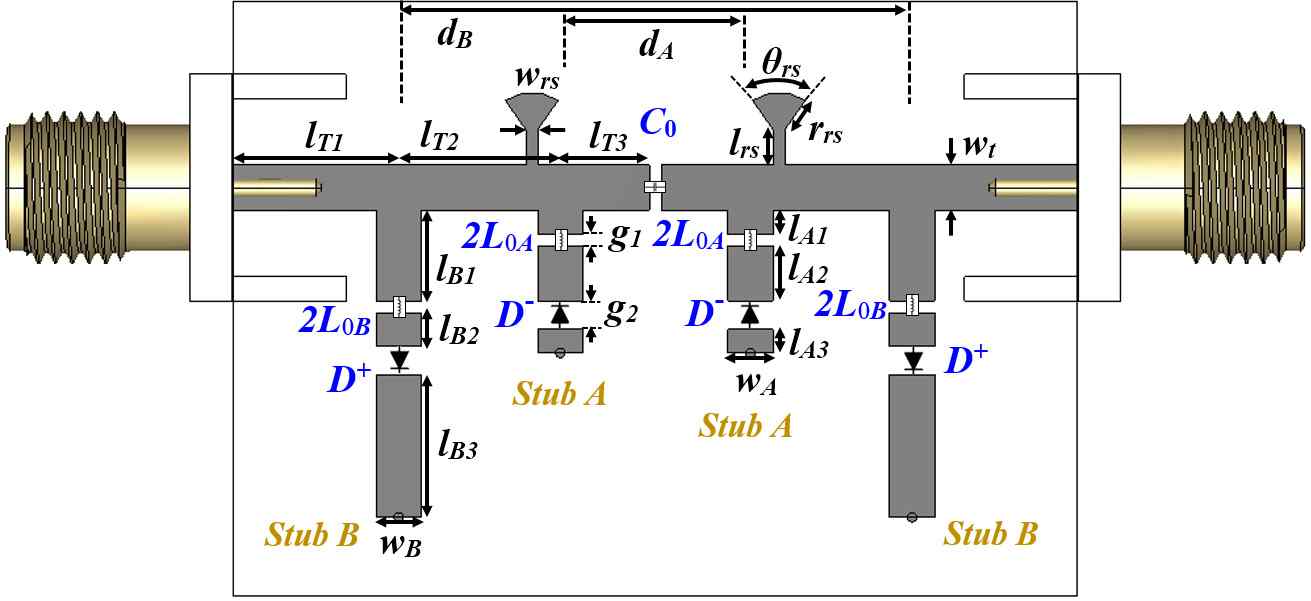}
	\caption{Layout configuration of the two-state reconfigurable NRI-TL metamaterial phase shifter. $\textcopyright$ 2017 IEEE.}
	\label{mtmf4}
\end{figure}

The proposed reconfigurable NRI-TL metamaterial phase-shifter shown in Figure \ref{mtmf4} resolves both these problems. The phase-shifting line consists of a microstrip transmission line with a series loading capacitor \textit{C${}_{0}$}, and two pairs of reconfigurable shorted stubs, namely \textit{Stub A} (\textit{l${}_{A1}$+l${}_{A2}$+l${}_{A3}$}) and \textit{Stub B} (\textit{l${}_{B1}$+l${}_{B2}$+l${}_{B3}$}), which host the shunt loading inductors \textit{2L}${}_{0}$\textit{${}_{A}$} and \textit{2L}${}_{0}$\textit{${}_{B}$}, and which are placed in shunt at fixed distances along the transmission line. The two main functions of these stubs are to facilitate the reconfigurability mechanism enabling them to switch between the two discrete shunt inductor loading values, and to aid in impedance matching. 

\begin{table}[htb]
	\centering
	\begin{tabular}{|cc|cc|cc|} \hline 
		parameter & Length & parameter & Length & parameter & Length \\ \hline 
		\textit{l${}_{T1}$} & 6.95 mm& \textit{l${}_{A1}$} & 1.00 mm & \textit{w${}_{A}$} & 1.90 mm \\ 
		\textit{l${}_{T2}$} & 6.80 mm& \textit{l${}_{A2}$} & 2.30 mm & \textit{w${}_{rs}$} & 0.50 mm \\ 
		\textit{l${}_{T3}$} & 3.75 mm & \textit{l${}_{A3}$} & 1.00 mm & \textit{g${}_{1}$} & 0.50 mm \\ 
		\textit{l${}_{B1}$} & 3.80 mm & \textit{l${}_{rs}$} & 1.49 mm & \textit{g${}_{2}$} & 1.20 mm \\ 
		\textit{l${}_{B2}$} & 1.40 mm & \textit{w${}_{t}$} & 1.90 mm & \textit{r${}_{rs}$} & 1.50 mm \\ 
		\textit{l${}_{B3}$} & 6.00 mm & \textit{w${}_{B}$} & 1.90 mm & \textit{$\theta$${}_{rs}$} & 70${}^\circ$ \\ \hline 
	\end{tabular}
	\caption{Dimensions of the Phase-Shifter Layout Shown in Figure \ref{mtmf4}}
	\label{mtmt1}
\end{table}

The metamaterial phase-shifter of Figure \ref{mtmf4} was first designed in the Keysight -- Advanced Design System (ADS) simulator to incur an insertion phase in \textit{State A} of \textit{$\phi$${}_{MTM\_A}$} = ${-}$15${}^\circ$ at 1 GHz using the two central stubs (\textit{Stub A}), while the two outer stubs (\textit{Stub B}) were turned off. Subsequently, the two outer stubs (\textit{Stub B}) were turned on, and the two inner ones (\textit{Stub A}) were turned off while keeping \textit{C}${}_{0}$ constant, in order to achieve an insertion phase in \textit{State B} of \textit{$\phi$${}_{MTM\_B}$} = +60${}^\circ$ at 1 GHz to achieve a phase difference of 75${}^\circ$. In doing so, a similar impedance matching degradation to the one shown in Figure \ref{mtmf3}(a) was observed. Thus, Quasi-Newton multivariable optimization was used to match the impedance of both \textit{State A }and \textit{State B }using following algorithm.\textit{ } 


\subsection{Optimization Algorithm}

Optimization workspace of Advanced Design System (ADS) was used as the optimization tool in this work. 
\noindent \textbf{Step 1:} Design a metamaterial unit cell on a 50 $\Omega$ transmission line with a length = 35 mm with \textit{$\phi$${}_{MTM\_A}$} = ${-}$15${}^\circ$ (\textit{State A}) at 1 GHz and formulate initial \textit{C}${}_{0}$ and \textit{L}${}_{0}$\textit{${}_{A}$} values.

\noindent \textbf{Step 2:} Design stub A with lengths L\_A1, L\_A2 and L\_A3 and gaps to host \textit{L}${}_{0}$\textit{${}_{A}$} and PIN diode with forward polarity.

\noindent \textbf{Step 3:} Calculate \textit{L}${}_{0}$\textit{${}_{B}$} for \textit{$\phi$${}_{MTM\_B}$} = +60${}^\circ$ (\textit{State B}) while \textit{C}${}_{0}$ is kept the same as defined in step 1.

\noindent \textbf{Step 4:} Design stub B with lengths L\_B1, L\_B2 and L\_B3 and gap to host \textit{L}${}_{0}$\textit{${}_{B}$} and PIN diode with reverse polarity.

\noindent \textbf{Step 5:} Evaluate optimized values of the \textit{optimization} \textit{variables} using Quasi-Newton multivariable optimization with the following simultaneous goals: 

1. {\textbar}S${}_{11}${\textbar} in state A $<$ ${-}$10 dB at 1 GHz

2. {\textbar}S${}_{11}${\textbar} in state B $<$ ${-}$10 dB at 1 GHz

3. \textit{$\phi$${}_{MTM\_A}$} = ${-}$15${}^\circ$$\pm$\textit{E${}_{A}$}${}^\circ$ in \textit{State A} 

4. \textit{$\phi$${}_{MTM\_B}$} = +60${}^\circ$$\pm$\textit{E${}_{B}$}${}^\circ$ in \textit{State B}

\begin{figure}[H]
	\centering
	\includegraphics[width=0.7\textwidth]{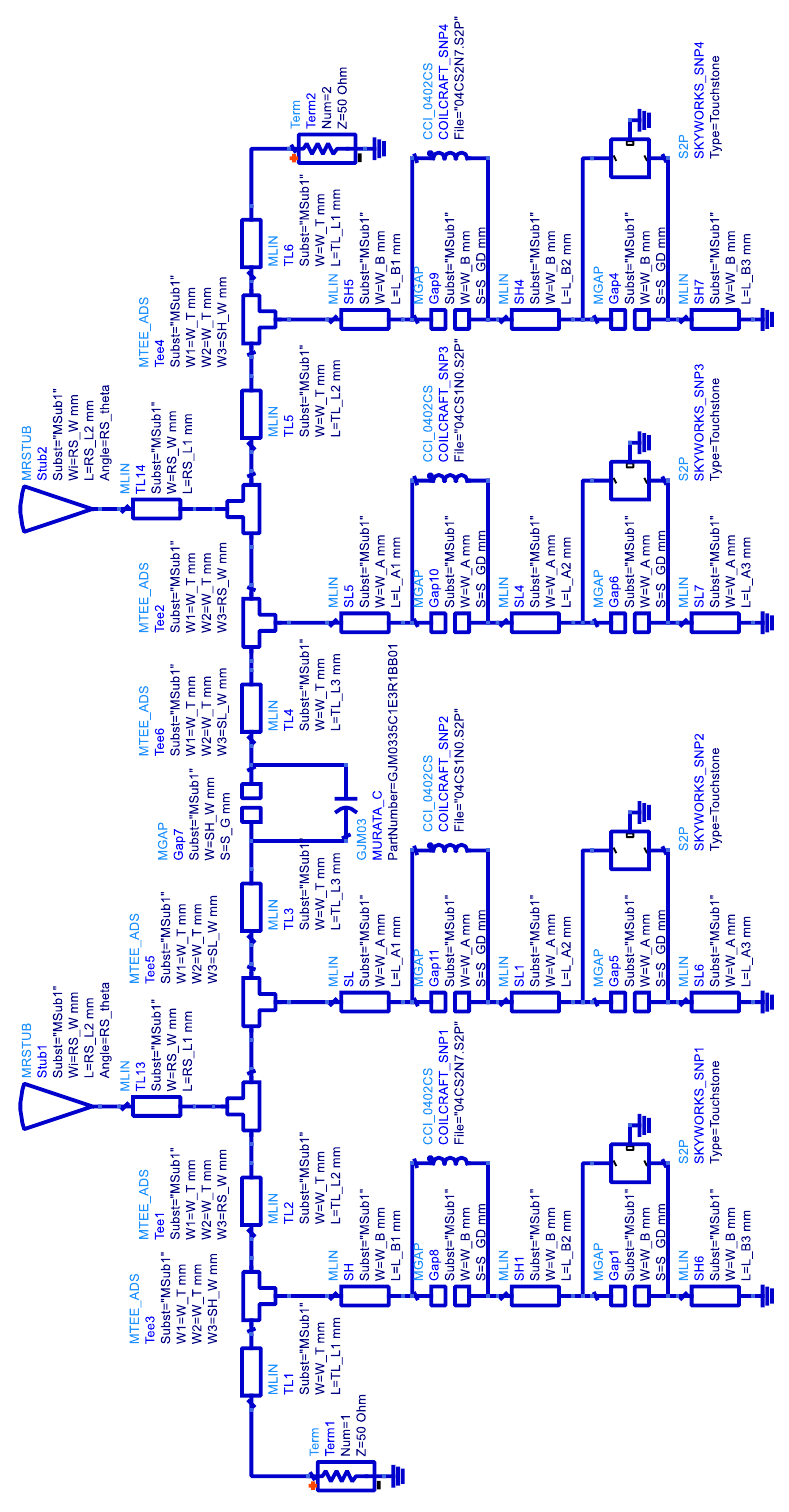}
	\caption{Schematic of the reconfigurable NRI-TL phase shifter in Keysight -- Advanced Design System (ADS). Realistic S-parameter models of the lumped LC components from Murata and Coilcraft were used, together with S-parameter models from Skyworks for the OFF and ON states of the PIN diodes. $\textcopyright$ 2017 IEEE.}
	\label{mtmf5}
\end{figure}

5. 2 $\times$ (TL\_L1+TL\_L2+TL\_L3) = 35 mm

(Where \textit{E${}_{A}$} and \textit{E${}_{B}$} are the tolerance factors in degrees with initial value = 0.5${}^\circ$) 

\noindent If the optimization criteria are satisfied go to the next sequential step. Else go to step 14.

\noindent \textbf{Step 6:} Define \textit{optimization} \textit{variables} L\_B1, LB\_2, LB\_3, L\_A1, L\_A2 and L\_A3. Go to step 5. If already optimized, go to next sequential step. 

\noindent \textbf{Step 7: }Define \textit{optimization} \textit{variables} TL\_L1, TL\_L2, TL\_L3. Repeat step 5 and 6. If already optimized, go to next sequential step. 

\noindent \textbf{Step 8: }Define \textit{optimization} \textit{variables} \textit{L}${}_{0}$\textit{${}_{A}$} and \textit{L}${}_{0}$\textit{${}_{B}$}. Repeat step 5, 6 and 7. 

\noindent \textbf{Step 9:} Replace the optimized values of \textit{L}${}_{0}$\textit{${}_{A}$} and \textit{L}${}_{0}$\textit{${}_{B}$} with closest discrete inductance value from the Coilcraft chip inductor library. Repeat step 5, 6 and 7.

\noindent \textbf{Step 10: }Define \textit{optimization} \textit{variables} \textit{C}${}_{0}$. Repeat step 5, 6 and 7.

\noindent \textbf{Step 11: }Replace the optimized value of \textit{C}${}_{0}$ with closest discrete capacitance value from the Murata chip capacitor library. Repeat step 5, 6 and 7.

\noindent \textbf{Step 12: }Define \textit{optimization variables} RS\_L1, RS\_L2, RS\_W, RS\_theta. Repeat step 5, 6 and 7.

\noindent \textbf{Step 13:} Decrease \textit{E${}_{A}$} and \textit{E${}_{B}$} values by 0.5${}^\circ$. Repeat steps 6-12. If already optimized in previous iteration, go to step 15. 

\noindent \textbf{Step 14:} Increase \textit{E${}_{A}$} and \textit{E${}_{B}$} values by 0.5${}^\circ$. Repeat step 6-12. 

\noindent \textbf{Step 15:} End

Upon successful completion of the optimization algorithm, the final values of the lumped components were \textit{C}${}_{0}$ = 3.10 pF, \textit{L}${}_{0}$\textit{${}_{A}$} = 0.55 nH, \textit{L}${}_{0}$\textit{${}_{B}$} = 1.40 nH, while the optimized geometrical dimensions of the phase shifter are enlisted in table \ref{mtmf4}. The resulted phase difference between states A and B is 73${}^{o}$.

\subsection{Realization}

To achieve reconfigurability, Skyworks PIN diode SMP1345 was used, which achieves ON and OFF states based on the biasing polarity. It is important to note that the polarity of the diode on\textit{ Stub A} D${}^{-}$ is opposite to the polarity of diode on \textit{Stub B} (D${}^{+}$). Negative biasing through the radial stubs (\textit{l${}_{rs}$},\textit{ w${}_{rs}$},\textit{ r${}_{rs}$}, and \textit{$\theta$${}_{rs}$}) along the top of the microstrip line in Figure \ref{mtmf4} turns on the two diodes D${}^{-}$, and turns off the two diodes D${}^{+}$, thus${}^{ }$defining \textit{State A}. Similarly, positive biasing through the radial stubs reverses this operation, consequently defining \textit{State B}. The DC ground for the differential diode biasing voltage is applied on the microstrip ground and is applied to the diodes through the grounding vias at the end of the stubs.

Surface-mount Murata chip capacitors (GJM03 series) and Coilcraft chip inductors (0402CS series), both with high Q-factors well above the operating frequency, were used in the realization of the phase shifter. It is important to mention that the achievable phase states rely on the available lumped-element component values in the Murata GJM03 series and the Coilcraft 0402CS series component libraries. The gaps \textit{g${}_{1}$} and \textit{g${}_{2}$} were designed keeping in mind the recommended soldering footprint for each lumped-element component and the diodes. 

\begin{figure}[t]
	\centering
	\begin{subfigure}{0.46\textwidth}
		\includegraphics[width=1\textwidth]{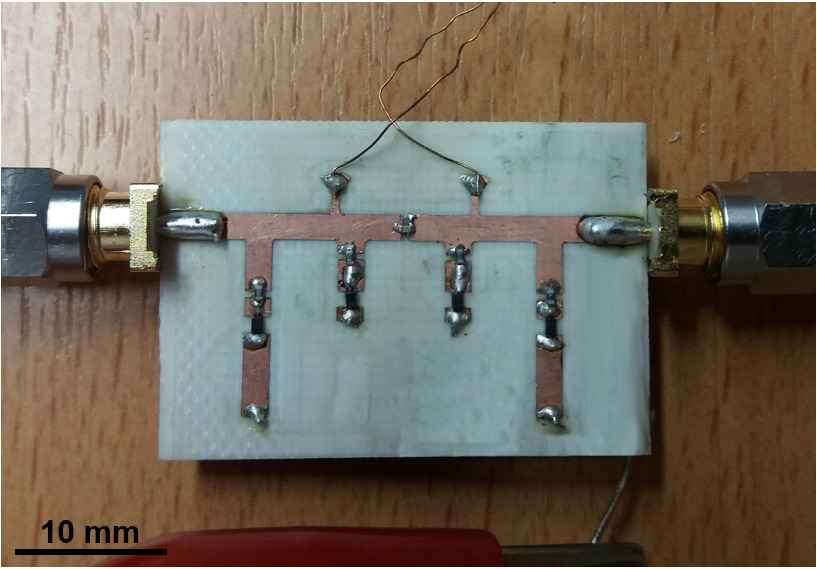}
		\caption{}
	\end{subfigure}
	\begin{subfigure}{0.48\textwidth}
		\includegraphics[width=1\textwidth]{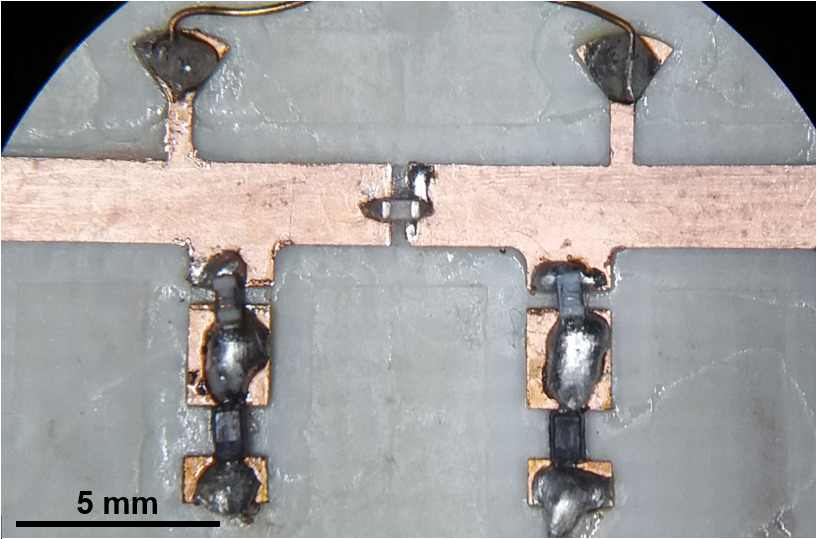}
		\caption{}
	\end{subfigure}
	\caption{(a) Fabricated prototype of the proposed reconfigurable NRI-TL phase shifter, (b) Zoomed-in view of the central unit cell. $\textcopyright$ 2017 IEEE.}
	\label{mtmf6}
\end{figure}

Finally, Keysight -- ADS was used to design and analyze the performance of the proposed reconfigurable NRI-TL phase shifter using realistic S-parameter models for the Murata capacitors, the Coilcraft inductors and the Skyworks diodes, in conjunction with a realistic model of the physical geometry of the microstrip lines and gaps on a Rogers RO4003C substrate ($\varepsilon$${}_{r}$ = 3.38, tan$\delta$ = 0.0027, \textit{h} = 0.813 mm), as can be seen in the full simulation models of Figures \ref{mtmf4} and \ref{mtmf5}. For a two-state insertion phase of \textit{$\phi$${}_{MTM\_A}$} = ${-}$15${}^\circ$ in \textit{State A} and of \textit{$\phi$${}_{MTM\_B}$} = +60${}^\circ$ in \textit{State B}, the final optimized set of loading values was determined to be \textit{C}${}_{0}$ = 3.1 pF, 2\textit{L}${}_{0}$\textit{${}_{A}$} = 1.1 nH, and 2\textit{L}${}_{0}$\textit{${}_{B}$} = 2.8 nH, corresponding to available discrete values in the Murata GJM03 and the Coilcraft 0402CS component series. It is worth mentioning here that the total shunt inductances for a particular state is a combination of the loading inductance (\textit{L}${}_{0}$\textit{${}_{A}$} or \textit{L}${}_{0}$\textit{${}_{B}$}), the internal inductance of Skyworks diodes (\textit{L${}_{PIN}$} = 0.7 nH) and the physical dimensions of the corresponding stubs (\textit{Stub A} or\textit{ Stub B}). The final layout dimensions of the phase shifter are also listed in \ref{mtmt1}.

The fabricated prototype of the proposed reconfigurable NRI-TL phase shifter is presented in Fig 6(a). First, the microstrip conductor and gap patterns shown of Figure \ref{mtmf4} were realized on the Rogers RO4003C substrate using a LPKF ProtoMat H100 milling machine. Holes of 0.4 mm diameter were drilled at the end of all four stubs for the placement of shorting vias. Then, the lumped components (\textit{C}${}_{0}$, 2\textit{L}${}_{0A}$ and 2\textit{L}${}_{0B}$) and diodes (D${}^{+}$ and D${}^{-}$) were soldered onto the microstrip transmission lines, as shown in the zoomed-in view of the central metamaterial unit cell in Figure \ref{mtmf6} (b). Finally, SMA connectors were attached for the S-parameter measurements. 

\begin{figure}[H]
	\centering
	\begin{subfigure}{0.6\textwidth}
		\includegraphics[width=1\textwidth]{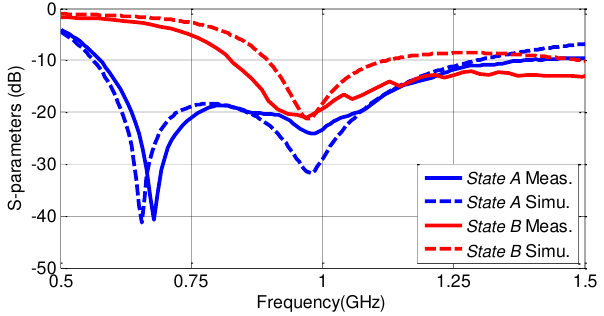}
		\caption{}
	\end{subfigure}
	\begin{subfigure}{0.6\textwidth}
		\includegraphics[width=1\textwidth]{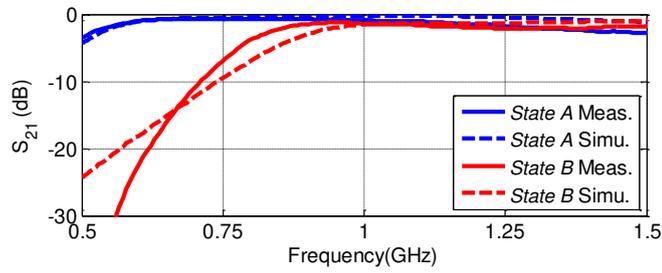}
		\caption{}
	\end{subfigure}
	\begin{subfigure}{0.6\textwidth}
		\includegraphics[width=1\textwidth]{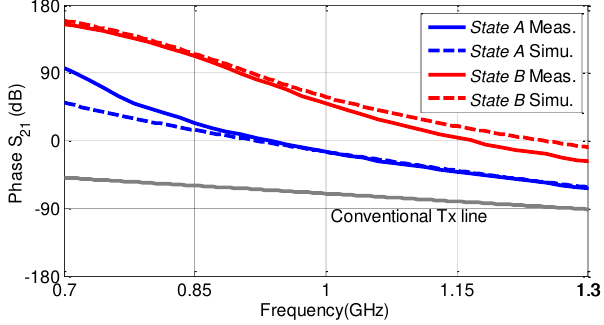}
		\caption{}
	\end{subfigure}
	\begin{subfigure}{0.6\textwidth}
		\includegraphics[width=1\textwidth]{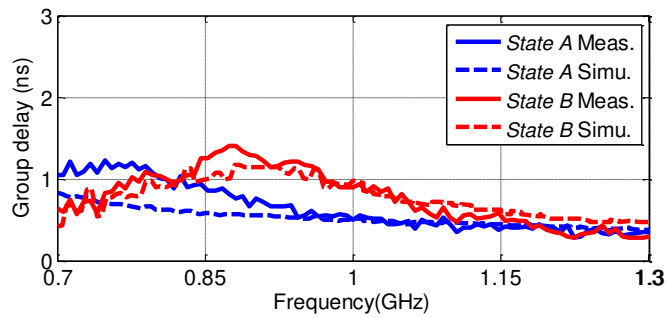}
		\caption{}
	\end{subfigure}
	\caption{(a) Simulated and measured responses of the reconfigurable NRI-TL metamaterial phase shifter. (a) {\textbar}S${}_{11}${\textbar}, (b) {\textbar}S${}_{21}${\textbar}, (c) Phase of S${}_{21}$, and (d) group delay. }
	\label{mtmf7}
\end{figure}

The simulated and measured results of the reconfigurable NRI-TL phase shifter are shown in Figure \ref{mtmf7}, where it can be observed that there is very good correlation between the two. The reflection coefficients for both S\textit{tate A} and \textit{State B} shown in Figure \ref{mtmf7}(a) is well below ${-}$10 dB at 1 GHz, indicating good impedance matching at the design frequency.

\subsection{Results and Discussion}

At 1 GHz, the simulated and measured {\textbar}S${}_{11}${\textbar} are ${-}$18.34 dB and ${-}$19.33 dB, respectively for \textit{State A}, while they are ${-}$29.05 dB and ${-}$23.38 dB, respectively for \textit{State B}. The advantages achieved by excluding the tuning mechanism from the direct signal path can be witnessed in terms of the low insertion losses, seen in Figure \ref{mtmf7}(b). The simulated and measured {\textbar}S${}_{21}${\textbar} are ${-}$1.53 dB and ${-}$1.43 dB, respectively for \textit{State A}, and ${-}$0.23 dB and ${-}$0.89 dB, respectively for \textit{State B}. In Figure \ref{mtmf7}(c), the phase response of an unloaded microstrip line is compared with the simulated and measured phase responses for both \textit{State A} and \textit{State B}. From Figure \ref{mtmf7}(a), (b) and (c), it can be observed that the reconfigurable NRI-TL metamaterial phase shifter retains its good impedance matching and transmission characteristics even though the insertion phase is significantly changed from \textit{State A} to \textit{State B}.


\begin{figure}[htb]
	\centering
	\includegraphics[width=1\textwidth]{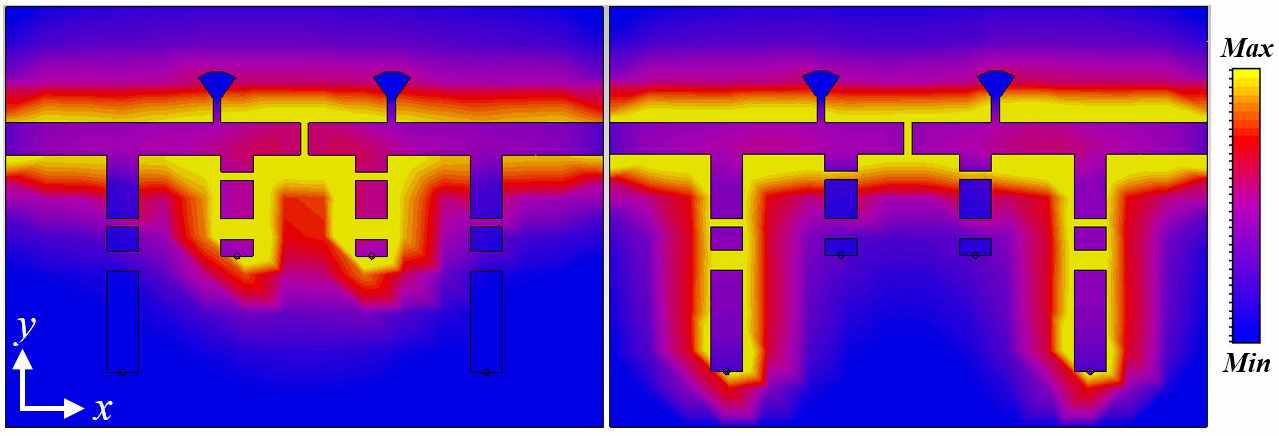}
	\caption{Normalized surface current distribution at 1 GHz on the conductors of the NRI-TL metamaterial phase shifter in \textit{State A} (left), and \textit{State B} (right).}
	\label{mtmf8}
\end{figure}

\begin{figure}[htb]
	\centering
	\includegraphics[width=0.6\textwidth]{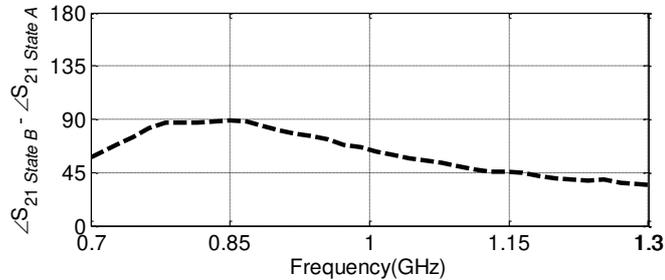}
	\caption{Measured phase difference between the two states. }
	\label{mtmf9}
\end{figure}

At 1 GHz, the ideal phase delay is 70${}^{\circ}$, which is compensated by \textit{$\phi$${}_{BW}$} = 55.5$^{\circ}$ and 128.5$^{\circ}$ for \textit{State A} and \textit{State B}, respectively, resulting in \textit{$\phi$${}_{MTM\_A}$} = ${-}$14.5${}^\circ$ and \textit{$\phi$${}_{MTM\_B}$} = +58.5${}^\circ$ at 1 GHz. The measured phase shift deviates by 2.4\% and 0.03\% from the predicted simulated value for \textit{State A} and \textit{State B,} respectively, which is well within the fabrication tolerances. 

The group delay of the reconfigurable NRI-TL phase shifter is shown in Figure \ref{mtmf7}(d), depicting good simulation and measurement agreement for both states. At 1 GHz, the measured group delay is 0.5 ns for \textit{State A}, and 0.9 ns for \textit{State B}, and these remain relatively flat in the range of 0.4 to 1.3 ns across the operating band shown. Figure \ref{mtmf8} shows the surface current distribution on the conductors of the reconfigurable NRI-TL phase shifter for \textit{State A} and \textit{State B}. It can be observed that the maximum surface current resides within the defined NRI-TL metamaterial unit cell for a particular state, while the un-used stubs carry an insignificant amount of surface current, resulting in very good isolation in both states. The measured phase difference between the two states for the frequency range of 0.75-1.25 GHz is shown in Figure \ref{mtmf9}, where it can be observed that it is relatively flat with a value of 73${}^\circ$ at 1 GHz that varies between 38.4${}^\circ$ to 89.1${}^\circ$ in the range shown. Finally, \ref{mtm2} summarizes the performance of the proposed reconfigurable NRI-TL metamaterial phase shifter and other phase shifters found in the literature, which compares quite favorably. It has the most compact size, 0.07 $\lambda$ x 0.06 $\lambda$, when normalized with respect to the central frequency wavelength, and the widest bandwidth (18.7\%) within which the S${}_{21}$ phase varies less than $\pm$${}_{ }$20$^{\circ}$. Albeit its compact and simple structure, one of the major disadvantages of the proposed architecture is the discrete phase shift that it can achieve. Other topologies that have achieved continuous phase tuning capabilities can be found in literature. An extension of this work would be to apply existing continuous phase tuning methods to the architecture proposed herein.

\begin{table}[]
	\centering
	\begin{tabular}{p{0.9in}p{0.7in}p{0.7in}p{0.7in}p{0.7in}p{0.7in}} \hline 
		Ref & \cite{choi2015novel} & \cite{abdalla2007printed} & \cite{damm2005compact} & \cite{michishita2011tunable} & This work \\ \hline 
		mm$\times$mm & 30.0$\times$68.0 & 10.8$\times$9.4 & 30.0$\times$40.0 & 85.5$\times$30.0 & 17.8$\times$21.6 \\ 
		$\lambda$ $\times$ $\lambda$ & 0.19$\times$0.09 & 0.09$\times$0.08 & 0.26$\times$0.20 & 0.57$\times$0.20 & 0.07$\times$0.06 \\ 
		Freq (GHz) & 0.9 & 2.5 & 2.0 & 2.0 & 1.0 \\
		IL (dB)\newline min-max & 4.0-5.6 & 0.55-1.1 & 2-10 & 0.56-0.69 & 0.89-1.43 \\ 
		Phase range\newline min-max & -22$^{\circ}$ to 30$^{\circ}$ & -40$^{\circ}$ to 34$^{\circ}$ & 51$^{\circ}$ to 184$^{\circ}$ & 313$^{\circ}$ to 355$^{\circ}$ & -14$^{\circ}$ to 59$^{\circ}$ \\ 
		Bandwidth \newline $\angle $S${}_{21 }$$\pm$${}_{ }$20$^{\circ}$ & 0.8-0.95\newline 13.5\% & 2.4-2.7\newline 11.8\% & 2.4-2.75\newline 13.6\% & 2-2.08\newline 3.9\% & 0.92-1.11\newline 18.7\% \\ 
		Technology & Micro-\newline fluids & MMIC & varactor diode & metallic perturber & PIN diode \\ \hline 
	\end{tabular}\\
	\caption{Performance comparison of Reconfigurable NRI-TL Metamaterial based phase shifters.}
	\label{mtmt2}
\end{table}


\section{A Compact NRI-TL Metamaterial Microstrip Crossover }

This work demonstrates a practical implementation of a miniaturized microwave crossover using negative-refractive-index transmission-line (NRI-TL) metamaterials. NRI-TL metamaterial theory has demonstrated that by loading a conventional transmission line with lumped LC elements, a large phase shift per-unit-length can be achieved. Using this feature of NRI-TL metamaterials, the miniaturization of a crossover has been achieved by replacing the conventional microstrip transmission lines in a three-section branch-line coupler crossover with NRI-TL metamaterials lines. A compact NRI-TL crossover with an area 47 times smaller than the area of the conventional counterpart is presented. Measured results demonstrate that the proposed NRI-TL Metamaterial crossover exhibits -10 dB {\textbar}S${}_{11}${\textbar} bandwidth of 32.67 \% and a {\textbar}3 dB{\textbar} {\textbar}S${}_{31}${\textbar} of 25.64 \% at around 1 GHz. 

\begin{figure}[t]
	\centering
	\includegraphics[width=0.9\textwidth]{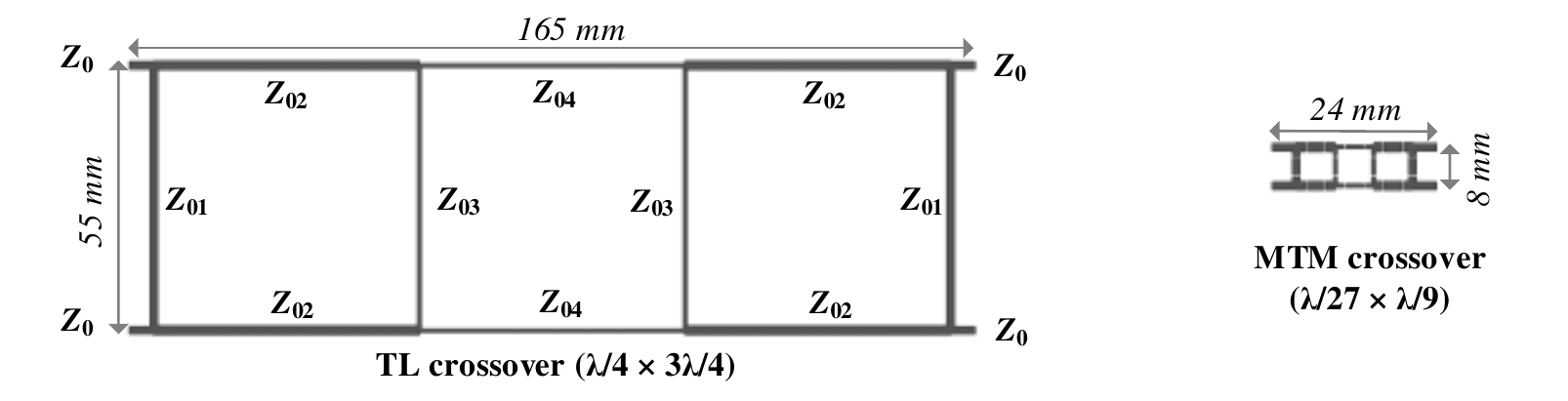}
	\caption{Footprint comparison between a microwave crossover using conventional transmission lines and a compact NRI-TL metamaterial microwave crossover.}
	\label{crossf1}
\end{figure}

Planar microwave circuits frequently use microwave transmission lines that intersect with each other. A crossover has been a well-known device suited for such intersection points after its first successful implementation by Weight \cite{choi2015novel}. Since then, a number of configurations of crossover devices have been presented to primarily increase the bandwidth of the device and to decrease the overall size \cite{wang2013broadband, wang2013broadband, markley2014ultra}. A broadband solution of a crossover has been presented using single and multi-section branch-line coupler configurations \cite{kholodniak2000wideband, yao2011microstrip}. Each branch in this configuration is set to incur a 90${}^\circ$ phase shift for the successful transmission of a signal from Port 1 to Port 3, and from Port 2 to Port 4. A $\lambda$/4 section of conventional transmission line is required in each branch of the crossover, which results in a physically large device (Figure \ref{crossf1}). For example, at 1 GHz, the size of a three-section crossover \cite{yao2011microstrip} is 55$\times$165 mm${}^{2}$, which is too large for most applications in modern microwave electronics. 

\begin{figure}[t]
	\centering
	\includegraphics[width=0.4\textwidth]{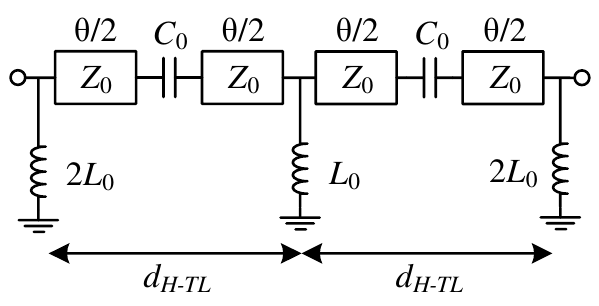}
	\caption{Two-stage NRI-TL metamaterial line formed by cascading two $\mathit{\Pi}$ unit cells for the implementation of each branch of the crossover.}
	\label{crossf2}
\end{figure}

\subsection{Implementation}

\begin{figure}[t]
	\centering
	\includegraphics[width=1\textwidth]{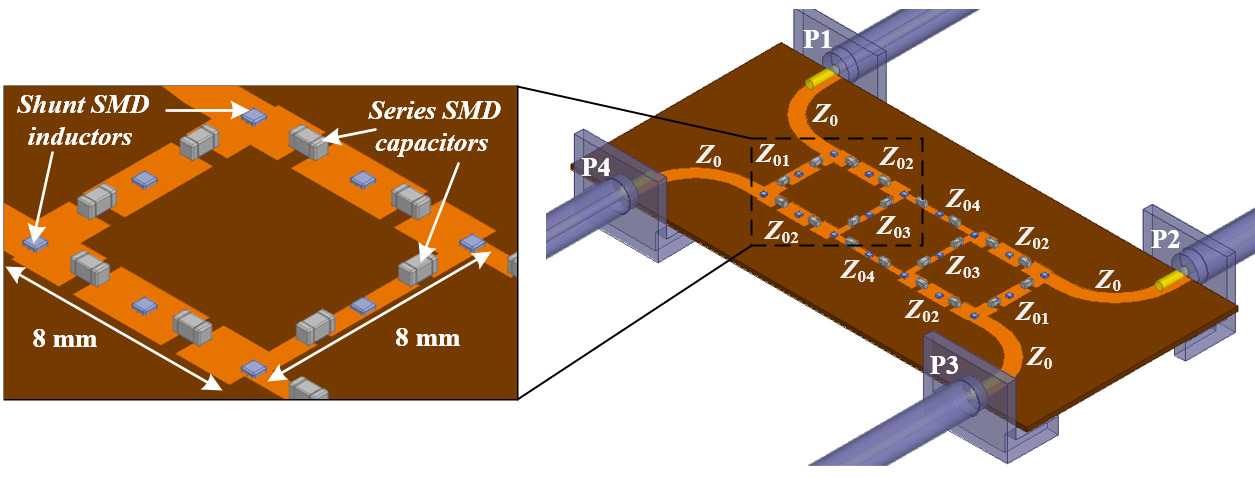}
	\caption{Full-wave EM simulation prototype of the three-section NRI-TL metamaterial crossover.}
	\label{crossf3}
\end{figure}

In this work, a practical implementation of a miniaturized three-section branch-line crossover using NRI-TL metamaterial lines has been achieved in a simple and realizable fashion. A new implementation of a three-section branch-line crossover has been employed using a two-stage NRI-TL metamaterial $\Pi$ unit cell \cite{antoniades2014transmission}. The approach also avoids an unrealizable overlapping of three capacitors at the junctions of each branch while using a T-unit cell (for example in \cite{antoniades2012compact}). A single NRI-TL metamaterial $\Pi$ unit cell consists of a host transmission line with characteristic impedance \textit{Z}${}_{0}$, series inductance per unit length \textit{L}, shunt capacitance per unit length \textit{C}, and length \textit{d}, which is loaded with a lumped-element series capacitor \textit{C}${}_{0}$ and two shunt inductors \textit{2L}${}_{0}$. A cascade of two metamaterial $\Pi$ unit cells to form a two-stage NRI-TL metamaterial line is presented in Figure \ref{crossf2}. The conventional host transmission line realizes a negative phase response, \textit{$\phi$${}_{H-TL}$}, while the series capacitive and shunt inductive loading forms a backward-wave (BW) line with a positive phase response\textit{ $\phi$${}_{BW}$}. Under effective medium conditions, the phase shift for \textit{N} NRI-TL metamaterial unit cells is given by [7]: 
\begin{equation} \label{crosse1}
\Phi _{MTM} \approx N(\phi _{MTM} )=N\left(-\omega \sqrt{LC} d+\frac{1}{\omega \sqrt{L_{0} C_{0} } } \right) 
\end{equation} 
This is valid under the impedance matching condition of \ref{impedance}. At the initial design phase, the impedance of each branch was calculated based on the three stage branch line coupler presented in [6]. The calculated characteristic impedance values of the transmission lines were \textit{Z}${}_{0}$ = \textit{Z}${}_{0}$\textit{${}_{1}$} = 50 $\Omega$, \textit{Z}${}_{0}$\textit{${}_{2}$} = 45 $\Omega$, and \textit{Z}${}_{0}$\textit{${}_{3}$} = \textit{Z}${}_{0}$\textit{${}_{4}$} = 81 $\Omega$. Each branch of the original crossover consisted of a 90${}^\circ$ microstrip transmission line, which was replaced with a compact two-stage NRI-TL metamaterial line, where each unit cell had an electrical length of \textit{$\theta$ }= 5${}^\circ$. A total number of 10 such branches formed the three-stage metamaterial crossover. The lumped components values for each unit cell were calculated using the guidelines presented in [7] and are enlisted in Table 1 along with the geometrical width of the host transmission line. The calculated \textit{L}${}_{0}$ and \textit{C}${}_{0}$ simultaneously need to satisfy the conditions of equation \ref{crosse1} and \ref{impedance}. In total, 20 surface-mount device (SMD) capacitors and 18 SMD inductors were used to implement the crossover, as can be seen in the diagram of Figure \ref{crossf3}.

\noindent The initial crossover was designed, simulated and evaluated in the Keysight -- Advanced Design System (ADS) simulator to operate at 1 GHz on a Rogers RT duroid 5880 substrate ($\epsilon$${}_{r}$ = 2.2, tan$\delta$ = 0.0004, \textit{h} = 0.787 mm). At first, ideal lumped-element components were used in each branch line to ensure the theoretical device operation. In the next stage, the ideal components were replaced with the closest available surface-mount component values from the Murata chip capacitor (GJM03 series) and Coilcraft chip inductor (0402CS series) component libraries. In the next and final stage, a full wave EM simulator was used to evaluate the device performance in a realistic environment. Figure \ref{crossf3} shows the geometrical configuration and component loading of the proposed device. It can be observed that due to the small size of the device, SMA connectors cannot be mounted directly to the ports of the crossover, therefore 50 $\Omega$ curved feed lines were added, and the SMA connectors were attached to these. The SMD capacitors were mounted in series along with the microstrip line trace, whereas the SMD inductors were placed vertically through the substrate. When assembling each of the individually-designed metamaterial branches to form the crossover, a shunt inductor is present at both ends of each branch, however, at each junction it is not practically feasible to place adjacently three separate inductors corresponding to separate branches in a very small area. Therefore, an equivalent parallel inductance has been evaluated (\textit{L}${}_{0}$\textit{${}_{1}$} {\textbar}{\textbar} \textit{L}${}_{0}$\textit{${}_{2}$} for a junction of 2 branches and \textit{L}${}_{0}$\textit{${}_{1}$} {\textbar}{\textbar} \textit{L}${}_{0}$\textit{${}_{2}$} {\textbar}{\textbar} \textit{L}${}_{0}$\textit{${}_{3}$} for a junction of 3 branches). The closest possible component value in the Coilcraft chip inductor (0402CS series) component library has been used to implement this new equivalent inductance value. Although the crossover was initially designed at a design frequency of 1 GHz, due to the limited and discrete range of component values available in the Murata and Coilcraft SMD component libraries, the final design was optimized at a slightly shifted design frequency of \textit{f}${}_{0}$ = 0.965 GHz.

\subsection{Fabrication and Measurements} 
\begin{figure}[t]
	\centering
	\begin{subfigure}{0.42\textwidth}
		\includegraphics[width=1\textwidth]{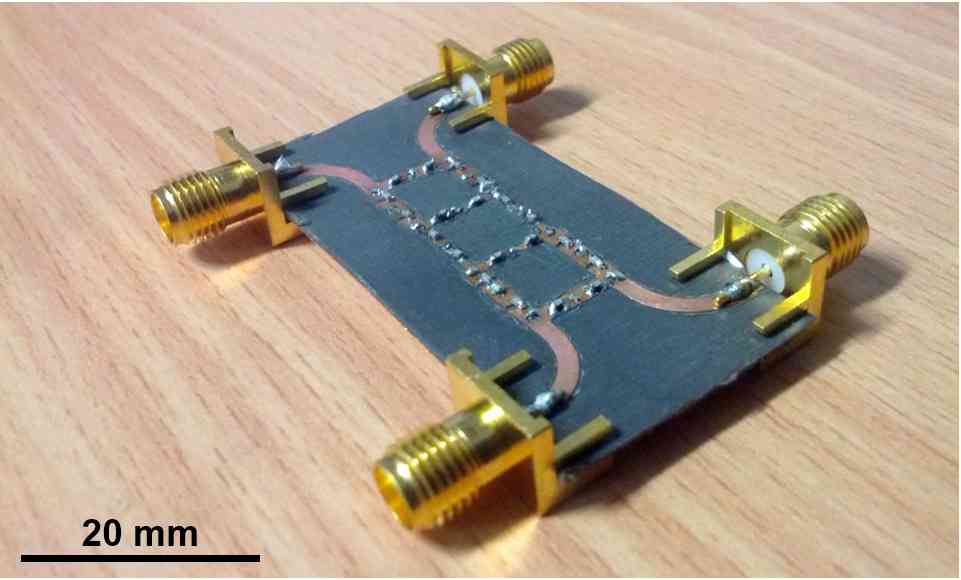}
		\caption{}
	\end{subfigure}
	\begin{subfigure}{0.42\textwidth}
		\includegraphics[width=1\textwidth]{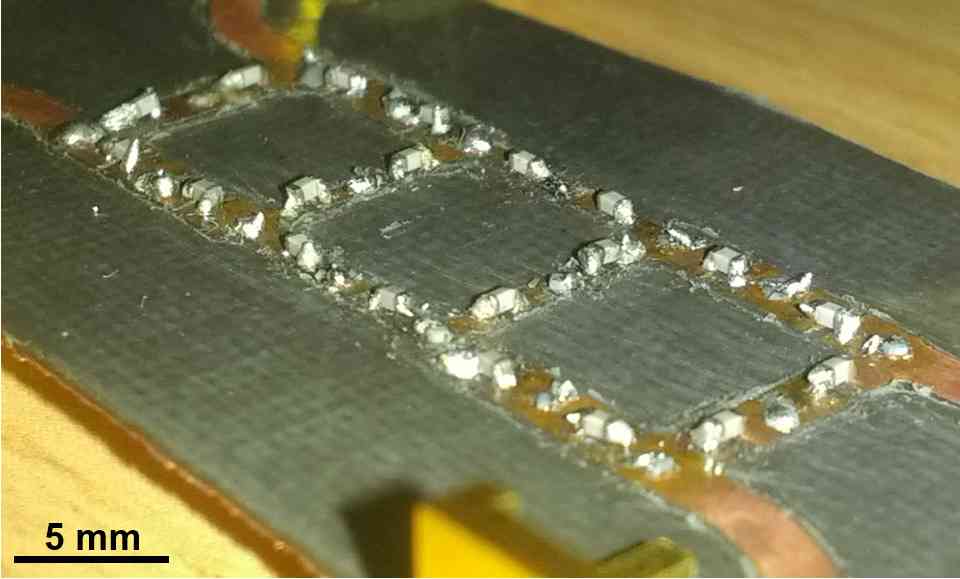}
		\caption{}
	\end{subfigure}
	\caption{Fabricated prototype of the compact NRI-TL metamaterial microwave crossover. (a) The device, (b) zoomed-in view of the soldered lumped-element components. $\textcopyright$ 2018 Wiley Periodicals, Inc.}
	\label{crossf4}
\end{figure}

\begin{figure}[H]
	\centering
	\includegraphics[width=0.8\textwidth]{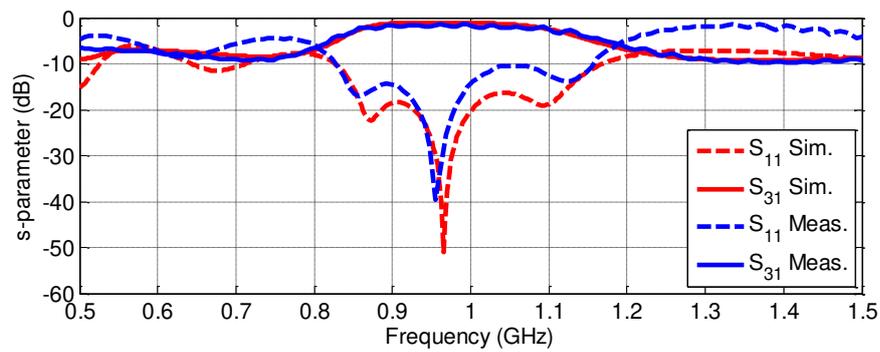}
	\caption{Simulated and measured S-parameter responses for the compact NRI-TL metamaterial microwave crossover. $\textcopyright$ 2018 Wiley Periodicals, Inc.}
	\label{crossf5}
\end{figure}

The microstrip conductor and gap patterns shown of Figure \ref{crossf3} were implemented on the Rogers 5880 substrate using a LPKF ProtoMat H100 milling machine. Holes of 0.4 mm diameter were drilled for the vertical placement of the SMD inductors. Then the lumped components were soldered onto the metallic traces. Figure \ref{crossf4} shows the fabricated prototype of the compact metamaterial crossover. To measure the performance of the crossover, the two ports of an Agilent E8363B vector network analyser were connected to Port 1 (P1) and Port 3 (P3), while two 50 $\Omega$ broadband loads were connected to the remaining two ports (P2 and P4) of the device. The reflection coefficient at Port 1 remains below -10 dB from 0.823 GHz to 1.155 GHz, as can be observed in Figure \ref{crossf5}, which amounts to a $\sim$26\% fractional bandwidth. At the design frequency of 0.965 GHz, the simulated and measured {\textbar}S${}_{11}${\textbar} of the device is -50.93 dB and -27.34 dB, respectively, indicating good impedance matching. The simulated and measured transmission coefficients {\textbar}S${}_{31}${\textbar} at the same frequency are -1.026 dB and -1.55 dB, respectively. A 3 dB transmission bandwidth can be defined as the bandwidth over which {\textbar}S${}_{31}${\textbar} does not fluctuate beyond {\textbar}3 dB{\textbar} from the value at \textit{f}${}_{0}$ = 0.965 GHz. For this device the simulated and measured 3 dB bandwidths were 26.53 \% (from 0.85 GHz to 1.11 GHz), and 25.64 \% (from 0.85 GHz to 1.10 GHz), respectively. Similar results were recorded when the crossover was connected in the opposite direction, i.e. for S${}_{33}$ and S${}_{13}$, indicating that the device is reciprocal. Also, analogous results were observed in the other crossover path, i.e. for transmission from P2 to P4 and in the opposite direction from P4 to P2. Since these results are almost identical to the ones presented in Figure \ref{crossf5}, they are omitted for brevity. The small discrepancies between the simulated and measured S-parameters can be attributed to the $\pm$5\% component tolerances and the soldering and component placement imperfections that can be observed in Figure \ref{crossf4}(b). These discrepancies can be reduced by efficient component placement and accurate soldering while in mass production. Nevertheless, the good reflection and transmission responses of the crossover at the design frequency indicate that the proposed device is well suited for use in next-generation microwave systems. 

\section{Chapter Conclusions}

A novel architecture of a reconfigurable phase-shifter based on NRI-TL metamaterial unit cells has been presented that is compact in size and exhibits low insertion loss characteristics. By modeling and optimizing two collocated NRI-TL metamaterial $\Pi$ unit cells loaded with the same series capacitance but different shunt inductances, two distinct phase states have been achieved. It has been shown that by decreasing the shunt inductance in an NRI-TL unit cell to achieve a higher phase shift, while using the same series capacitance, disturbs the matching of the unit cell severely. Good matching has been achieved in both phase states by adjusting the length and positions of the inductively-loaded shorted stubs. The reconfigurable mechanism has been achieved by placing PIN diodes on four shorted stubs, enabling the device to operate in two phase states based on the biasing polarity. This also avoids the presence of lossy components in the direct signal path. The proposed device was fabricated on a 35.5$\times$25 mm${}^{2}$ Rogers RO4003C substrate and the measured results show good agreement with the simulations. The useable bandwidth of the proposed device for which the {\textbar}S${}_{11}${\textbar} of both states simultaneously remains below $\mathrm{-}$10 dB is 0.46 GHz (46\%), from 0.83 GHz to 1.29 GHz. In State A, the phase shifter achieves a phase shift of $\mathrm{-}$14.5${}^\circ$, while in State B +58.5${}^\circ$, resulting in a measured differential phase shift of 73${}^\circ$ at 1 GHz. The measured {\textbar}S${}_{21}${\textbar} for State A is $\mathrm{-}$1.43 dB, and for State B $\mathrm{-}$0.89 dB. The realized phase difference is representative, and the same metamaterial phase-shifting architecture can also be used to achieve any other arbitrary phase shifting value. Thus, the proposed reconfigurable NRI-TL metamaterial phase-shifter is ideal for use in feeding networks of next-generation antenna array systems for pattern and polarization diversity applications. Also, the capability to achieve high phase differences while maintaining compactness makes it well suited for use in massive MIMO systems. 
An effective and practical implementation of a compact NRI-TL metamaterial microwave crossover is presented in this work. The overall footprint has been reduced significantly by a factor of 47, while the signal reflection and transmission characteristics of the device are maintained at a good level. The design frequency of the proposed device around 1 GHz has been selected to enable rapid prototyping using commercially-available SMD components, however, the same design procedure that has been presented can be used to design NRI-TL metamaterial crossovers at higher frequencies for use in future microwave systems such as 5G and beyond. Thus, future work would involve the replacement of SMD components with inter-digital capacitors and printed inductors, which can be investigated to further decrease the cost and losses of the device.



\chapter{Metasurface Based Miniaturized Wearable Antenna} \label{chp2}
\ifpdf
\graphicspath{{Chapter2/Chapter2Figs/PNG/}{Chapter2/Chapter2Figs/PDF/}{Chapter2/Chapter2Figs/}}
\else
\graphicspath{{Chapter2/Chapter2Figs/EPS/}{Chapter2/Chapter2Figs/}}
\fi

In recent years synthesizing new materials with meta-physical properties has been focused by a number of research efforts. These metamaterials are designed for better performance of antennas by making use of periodic structures. Special attention has been given to high impedance surfaces. A specific form of these surfaces turn out to have frequency bands (bandgaps) inside which surface waves are reflected rather than being propagated, and they are therefore such specific metasurfaces are referred to as electromagnetic bandgap (EBG) surfaces. This chapter presents a planar monopole backed with a 2$\boldsymbol{\times}$1 array of Electromagnetic Band Gap (EBG) structures. The reflection phase of a single EBG unit cell has been studied and exploited towards efficient radiation of a planar monopole antenna, intended for wearable applications. The shape of the EBG unit cell and the gap between the ground and the EBG layer are adjusted so that the antenna operates at 2.45 GHz. The proposed antenna retains its impedance matching when placed directly upon a living human subject with an impedance bandwidth of 5\%, while it exhibits a measured gain of 6.88 dBi. A novel equivalent array model is presented to qualitatively explain the reported radiation mechanism of the EBG-backed monopole. The proposed antenna is fabricated on a 68$\boldsymbol{\times}$38$\boldsymbol{\times}$1.57 mm${}^{3}$ board of semi-flexible RT/duroid 5880 substrate. Detailed analysis and measurements are presented for various cases when the antenna is subjected to structural deformation and human body loading, and in all cases the EBG-backed monopole antenna retains its high performance. The reported efficient and robust radiation performance with very low specific absorption rate (SAR), the compact size, and the high gain, make the proposed antenna a superior candidate for most wearable applications used for off-body communication.

\section{Introduction to Metasurfaces}

Metasurfaces (MTSs) are the two-dimensional equivalents of volumetric metamaterial engineered to achieve extraordinary electromagnetic properties in 3-dimensions. Some recent studies have combined the reconfigurablity with metasurfaces to achieve multi-dimensional benefits in an attempt to achieve stronger fields, smaller size and increased controllability. The various tuning mechanisms of metasurfaces at microwave and photonic level that have been demonstrated in the literature can be divided into three general categories; 1) electronic switching 2) microfluidics and 3) photonics.

With the concise planar structure, Metasurface can be used in the design of planar antennas with improved performances. In the design of antennas, operating frequency, polarization and radiation pattern are the most important characteristics to be considered. With the rapid development of different wireless communications systems, it is highly desirable if one or more of the above-mentioned antennas characteristics can be reconfigured. In very recent studies, anisotropic metasurfaces using mechanically controlled reconfiguriblity is used to achieve frequency tuning from 2.55 GHz to 3.45 GHz \cite{majumder2016frequency} and polarization tuning is achieved at 4 GHz \cite{kandasamy2015low}. Similarly, piezoelectric actuators are used at 15 GHz at achieve frequency tuning in \cite{mavridou2016dynamically} while varactors are used for polarization reconfiguriblity in \cite{scarborough2016compact}. Such vast areas of research with variety of tuning mechanism options have opened up endless possibilities of metasurfaces in antenna applications.

Metasurface made of carefully designed discrete metallic or dielectric elements with spatially varying characteristics across its surface can direct the flow of electromagnetic radiation in way similar to optical holograms. Reconfigurable metasurfaces using microfluidics presented in \cite{zhao2007electrically, zhang2008magnetic} using nematic liquid crystals demonstrated magnetic resonance frequency tuning, while most recently, microfluidic split-ring resonators (MF-SRR) fabricated inside a flexible elastomeric material \cite{kasirga2009microfluidics} opened up directions toward switchable metamaterials and reconfigurable devices such as filters, switches, and resonators. Although a lot of work has not being put forward in this direction, yet substantial efforts are now focused on developing metasurfaces with switchable and reconfigurable capabilities in the microwave and optical regions of the spectrum, thus making reconfigurable photonic devices controllable by external signals a realistic possibility.

The latest developments have shown that optical metasurfaces comprising a class of optical metasurfaces with a reduced dimensionality can exhibit exceptional abilities for controlling the flow of light to achieve the anomalously large photonic density of states and provide super-resolution imaging. Tuning capability in planar photonics technology is expected to facilitate new physics and enhanced functionality for devices \cite{gwo2016plasmonic}. This technology have enabled new applications in imaging, sensing, data storage, quantum information processing, and light harvesting. Metasurface represented by a patterned metal-dielectric layer that is very thin compared with the wavelength of the incident light and is typically deposited on a supporting substrate \cite{kildishev2013planar}. Photonic components with adjustable parameters optically reconfigured to result in the phase change \cite{wang2016optically} or reflectivity/absorption of metasurface \cite{waters2015optically} has been demonstrated recently. The functionality of a device based on such a metasurface depends directly on the effective, surface-confined, optical dispersion. 

\subsection{Possible application of Metamaterial and Metasurfaces}

Metasurfaces has been practically used in wavefront shaping and beam forming for 4G communication. Impedance matched metasurface to free space are able to fully control the phase of the transmitted light, also used for beam deflector and a flat lens with high transmission efficiency for efficient wireless signal diversity. The metasurfaces for circularly polarized incident wave can control the component of the circularly polarized transmission with the opposite handedness. They can also helo in phase jumping associated with polarization change created by using anisotropic, subwavelength scatterers with identical geometric parameters but spatially varying orientations for wearable antenna applications. Some studies has shown a linear-to-circular polarization conversion for wireless signal diversity and linear polarization rotation for wireless signal diversity. Polarization sensitive metasurfaces that display asymmetric transmission with respect to the direction of wave propagation can be used for radar communication. Metasurfaces based on Dielectric resonators for wearable and bio-compatible implantable applications. Directional scattering metasurfaces can be used for indoor wireless applications. Beam forming and wavefront control enabled by dielectric metasurfaces can be implemented for wearable applications. The coupling between free space and surface waves using metasurfaces can also be exploited for low profile antennas. Graphene hybrid metasurfaces can be used for next generation material sciences. Last, but not least, resonance switchable and frequency tunable metasurfaces seem to show enormous potential in 5G transmitters and base stations.

Current reconfigurable beam shifters uses switchable feeding network for beam forming of large antenna arrays. It involves the inherent transmission loss of power from the network of switches. This transmission loss is proportional to the number of switches involve to achieve a particular phase difference. Reconfigurable metasurfaces designed beneath the radiating antenna can reflect the incident waveform in the given direction. Some possible innovations in this domain can be: 

Reconfigurable metasurface using resonators (split ring, L-shaped) with electronic switches (PIN, varactor, MEMS) to define a tunable reflection characteristics for beam forming (inkjet printer and laser milling machine)
Reconfigurable metasurface using resonators (split ring, L-shaped) with fluid channels enabling microfluids with different material properties to create variations of reflection surface for beam forming (inkjet printer and 3D printer)
Reconfigurable merasurface created by inkjet printed or 3D printed substrate tunable using incident light of given wavelength to alter the reflection properties of metasurface for beamforming. (inkjet printer and 3D printer).

Polarization is one of the most important factor in microwave and optical signals. Reconfigurable microwave and photonics metasurfaces can be engineered for polarization tuning in following innovative ways: 
Inkjet printed resonators (L-shaped, split ring) on substrate with microfluidic channels placed in the direction of propagation of microwave. Material properties of fluids changing the characteristics of metasurfaces to alter the polarization of microwave and optical signal (laser milling machine and inkjet printer)
3D printed substrate with hollow space for inkjet printing of photoelectric substrate placed in the direction of propagation of high frequency microwave signals. Dispersion properties of the substrates tuned by another incident wave to alter to polarization properties of the metasurface. (inkjet printer, laser milling machine and 3D printer).

\section{2$\times$1 EBG-backed Planar Monopole Antenna Design}

\noindent With the rapid development in communication systems in the past few years, the area of wireless body area networks (WBAN) has grown significantly, supporting a large number of applications, including personalized health care systems, patient monitoring systems, rescue systems, battle field survival, and wearable gaming consoles \cite{hao2005antennas}, \cite{roblin2011antenna} . Several frequency bands have been allocated for these applications to commercialize WBAN communication systems, which include the Medical Implantable Communication Systems (MICS) band (402-405 MHz), Industrial Scientific and Medical (ISM) band (2.40-2.48 GHz) and Ultra Wideband (UWB). For optimum performance, the antennas used for WBAN applications are required to be compact, mechanically robust, lightweight and preferably comfortable while being worn. It has been reported, that the performance of an antenna may degrade significantly while operating in close proximity with the human body \cite{hao2005antennas,roblin2011antenna}. This occurs because the antenna's surface currents are affected by the near field coupling with the body, which in turn affects the input impedance matching of the antenna. Specifically, for narrow-band operation, the dominant effect of the body proximity is a shift of the resonance frequency, which causes a mismatch at the designed frequency, resulting in a significant degradation of the total efficiency (P${}_{radiated}$/P${}_{incident}$). Designing a narrowband wearable antenna with a high total efficiency can be a challenging task, especially when it is also expected for the antenna to have low-profile, conformal and lightweight characteristics \cite{roblin2011antenna}. At the same time, the effect of wearable antennas on the human body in terms of maximum allowable specific absorption rate (SAR) needs to be addressed \cite{fields1997evaluating}. So far a number of configurations have been investigated as potential candidates for wearable antennas including fractal \cite{karimyian2015super}, inverted-F \cite{el2014novel}, planar monopoles \cite{jiang2014compact,kim2012monopole,poffelie2016high}, magneto electric dipoles \cite{yan2015wearable}, cavity backed \cite{agneessens2014compact}, \cite{yun2015folded}, and stacked microstrip antennas \cite{sheikh2013directive}. 

The microstrip antenna presented in \cite{sheikh2013directive}, and the cavity backed antennas \cite{agneessens2014compact}, \cite{yun2015folded}, can be considered good potential candidates for wearable applications, however they do not exhibit conformal characteristics. Several textile-based conformal antennas, including a wearable magneto electric dipole \cite{yan2015wearable}, a fully grounded microstrip \cite{samal2014uwb}, textile fractal \cite{karimyian2015super} and textile antenna based on Substrate-Integrated Waveguide (SIW) technology \cite{yan2015dual} are proposed to be pliable for off-body communication, however they have a relatively large footprint. A recent investigation on a compact conformal inverted-F wearable antenna has been presented \cite{el2014novel}; however, due to its near-omnidirectional radiation properties, a significant amount of energy is directed towards the human body. In an attempt to direct the antenna radiation away from the body for off-body communication with a BAN base station like the one presented in \cite{chamaani2015miniaturized}, it is always desirable to have a full ground plane. The presence of a ground plane also increases the isolation between the wearable antenna and the body, thus resulting in lower SAR values \cite{trajkovikj2015diminishing}. Other than conventional full or extended ground planes \cite{sheikh2013directive,samal2014uwb,koohestani2014novel,jiang2016design}, periodically loaded configurations using high impedance surfaces \cite{mohamed2015wideband,folayan2009dual}, artificial magnetic conductors and electromagnetic/photonic band gap structures \cite{jiang2014compact,kim2012monopole,mohamed2015wideband,raad2013flexible,kim2012wearable,zhu2009dual} have been investigated to introduce improved isolation between the radiating wearable antenna and living human tissues. However, these configurations still suffer from frequency shifts due to either bending, or crumpling, or they have relatively large lateral sizes. 
In this paper, we present the novel design of a very compact EBG-backed planar monopole antenna and the experimental realization of a conformal low-profile wearable antenna operating at 2.45 GHz in the ISM band. It is worth mentioning here that contrary to recently reported wearable antennas operating in the UHF \cite{trajkovikj2015diminishing,mohamed2015wideband}, ultra wideband (UWB) \cite{sheikh2013directive,samal2014uwb,koohestani2014novel} and super wideband \cite{karimyian2015super} frequency bands, this paper primarily focuses on the implementation, of a low profile, conformal, highly efficient, narrowband wearable antenna for off-body communication in wireless body area networks. The proposed antenna shown in Figure \ref{ebg1} consists of a radiating monopole, backed by a very compact 2$\times$1 cell EBG structure exhibiting controllable reflection phase, which radiates with high directivity and high efficiency even when it operates adjacent to living human body. The finite size ground plane provides high isolation between the antenna and the living human body tissues beneath it. At the same time, the ground plane in combination with the radiating monopole backed with an EBG surface contributes towards significantly higher gain compared to conventional planar monopole antennas. The antenna design details are presented in section II with a novel array model that provides a qualitative explanation of the radiation mechanism, while section III presents the experimental results of the fabricated prototype. Finally, section IV presents the performance of the antenna when it is subjected to structural deformation and placed on a human body.


\subsection{ Antenna Configuration and EBG Structure Selection}

\begin{figure}[htb]
	\centering
	\includegraphics[width=1\textwidth]{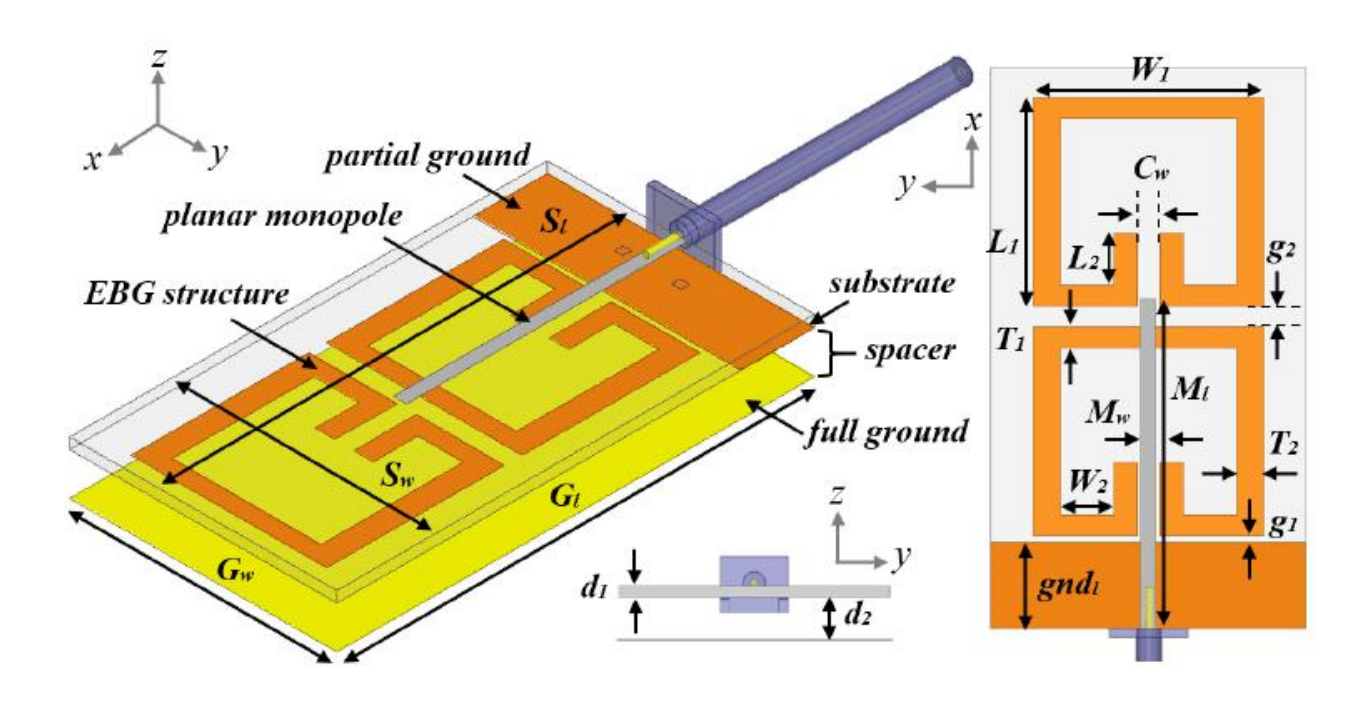}
	\caption{(a) Configuration of the 2$\times$1 EBG-backed planar monopole antenna with full ground (b) Top view of the monopole and finite sized EBG layer. Antenna is fabricated on semi flexible Rogers RT/duroid 5880 substrate with $\epsilon_r$ = 2.2 and tan$\delta$ = 0.0009. $\textcopyright$ 2016 IEEE. }
	\label{ebg1}
\end{figure}

The antenna schematic shown in Figure \ref{ebg1} consists of a radiating monopole, backed by a 2$\times$1 array of EBG cells. The top planar monopole and the partial ground plane are printed on the opposite sides of a substrate making a standalone planar monopole antenna. A customized EBG structure is printed on the same side of the substrate where the partial ground plane is printed to form an EBG surface, hence utilizing the available space and reducing the overall size of the antenna. The full ground plane beneath the EBG structure is separated by a thin foam spacer.

\begin{figure}[htb]
	\centering
	\begin{subfigure}{0.26\textwidth}
		\includegraphics[width=1\textwidth]{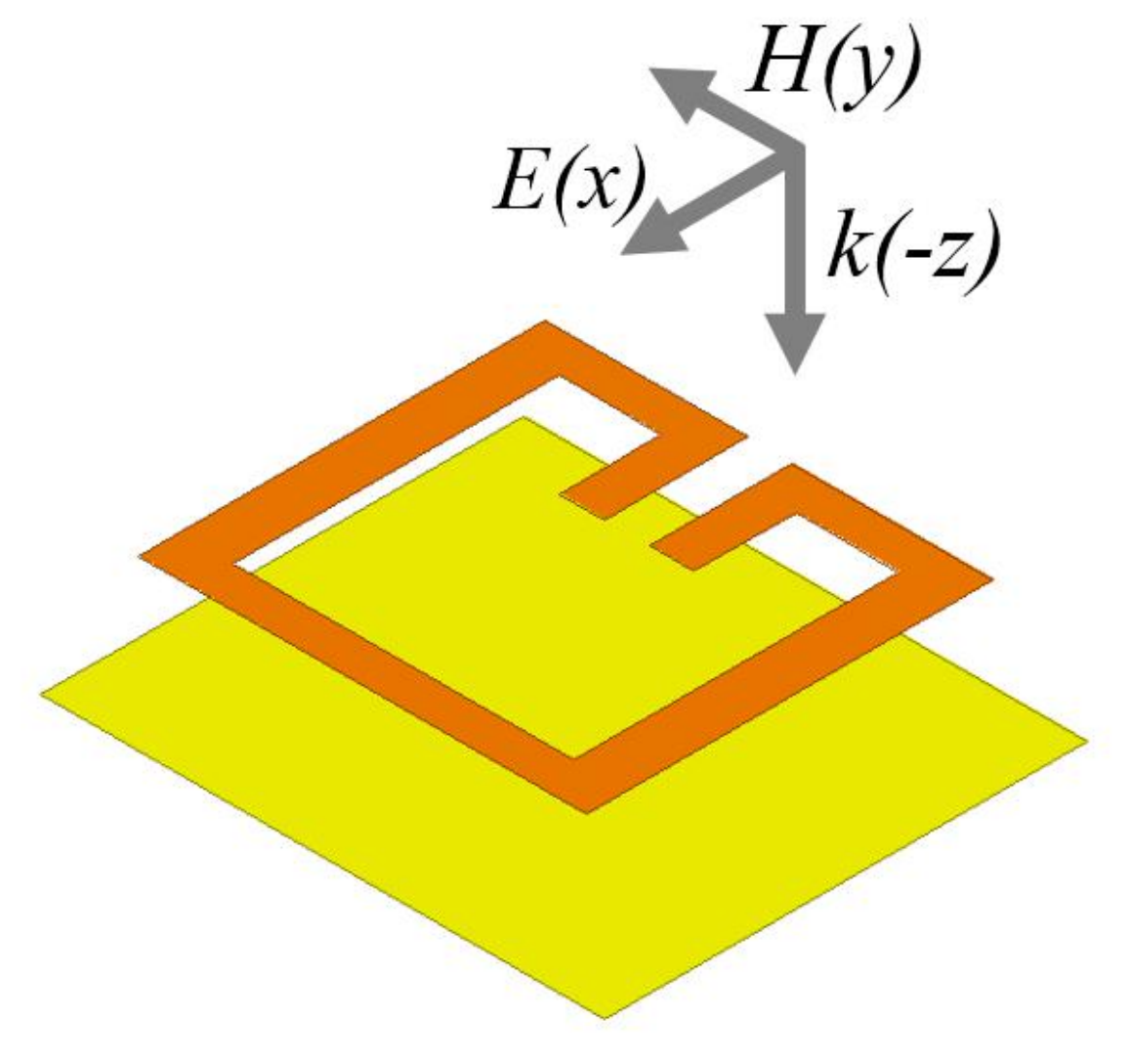}
		\caption{}
	\end{subfigure}
	\begin{subfigure}{0.6\textwidth}
		\includegraphics[width=1\textwidth]{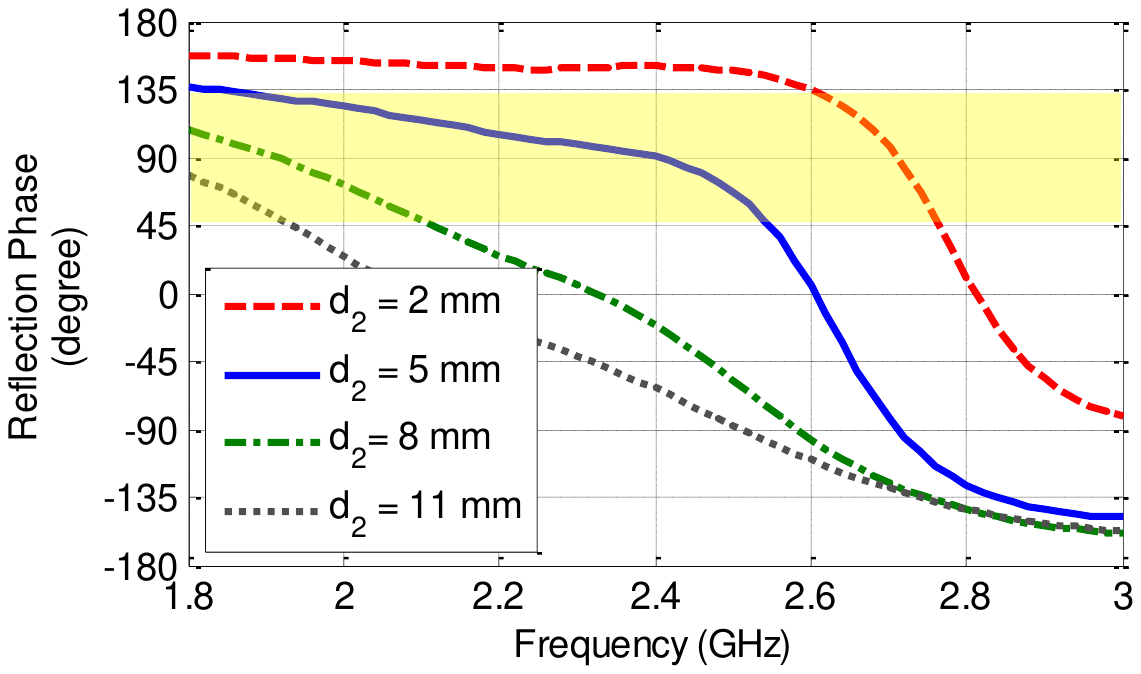}
		\caption{}
	\end{subfigure}
	\begin{subfigure}{0.35\textwidth}
		\centering
		\includegraphics[width=1\textwidth]{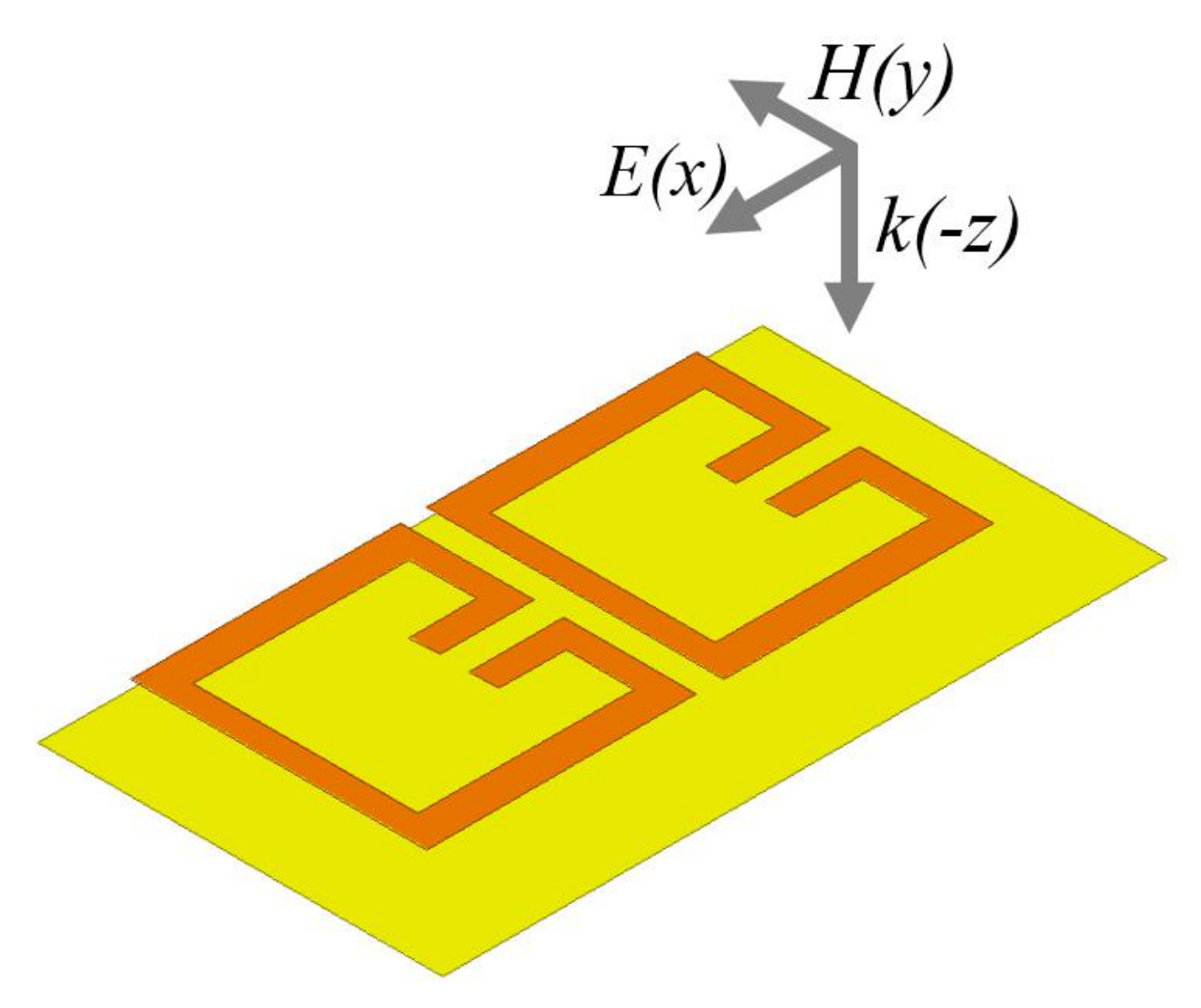}
		\caption{}
	\end{subfigure}
	\begin{subfigure}{0.6\textwidth}
		\centering
		\includegraphics[width=1\textwidth]{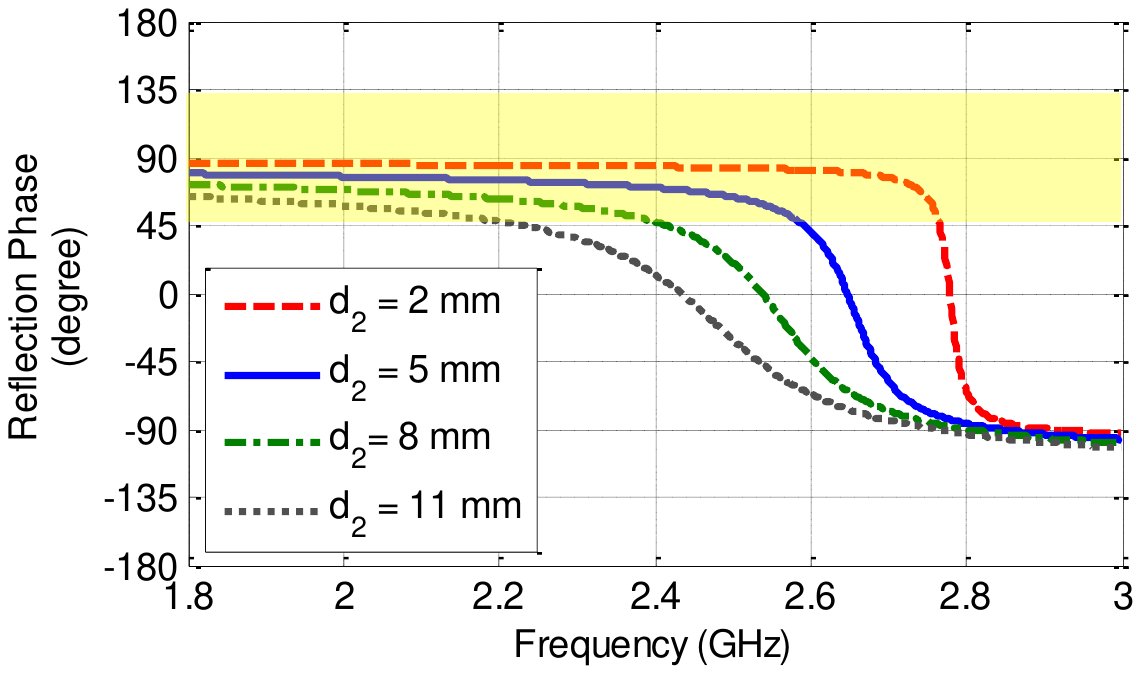}
		\caption{}
	\end{subfigure}
	\caption{(a) Configuration of a single unit cell consisting of the EBG element with metallic backing under plane wave excitation. (b) Reflection phase of an infinite array of the EBG unit cells. (d) Configuration of a 2×1 array of EBG cells under plane wave excitation (d) Reflection phase of the 2×1 array of EBG cells}
	\label{ebg2}
\end{figure}

At the initial design stage, the EBG structure has been designed so that the surface wave frequency bandgap of the EBG layer overlaps the matched frequency band of the planar monopole, to form a combined operating frequency band for the antenna. The EBG unit cell consists of a compact split squared ring resonator with an elongated length \textit{L${}_{2}$}${}_{ }$and a coupling gap \textit{C${}_{w}$} between the two symmetrical parts of the structure \cite{smith2004metamaterials}. Note that the theoretical resonance of split ring is different from full ring resonators e.g. \cite{abbasi2015humidity, arainreconfigurable, arain2016square, arain2017wideband}.The EBG structure is symmetric along the \textit{x}-axis\textit{ }as shown in Figure \ref{ebg1}(b) and \ref{ebg2}(a). The adapted shape of the single EBG cell \cite{tang2010novel} and the air gap between the EBG layer and the ground plane depict built-in anisotropic behavior, having a strong LC resonance along the \textit{x}-axis \cite{eleftheriades2005negative}. This implies that the EBG structure exhibits a different electric response for a wave polarized parallel to the \textit{x}-axis as compared to a wave polarized parallel to the \textit{y}-axis, when the wave is propagating along the \textit{z}-axis. 

\begin{table} [htb]
	\centering
	\begin{tabular}{cccc} \hline 
		parameter & Length (mm) & parameter & Length (mm) \\ \hline 
		\textit{S${}_{w}$} & 38.00 & \textit{W${}_{1}$} & 27.83 \\ 
		\textit{S${}_{l}$} & 68.00 & \textit{W${}_{2}$} & 06.40 \\ 
		\textit{G${}_{w}$} & 38.00 & \textit{L${}_{1}$} & 25.30 \\ 
		\textit{G${}_{l}$} & 68.00 & \textit{L${}_{2}$} & 06.32 \\ 
		\textit{d${}_{1}$} & 01.57 & \textit{M${}_{l}$} & 40.00 \\ 
		\textit{d${}_{2}$} & 05.00 & \textit{M${}_{w}$} & 01.80 \\ 
		\textit{T${}_{1}$} & 02.53 & \textit{g${}_{1}$} & 00.75 \\ 
		\textit{T${}_{2}$} & 03.34 & \textit{g${}_{2}$} & 02.50 \\ 
		\textit{C${}_{w}$} & 02.78 & \textit{gnd${}_{l}$} & 10.50 \\ \hline 
	\end{tabular}
	\caption{Dimensions of the EBG-backed planar monopole of Figure \ref{ebg1}.}
	\label{ebgtable1}
\end{table}

Ansoft HFSS, based on the finite element method (FEM), is used to simulate the electromagnetic response of the EBG structure. Figure \ref{ebg2}(b) shows the reflection coefficient phase when a single EBG unit cell is illuminated with a linearly polarized plane wave that is polarized along the \textit{x}-axis. The reference plane has been de-embedded up to the point where the linearly polarized planar monopole will be located. The structural parameters of the EBG structure, namely, \textit{L${}_{1, }$L${}_{2}$, W${}_{1}$, W${}_{2}$, T${}_{1}$, T${}_{2}$ }and \textit{C${}_{w}$} were optimized to define the EBG surface wave frequency band gap. The designated reflection phase for the EBG structure operation agrees well with the work presented in \cite{yang2003reflection,akhoondzadeh2007wideband}, which argues that it is advantageous in terms of the antenna impedance matching to work in the region where the reflection phase of the EBG plane is 90${}^\circ$$\pm$45${}^\circ$. The defining parameter for this 90${}^\circ$$\pm$45${}^\circ$ bandwidth is the gap between EBG structure and ground plane (\textit{d${}_{2}$}) which provides direct control over the reflection phase of the EBG structure (see Figure \ref{ebg2}(b)). In \cite{yang2003reflection} it was also concluded that the operational EBG bandwidth increases as the contour of the reflection phase in the transition region becomes smoother, a characteristic that also depends on the parameter \textit{d${}_{2}$}.

\subsection{ Antenna Miniaturization}

The work presented in \cite{kim2012monopole} describes an inkjet printed planar monopole backed by a 4$\times$3 array of similar planar split ring structures \cite{tang2010novel} operating at a $\sim$0${}^\circ$ phase reflection bandwidth namely at the Artificial Magnetic Conductor (AMC) band. The dimensions of the antenna in \cite{kim2012monopole} are rather large for wearable applications. Also, the use of a paper substrate further decreases the robustness of the antenna in adverse environmental conditions. These seemingly disadvantageous characteristics motivated the design and miniaturization of an EBG array backing a printed monopole, on a more robust and semi-flexible substrate. In this section, the miniaturization from a 4$\times$3 to 2$\times$1 EBG array is discussed stepwise in order to obtain some engineering design guidelines. The miniaturization presented in Figure \ref{ebg3} and summarized in Table \ref{ebgtable2} consists of the following steps: 

\begin{figure}[htb]
	\centering
	\begin{subfigure}{0.65\textwidth}
		\includegraphics[width=1\textwidth]{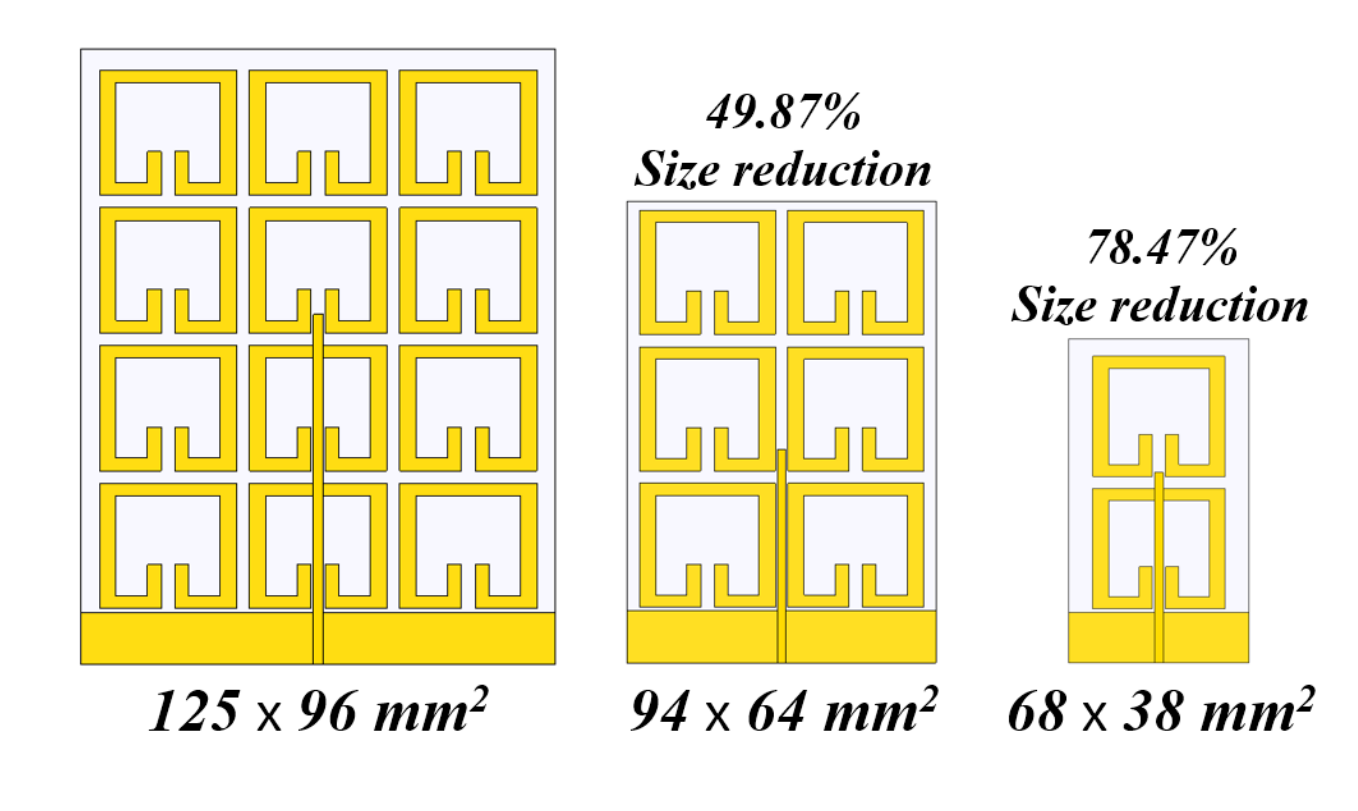}
		\caption{}
	\end{subfigure}
	\begin{subfigure}{0.8\textwidth}
		en		\includegraphics[width=1\textwidth]{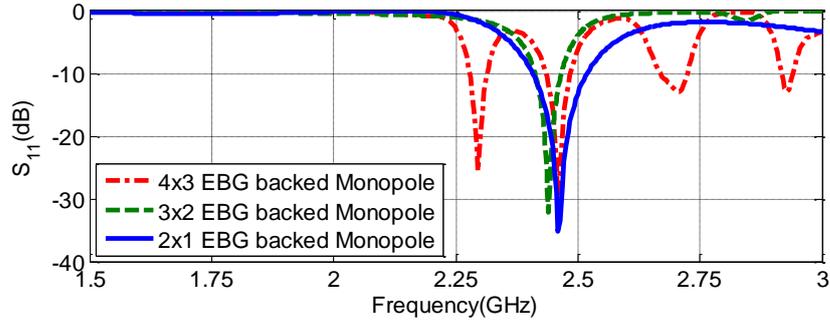}
		\caption{}
	\end{subfigure}
	\caption{Miniaturization of antenna (a) Return loss response of the steps with percentage size reduction (b) Reflection coefficient when the antenna is optimized in the 2.45 GHz ISM band.}
	\label{ebg3}
\end{figure}

\textit{Step 1: }A radiating monopole backed by a 4$\times$3 EBG array was designed on a Rogers RT/duroid 5880. Note that the ground plane, backing the EBG structure, was reduced to 125$\times$96 mm${}^{2}$ compared to 150$\times$120 mm${}^{2}$ in \cite{kim2012monopole}, hence decreasing the overall footprint of the antenna at this initial design phase. The reflection coefficient and radiation efficiency of a planar monopole backed with 4$\times$3 EBG array is mainly determined by three parameters (see Figure \ref{ebg1}): the monopole length (\textit{M${}_{l}$}), the monopole width (\textit{M${}_{w}$}) and the gap between the monopole and the bottom ground plane (\textit{d${}_{2}$}). The EBG reflection phase band (controlled primarily by \textit{d${}_{2}$}) was matched with the resonance of the planar monopole by optimizing the parameters\textit{ M${}_{l}$ }and \textit{M${}_{w}$}. The optimized reflection coefficient of the 4$\times$3 EBG-backed monopole is presented in Figure \ref{ebg3}(b) whereas the optimized parameters and results are listed in Table \ref{ebgtable2}.

\textit{Step 2: }Since most of the radiating fields reside in the center of the 4$\times$3 EBG array, the array matrix was decreased from 4$\times$3 to 3$\times$2, targeting a $\sim$50\% size reduction. The initial reflection coefficient was poor in the ISM band, but it was subsequently optimized so that the antenna radiated at a higher efficiency. The noticeable effect of this miniaturization was the small decrease in the peak gain of the antenna, from 9.78 dBi to 8.97 dBi, when the array size was reduced from 4$\times$3 to 3$\times$2 \cite{abbasi2016high}. 

\textit{Step 3: }Based on the same observation of weak radiating fields along the edges of the antenna, miniaturization from a 3$\times$2 to a 2$\times$1 EBG array was achieved. Re-optimization of the antenna dimensions led to the set of parameters listed in Table \ref{ebgtable2} that resulted in efficient radiation of the antenna in the ISM band, from 2.40 -- 2.52 GHz, with a gain of 6.88 dBi at 2.45 GHz. The final proposed structure, from Step 3, shows a significant size reduction of the proposed antenna, compared to previously reported EBG- and AMC-backed designs \cite{jiang2014compact,kim2012monopole,mohamed2015wideband,raad2013flexible,kim2012wearable,zhu2009dual,yang2005novel,yan2014low}.

\begin{table}
	\centering
	\begin{tabular}{c|ccc} \hline 
		& \textbf{Step 1} & \textbf{Step 2} & \textbf{Step 3} \\ \hline 
		\multicolumn{4}{c}{Optimized dimensions of parameters (mm)} \\ \hline 
		\textit{S${}_{w}$} & 96.00 & 64.00 & 38.00 \\ 
		\textit{S${}_{l}$} & 125.00 & 94.00 & 68.00 \\ 
		\textit{d${}_{2}$} & 03.25 & 04.00 & 06.00 \\ 
		\textit{M${}_{l}$} & 70.50 & 43.50 & 40.00 \\ 
		\textit{M${}_{w}$} & 03.80 & 01.60 & 01.80 \\ \hline 
		\multicolumn{4}{c}{Optimized results} \\ \hline 
		Reflection Coefficient & -32 dB & -31 dB & -34 dB \\ 
		Bandwidth & 2.43-2.48 GHz & 2.41-2.46 GHz & 2.40-2.52 GHz \\ 
		Radiation Efficiency & 95\% & 96\% & 76\% \\ 
		Gain Max. & 9.78 dBi & 8.97 dBi & 6.88 dBi \\ \hline 
	\end{tabular}
	\caption{Summary of Miniaturization.}
	\label{ebgtable2}
\end{table}

\subsection{ Parametric analysis}

\begin{figure}[t]
	\centering
	\includegraphics[width=0.7\textwidth]{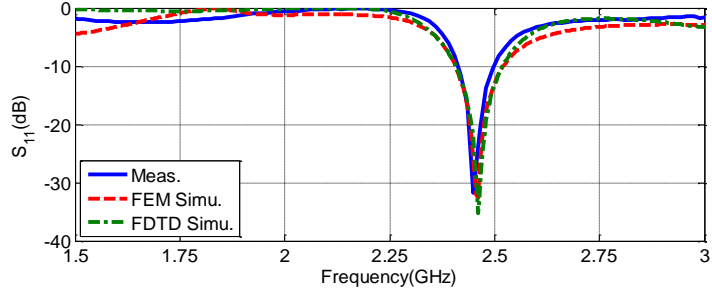}
	\caption{Comparison of 2$\times$1-EBG-backed monopole measured S${}_{11}$ with simulated S${}_{11}$ using finite element method (FEM) and simulated using finite difference time domain method (FDTD).}
	\label{ebg4.1}
\end{figure}

\begin{figure}[H]
	\centering
	\begin{subfigure}{0.45\textwidth}
		\includegraphics[width=1\textwidth]{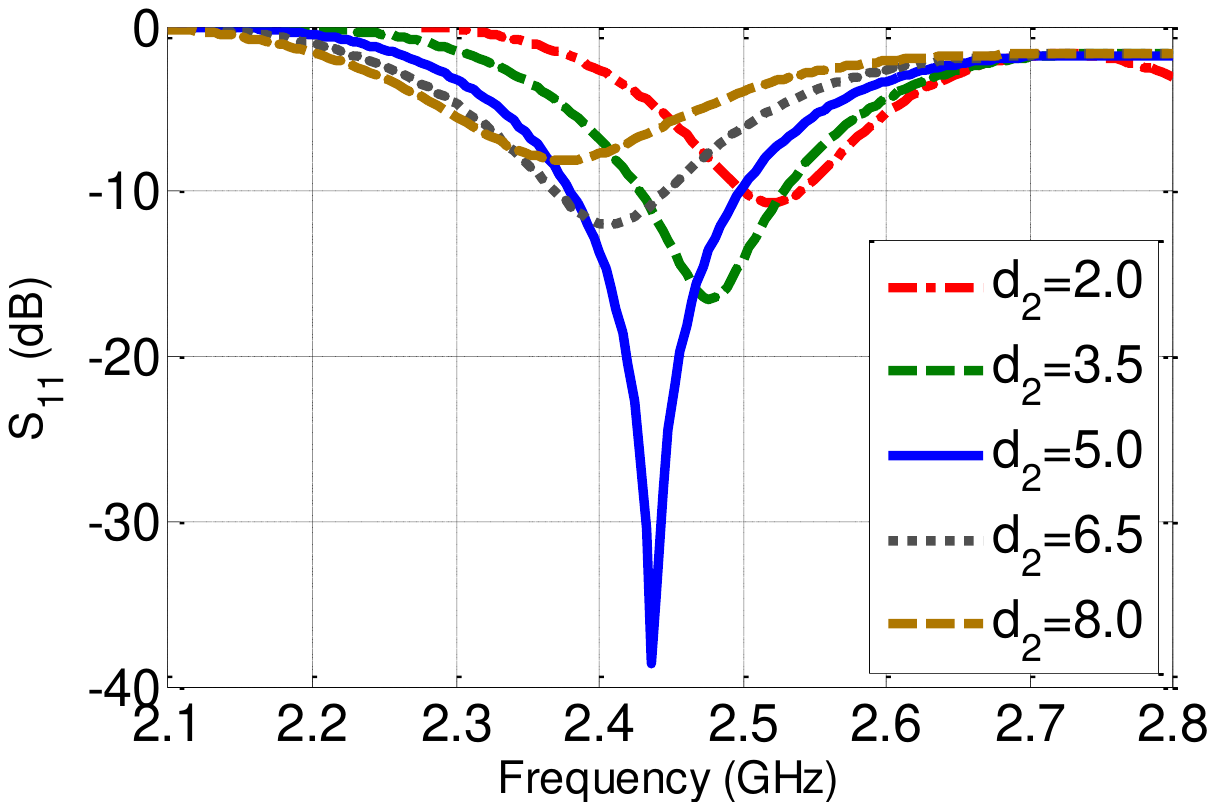}
		\caption{}
	\end{subfigure}
	\begin{subfigure}{0.45\textwidth}
		\includegraphics[width=1\textwidth]{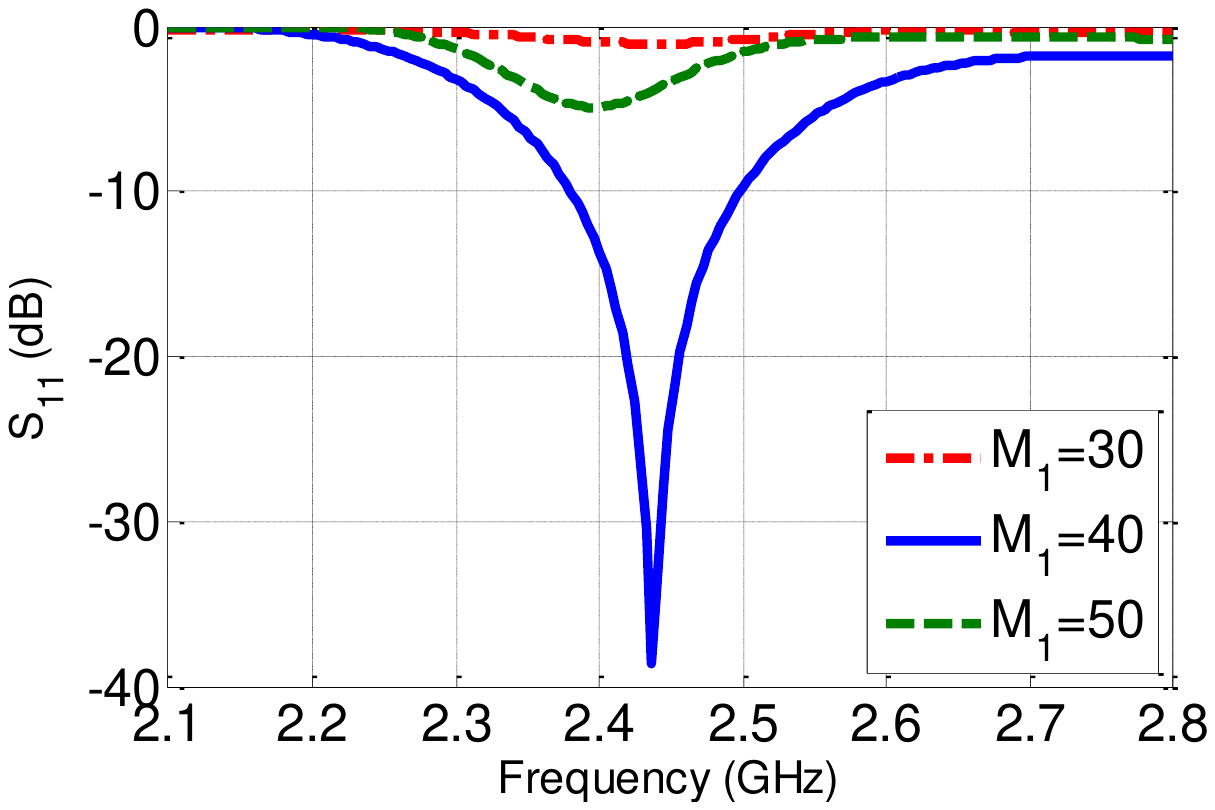}
		\caption{}
	\end{subfigure}
	\begin{subfigure}{0.45\textwidth}
		\includegraphics[width=1\textwidth]{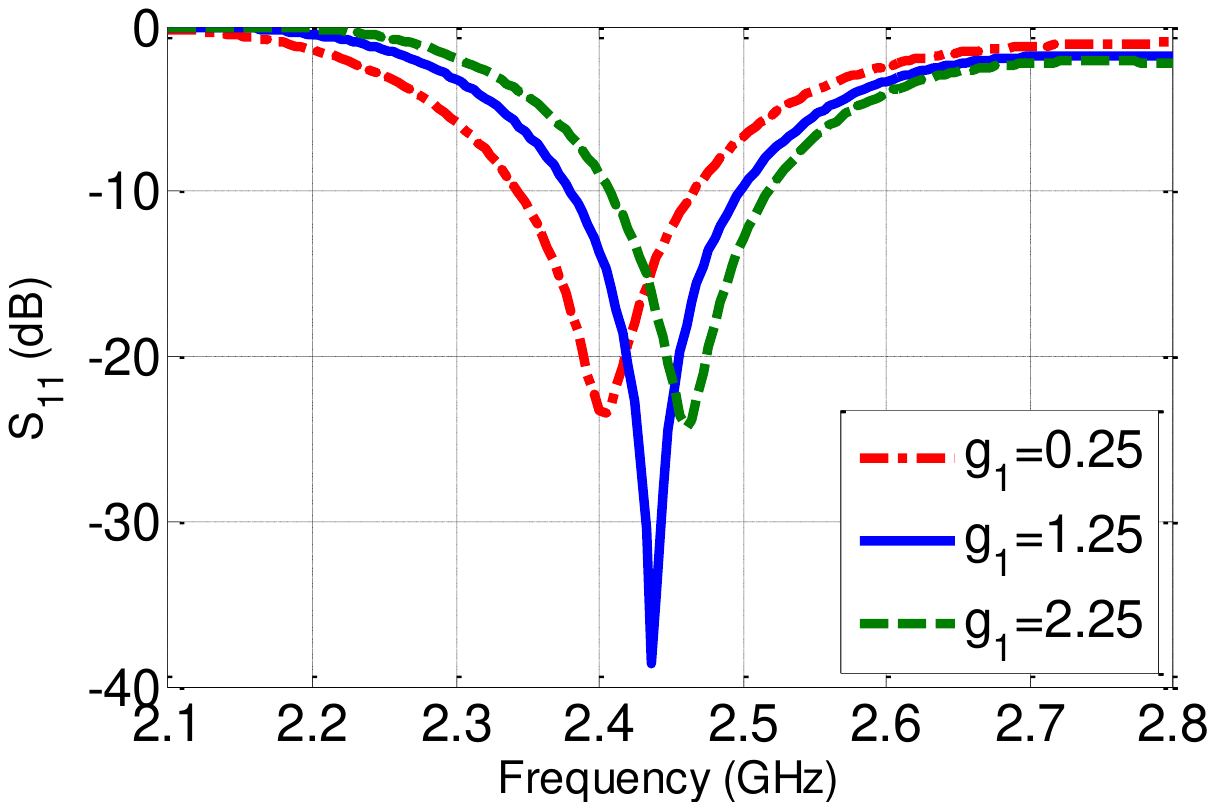}
		\caption{}
	\end{subfigure}
	\begin{subfigure}{0.45\textwidth}
		\includegraphics[width=1\textwidth]{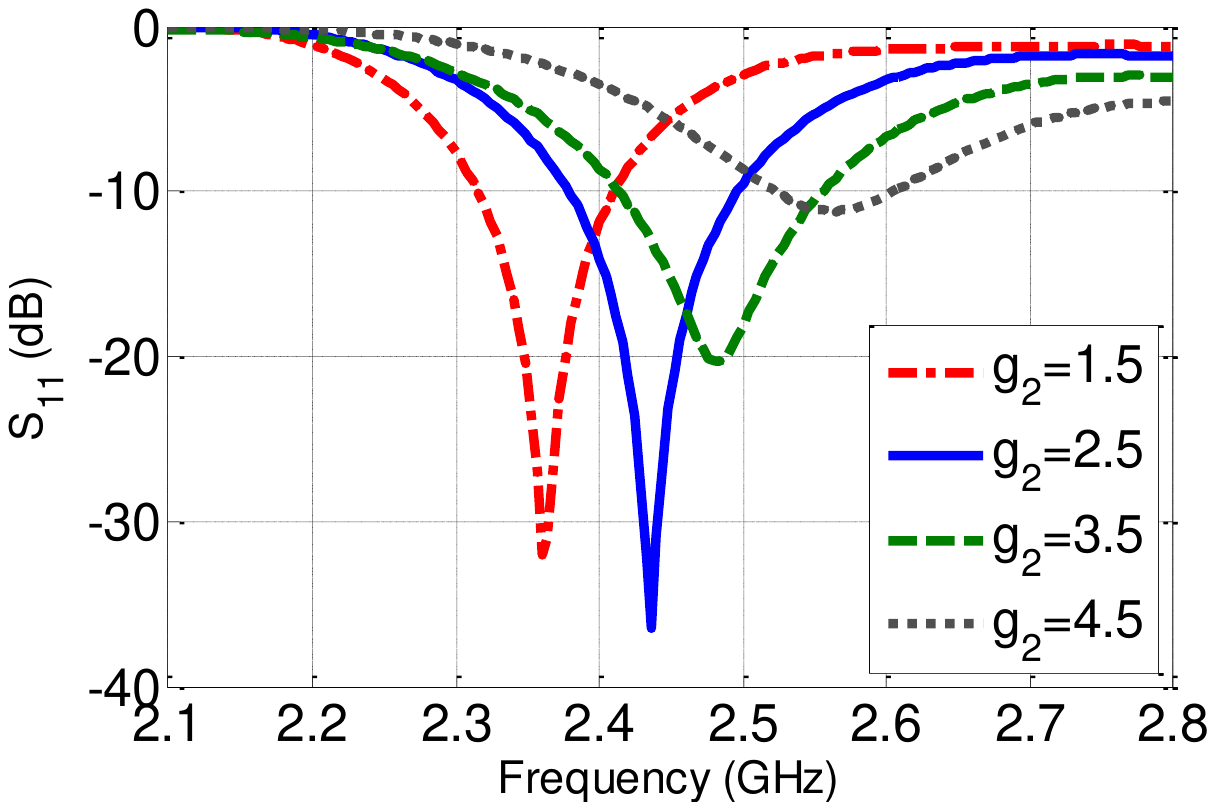}
		\caption{}
	\end{subfigure}
	\caption{Parametric study of: (a) monopole length \textit{M${}_{1}$, }(b) gap between EBG and ground plane \textit{d${}_{2}$} (c) gap between RF ground and first EBG cell \textit{g${}_{1}$} and (d) gap between EBG cells \textit{g${}_{2}$} all in mm. }
	\label{ebg4}
\end{figure}

The simulated and measured reflection coefficient of the fabricated 2$\times$1 EBG-backed monopole is shown in Figure \ref{ebg4.1}. The 90${}^\circ$$\pm$45${}^\circ$ phase bandwidth available for the operation of the optimized EBG layer is considerably wider than the S${}_{11}$ $<$ -10 dB bandwidth of a conventional planar monopole. To further appreciate this point, let us observe the effect of gap \textit{d${}_{2}$} on the reflection phase in Figure \ref{ebg2}(d) compared to the effect on the reflection coefficient in Figure \ref{ebg4}(a). It is evident that, the input impedance frequency band overlaps with the surface-wave frequency band of the EBG structure for a wide range of \textit{d${}_{2}$ }values\textit{ }(i.e. around 3 mm). This operation also corroborates the theoretical studies presented in \cite{yang2003reflection,mosallaei2004antenna}. This fact makes the proposed antenna rather robust and immune to gap size (\textit{d${}_{2}$}) changes, and as a result its operation is not significantly disturbed when the gap is perturbed. This robustness feature is always desirable for a wearable antenna because it is often subjected to mechanical stress, which would tend to change the value of \textit{d${}_{2}$}. The length of the planar monopole (\textit{M${}_{l}$}) significantly affects the impedance matching of the antenna, whereas, the gap \textit{g${}_{1}$} controls the resonance frequency of the antenna as shown in Figure \ref{ebg4}(b) and (c) respectively. The variation of the gap between two consecutive EBG cells, \textit{g${}_{2}$}, affects both the impedance matching and the antenna resonance frequency and is the most sensitive design parameter. As the \textit{g${}_{2}$} increases, the frequency resonance shifts towards higher frequencies as reported also in \cite{yang2003reflection}.


\subsection{ Radiation performance and mechanism}

\begin{figure}[htb]
	\centering
	\begin{subfigure}{0.8\textwidth}
		\includegraphics[width=1\textwidth]{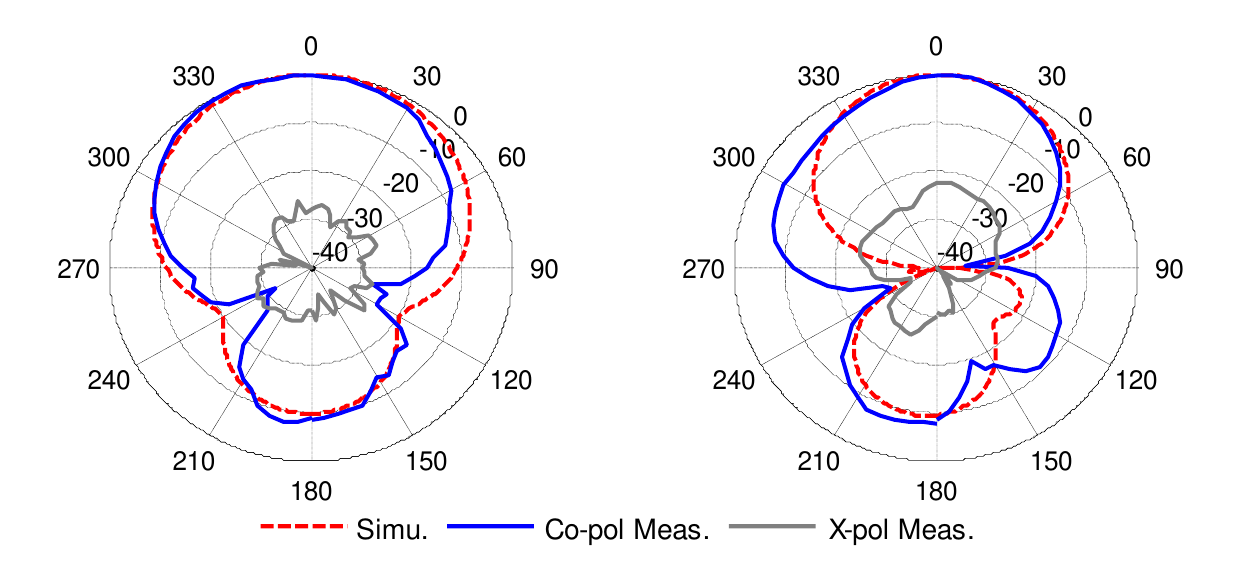}
		\caption{}
	\end{subfigure}
	\begin{subfigure}{0.7\textwidth}
		\includegraphics[width=1\textwidth]{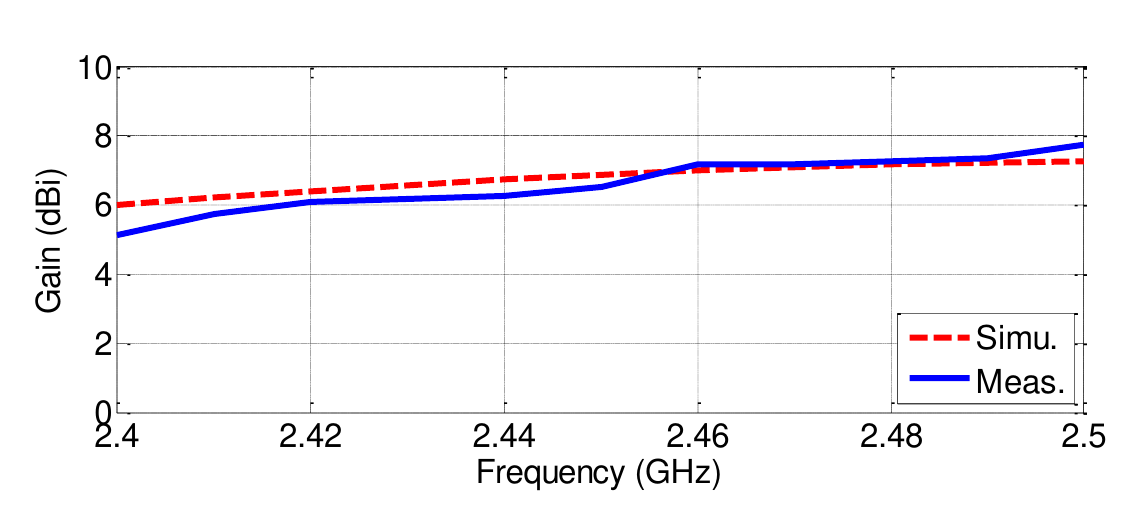}
		\caption{}
	\end{subfigure}
	\caption{(a) Simulated and measured \textit{y-z} plane (left) and \textit{x-z} plane (right) normalized radiation patterns of the 2$\times$1 EBG-backed monopole antenna at 2.45 GHz. (b) Simulated and measured peak gain of the 2$\times$1 EBG-backed monopole}
	\label{ebg5}
\end{figure}

The proposed antenna has the maximum directivity along the positive\textit{ z}-axis as shown in Figure \ref{ebg5}. The main radiation lobe has a {94}${}^\circ$ and {62}${}^\circ$ half-power beam width (HPBW) in the \textit{y-z} plane and \textit{x-z} plane, respectively. The radiation pattern depicts relatively low radiation in the backward direction, at the location of human tissue when the antenna is placed directly on the human body. This factor also decreases the maximum SAR value of the antenna which is a desired characteristic for any wearable antenna.

\begin{figure}[H]
	\centering
	\begin{subfigure}{0.6\textwidth}
		\includegraphics[width=1\textwidth]{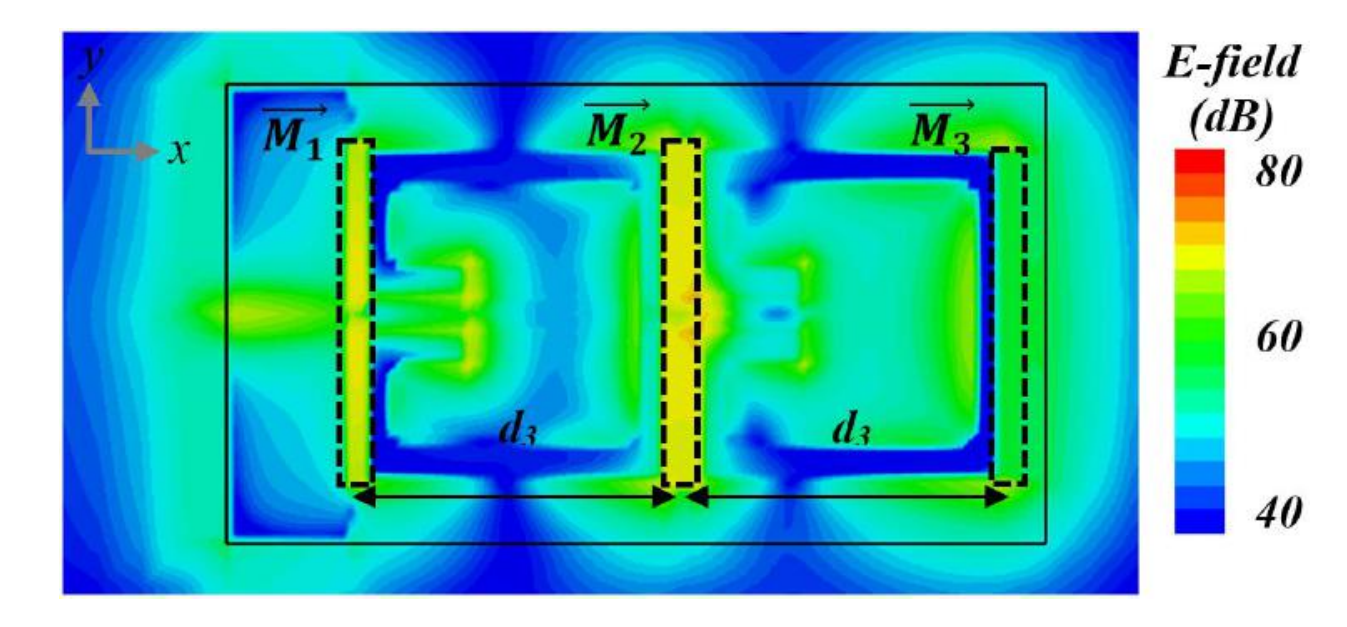}
		\caption{}
	\end{subfigure}
	\begin{subfigure}{0.7\textwidth}
		\includegraphics[width=1\textwidth]{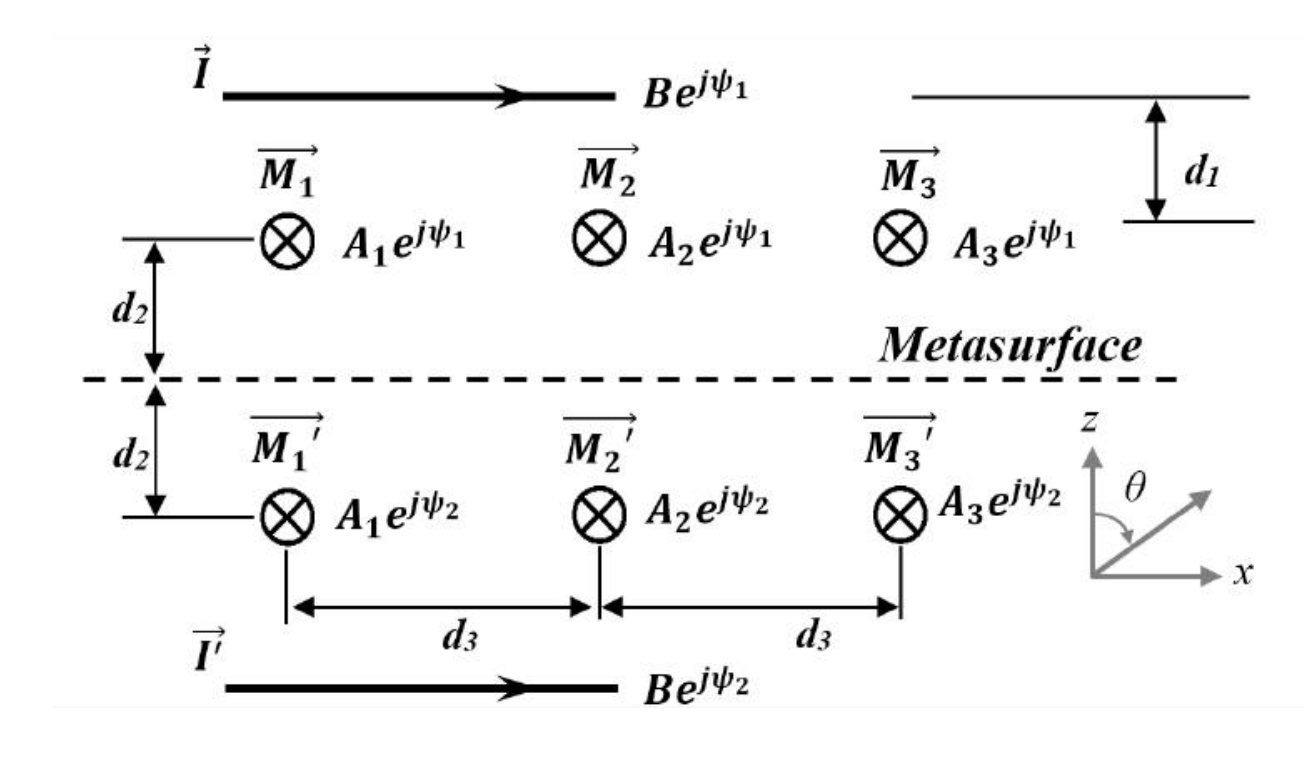}
		\caption{}
	\end{subfigure}
	\begin{subfigure}{0.6\textwidth}
		\includegraphics[width=1\textwidth]{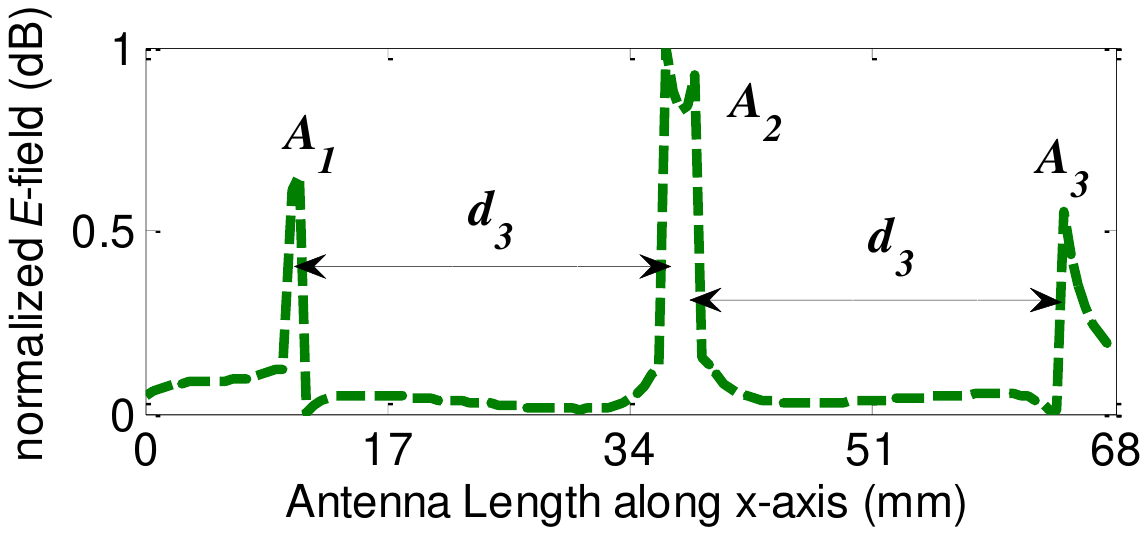}
		\caption{}
	\end{subfigure}
	\begin{subfigure}{0.35\textwidth}
		\includegraphics[width=1\textwidth]{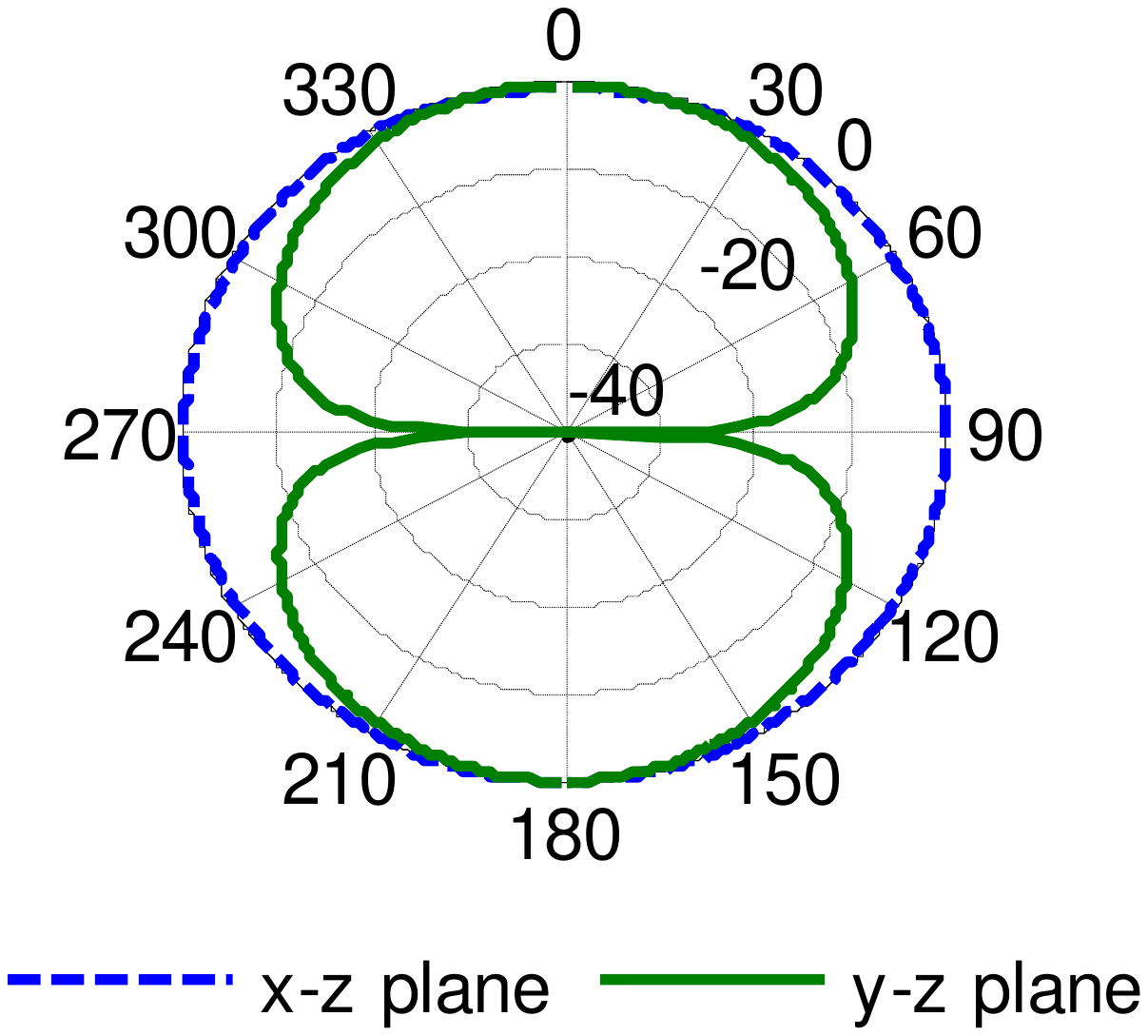}
		\caption{}
	\end{subfigure}
	\caption{(a) Electric field distribution on EBG surface depicting three magnetic current sources at 2.45 GHz (b) Equivalent eight-element array model consisting of two electric current sources and six magnetic current sources, (c) Simulated electric field on the EBG layer along the \textit{x}-axis, which is used to model the magnitude of the magnetic current sources, (d) Simulated electric field pattern of a $\sim$$\lambda$/4 magnetic current source.}
	\label{ebg6}
\end{figure}

The directional radiation patterns of the proposed antenna are caused not only by the electric monopole but also by the surface currents induced on the metallic surface of the EBG structure, placed over a full ground plane that creates an array of slots. These slots can be modeled as equivalent magnetic current sources, which form a centrally-excited 3$\times$1 array. To further elaborate on this, the electric field plotted on the substrate directly beneath the EBG surface is presented in Figure \ref{ebg6}(a), where the locations of the equivalent magnetic current sources are identified in dotted rectangles. These radiating slots can be replicated by placing three independent magnetic current sources ($\vec{M_{{1}}}\mathrm{,\ }\vec{M_{{2}}}\mathrm{,\ }\vec{M_{{3}}}$) at the same location. Considering the combined effect of the ground plane and the EBG layer, it can be assumed that the resultant radiation pattern is radiated by a virtual eight-element array as shown in Figure \ref{ebg6}(b).

The eight-element array consists of 2x1 elements excited by electric current sources and 3x2 elements excited by magnetic current sources. In Figure \ref{ebg6}, $\vec{M_{\mathrm{1}}}\mathrm{,\ }\vec{M_{\mathrm{2}}}$ and $\mathrm{\ }\vec{M_{\mathrm{3}}}$ represent the independent magnetic current sources, while $\vec{{M'_1}}$, $\vec{{M'_2}}$ and $\vec{{M'_3}}$ indicate their images upon reflection on the metasurface that causes a 90${}^\circ$ phase difference between the radiating elements and their images. Similarly, $\vec{I}$ represents the electric current on the planar monopole and $\vec{I'}$ represents its image. For the analysis, the current on each element is expressed in complex form $A_{\mathrm{1}}e^{j{\psi }_{\mathrm{1}}}$,$A_{\mathrm{2}}e^{j{\psi }_{\mathrm{1}}}$, $A_{\mathrm{3}}e^{j{\psi }_{\mathrm{1}}}$, $A_{\mathrm{1}}e^{j{\psi }_{\mathrm{2}}}$, $A_{\mathrm{2}}e^{j{\psi }_{\mathrm{2}}}$, $A_{\mathrm{3}}e^{j{\psi }_{\mathrm{2}}}$, $Be^{j{\psi }_{\mathrm{1}}}$ and $Be^{j{\psi }_{\mathrm{2}}}$, where $A_{\mathrm{1}}$, $A_{\mathrm{2}}$, $A_{\mathrm{3}}$ and $B$ represent the magnitudes of each current source and \textit{$\psi$${}_{1}$} and \textit{$\psi$${}_{2}$} represent the phases of the real and image sources, respectively. The magnitude excitations $A_i$ and $B$ can be approximated from the simulated electric field distribution presented in Figure \ref{ebg6}(c) and are normalized with respect to the maximum value$\ A_{\mathrm{2}}$. Based on the same distribution, the normalized value of the electric current magnitude is $B$ = 1.00 and the estimated relative magnetic current magnitudes are$\ A_{\mathrm{1}}=0.65$, $A_{\mathrm{2}}=1.00$ and$\ A_{\mathrm{3}}=0.55$. Based on the time domain animation of the \textit{E}-field distribution they radiate in phase with each other. Hence, all three magnetic current sources above the metasurface have the same phase, \textit{$\psi$${}_{1}$}, while the image sources below the metasurface have a phase of \textit{$\psi$${}_{2}$}. 

Note that the simulations to evaluate aforementioned quantities were based on FDTD method with a high accuracy. The FDTD mesh cells defined on the EBG structure were intensified to a level where three mesh cells defined the are between the first EBG structure and the RF ground plane. This added up the complexity of the simulation setup specifically designed for accurate data matrix of the magnetic current sources, both in terms of magnitude as well as the spacial phase values. For cross verification, similar simulation was performed using FEM when the antenna port was excited using waveguide port. Both setups agreed well up to one significant figure of normalized electric field.

\begin{figure}[htb]
	\centering
	\includegraphics[width=0.8\textwidth]{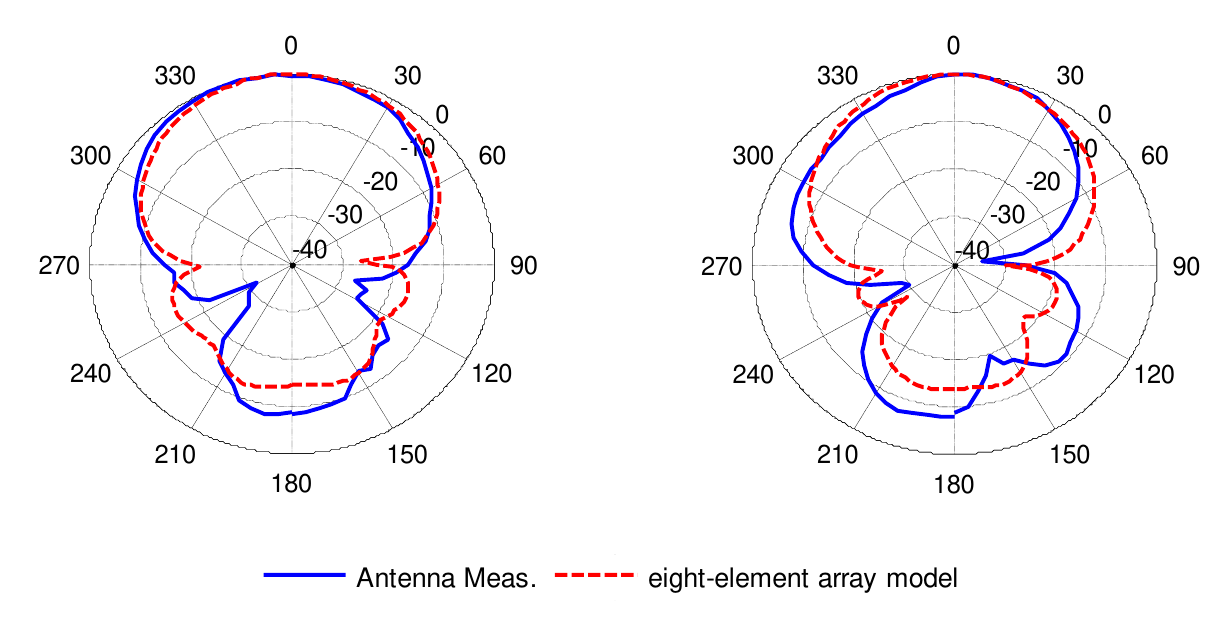}
	\caption{Electric field patterns comparison between of 2$\times$1 EBG-backed monopole antenna and the eight-element array model along the \textit{y-z} plane (left) and along the \textit{x-z} plane (right).}
	\label{ebg7}
\end{figure}

As explained in Section II.A and Figure \ref{ebg2}(b), the reflection phase exhibited by the grounded EBG metasurface is selected to be 90${}^\circ$, meaning that the phase difference between \textit{$\psi$${}_{1}$} and \textit{$\psi$${}_{2}$ }is 90${}^\circ$ as well. It is known that the total electric field of multiple independent radiating elements is equal to the product of the field of a single source and the array factor, so the behavior of the six-element array is determined by the array factor for six magnetic current sources given by: 

\begin{equation}
{AF}_M={AF}_h\ \times {AF}_v\ 
\end{equation}

Here, ${AF}_h$ is the array factor for the three-element magnetic current sources distributed along \textit{x}-axis, and ${AF}_v\ $is the array factor for the two sets of three-element linear arrays, distributed along the \textit{z}-axis. The six magnetic currents which are aligned along the \textit{y}-axis have omni-directional radiation patterns along the \textit{x-z} plane and are linearly polarized with an \textit{E${}_{\phi}$} component (assuming that the magnetic currents are oriented parallel with the \textit{z}-axis) to be the dominant polarization. The electric field derived for a magnetic current source is presented in \cite{nikolaou2006pattern}:

\begin{equation}
E(\theta )\cong -j\frac{M_oe^{-jkr}}{2\pi r}\frac{{\mathrm{cos} \left(\frac{kl}{2}{\mathrm{cos} \theta \ }\right)\ }-{\mathrm{cos} \left(\frac{kl}{2}\right)\ }}{{\mathrm{sin} \theta \ }}
\end{equation}

Where $k$ is the wavenumber, $r$ is the distance from the source centre, and $M_o$ is the amplitude of the magnetic current excitation. Considering the ``global'' axes presented in Figure \ref{ebg6}(b), this coincides with the \textit{E${}_{\theta}$} polarization. The polarization of the radiated field is in agreement with the measured co-polarization. However, for the calculation of the \textit{E}-field along the \textit{y-z} plane, the normalized distribution of the individual magnetic currents has to be considered. The normalized array factor pattern is calculated for an arrangement replicating the radiating mechanism of the proposed 2$\times$1 EBG-backed monopole, where, \textit{d${}_{2}$} = 5.8 mm ($\sim$$\lambda$/20) and \textit{d${}_{3}$ =} 26.7 mm ($\sim$$\lambda$/4). The calculated array factor for the magnetic current sources, ${AF}_M$, is then superimposed with the radiation pattern of the remaining 2$\times$1 monopole elements represented by $\vec{I}$ and $\vec{I'}$ to form a radiation pattern for an eight-element array model. It should be noted that despite the fact that the orientation of the 2$\times$1 electric currents is orthogonal with the orientation of the 3$\times$2 magnetic currents, their dominant radiated \textit{E}-field polarization coincides because they are excited by electric and magnetic sources respectively. In both Figures \ref{ebg7}(a) and (b), the solid lines show the co-polarized measured radiation patterns of the 2$\times$1 EBG-backed monopole, whereas the dotted lines show the calculated \textit{E}-field pattern of the combined eight-element current source array model. The location of the nulls in Figure \ref{ebg7}(a) for the eight-element model is exactly at 90${}^\circ$ and 270${}^\circ$, which do not coincide exactly with the nulls in the measured radiation pattern (blue solid contour). This discrepancy is mostly because of the approximation of the limited size ground plane \textit{G${}_{l}$}$\times$\textit{G${}_{w}$} with an ideal infinite ground plane for the eight-element array. Other than the slight relocation of nulls, it can be observed that the suggested radiation mechanism of the eight-element array is in very good agreement with the measured radiation patterns of the fabricated 2$\times$1 EBG-backed monopole. This verifies the validity of the presented eight-element array model presented in Figure \ref{ebg6}(b) and also demonstrates the effect of the metasurface that causes the intended 90${}^{o}$ phase difference between the physical radiators and their images. 

\subsection{ Experimental Results}

\begin{figure}[htb]
	\centering
	\begin{subfigure}{0.19\textwidth}
		\includegraphics[width=1\textwidth]{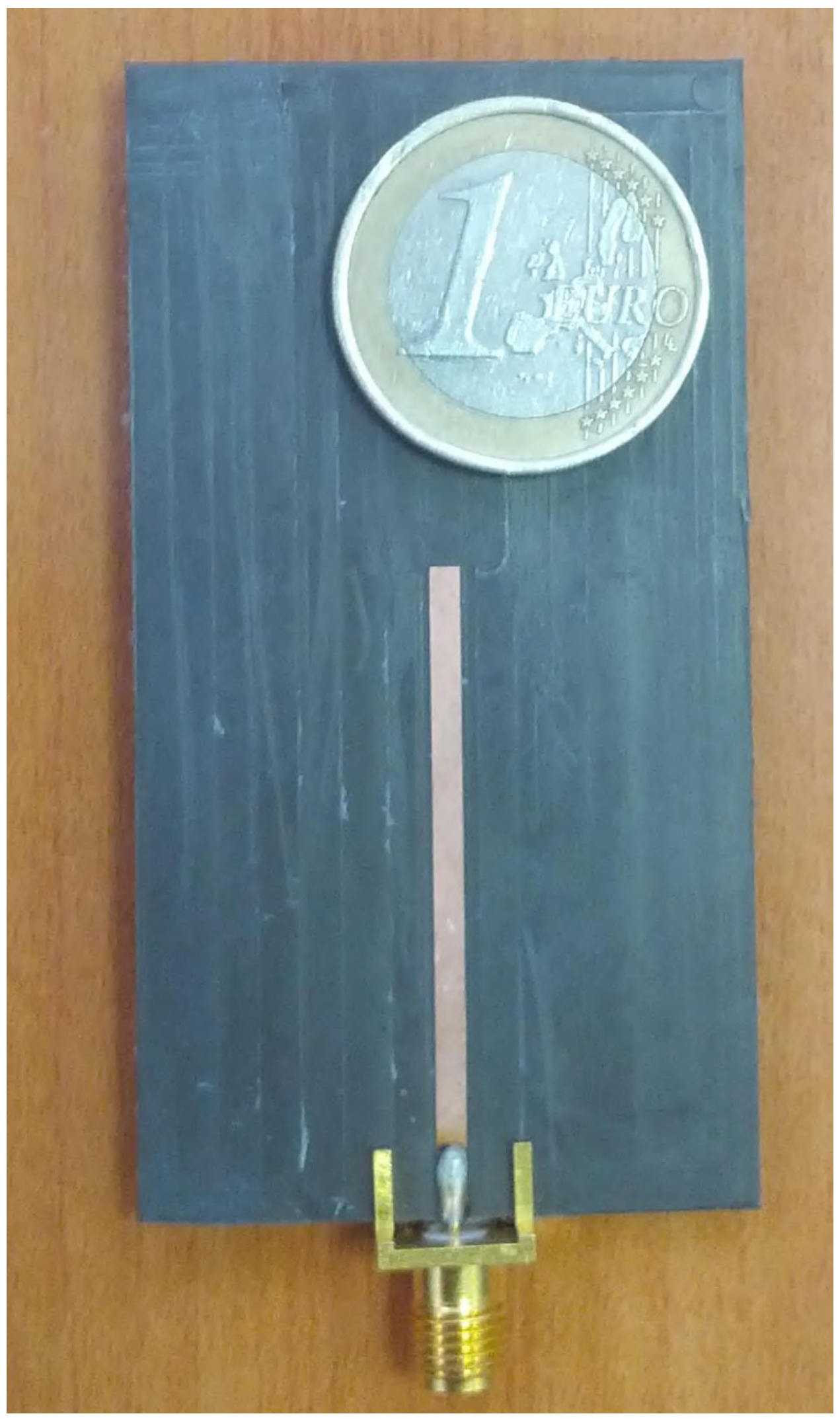}
		\caption{}
	\end{subfigure}
	\begin{subfigure}{0.2\textwidth}
		\includegraphics[width=1\textwidth]{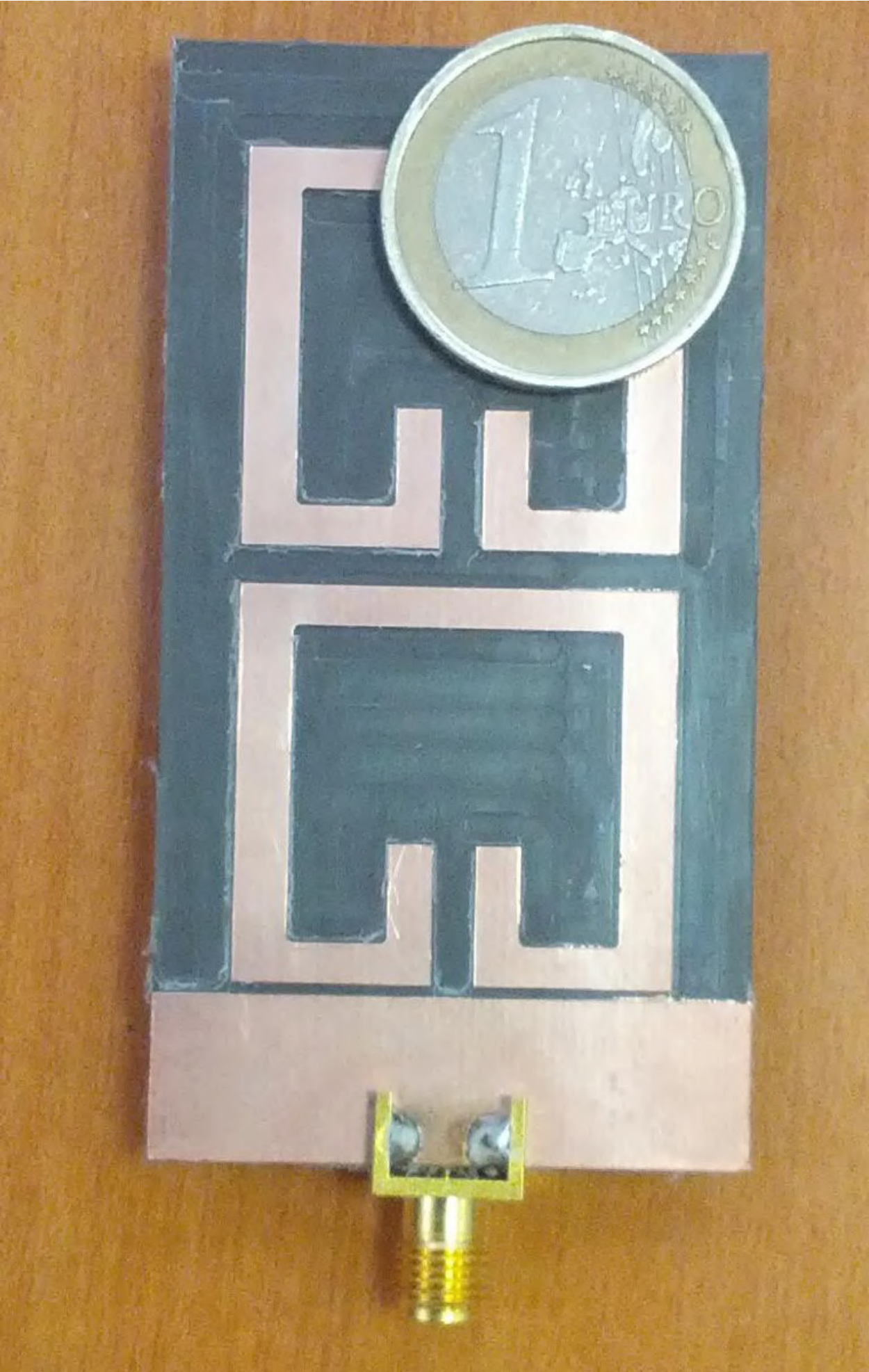}
		\caption{}
	\end{subfigure}
	\begin{subfigure}{0.51\textwidth}
		\includegraphics[width=1\textwidth]{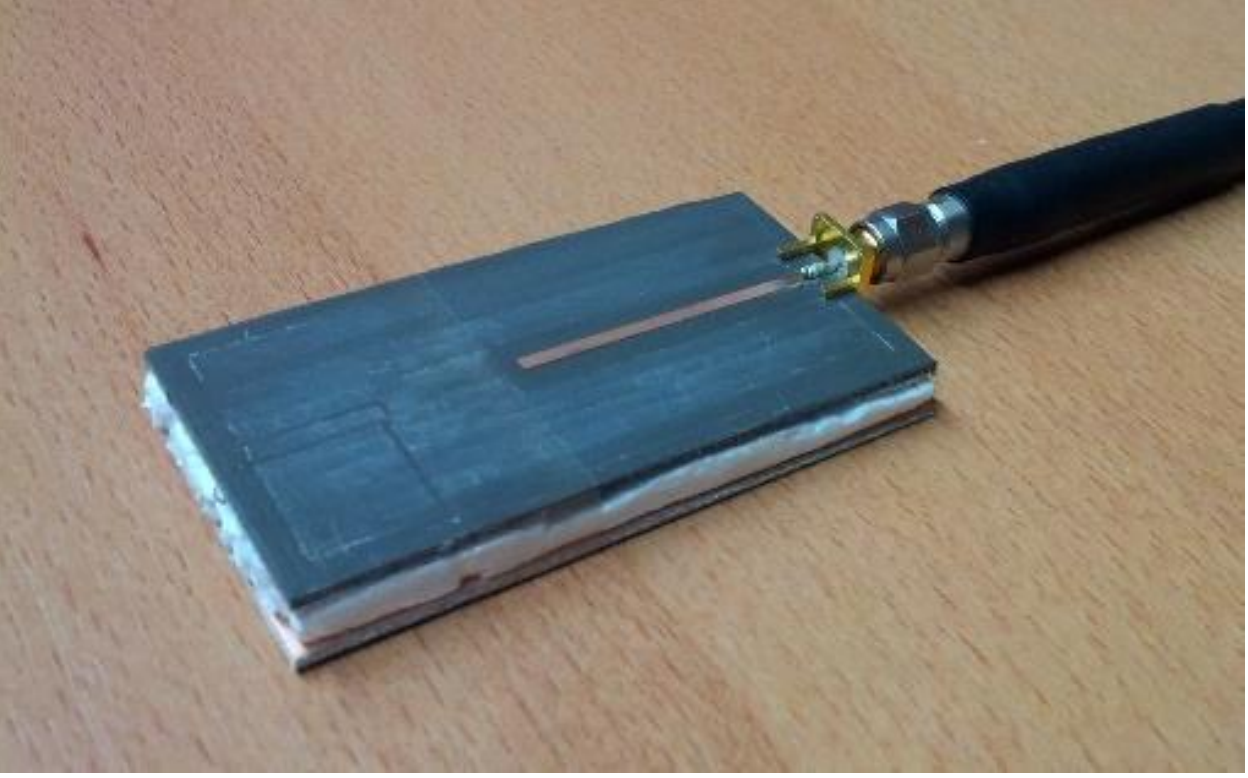}
		\caption{}
	\end{subfigure}
	\begin{subfigure}{0.47\textwidth}
		\includegraphics[width=1\textwidth]{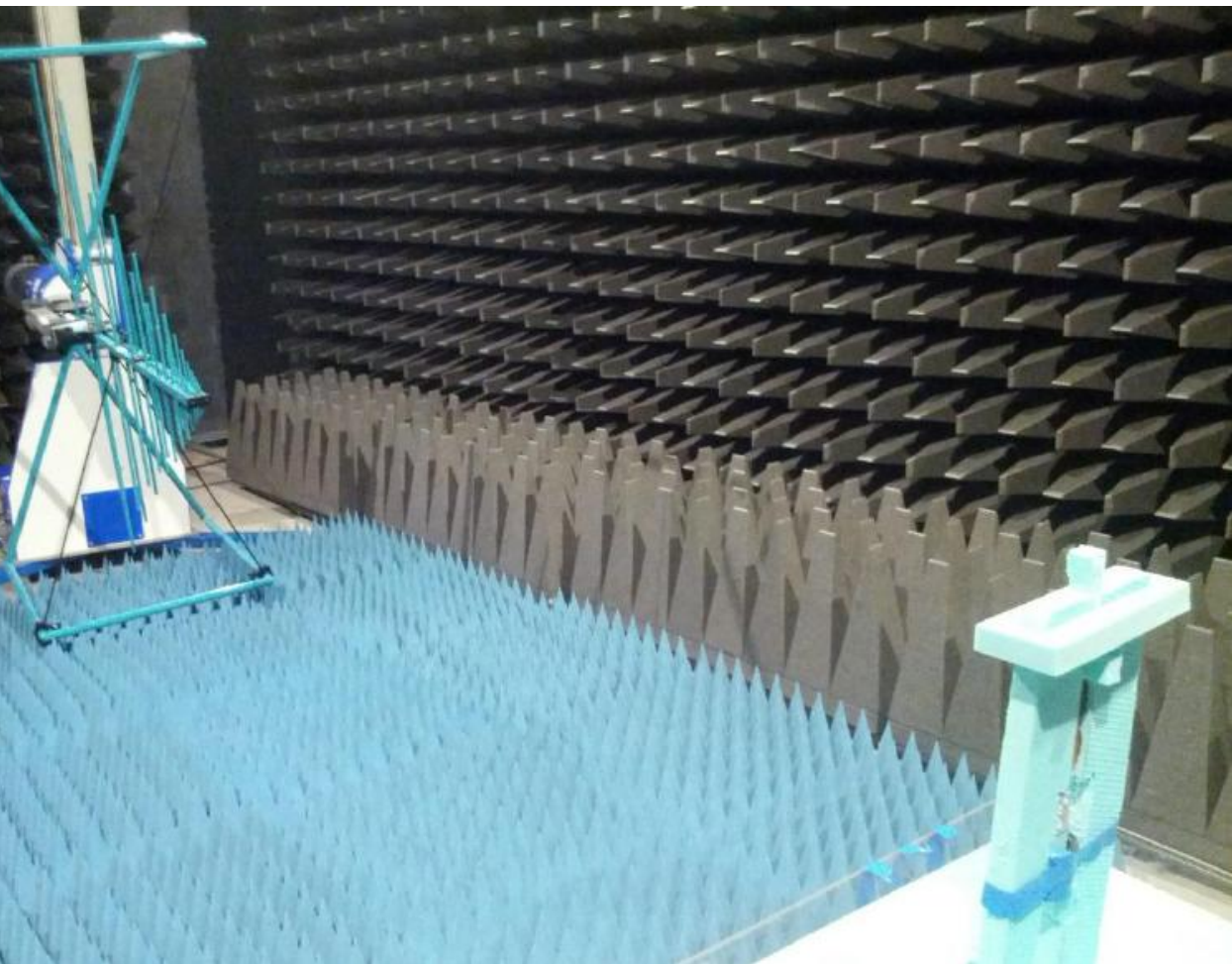}
		\caption{}
	\end{subfigure}
	\begin{subfigure}{0.45\textwidth}
		\includegraphics[width=1\textwidth]{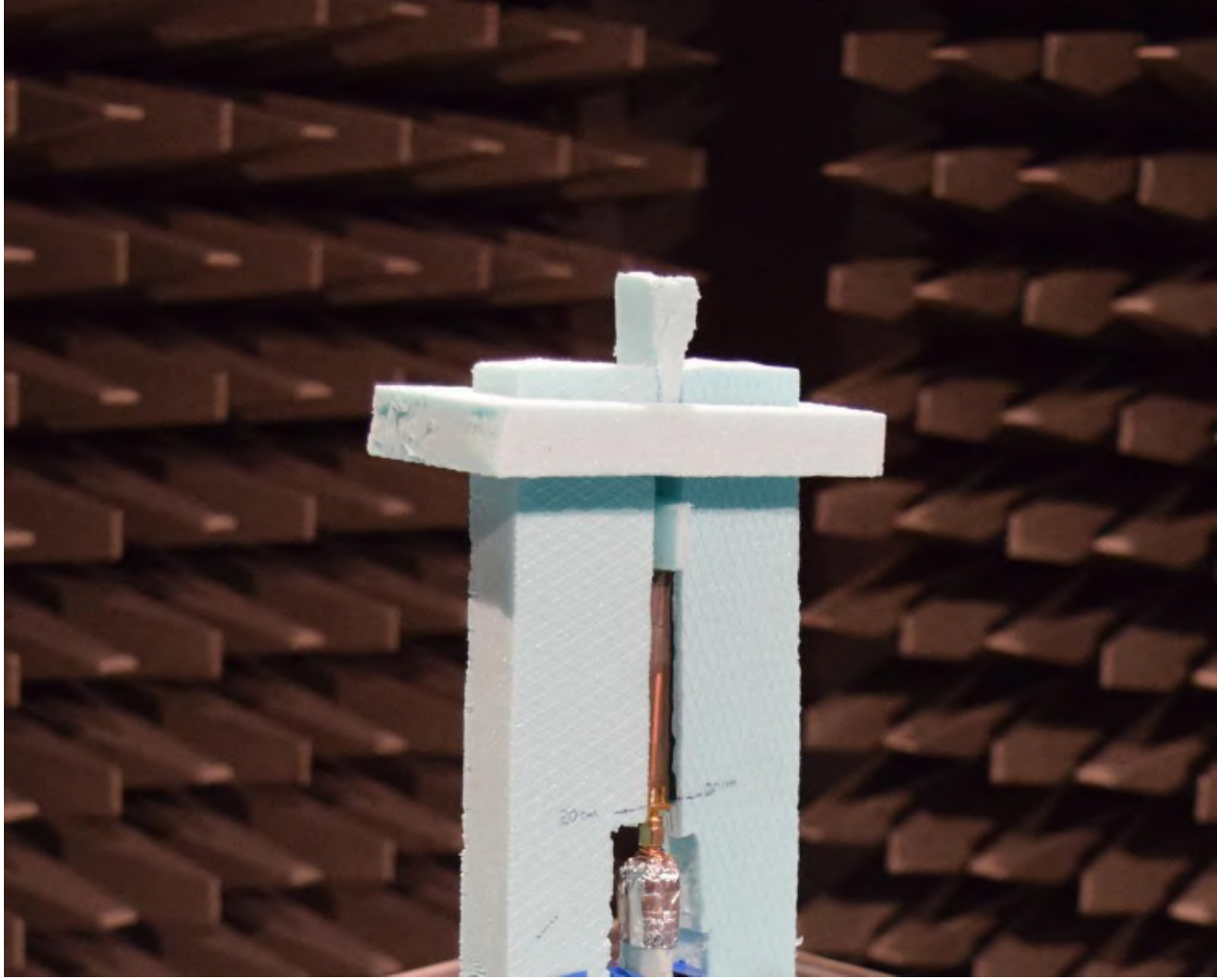}
		\caption{}
	\end{subfigure}
	\caption{Photographs of the fabricated antenna (a) on the planar monopole side (front side) (b) on the 2$\times$1 EBG plane side (back side) (c) assembled integrated antenna connected to a VNA (d) Radiation pattern measurement setup in an anechoic chamber and (e) antenna supported by low-loss Styrofoam over a rotating platform connected to an SMA connector and a balun in the anechoic chamber. $\textcopyright$ 2016 IEEE.}
	\label{ebg8}
\end{figure}

The 2$\times$1 EBG-backed monopole was fabricated on an LPKF ProtoMat H 100 milling machine. Styrofoam with the prescribed thickness was added as a spacer between the EBG layer and the finite ground plane (Figure \ref{ebg8}). An Agilent E8363B network analyzer was used to characterize the reflection coefficient of the antenna in free space. As shown in Figure \ref{ebg4}(a), good agreement can be seen between measurements and simulated predictions not only in terms of resonance position, but also with respect to the 10 dB bandwidth. The measured reflection coefficient of the 2$\times$1 EBG-backed monopole antenna has a -10 dB bandwidth ranging from 2.40 to 2.50 GHz, which is slightly narrower than the simulated response. This small discrepancy between the measurement and simulation can be related to the additional Ohmic losses normally generated while soldering the SMA connector on the fabricated prototype.


The measured co-polarized and cross-polarized radiation patterns compared to the simulations are shown in Figure \ref{ebg5}. The measurements were taken in an anechoic chamber at the Idvorski Laboratories Serbia, co-owned by the School of Electrical Engineering, University of Belgrade. The antenna was placed on an automatic rotating platform (Figure \ref{ebg8}(d) and (e)) and gain patterns were measured at 11 discrete frequency points from 2.4 to 2.5 GHz as shown in Figure \ref{ebg5}(c). Simulation and measurements are in good agreement. It can be observed that the antenna retains a reasonable directional radiation pattern in the \textit{+z }hemisphere, confirming the simulation predictions. In the \textit{x-z} plane radiation pattern, the measured HPBW agrees well with the simulated predictions, which is around 93${}^\circ$. In the \textit{y-z} plane radiation pattern, the measured HPBW of the 2$\times$1 EBG-backed monopole antenna is 56${}^\circ$ (slightly narrower than the simulated HPBW) and a shift in the maximum directivity is also observed, but predominantly this was due to fabrication and measurement setup imperfections. Nevertheless, the measured directional radiation patterns confirm the reliable radiation performance of the proposed wearable antenna.

\section{ Analysis of Antenna for Wearable Applications }

\subsection{ Effects of Structural Deformation}

\begin{figure}[htb]
	\centering
	\begin{subfigure}{0.7\textwidth}
		\includegraphics[width=1\textwidth]{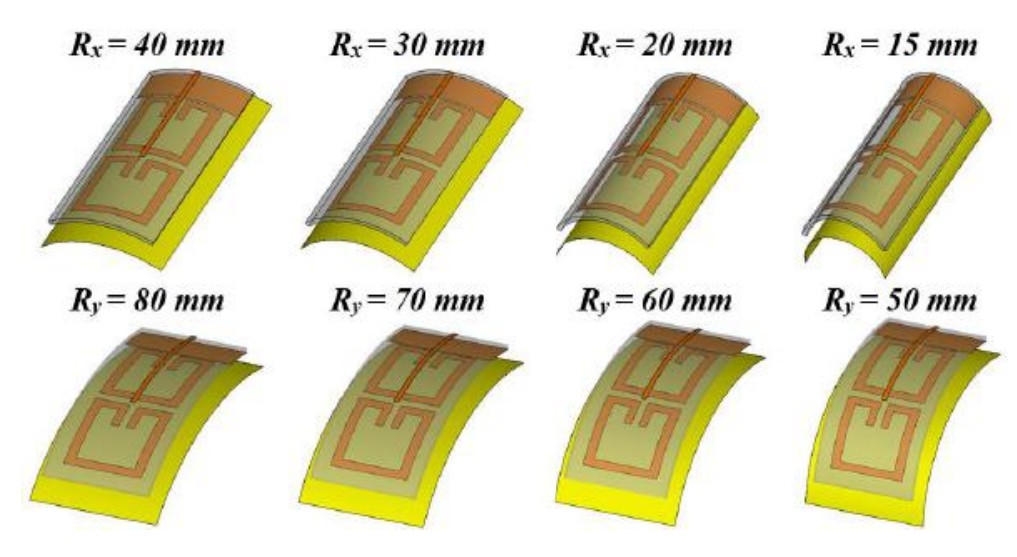}
		\caption{}
	\end{subfigure}
	\begin{subfigure}{0.2\textwidth}
		\includegraphics[width=1\textwidth]{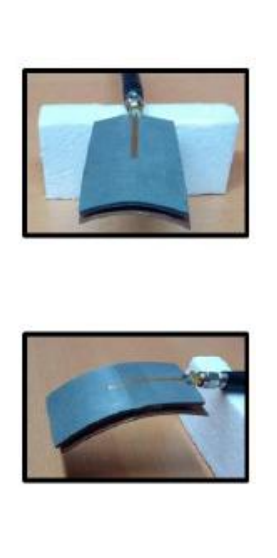}
		\caption{}
	\end{subfigure}
	\caption{(a) Structurally deformed assembled 2$\times$1 EBG-backed monopole antenna with different values of the curvature radius, ranging from \textit{R${}_{x}$} = 15 mm to 40 mm along the \textit{x}-axis and from \textit{R${}_{y}$} = 50 mm to 80 mm along the \textit{y}-axis (b) Photograph of deformed antenna. $\textcopyright$ 2016 IEEE.}
	\label{ebg9}
\end{figure}

In BAN applications, the wearable antennas are expected to be deformed or conformed during operation. Before investigating the performance in wearable scenarios, we first examined the antenna performance under structural deformation in free space to ensure its reliability \cite{antoniades2017conformai}. The parameters ``\textit{R$_x$}'' and ``\textit{R$_y$}'' are used to represent the bending radii of the antenna along the \textit{x}-axis and \textit{y}-axis respectively. As shown in Figure \ref{ebg9}(a), the assembled antenna with four different radii of curvature values, namely 15, 20, 30, and 40 (mm) along the \textit{x}-axis and 50, 60, 70 and 80 (mm) along the \textit{y}-axis have been studied. A viably deformed fabricated prototype is also measured to confirm the accuracy of the simulation prediction as shown in Figure \ref{ebg9}(b). 

\begin{figure}[H]
	\centering
	\begin{subfigure}{0.7\textwidth}
		\includegraphics[width=1\textwidth]{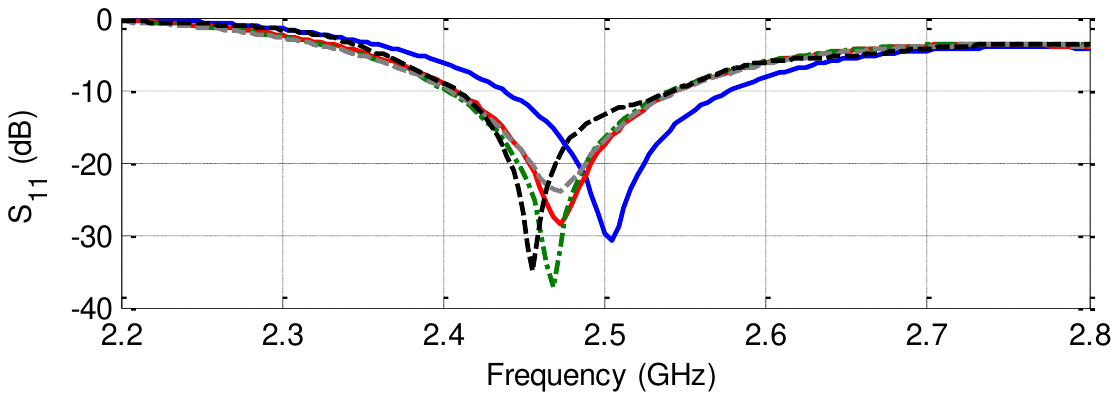}
		\caption{}
	\end{subfigure}
	\begin{subfigure}{0.8\textwidth}
		\includegraphics[width=1\textwidth]{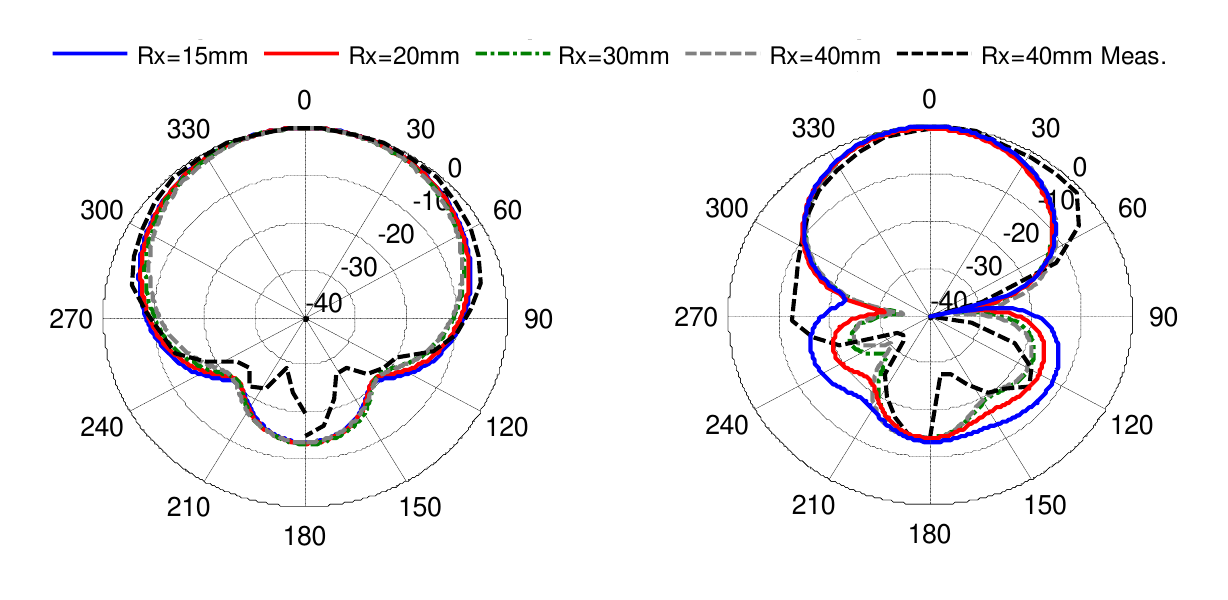}
		\caption{}
	\end{subfigure}
	\caption{(a) Simulated and measured reflection coefficient. (b) Radiation patterns of the antenna in the \textit{y-z} (left) plane and \textit{x-z plane} (right) for all deformation cases in the \textit{x}-axis.}
	\label{ebg10}
\end{figure}

The chosen curvature radii are reasonable representations for the radii of different sizes of human arms and legs where both vertical and horizontal antenna loading is possible. Figure \ref{ebg10}(a) and 11(a) show the simulated and measured reflection coefficients of the antennas with the four bending radii values. As it can be seen, the resonance frequency of the antenna is well maintained below -10 dB for all selected values of \textit{R${}_{x}$} and \textit{R${}_{y}$}. In the case of the \textit{x}-axis bending deformation, the frequency shift of less than 20 MHz can be observed when \textit{R${}_{x}$} is decreased from 40 to 20 mm, which is negligible. While looking at the near-extreme deformation where \textit{R${}_{x}$}=15 mm, a 38 MHz shift in the resonance frequency towards higher frequencies is observed. On the other hand, for the values of \textit{R${}_{y}$} corresponding to deformation along the \textit{y}-axis with a decrease in the bending radius, consequent degradation of the impedance matching is observed. Particularly when \textit{R${}_{y}$ }is decreased from 60 to 50 mm, the bandwidth decreases from 2.36--2.51 GHz when \textit{R${}_{y}$} = 60 mm to 2.36--2.47 GHz when \textit{R${}_{y}$} = 50 mm. This can be attributed to the fact that the radiated fields from the 2$\times$1 EBG-backed monopole antenna are provided mainly by the array of EBG structures and the ground plane. 


\begin{figure}[H]
	\centering
	\begin{subfigure}{0.7\textwidth}
		\includegraphics[width=1\textwidth]{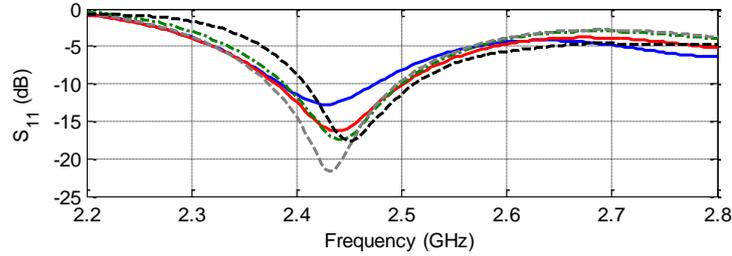}
		\caption{}
	\end{subfigure}
	\begin{subfigure}{0.8\textwidth}
		\includegraphics[width=1\textwidth]{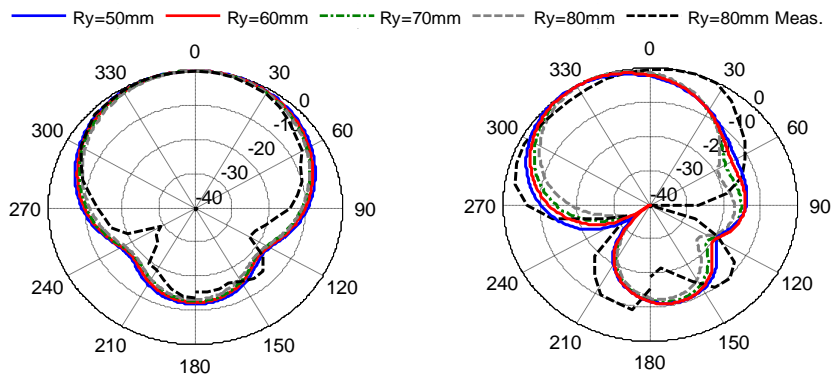}
		\caption{}
	\end{subfigure}
	\caption{(a) Simulated and measured reflection coefficient. (b) Radiation patterns of the antenna in the \textit{y-z} plane (left) and \textit{x-z plane} (right) for all deformation cases in the \textit{y}-axis.}
	\label{ebg11}
\end{figure}

The simulated and measured radiation pattern for all the cases demonstrated in Figure \ref{ebg9} are presented in Figure \ref{ebg10}(b) and (c) for \textit{x}-axis bending deformation and in Figure \ref{ebg11}(b) and (c) for \textit{y}-axis bending deformation. The gain comparison for the same cases is summarized in Table \ref{ebgtable3}, confirming the high gain response of the proposed antenna even when it is deformed significantly in either the \textit{x} or \textit{y}-axis. 

The antenna efficiency is calculated based on the measured gain and simulated directivity because the measurement facility did not allow 3D radiation pattern measurements for the calculation of the measured directivity.\textit{ }The measured radiation patterns for the tested antennas are presented in Figure \ref{ebg10} and \ref{ebg11} and were taken at 2.45 GHz. Because of the consistency of the simulated radiation patterns they were verified with a single measurement of the deformed antenna per axis. The measured results are in good agreement with the extensive simulation predictions and some discrepancies mostly evident in the back lobes of the antenna can be related to the measurements setup limitations. Overall, the proposed antenna performance has been shown to be robust to structural deformation along both the \textit{x}-axis and \textit{y}-axis and is compared favorably to several previously reported designs where degradation in impedance matching and/or significant band shifting were observed \cite{zhu2009dual,sundarsingh2014polygon,hu2016bending,4145071,velan2015dual}.

\begin{table}
	\centering
	\begin{tabular}{c|ccccc} \hline 
		\multicolumn{6}{c}{Bending along \textit{x}-axis} \\ \hline 
		\textit{R${}_{x}$} & 15 & 20 & 30 & 40 & 40 Meas. \\ 
		Gain (dBi) & 5.31 & 6.78 & 7.20 & 7.30 & 6.70 \\ 
		Radiation Efficiency & 58.3\% & 77.8\% & 80.5\% & 78.8\% & 67.48\% \\ \hline 
		\multicolumn{6}{c}{Bending along y-axis} \\ \hline 
		\textit{R${}_{y}$} & 50 & 60 & 70 & 80 & 80 Meas. \\ 
		Gain (dBi) & 5.45 & 6.13 & 6.63 & 6.86 & 6.40 \\ 
		Radiation Efficiency & 61.0\% & 65.3\% & 69.7\% & 71.5\% & 64.1\% \\ \hline 
	\end{tabular}
	\caption{Comparison of Gain and Radiation Efficiency for different values of bending radii.}
	\label{ebgtable3}
\end{table}

\subsection{ Human body loading}

\begin{figure}[htb]
	\centering
	\includegraphics[width=0.8\textwidth]{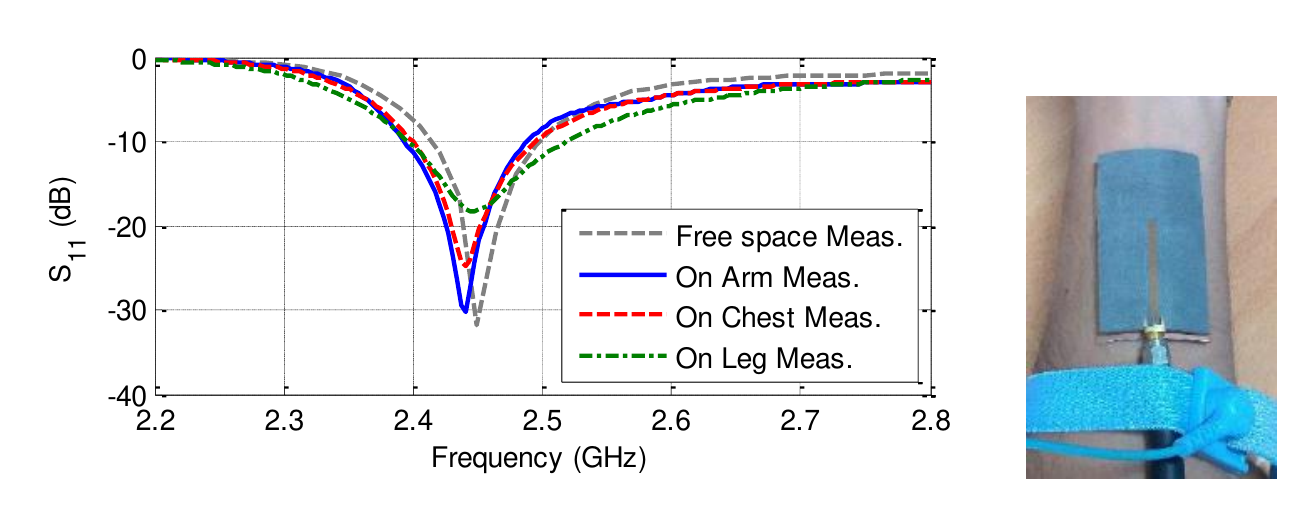}
	\caption{Measured reflection coefficient of the assembled 2$\times$1 EBG-backed monopole antenna placed on different parts of the human body, and a photograph of the antenna placed on the arm. $\textcopyright$ 2016 IEEE.}
	\label{ebg12}
\end{figure}

The EBG structure in the proposed design provides less backward radiation, hence the antenna should be tolerant to human body loading. To verify this, a series of experiments were performed where the fabricated antenna was directly placed on different parts of the body including the arm, leg and chest, while the antenna was connected to the VNA. The measured S${}_{11}$ plots presented in Figure \ref{ebg12} show a very stable reflection coefficient maintained for all the measured cases of human body loading.

\begin{figure}[H]
	\centering
	\begin{subfigure}{0.29\textwidth}
		\includegraphics[width=1\textwidth]{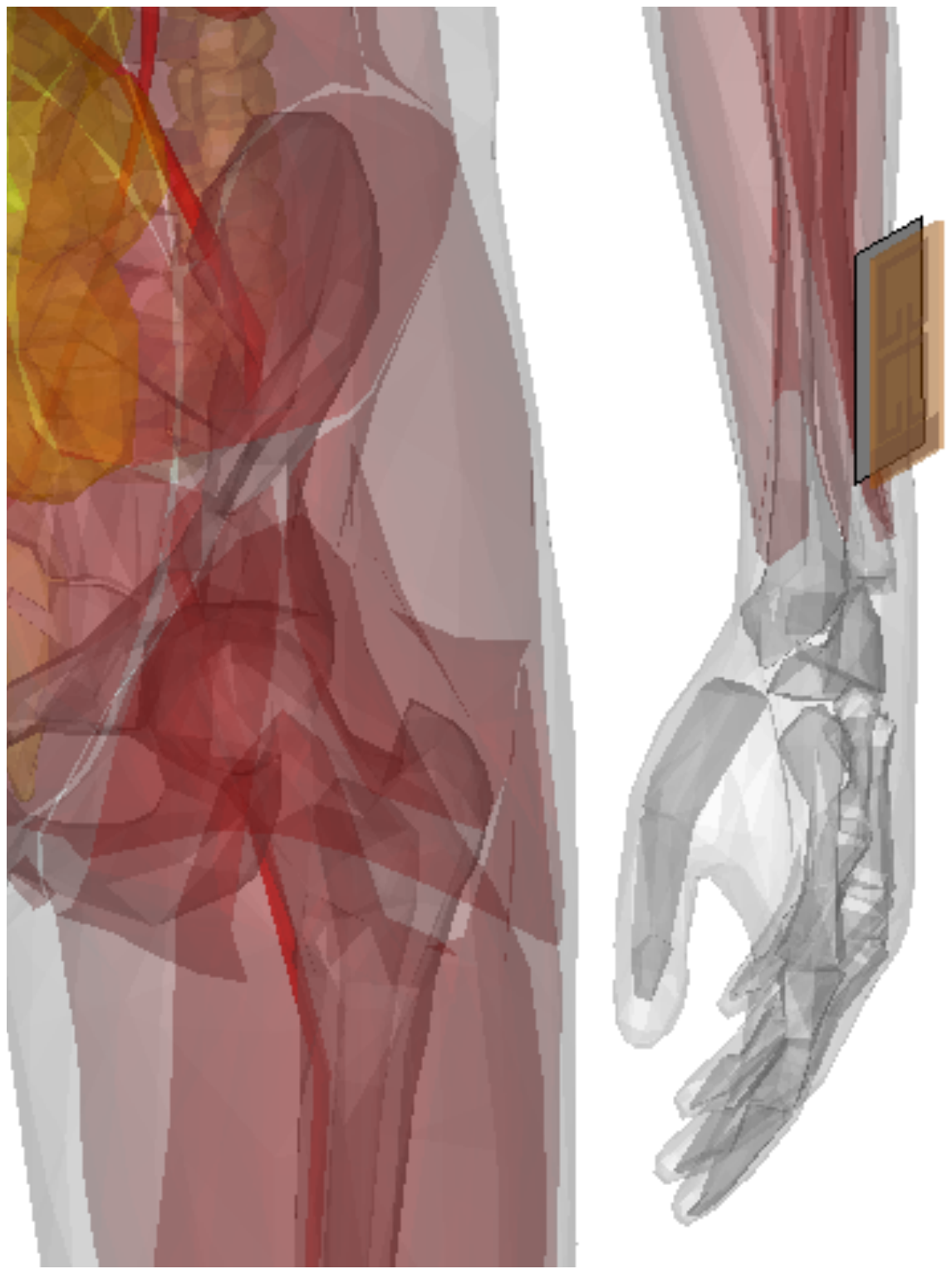}
		\caption{}
	\end{subfigure}
	\begin{subfigure}{0.41\textwidth}
		\includegraphics[width=1\textwidth]{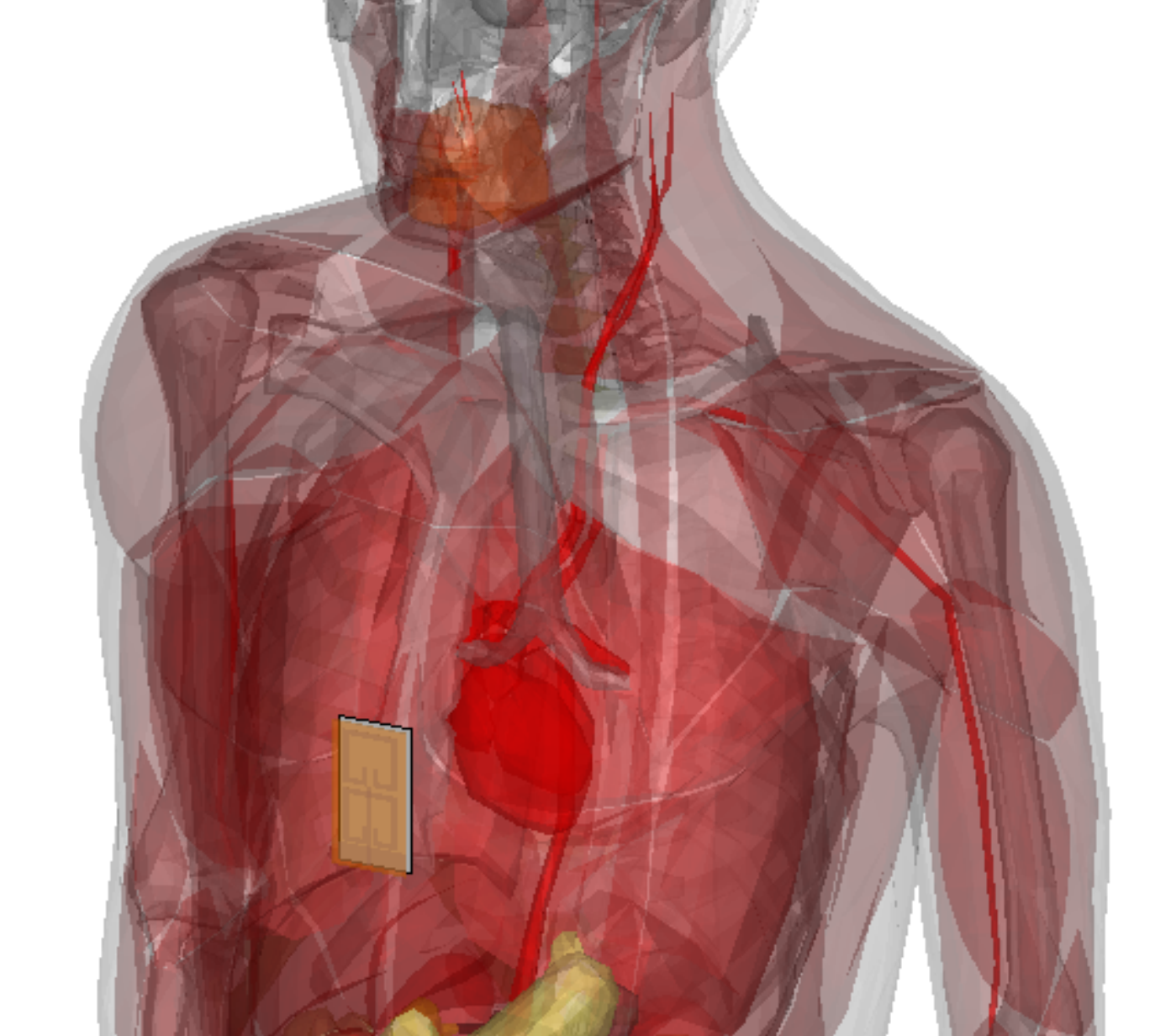}
		\caption{}
	\end{subfigure}
	\begin{subfigure}{0.27\textwidth}
		\includegraphics[width=1\textwidth]{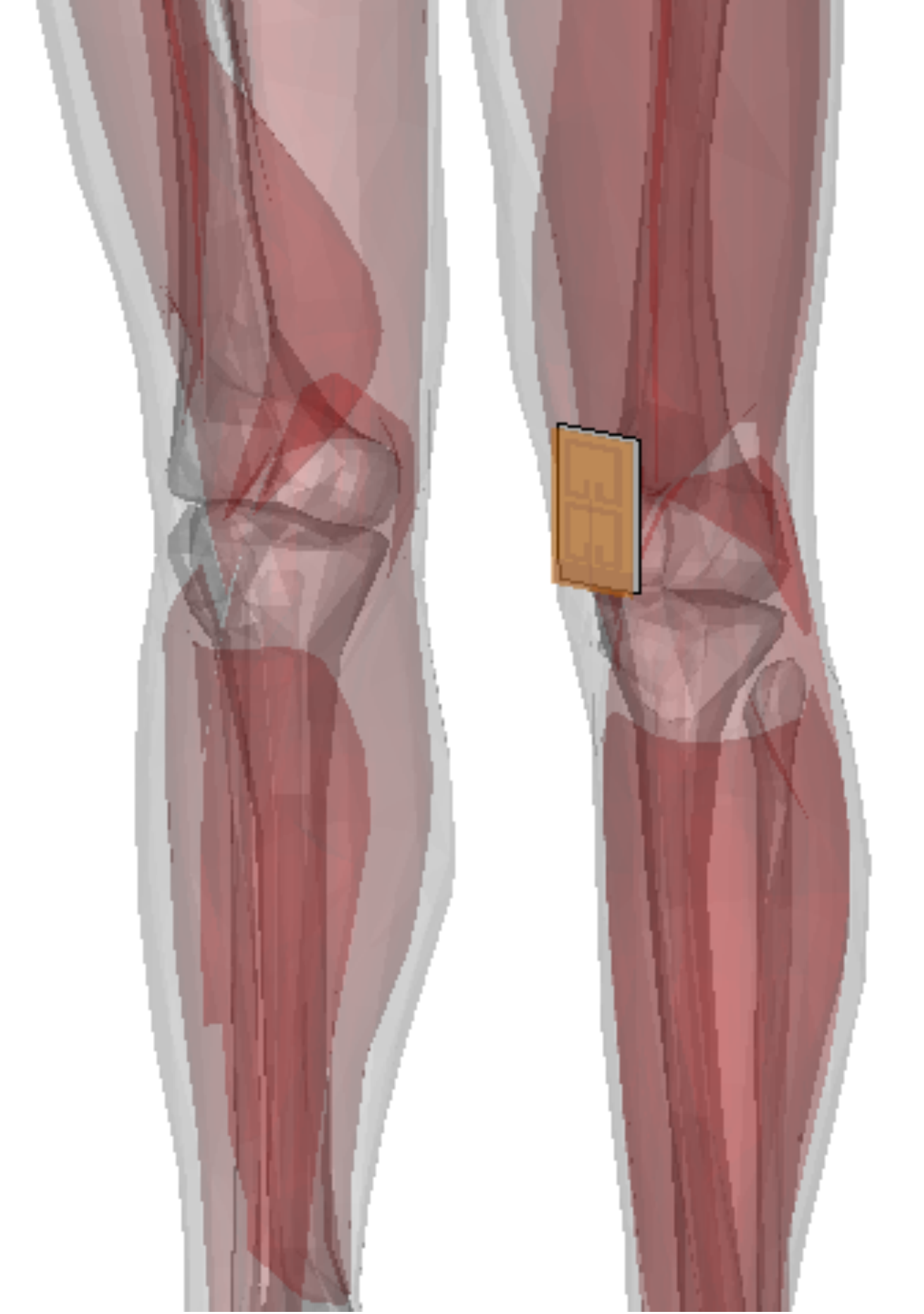}
		\caption{}
	\end{subfigure}
	\begin{subfigure}{0.8\textwidth}
		\includegraphics[width=1\textwidth]{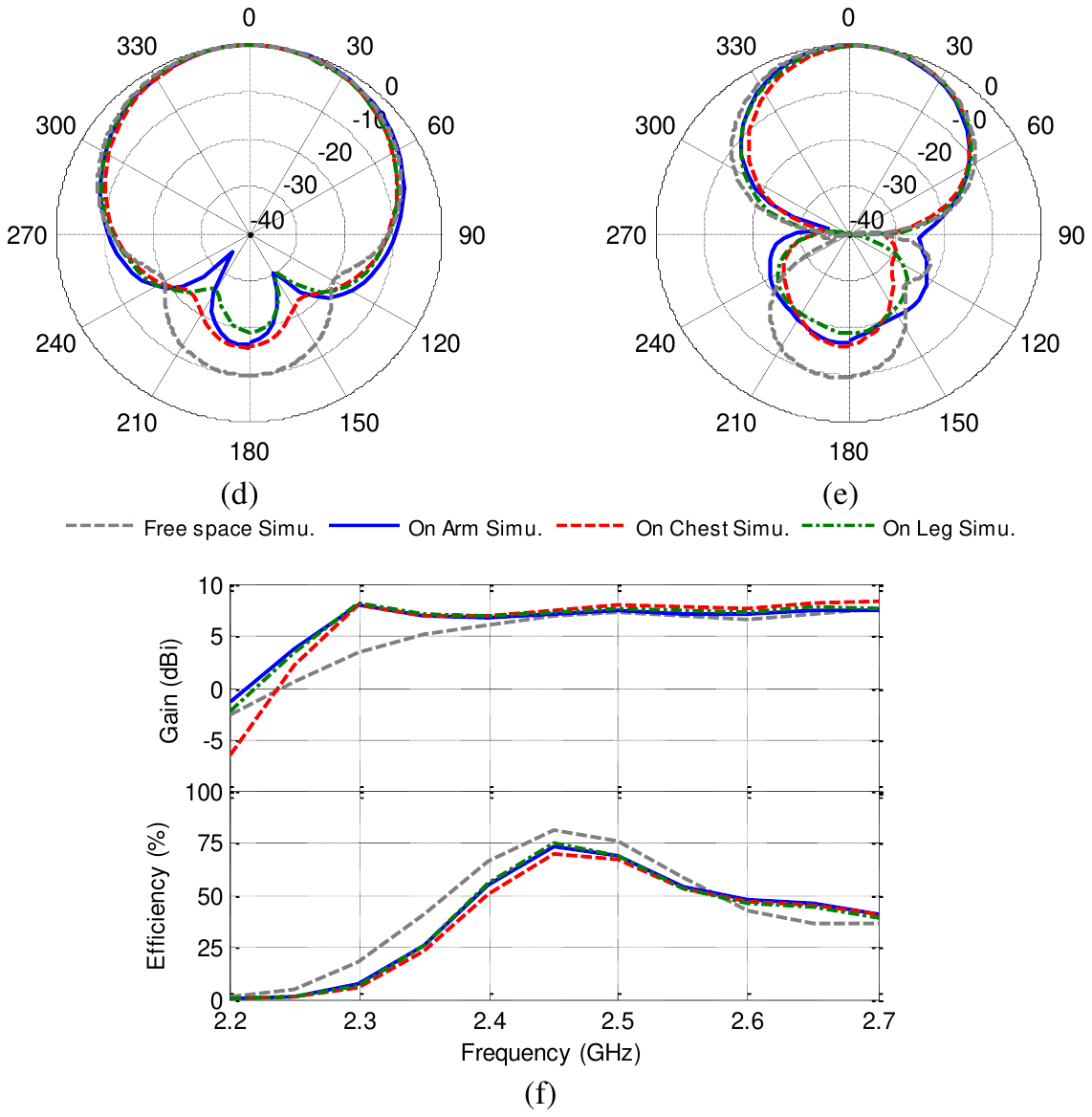}
		\caption{}
	\end{subfigure}
	\begin{subfigure}{0.75\textwidth}
		\includegraphics[width=1\textwidth]{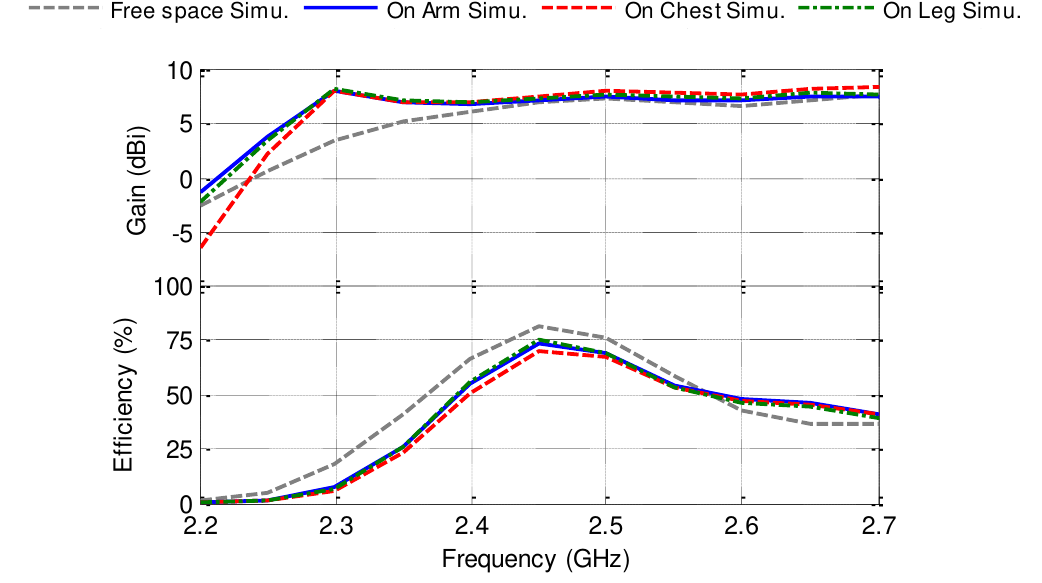}
		\caption{}
	\end{subfigure}
	\caption{2$\times$1 EBG-backed monopole antenna placed on the (a) arm, (b) chest and (c) leg of a realistic human body numerical phantom to evaluate the radiation characteristics presented in terms of (d) radiation patterns at 2.45 GHz along the \textit{y-z} axis (left), along the \textit{x-z} axis (right). (e) Antenna gain and efficiency. $\textcopyright$ 2016 IEEE.}
	\label{ebg13}
\end{figure}

When placed on the leg, the -10 dB bandwidth of the antenna broadens from 2.41-2.50 GHz to 2.39-2.53 GHz, where the impedance matching is still well maintained throughout the band. A full wave EM simulation with detailed realistic human body phantom \cite{ansoft1994ansoft, chow2013implantable} was used to evaluate the radiation performance of the antenna when operating in close proximity to the human body. To match the measurement cases presented in Figure \ref{ebg12}, the antenna was placed on the arm, chest and leg of a numerical phantom as presented in Figure \ref{ebg13}(a) to (c). After readjusting the on-body antenna orientation the simulated radiation patterns are compared to the measurements presented in Figure \ref{ebg5}(a) and (b). In addition, the simulated radiation patterns, gain and efficiency for all the cases is presented in Figure \ref{ebg13}(d) to (e) respectively. It is evident that the front-to-back ratio of antenna for all three cases increases because the body behaves as an extension of the ground plane and it directs the radiation away from the body. This increase in directivity impacted directly the peak gain of the antenna. The efficiency of the antenna remained above 70\% and the gain of antenna remained within the range of 7.07 to 7.43 dBi for all human body loading cases. 


\subsection{ Specific Absorption Rate (SAR) analysis}

\begin{figure}[htb]
	\centering
	\includegraphics[width=0.8\textwidth]{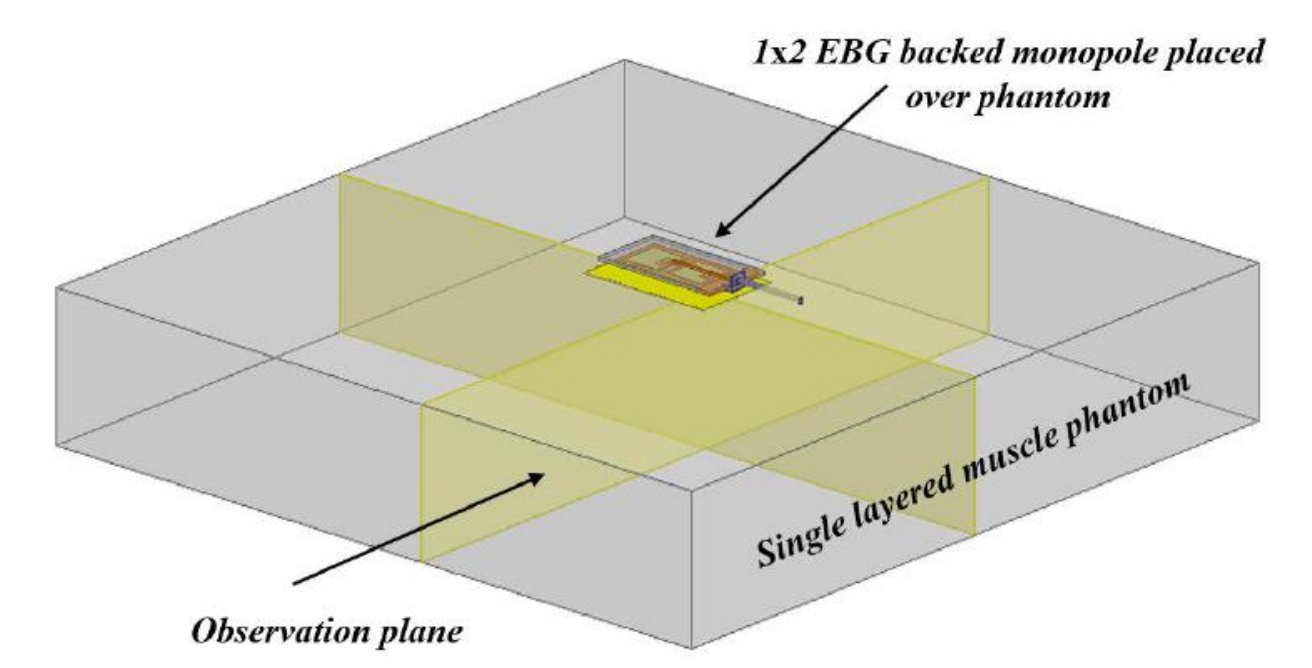}
	\caption{Single-layer phantom for SAR computation and 2$\times$1 EBG-backed monopole mounted on top of it.}
	\label{ebg14}
\end{figure}

Since the human body is always in the near-field of the wearable antenna, it is critical to pay adequate attention to the amount of radiation entering the human body. Near-field non-radiated power is stored around a radiating antenna and most of the energy produced in any lossy material residing close to an antenna is due to its reactive near fields. Similar conclusions have been drawn in a recent study \cite{velan2015dual}. Despite the fact that the proposed antenna operates at a comparatively narrow frequency band, having thus lower near field power density, the reactive near field is still a major contributor to the SAR. According to the FCC specifications, SAR values must be no greater than 1.6 W/kg averaged over 1 g of tissue. The SAR evaluation setup in this work consists of a simplified single-layer phantom mimicking the muscle tissue characteristics \cite{jiang2014compact}. The size of the phantom is 400$\times$400$\times$80 mm${}^{3}$,${}^{ }$large enough to ensure \textit{$\lambda$}/4 margin between antenna and phantom edges. Two SAR observation planes were introduced across the phantom box in \textit{x-z} and \textit{y-z} planes in the near field region for the SAR (see Figure \ref{ebg14}). The feeding structure of the antenna is a coaxial cable with an excitation port at its end, and the antenna is placed 2 mm above the phantom. For an input power of 0.5 W (rms), the SAR value is observed to be 0.244 W/kg averaged over 1 g of tissue, which falls well within the FCC specifications. 

\begin{figure}[htb]
	\centering
	\begin{subfigure}{0.7\textwidth}
		\includegraphics[width=1\textwidth]{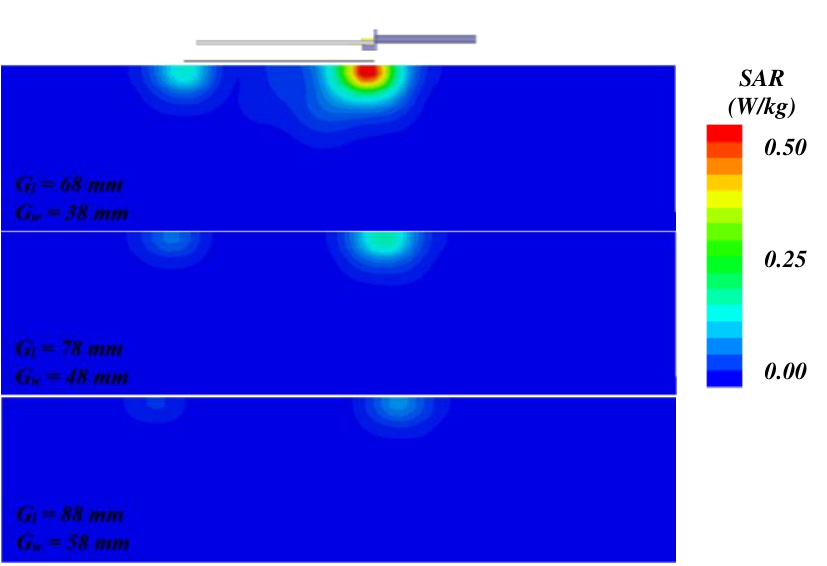}
		\caption{}
	\end{subfigure}
	\begin{subfigure}{0.7\textwidth}
		\includegraphics[width=1\textwidth]{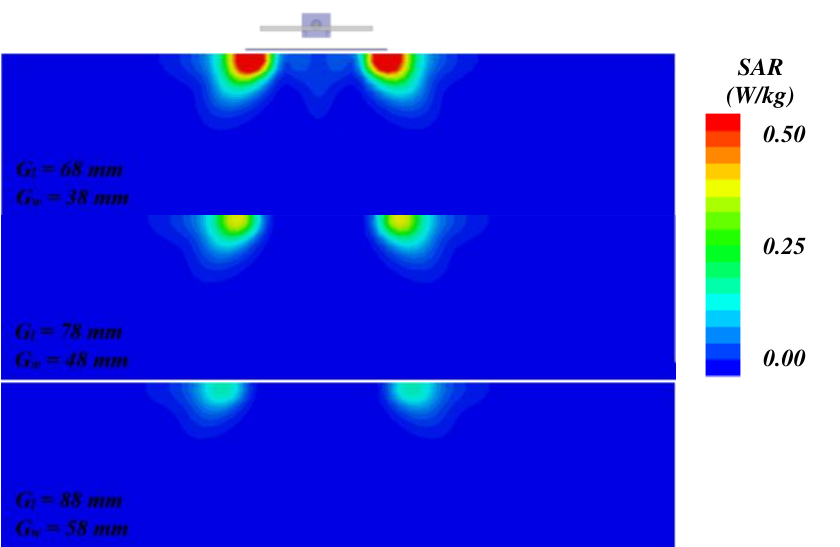}
		\caption{}
	\end{subfigure}
	\caption{SAR evaluation comparison for the 2$\times$1 EBG-backed monopole with extensions in the ground plane (a) along \textit{x-z} evaluation plane (b) along \textit{y-z} evaluation plane [all the plots are averaged to the maximum value of 0.5 W/kg]. $\textcopyright$ 2016 IEEE.}
	\label{ebg15}
\end{figure}

It was demonstrated in a study performed in \cite{trajkovikj2015diminishing} that the SAR of the antenna can be directly controlled by increasing the size of the ground plane. A similar evaluation of the proposed antenna was performed in which the ground dimensions\textit{ G${}_{l}$}$\times$\textit{G${}_{w}$} were varied to further decrease the SAR value. An identical environment in terms of operating frequency, phantom size, phantom structure, feeding, and position of the antenna was used and the SAR was estimated over the observation planes shown in Figure \ref{ebg15}. When the ground dimensions\textit{ G${}_{l}$}$\times$\textit{G${}_{w}$} were increased from 68$\times$38 mm${}^{2}$ to 78$\times$48 mm${}^{2}$ (144\%) and 88$\times$58 mm${}^{2 }$(197\%) respectively, while keeping all other parameters constant, a significant drop in SAR was observed, as can be witnessed from the 2D plots on the observation planes shown in Figure \ref{ebg15} (a) and (b). Note that all observation planes are uniformly color coded from 0 to 0.5 W/kg. It was discussed previously in sections I and II how the grounded EBG plane directs the radiation of the antenna towards the positive \textit{z}-axis, however some of the diffracted fields from the edges of the ground plane were still seen to radiate towards the body. This effect can be controlled by increasing further the dimensions of the ground plane, giving an additional control parameter to the proposed antenna over the SAR. In conclusion, whenever it is possible, and allowed by the application, the size of the ground plane backing the EBG should be increased to further decrease the maximum SAR.

\section{ Chapter Conclusions}

A compact semi-flexible 2$\times$1 EBG-backed planar monopole antenna for wearable applications has been proposed and tested experimentally. The proposed antenna exhibits very good size and gain characteristics compared to most recently reported configurations listed in Table \ref{ebgtable4}, which are intended for wearable applications. The EBG array in the proposed structure not only serves to isolate the antenna from the human body, but also contributes towards enhanced radiation efficiency. The fabricated antenna performs in good agreement with the simulated structure. It has a 4.8\%, $\mathrm{-}$10 dB, fractional bandwidth at the ISM band and a measured gain of 6.88 dBi, far greater than the gain of a conventional planar monopole antenna. The measured radiation pattern characteristics have been qualitatively explained using an eight-element array model consisting of six magnetic current sources and two electric current sources. Full-wave simulations and experimental measurements further revealed the robustness of the antenna to structural deformation and human body loading effects. The metallic sheet backing the EBG structure not only greatly reduces the SAR levels inside the body, but also provides direct control of its peak value. This, in addition to its low-weight, compact dimensions and ease in fabrication makes the proposed antenna very well suited for body-worn applications where the antenna can be integrated with systems for wearable biosensors or medical monitoring.

\begin{table}
	\centering
	\begin{tabular}{ccccc} \hline 
		Ref. & Size ($mm^{2}$) & Bandwidth & Gain (dBi) & Size comp. \\ \hline 
		\cite{velan2015dual} & 150$\times$150 & \begin{tabular}[c]{@{}l@{}}1.8 GHz (10.92\%)\\ 2.45 GHz (5.08\%)\end{tabular} & - & 870.7\% \\ 
		\cite{zhu2009dual} & 120$\times$120 & \begin{tabular}[c]{@{}l@{}}2.45 GHz (4\%)\\ 5.50 GHz (16\%)\end{tabular} & 6.4, 7.6 & 557.3\% \\ 
		\cite{kim2012monopole,kim2012wearable} & 127$\times$87 & 2.45 GHz (4.08\%) & 0.86 & 553.4\% \\ 
		\cite{4145071} & 110$\times$130 & 2.45 GHz (6.18\%) & - & 427.6\% \\ 
		\cite{yan2014low} & 100$\times$100 & \begin{tabular}[c]{@{}l@{}}2.45 GHz (12\%)\\ 5.50 GHz (16\%)\end{tabular} & 2.5, 4.0 & 386.9\% \\ 
		\cite{raad2013flexible} & 65.7$\times$65.7 & 2.45 GHz (18\%) & 4.8 & 167\% \\ 
		\cite{jiang2014compact} & 62$\times$42 & 2.40 GHz (4.63\%) & 6.2 & 100.7\% \\ 
		This work & 68$\times$38 & 2.45 GHz (4.88\%) & 6.88 & 100\% \\ \hline 
	\end{tabular}
	\caption{Comparison of Proposed Compact Antenna with EBG or AMC backed Planar Antennas.}
	\label{ebgtable4}
\end{table}


%

\chapter{Electrically Small Implantable Antenna}\label{chp3}

\ifpdf
\graphicspath{{Chapter3/Chapter3Figs/PNG/}{Chapter3/Chapter3Figs/PDF/}{Chapter3/Chapter3Figs/}}
\else
\graphicspath{{Chapter3/Chapter3Figs/EPS/}{Chapter3/Chapter3Figs/}}
\fi


This work discusses the design method towards the implementation of a compact, stacked, implantable antenna for biotelemetry applications. The proposed antenna consists of three stacked layers, printed on high permittivity grounded substrate. The bottom layer of the antenna above the ground plane, contains a meandered structure with a symmetrically placed T-shaped slot. The middle layer consists of two U-shaped radiators and an M-shaped metallic segment is further added on the top layer to further enhance the radiation efficiency of the antenna. These three stacked layers, form a symmetrical closed loop structure resonating on its fundamental resonant mode, at 403 MHz, for the Medical Implant Communication Service (MICS) band. A shifted higher order mode of the same closed loop structure, along with the negative currents on ground plane enables wide band operation, at 2.45 GHz, for the industrial, scientific, and medical (ISM) band. To resolve the constraints associated with implanting an antenna in human body that usually results in detuning and impedance mismatch, the antenna’s most important radiating sections were thoroughly investigated. A fully parametrized solution is proposed that makes the antenna a good candidate for a device, implanted at several different areas of the human body with potentially different electrical properties with consequently different detuning effect. For further investigation, the particle swarm algorithm was implemented to optimize the antenna’s performance while operating inside a compact, 23x23x5mm$^{3}$ block of human skin, equivalent phantom. The simulated performance of the proposed prototype antenna, indicates that it can be used for either \textit{in-vitro} or \textit{in-vivo} operations. 
Any wireless communication between implantable antenna and an external off-body antenna suffers significant link loss due to the lossy nature of living human tissues. This link degradation is normally unpredictable and varies for different parts of the human body. To study an implantable communication scenario, complete setup is generally simulated using a full wave electromagnetic simulator. Such simulation requires human body phantoms having electrical properties such as permittivity and bulk conductivity, or even mass density similar to that of real body tissues. It has been observed that geometry of such phantoms is usually chosen as simple as possible to decrease the required computational resources and the overall time for the completion of the simulation. It is difficult to predict how much error is expected when realistic phantom is replaced by such geometrically simpler phantom. This study focuses on a comparison of the usefulness of using realistic human body phantoms over geometrically simple body phantoms for through-body communication simulation. To study this comparison, an implantable antenna, placed inside a human stomach model, is set to communicate with an off-body antenna at Medical Implant Communication Service (MICS) band. Keeping the distance between implantable antenna and off-body antenna constant, the realistic human body phantom is then replaced by two types of layered body phantoms with different complexity. Results in the form of calculated s-parameters and phantom complexity metrics are further investigated to conclude the study. 

\section{Design Challenges}
Biotelemetry enables full duplex wireless communication between implantable antennas with on-body receiver antennas. Designing a wearable on-body antenna is comparatively easier because of its placement in an environment where body is normally in one direction and air in the other. This, allows the designers to control the radiation pattern of the antenna towards the human body by introducing a reflector or a cavity. On the other hand, designing an implantable antenna is a more challenging task due to a number of factors. First and far most challenge is the detuning factor and impedance mismatching that occurs when the antenna is placed inside human body. This is due to high water content and high conductive human tissues. Moreover detuning and mismatching are normally unpredictable and vary from tissue to tissue because different tissues have different electrical properties. The second major problem that arises while designing implantable antenna is its size. Since, MICS band at 403 MHz is assigned for implantable devices for on-body devices communication, the antenna effective aperture area at this frequency is rather large. However, the requirement is that the antenna integrated with the implantable device, along with power unit, should be as small as possible since the device is expected to be implanted inside the human body. A significant amount of work has already been done where different antenna design topologies have been proposed by researchers. To test the performance of an antenna in \textit{in-vivo} environment, simulations are generally carried out inside simple layered phantoms \cite{gemio2010human,permana2013hermetic} or in realistic body phantoms \cite{yilmaz2012patch,alrawashdeh2013flexible}. Electrical properties of these phantoms are either modeled using Cole-Cole model \cite{permana2013hermetic,karacolak2009electrical} or multi-pole Debye model \cite{noroozi2012three}. Some researchers \cite{karacolak2009electrical,karacolak2010vivo,yilmaz2012patch} also measured the electrical properties of human tissues (skin, fat, muscle) using dielectric probe kit or impedance analyzer and then used these properties in simulations to test the performance of their proposed solution. To achieve compactness, researchers have tried different methods and several types of antennas, including spiral \cite{permana2013hermetic,merli2011design,kim2006planar}, meandered patch \cite{alrawashdeh2013flexible,karacolak2009electrical,karacolak2010vivo,merli2011design,kim2006planar}, folded square IFA \cite{vidal2012detuning}, 2D and 3D spherical \cite{abadia20093d,merli2011design} and multilayered \cite{huang2011rectenna} configurations. All these methods reduce the antenna size by increasing the length of the current flow path on the same plane (meandered and spiral) or in all three dimensions (3D spiral, 3D ring and stacked). In addition, a shortening pin between ground and patch may also effectively reduce the required physical dimensions of an implantable antenna \cite{kiourti2012review} and has been widely used. It has been observed that antennas designed for a specific tissue are not expected to perform equally well in any substantially different tissue surrounding. In this work, we propose a solution which not only covers MICS and ISM bands for implantable applications, but additionally it can be customized to a random tissue environment. The behavior of the antenna is thoroughly studied and a range of geometric parameters are presented for best antenna performance. 

\begin{figure}[htbp]
	\centering
	\includegraphics[width=0.9\textwidth]{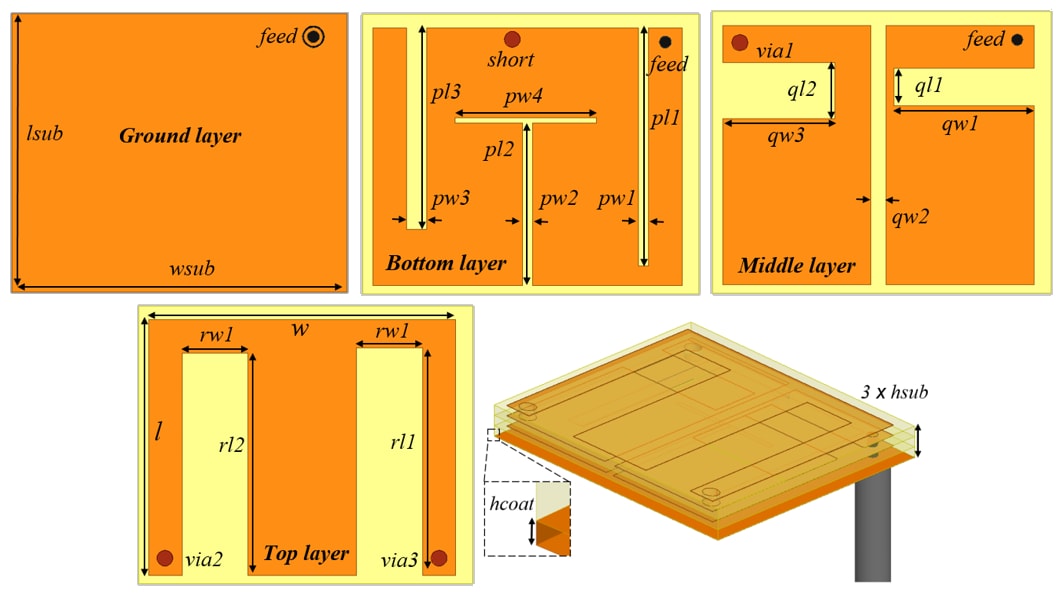}
	\caption{The geometry of proposed Implantable antenna. with \textit{lsub} = 15mm, \textit{hsub} = 0.635mm, \textit{wsub} = 18mm, \textit{hcoat} = 17$\mu$m, \textit{l} = 13.8mm and \textit{w} = 16.56mm. $\textcopyright$ 2015 PIERS}
	\label{Figure11}
\end{figure} 

\begin {table} [htb]
.\caption {Antenna optimization parameters in skin phantom (all values in mm)}
\centering
\begin{tabular}{ c|cccc }
	\hline
	\textbf{Parameter} & \textbf{min. value} & \textbf{max. value} & \textbf{seed. value} & \textbf{optimized} \\ \hline
	\textit{pl1}	& 08.60	& 13.60	& 10.60	& 12.75\\
	\textit{pl2}	& 01.00	& 13.00	& 01.00	& 08.70\\
	\textit{pl3}	& 08.60	& 13.60	& 10.60	& 10.80\\
	\textit{pw1}	& 00.50	& 02.00	& 01.00	& 00.54\\
	\textit{pw2}	& 00.50	& 00.70	& 01.00	& 00.50\\
	\textit{pw3}	& 00.50	& 02.00	& 01.00	& 01.17\\
	\textit{pw4}	& 04.00	& 08.00	& 06.00	& 07.56\\
	\textit{ql1}	& 03.00	& 09.00	& 06.00	& 03.00\\
	\textit{ql2}	& 03.00	& 09.00	& 06.00	& 06.00\\
	\textit{qw1}	& 05.00	& 07.50	& 06.00	& 07.46\\
	\textit{qw2}	& 00.20	& 00.90	& 00.30	& 00.81\\
	\textit{qw3}	& 05.00	& 07.50	& 06.00	& 05.90\\
	\textit{rl1}	& 11.00	& 13.00	& 12.00	& 12.30\\
	\textit{rl2}	& 11.00	& 13.00	& 12.00	& 12.00\\
	\textit{rw1}	& 03.00	& 06.00	& 04.50	& 03.50\\
	\textit{rw2}	& 03.00	& 06.00	& 04.50	& 03.50\\ \hline
	
\end{tabular}
\label{Table 1:}
\end{table}

\section{Multi-layered Implantable Antenna}
Figure \ref{Figure11} shows the geometry of the proposed layered implantable antenna. The overall dimensions of the antenna are 15x18x1.97mm$^{3}$, and the antenna consists of three copper coated substrate layers. In this study, Roger's RO3210 as substrate was used since its electrical properties resemble with those of biocompatible substrate ceramic alumina ($\epsilon_r$=9.4 and tan$\delta$=0.006). RO3210 was preferred because it can be easily available and is compatible with conventional fabrication techniques making in-house prototype fabrication possible. The proposed antenna consists of the ground layer and three stacked patch layers as shown in Figure \ref{Figure11} The bottom layer of the antenna, fed at the upper right corner, comprises a patch with identical slits having length and width \textit{pl1, pl3, pw1} and \textit{pw3} respectively. These slots along with the T shaped slot in the middle of the patch introduce a meandered structure used for size reduction. To further decrease the size of the antenna, a short pin is introduced on the bottom layer as shown in Figure \ref{Figure11}. The middle layer of the antenna comprises of two U-shaped coupling resonators with a capacitive coupling gap \cite{abbasi2014board}. The right U-shaped resonator is fed at the upper right corner with extended coaxial feed whereas the left U-shaped resonator is fed by via1 connecting the bottom layer and the middle layer at the upper left corner, exactly opposite to the feed point. To further increase the resonance capacity of the antenna, an M-shaped, third layer is introduced on the top layer of the substrate. Via2 and via3 connect the middle with the top layer at lower right and lower left corners respectively. The M-shaped top layer consists of a radiating patch with two identical slits, having dimensions \textit{rl1, rl2, rw1} and \textit{rw2} respectively. $\lambda/4$ closed loop structure, resonating at the fundamental resonant mode at 403 MHz covers the MICS band and a higher order harmonic of the same resonant closed loop structure, was adjusted to operate at 2.45 GHz ISM band.

\begin{figure} [htb]
\centering
\begin{subfigure}{0.6\textwidth}
	\includegraphics[width=1\linewidth]{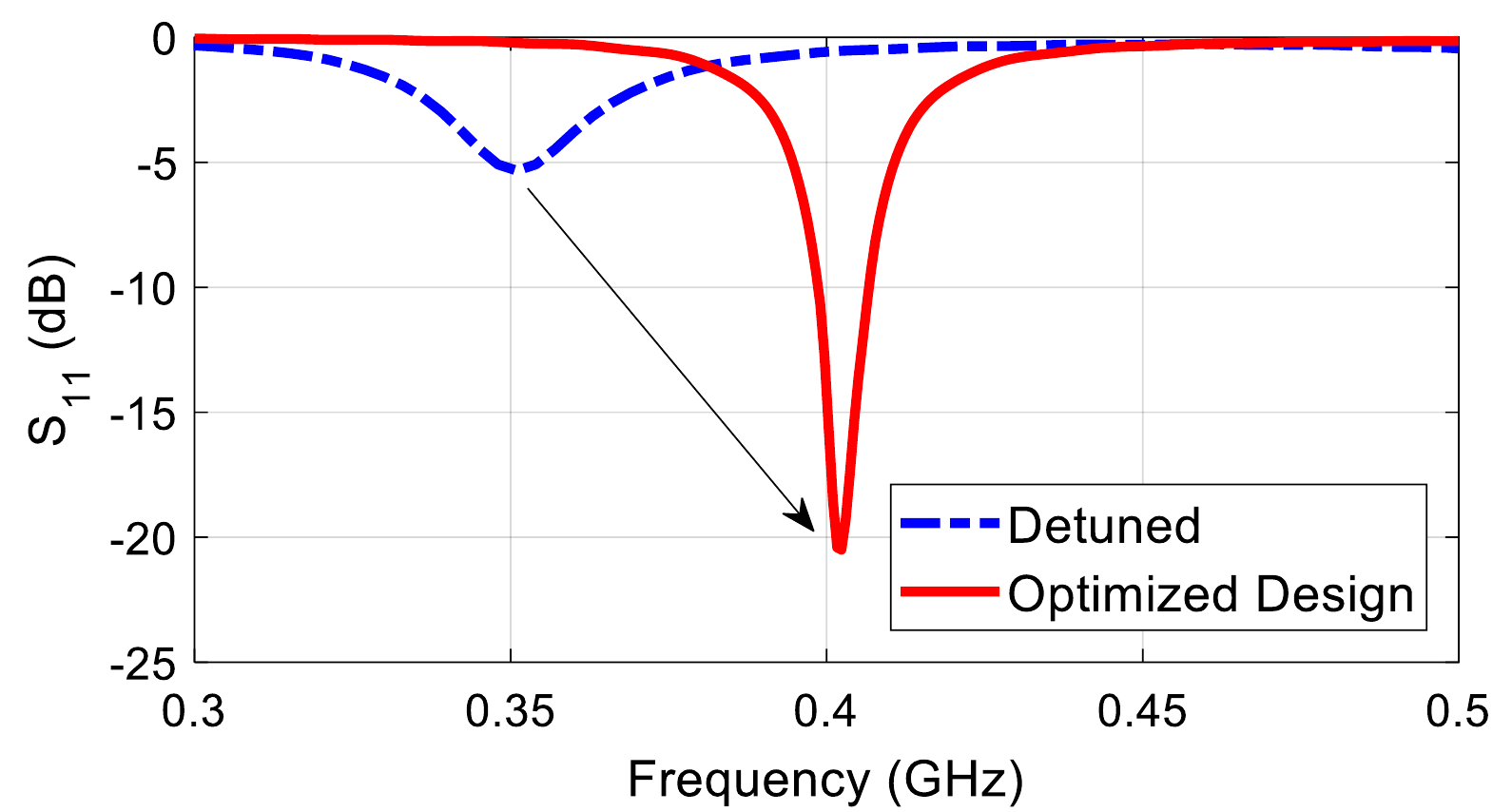} 
	\centering
	\caption{}
	\label{Figure21a}
\end{subfigure}
\begin{subfigure}{0.6\textwidth}
	\includegraphics[width=1\linewidth]{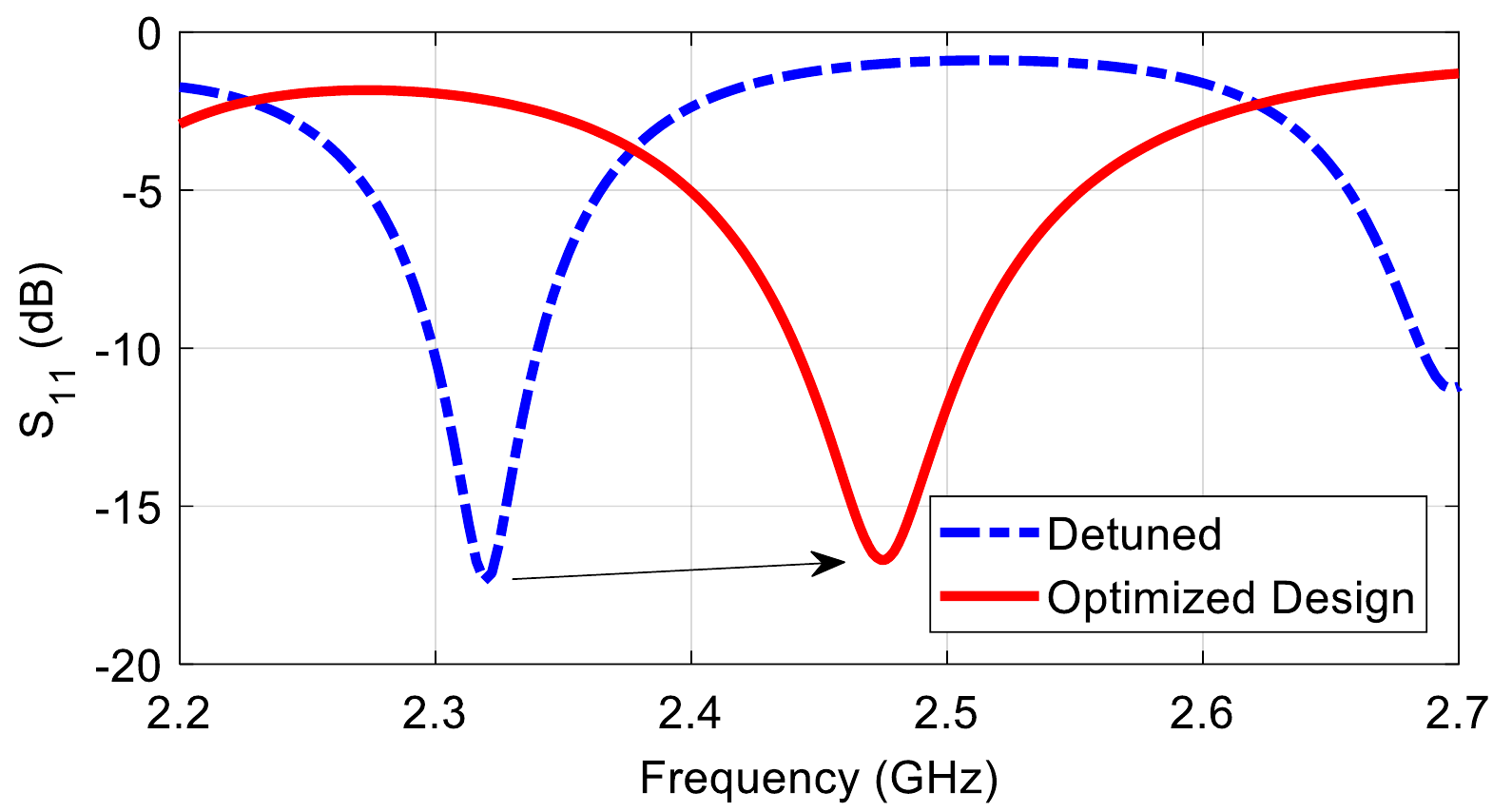} 
	\centering
	\caption{}
	\label{Figure21b}
\end{subfigure}
\caption{Detuning of antenna when placed in 23x23x5 mm$^3$ human skin phantom and optimization using particle swarm algorithm at (a) MICS band and (b) ISM band}
\label{Figure21}
\end{figure}

The effective current path in this antenna comprises of a meandered patch on the bottom layer, the left U-shaped patch of the middle layer, the M-shaped patch of the top layer and the right U-shaped patch of the middle layer. This current path makes an approximately $\lambda/4$ closed loop structure, resonating at the fundamental resonant mode at 403 MHz covering the MICS band. Furthermore, a higher order harmonic of the same resonant closed loop structure, was adjusted to operate at 2.45 GHz ISM band, resulting in a dual band operation using the same structure. To further enhance the radiation efficiency of the proposed antenna at ISM band, the effective length of the meandered patch on the bottom layer, that depends upon its coupling with the ground plane was further optimized. The next step was to study the behavior of the antenna in different tissue surroundings, which is equivalent to implanting the antenna at different body parts, and to identify various antenna tuning parameters simultaneously, for MICS and ISM bands. This study resulted in a fully parametrized antenna model with a minimum and maximum defined range of geometrical parameters that may adjust the effective wavelength of the $\lambda/4$ closed loop resonator, and the meandered patch of the bottom layer. One case is presented in this paper where the antenna is placed in 23x23x5 mm$^{3}$ block of human skin phantom. Note that measured electrical properties of skin tissue reported in \cite{karacolak2009electrical}, were used for this test. As predicted, significant detuning was observed both at MICS and ISM bands as shown in Figure \ref{Figure21a}. 

Results shown in Figure \ref{Figure21} (a), indicate that, when the antenna was placed in skin phantom, a resonant frequency shift of around 50 MHz towards lower frequencies, was observed for the MICS band. Significant degradation in impedance matching can also be observed in Figure \ref{Figure21} (a). The frequency downshift was raised up to 130 MHz at the ISM band as can be seen in Figure \ref{Figure21} (b). In order to tune the antenna dimensions on both targeted frequency bands simultaneously, particle swarm algorithm in combination with FDTD simulation was used. The resulting set of optimized values of the antenna parameters are listed in the fourth column of Table 1, where the third column shows the seed values for the particle swarm optimizer. The watch function to optimize the antenna in given tissue environment is set to be: 

\begin{equation}
f=max(|S_{11}(dB)_{403MHz}|, |S_{11}(dB)_{2.45GHz}|)
\label{eq1}
\end{equation}

It took 20 iterations for the particle swarm algorithm to converge and tune both minima at the desired frequency bands. Impedance bandwidth of 9 MHz (from 398 MHz to 407 MHz) covering the MICS band and 72 MHz (from 2438 MHz to 2510 MHz) covering the ISM band, were achieved as a result of optimization. Note that return loss minima can be further tuned at the desired frequencies at a cost of increased iterations of the optimization algorithm. Columns 2 and 3 of Table 1 list the range of controlling geometric parameters. The particle swarm algorithm requires both a rang as well as a seed value of a geometrical parameter optimization to be initiated. Values listed in Table 1 can be used for this purpose. The expected result should be an optimized set of values for which the customized antenna operates simultaneously at both MICS and ISM bands and this should be feasible for any given random tissue surrounding environment.


\section{Numerical Phantom Selection for Implant Study}

Implantable devices for biomedical applications play vital role in building comprehensive telemedicine wireless communication networks. Before an implantable device is proposed, it is usually thoroughly simulated at system level, using numerical methods in order to study the electromagnetic propagation loss when the DUT is placed inside human body. Computerized MRI scanner deduced, detailed body phantoms are finding increasingly important role in the attempt to test the performance of such implantable devices on system level. A number of 3D phantoms, deduced from MRI scans of living subjects \cite{seo2014electrical,motovilova2015mri}, depicting indirectly measured electrical properties of human tissues, are widely used, for case to case based, implantable device, design problems. These models generally have frequency-dependent electrical properties, and they are similar in shape to realistic human body parts and thus geometrically complex. Example of such complex phantoms for medical imaging and implantable applications are presented in \cite{veress2006normal}. Alternatively another simpler approach is widely used, where dense material blocks or homogeneous layers with constant electrical properties similar to those of predominant human tissues are built and used in EM simulations for implantable devices problems \cite{huang2015considerations,nawaz2015body}. These mainly homogeneous blocks, consist of simple geometric shapes which result in low mesh complexity, during modeling in 3D CAD tools, and are therefore far less resources hungry when simulated using full wave EM solvers. The main problem limiting the use of layered homogeneous models, is related to their accuracy and consequently their reliability. This study aims to assess the reliability of such simpler models in comparison with the more complex structures of realistic body-mimicking models. Towards this end, a theoretical scenario is discussed through extensive simulations, in which the communication between an implantable device, located inside a human stomach model, and an off-body device placed near the abdomen is investigated. A novel, compact, multilayered, implantable antenna, operating at 403 MHz, is used to establish a through-body communication scenario with a near-body transceiver’s antenna. Keeping in mind the detuning effect of implantable and near-body antennas, both antennas are matched properly with respect to the surrounding human tissue models. Three different phantom models are compared in the proceeding section and the simulated results are further investigated.

\begin{figure}
\centering
\includegraphics[width=0.7\textwidth]{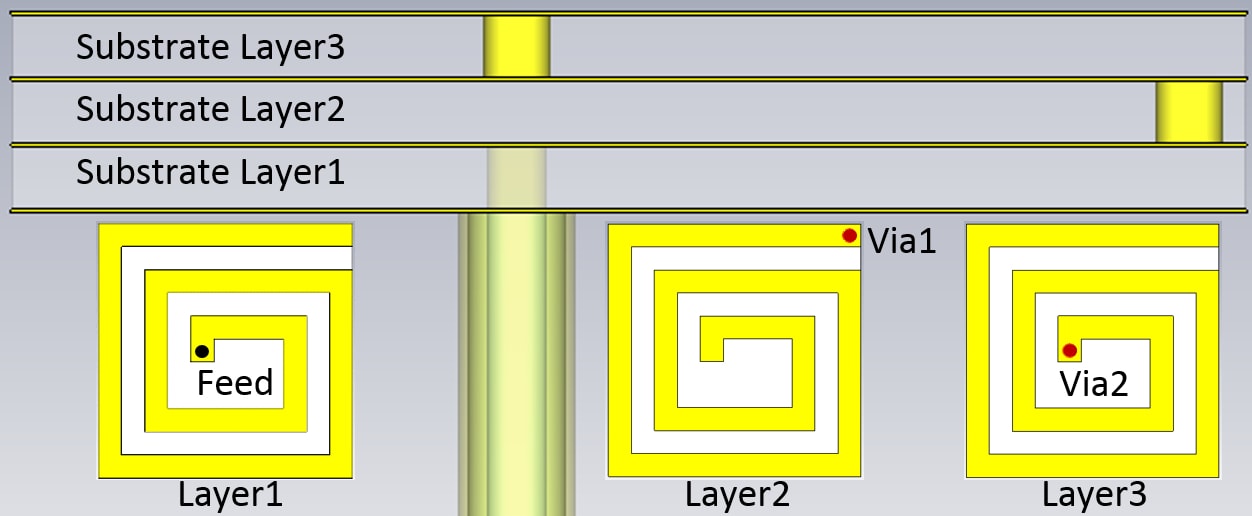}
\caption{Geometry of implantable antenna operating at MICS band. $\textcopyright$ 2015 PIERS.}%
\label{f1}
\end{figure}

\begin{table}
\centering
\begin{tabular}{ p{4cm}|cc}
	\hline
	\textbf{Tissue} & \textbf{$\epsilon_r$} & \textbf{$\sigma (S/m)$} \\ \hline
	1-Stomach						&	67.51		&	1.00\\
	2-Abdominal Muscle	&	57.15		&	0.79\\
	3-Body Average			&	35.00		&	0.60\\ 
	4-Gallbladder				&	61.25		&	1.13\\
	5-Large Intestine		&	62.63		&	0.95\\
	6-Liver							&	51.27		&	0.65\\
	7-Pancreas					&	61.56		&	0.87\\
	8-Small Intestine		&	66.19		&	1.90\\
	9-Spleen						&	63.27		&	1.02\\\hline
\end{tabular}
\caption{Electrical properties of human organs at MICS band.}
\label{tprop}
\end{table}

\begin{figure}[htb]
\centering
\begin{subfigure}{0.4\textwidth}
	\includegraphics[width=1\textwidth]{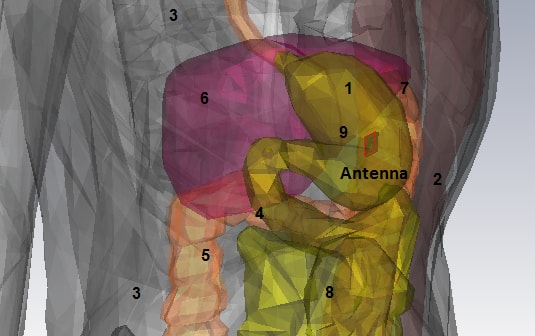}
	\caption{Case1}
\end{subfigure}
\begin{subfigure}{0.234\textwidth}
	\includegraphics[width=1\textwidth]{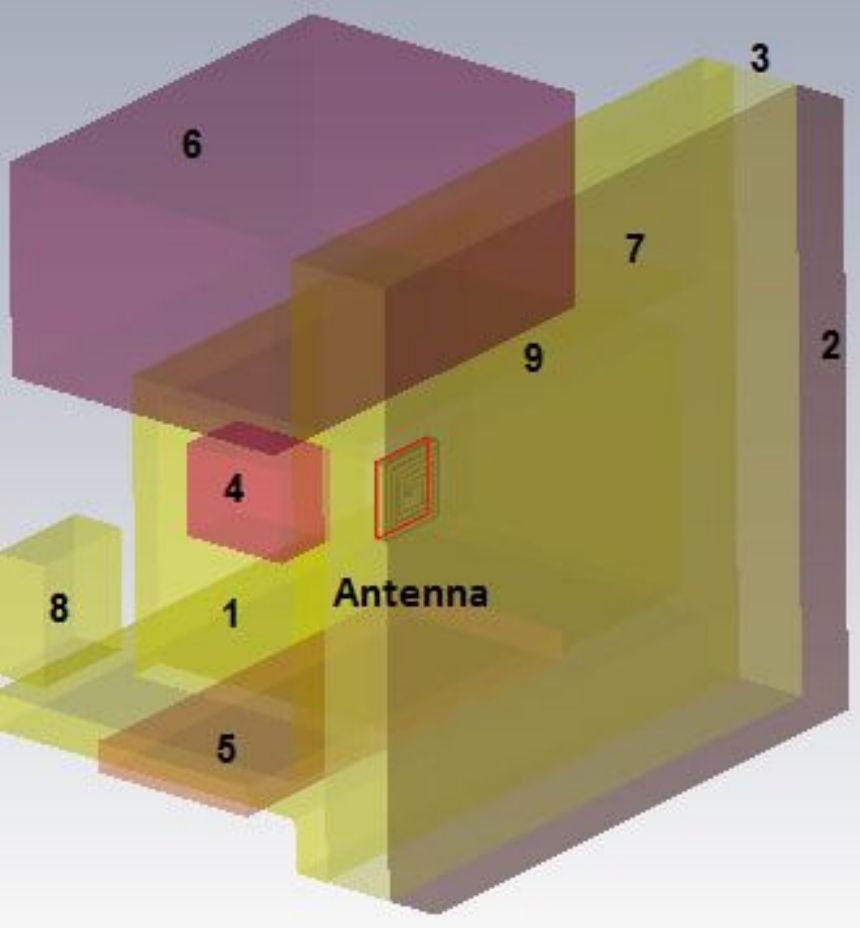}
	\caption{Case2}
\end{subfigure}
\begin{subfigure}{0.195\textwidth}
	\includegraphics[width=1\textwidth]{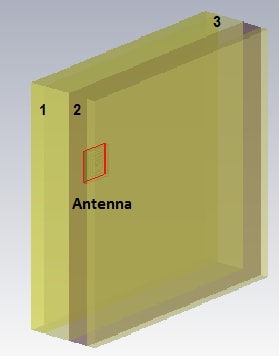}
	\caption{case3}
\end{subfigure}
\caption{Implanting antenna in (a) realistic human phantom i-e Case1 (b) complex layered phantom i-e Case2 and (c) simple layered phantom i-e Case3. $\textcopyright$ 2015 PIERS}
\label{f2}
\end{figure}

\subsection{Simulation Setup}
The proposed implantable setup in order to investigate through body communication consists of an implantable antenna shown in Figure \ref{f1}, placed inside the stomach of a realistic body phantom. The used antenna consists of three substrate layers of Rogers RO3210 (h=0.635mm, $\epsilon_r$=9.4 and tan$\delta$=0.006) coated with copper having spiral and meandered configuration. Three spiral layers having 1mm thick copper strip, connected to each other through via1 and via2, realize a radiating resonance at 403 MHz, MICS band. These layers are stacked to form a 3D meandered patch, in an attempt to achieve the desired compactness for an implantable antenna. The antenna used in this study is designed and matched in accordance to its operation inside stomach of a realistic human body phantom. On the other hand, a wideband antenna having directional radiation pattern with realizable gain ranging from 4 dB to 4.5 dB at MICS band is placed near the abdominal region of the phantom, to operate as an off-body transceiver antenna. Keeping the location and operation of implantable and off-body antennas constant, three types of phantoms, representing human tissues have been studied as described in Figure \ref{f2}. The properties of human tissues considered in this study are listed in Table 1 and were presented in \cite{sani2009numerical,kiourti2012review}. Note that the average human tissue properties at MICS band \cite{kiourti2012review} ($\epsilon_r$=35, $\sigma$=0.60 $S/m$) have been assigned to all other organs of the body phantom that are not listed in Table \ref{tprop}. As shown in Figure \ref{f2}, Case1 represents a realistic human phantom model that can be considered as a reference phantom. Case2 represents a geometrically simpler phantom model where dense material blocks are placed at the same location where real human tissues are located. This layered phantom represents a comparatively simpler model than that of the realistic body phantom presented in Case1. The simplest phantom model is presented in Case3 in which only those tissues which are in the LOS path of the implantable to off-body antenna communication link, are modeled for the simulation setup. These tissues are represented by dense homogeneous material layers with the implantable antenna placed inside the equivalent stomach layer. 

\subsection{Results and Discussion}
\begin{figure}[htbp]
\centering
\includegraphics[width=0.8\textwidth]{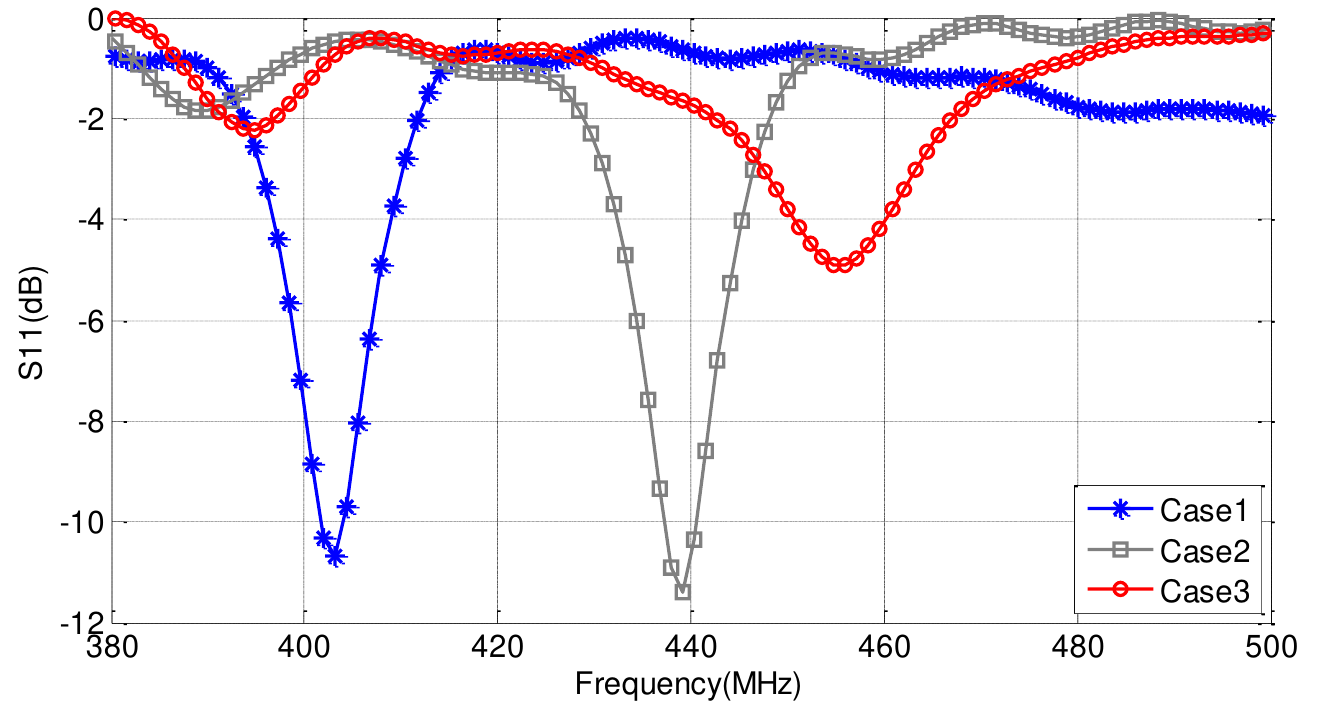} 
\caption{Return loss comparison of implantable antenna for all three cases.}
\label{f3}
\end{figure}

\begin{figure}[htbp]
\centering
\includegraphics[width=1\textwidth]{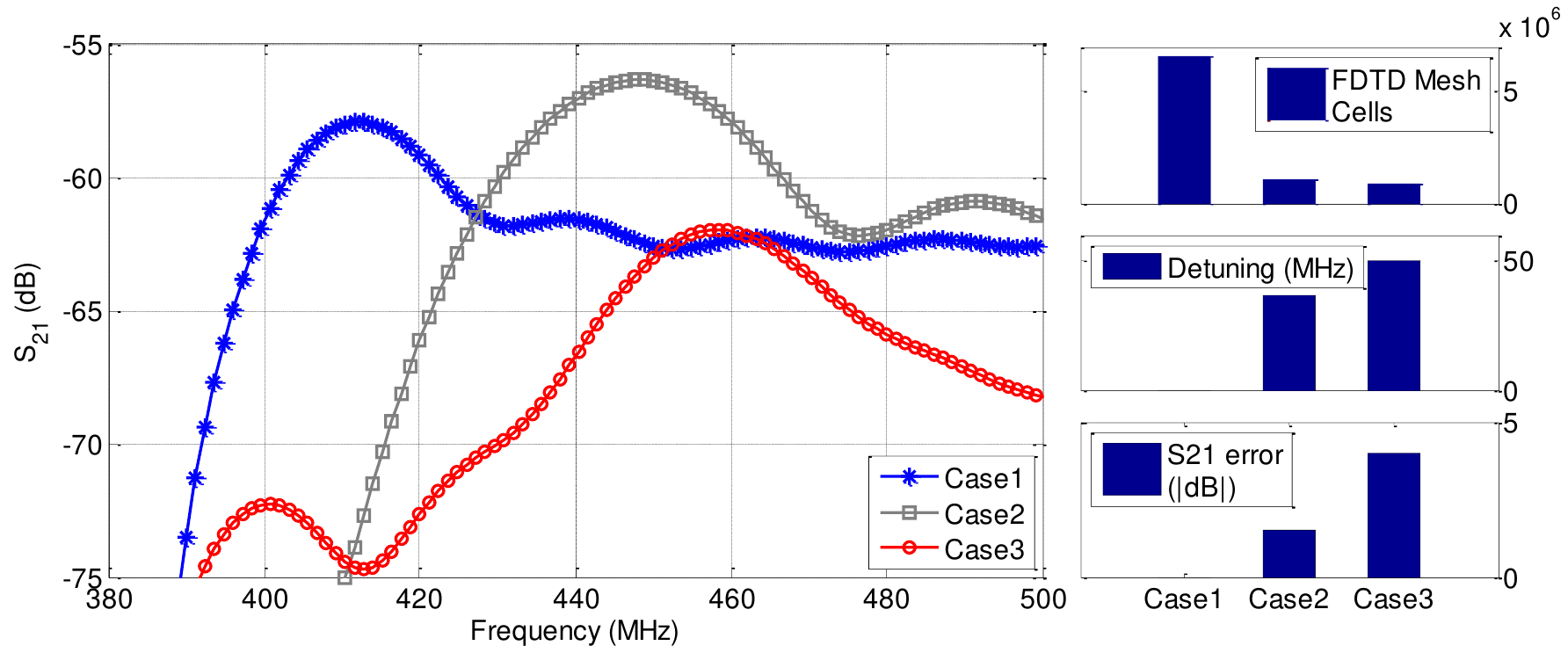} 
\caption{S$_{21}$ comparison for implantable antenna (left), comparison of FDTD mesh cell count (Top right), comparison of detuning in MHz (middle right) and comparison of error observed in S-parameter for all three cases.}
\label{f4}
\end{figure}

To study the performance of each phantom, Finite Difference Time-Domain (FDTD) solver with Finite Integration Perfect Boundary Approximation (FPBA) hexahedral mesh was used. Considering only the organs listed in Table 1, full wave simulation was performed for all three cases. The number of pulses to define steady state convergence criteria was 20, and it was carefully selected to ensure accuracy up to 1$\%$. Simulation parameters including adaptive mesh refinement were kept constant for all three cases. It was observed that a total number of FDTD mesh cells was 6.573952x10$^6$ for Case1 as compared to that of 1.095438x10$^6$ and 0.898688x10$^6$ mesh cells generated in Case2 and Case3 respectively. In Case1, the antenna is placed inside stomach where it is surrounded by not only stomach tissue but also by other high water content and high conductivity tissues (like gallbladder, small intestine etc). Intentionally the antenna is matched for this implant as evident by the simulated return loss shown in Figure \ref{f3}. In Case2, realistic body organs are replaced by dense material blocks representing comparatively simpler models of human organs. Mesh cell count to simulate Case2 setup decreases but with the cost of significant detuning of resonant frequency from 403 MHz to 440 MHz. This shift is much greater than the total bandwidth assigned to MICS band i-e 3MHz (402 MHz-405 MHz \cite{kiourti2012review}) resulting in operation of the implantable antenna outside the designated, licensed, MICS band. Moreover, when this phantom is further replaced by the second even simpler layered phantom, widely used to test the performance of implantable antennas \cite{huang2015considerations}, the resonance frequency further shifted towards higher frequencies. S$_{11}$ for Case3 shown in Figure \ref{f4} depicts a detuning of 50 MHz, and a significant impedance mismatch. Simulation results indicate that not only the electrical properties, but also the location and even the very geometry of high conductive tissues play a significant role in through-body communication. Disregarding these tissues weill introduce a significant error in s-parameter calculation for any similar implantable simulation setup.

\section{Practical Realization}

\begin{figure}[htb]
\centering
\begin{subfigure}{0.35\textwidth}
	\includegraphics[height=1\textwidth]{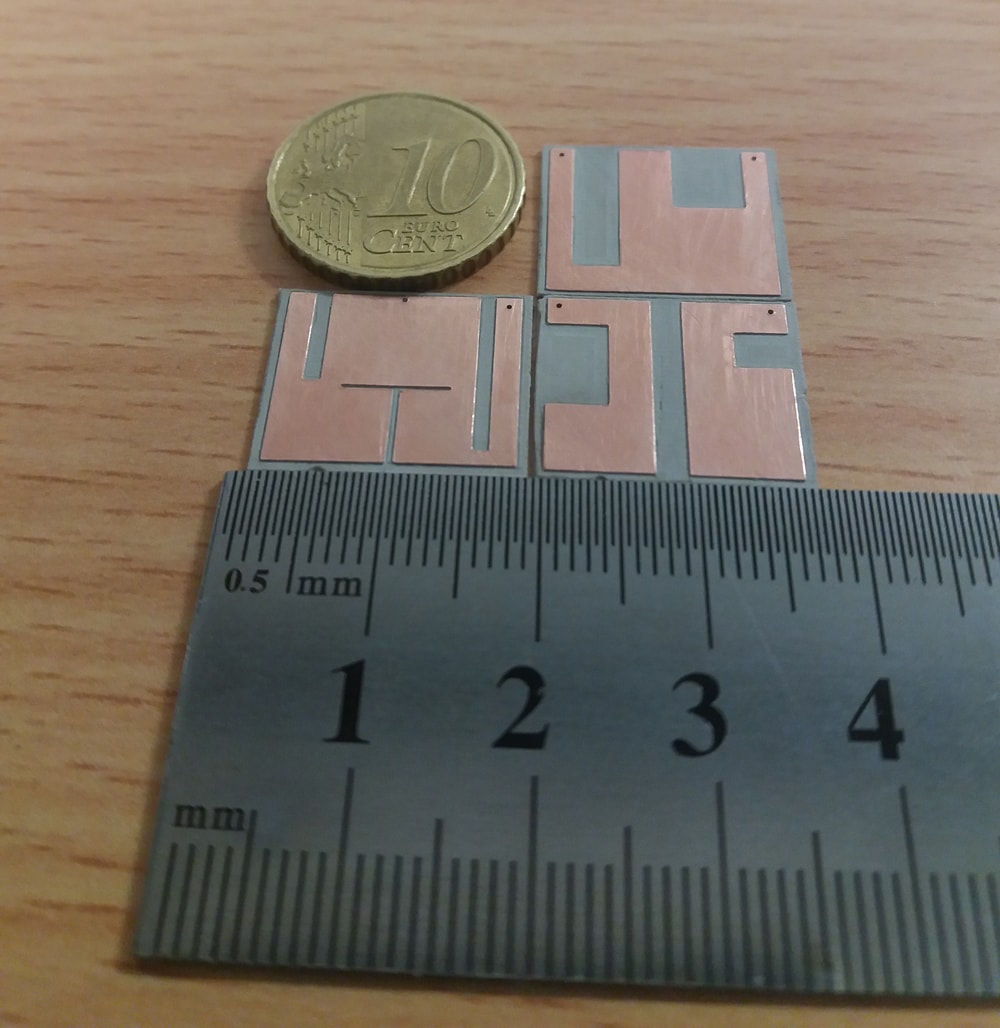}
	\caption{}
\end{subfigure}
\begin{subfigure}{0.35\textwidth}
	\includegraphics[height=1\textwidth]{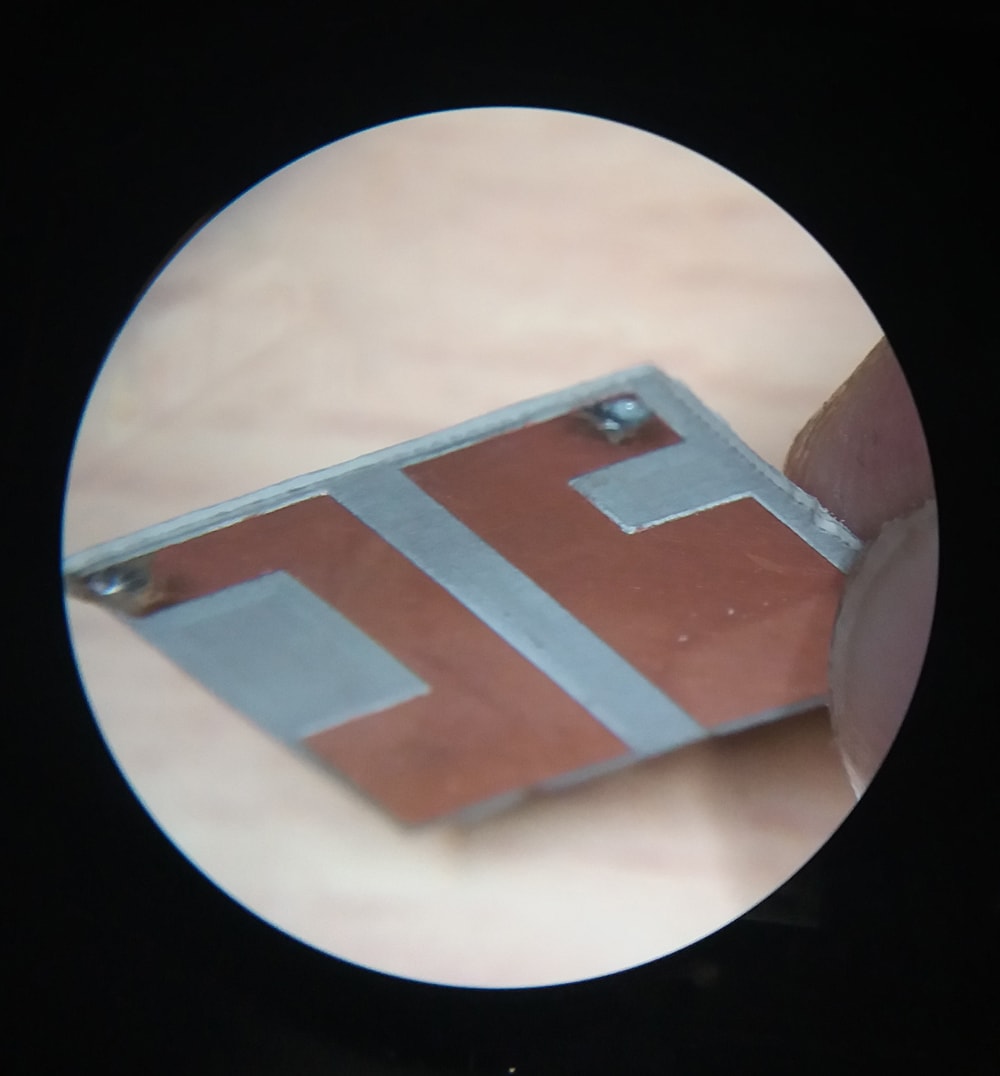}
	\caption{}
\end{subfigure}
\begin{subfigure}{0.6\textwidth}
	\includegraphics[width=1\textwidth]{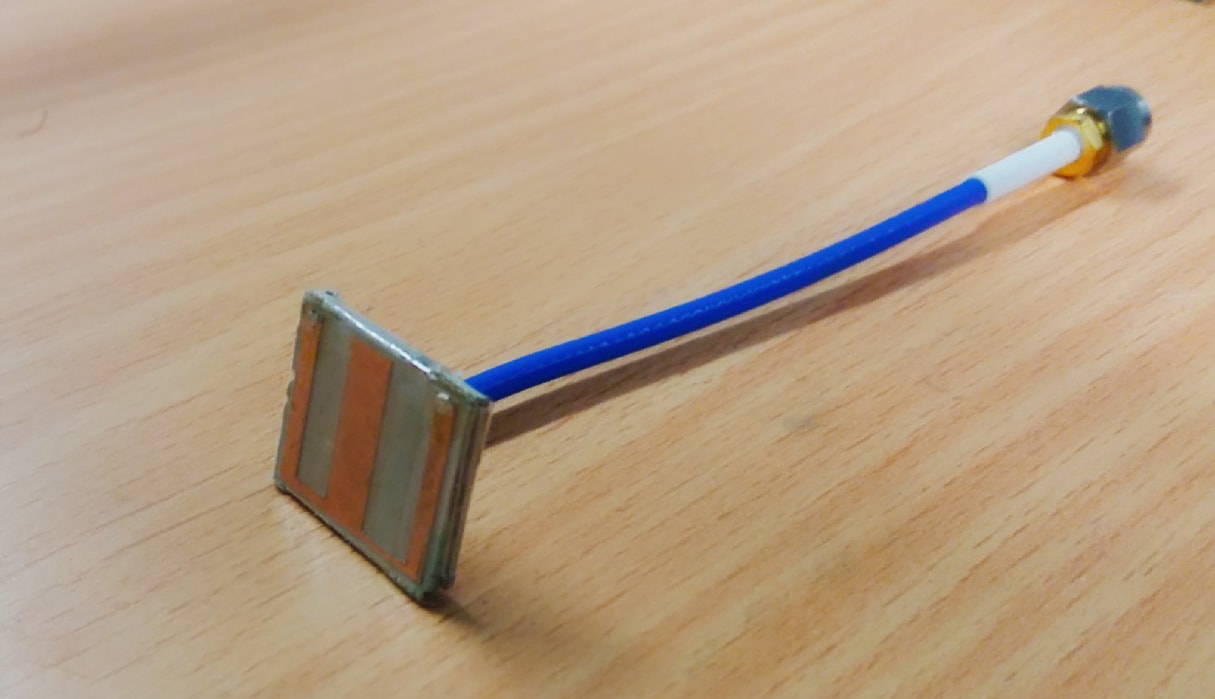}
	\caption{}
\end{subfigure}
\caption{Fabricated prototype with implantable antenna (a) fabricated layers after milling and cutting by milling machine (b) microscopic photograph of middle layer showing soldered vias (c) photograph of final implantable antenna prototype. $\textcopyright$ 2018 IEEE.}
\label{fabrication}
\end{figure}

As discussed in the previous sections, the implantable antenna need to be designed for a specific tissue environment and the variables mentioned in Table \ref{Table 1:} should be re-adjusted. The material properties of mined meat resemble that of human muscle tissue so the antenna was readjusted to ensure its resonance tested in full wave EM simulator while placed inside a 1.548$\times$1.548$\times$ 4.445 mm$^2$ minced pork meat. The antenna resonance and the radiation properties were carefully optimized. Using LPKF protomate h100 milling machine, the antenna traces were fabricated on a 0.635 mm sheet of Roger's TO 3210 substrate. The substrate is very soft and need extra care to handle while placed inside the milling machine. The process consisted of two major steps. First, the substrate sheet was placed in vacuum suction panel of the machine and drill holes were created for the via wires to pass through the substrate. In the next step, the excessive copper was removed from the substrate in such a way that the milling routing need not to go too deep, but scratch the copper evenly throughout the board. The removal of the substraet sheet from the vacuum suction panel was done without bending the antenna sheets. The antenna layers are presented in Figure \ref{fabrication}(a) along with a size comparison. When the layers were ready, first, the via wires of 0.35 mm
radius were passed from the drilled holes. The wires were sold at the top and bottom of each layer's copper coating, while the bulges of the solder were squeezed manually to ensure they stay at the same level with the layer. Figure \ref{fabrication}(b) shows the microscopic picture taken in a prospective view to show the accuracy of the soldering and leveling of via wires. A low loss and highly shielded coax cable from Amphenol RF$\textsuperscript{\textregistered}$ (Model: 523-095-902-451-005 and Manufacturing Part: 095-902-451-005) was used to feed the antenna. The choice of the cable was made keeping in mind the leaky wave and standing wave possibility since the antenna ground plane is electrically small. When they layers were stacked together, a few drops of low loss epoxy (super glue) were used to stick from each other. A very careful taping of the entire implantable antenna was done and the final fabricated antenna is shown in Figure \ref{fabrication}(c).

\subsection{Measurements within Tissue}
\begin{figure}[htb]
\centering
\begin{subfigure}{0.35\textwidth}
	\includegraphics[height=1\textwidth]{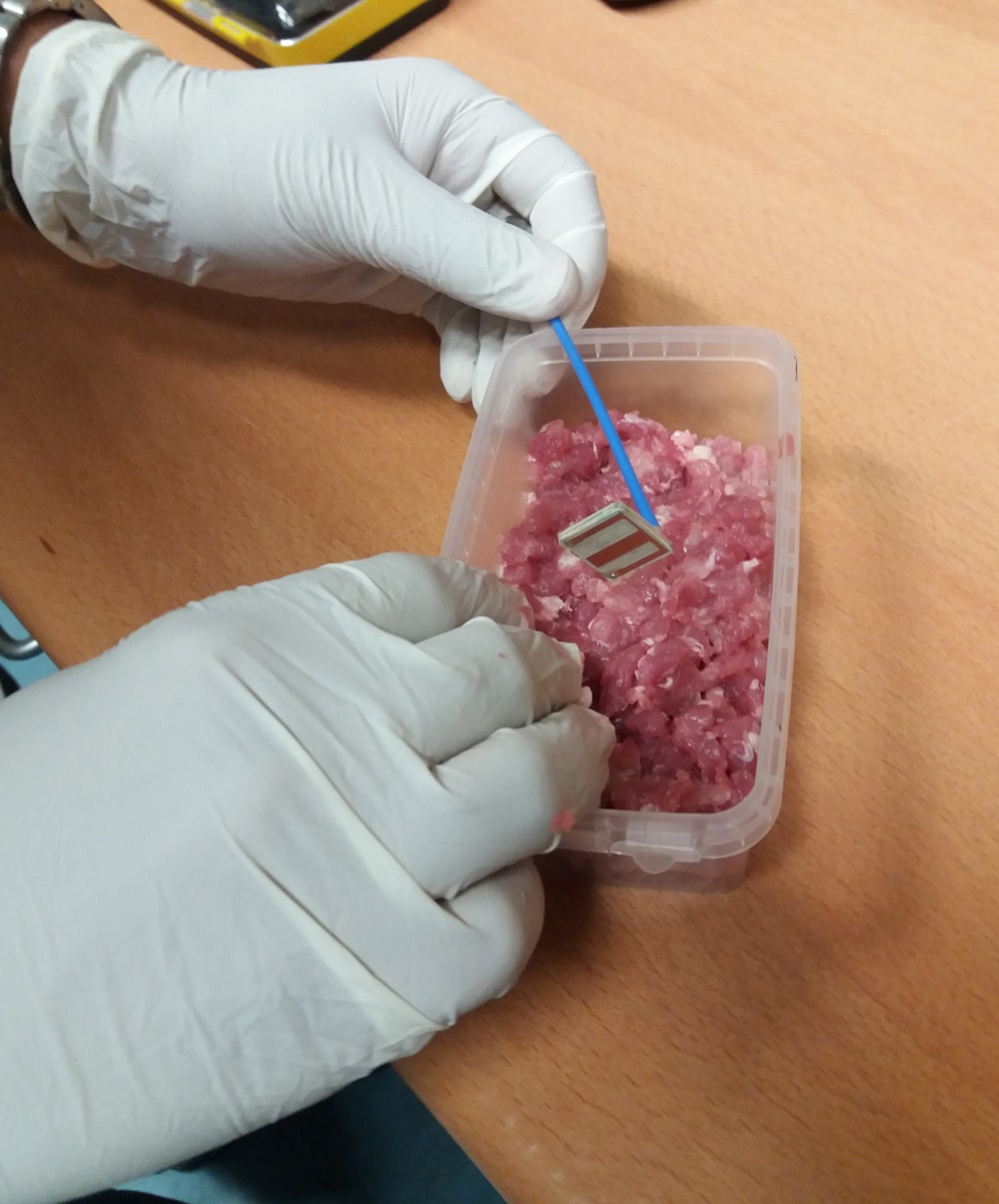}
	\caption{}
\end{subfigure}
\begin{subfigure}{0.35\textwidth}
	\includegraphics[height=1\textwidth]{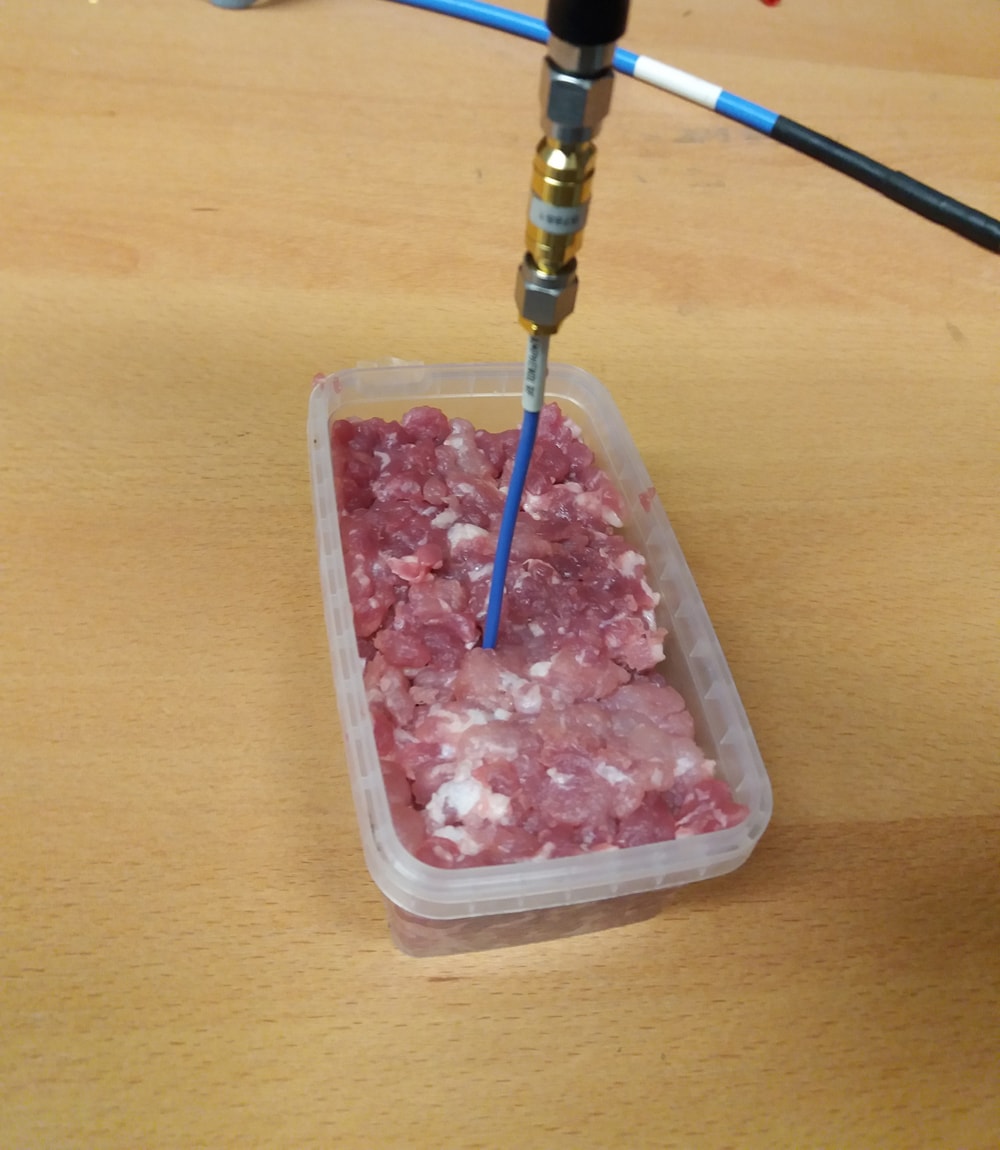}
	\caption{}
\end{subfigure}
\caption{Measurement setup of minced pork meat implant to test the performance of implantable antenna. $\textcopyright$ 2018 IEEE.}
\label{measurement}
\end{figure}

\begin{figure}[htbp]
\centering
\includegraphics[width=1\textwidth]{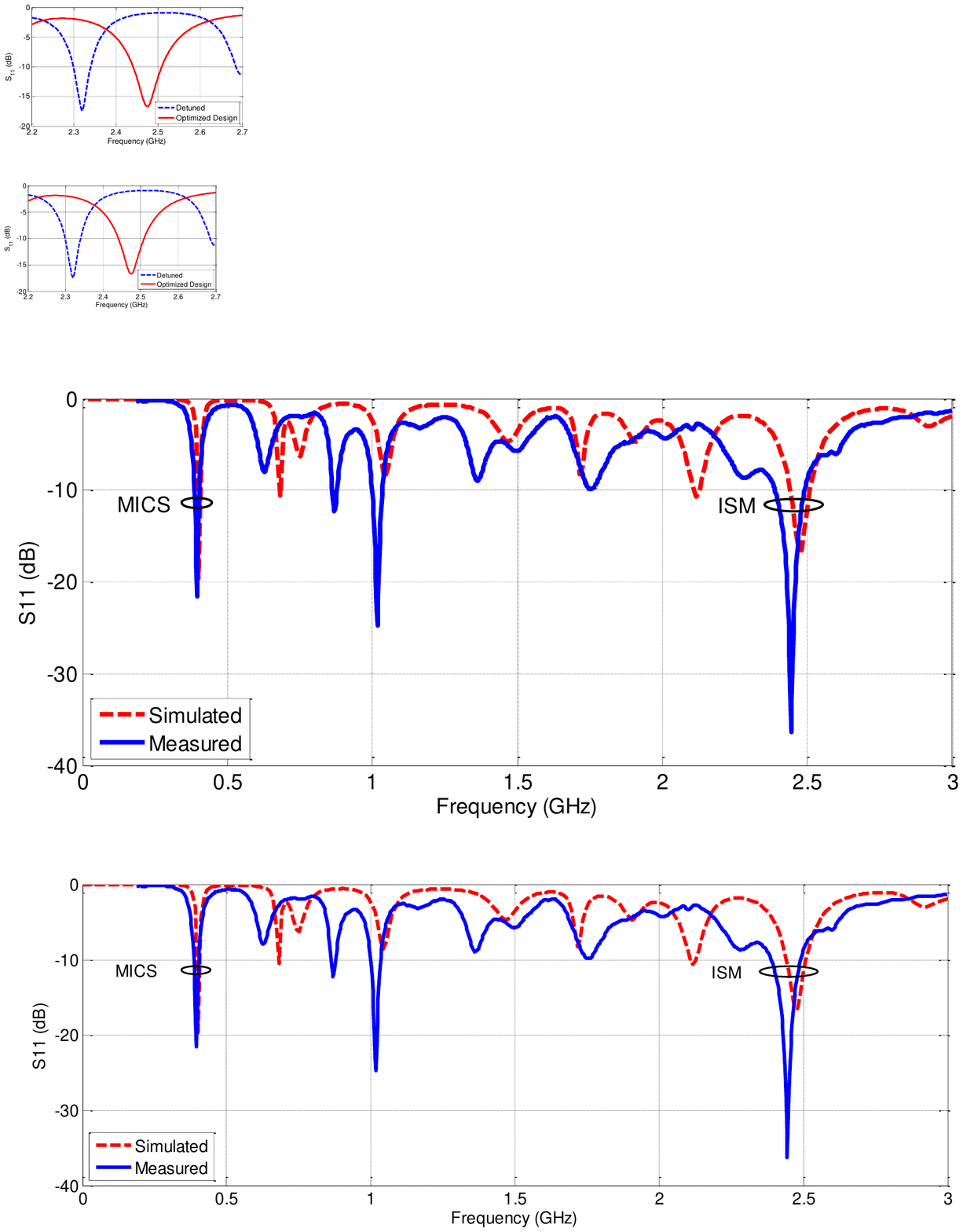} 
\caption{Comparison between simulated and measured results for the antenna prepared for minced pork meat implant.}
\label{final}
\end{figure}

Initial antenna testing were done by directly connecting it to the VNA and placing it in the hand just to test that the whole fabrication process goes well. In the next step, minced pork meat placed in a plastic container ($\epsilon_r$ is roughly equal to 1.5). Since the $\epsilon_r$ of the antenna substrate is much higher and the distance between the boundary of container is higher compared to the antenna size, the thin plastic container with low $\epsilon_r$ was not taken into consideration while simulations. The antenna was carefully placed in the middle half, and the was fully covered with minced meat to form a block where the antenna is approximately in the middle. The cable connecting the antenna to the VNA was carefully attached to a stand to ensure proper isolation between the AUT and the measurement equipment. Figure \ref{measurement} shows the pictures taken during the measurement process. The results presented in Figure\ref{final} indicates a good agreement between the simulated and the measurement results. This ensures the proper fabrication and measurement process in which even minor details like milling bit selection and antenna layering selection was carefully monitored. Some of the non-radiating poor resonances occur in between MICS and ISM band. These resonances appear commonly in wideband electrically small antennas \cite{hansen2006electrically} but due to their non-radiating nature, they do not have a major impact in radiation efficiency of the antenna for the desired bands. The fabricated antenna retains the predicted impedance bandwidth, both for ISM and MICS bands. 

\section{Chapter Contributions}
This chapter describes a miniaturized implantable antenna with a meandered and stacked structure. By varying the length and width of each parameter listed in Table \ref{Table 1:}, the antenna is excited at two resonant frequencies, at MICS and ISM bands. It was demonstrated that the proposed antenna can be tuned using the particle swarm algorithm, to operate adequately when placed inside several different human tissue models. Antenna testing within dead tissue, to verify the practical implementation in a real life implant scenario was discussed. Possible bottle necks while fabricating and testing of the antenna were also presented. In this chapter a realistic body phantom, a complex body phantom and a widely used simple layered phantom are compared in a through-body communication scenario at the MICS band. Complexity of these phantoms in terms of full wave simulator meshing density and resulting s-parameters accuracy were discussed. It was observed that when a realistic phantom was replaced by widely used simpler phantom models, significant detuning of the resonant frequency of the implantable antenna occurred. Due to the narrow bandwidth of MICS band, this detuning can be considered as “unacceptable” for such simulation setups. Further simulation results indicated that the additional computational requirements (computational resources and time) to simulate a realistic body phantom model is only 6 to 7 times the computational requirements, needed to simulate simple layered phantoms. Based on this observation, it is recommended to use realistic body phantom rather than layered phantoms for simulation setups created for implantable antennas, operating in the MICS band.

\chapter{Through- and across-body communication}

\ifpdf
\graphicspath{{Chapter4/Chapter4Figs/PNG/}{Chapter4/Chapter4Figs/PDF/}{Chapter4/Chapter4Figs/}}
\else
\graphicspath{{Chapter4/Chapter4Figs/EPS/}{Chapter4/Chapter4Figs/}}
\fi
In health care sector, a reliable connectivity between the Body Centric Wireless Communication (BCWC) sensor nodes is vital to patient's health monitoring. Investigation in the previous chapters revealed that the radiators for wearable and implantable nodes need to work efficiently even in the presence of human body. For a wearable antenna maintaining an off-body communication, high directivity in the direction away from the body is the most important factors as discussed in Chapter \ref{chp2}. To achieve this goal, a real or virtual isolation from human body is preferable, which reduces the detuning and mismatching factor due to body. For an implantable antennas, the directivity is of a secondary importance, while the matching when an antenna is covered in a given tissue is vital as discussed in Chapter \ref{chp3}. For a wearable sensor antenna intended for an efficient communication with an implantable antenna or another node present on any part of the body, both, the directivity as well as impedance miss-matching need equal consideration. In this chapter, a wearable antenna for BCWC sensor node (referred to in this chapter as wearable sensor antenna) is discussed, designed to work reliably for a communication with an implantable. This completes the link between body implant and off-body receiver, communicating with each other via wearable node. Moreover, the connectivity between multiple wearable antenna nodes placed around human body is discussed in the light of creeping-wave theory.

\section{Introduction}

With ongoing population growth and the associated rising health concerns, the idea of transforming wired medical facilities into wireless biomedical sensors has been increasingly studied in recent years. For an implantable sensor, physiological variables such as glucose, oxygen and pH levels, and device parameters such as battery life and mode of operation are considered to be essential readings. When a patient is in critical condition in intensive care, these variables are required to be monitored either continuously, or at very frequent intervals. The biggest challenge in ensuring a stable wireless communication between an implantable sensor and an on/off-body sensor lies in the fact that the RF signal needs to propagate through the body. The impedance boundaries of the biological tissues and the electromagnetic wave absorption result in a significant reduction of signal integrity. Furthermore, the performance of a newly designed sensor antenna is difficult to be predicted in real life conditions due to the complicated heterogeneous time-changing electrical properties of the living tissues. Some studies have shown experimental results of implantable antennas placed in either rat skin \cite{jin2013injectable,karacolak2009electrical,karacolak2010vivo}, pork meat \cite{surapan2016design,xu2015miniaturized}, or skin-mimicking material \cite{yang2017circularly} communicating with an external antenna. Recently, successful communication between an implantable antenna and a smart phone has been shown in \cite{karnaushenko2015compact}. With advancements in simulation technology, new methods to investigate through-body communication scenarios have been established, where the design of biomedical sensors, was carried out in the presence of a simple, or high resolution, numerical body model \cite{segars2009mcat,chow2013implantable,kiourti2012review}. In this work, the influence of the body on the performance of through-body communication between a wearable and an implantable antenna, is investigated. 


\subsection{Wearable Sensor Antenna for Through-body communication}

\begin{figure}[htb]
	\centering
	\begin{subfigure}{0.4\textwidth}
		\includegraphics[width=0.6\linewidth]{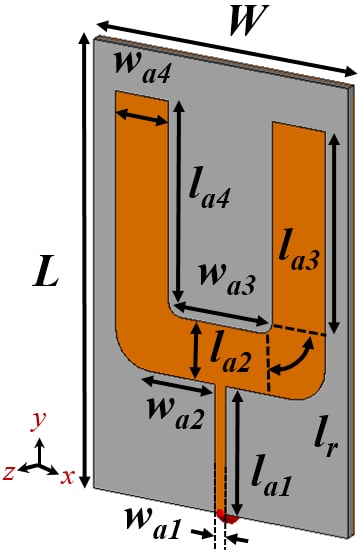} 
		\centering
		\caption{}
	\end{subfigure}
	\begin{subfigure}{0.5\textwidth}
		\includegraphics[width=1\linewidth]{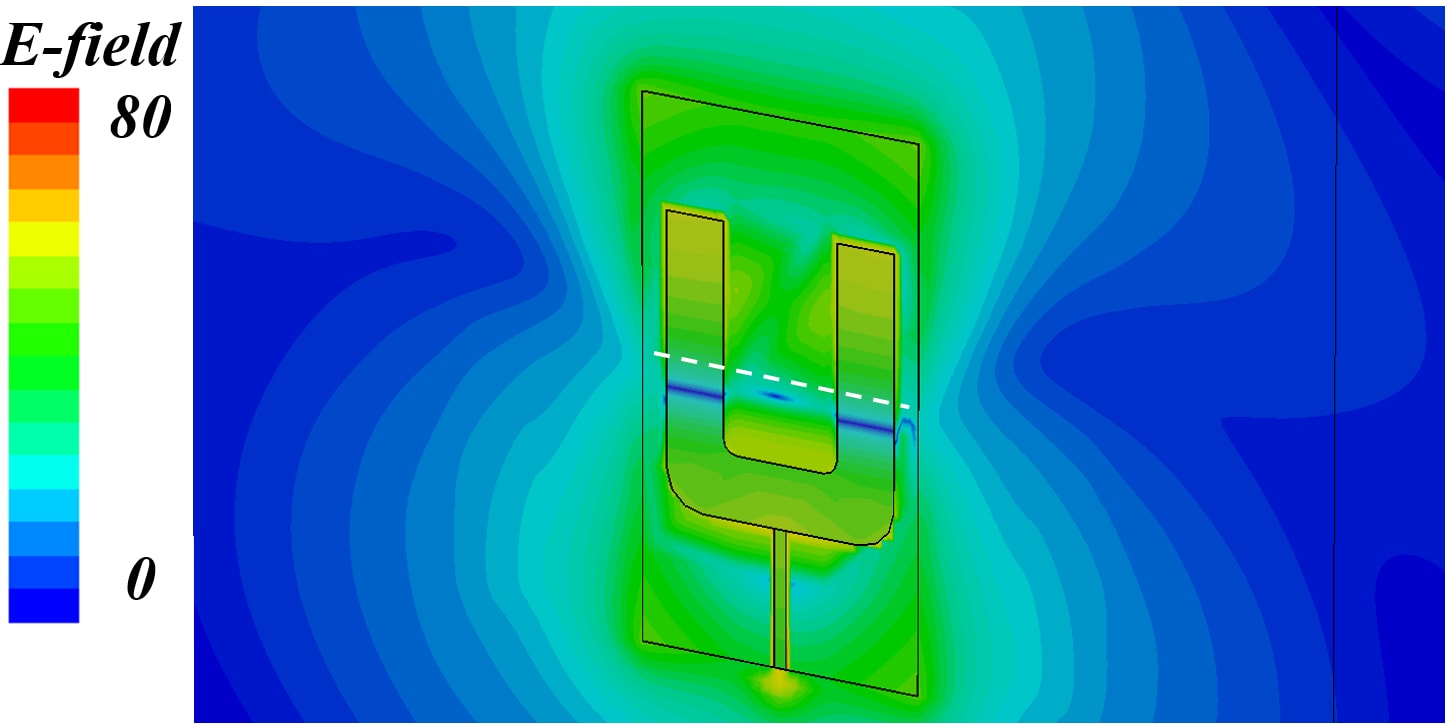} 
		\centering
		\caption{}
	\end{subfigure}
	\caption{(a) Schematic of a wearable sensor antenna when \textit{L} = 70, \textit{W} = 38, \textit{$w_{a1}$} = 2.05, \textit{$w_{a2}$} = 10.25, \textit{$w_{a3}$} = 7.33, \textit{$w_{a4}$} = 8.34, \textit{$l_{a1}$} = 22.60, \textit{$l_{a2}$} = 10.50, \textit{$l_{a3}$} = \textit{$l_{a4}$} = 32.89 (all parameters are in mm) (b) Normalized \textit{E}-field distribution. }
	\label{tb1}
\end{figure}

An antenna designed for operation in free space, it will not radiate efficiently when it operates in close proximity to human body. This phenomenon is more intense when the antenna is part of a wearable sensor establishing communication with an implantable antenna placed inside the body. Since the main beam direction of the wearable sensor antenna (which can be a part of textile \cite{shahid2012textile, nawaz2015body}) needs to be towards the body, reflections from the body can affect the radiation performance of the antenna. A wearable sensor antenna inspired by recent development in efficient radiators for microwave imaging \cite{rahman2016electromagnetic} is presented. Some of these antennas are designed on inflexible substrates that are ill adapted to the body curvatures, however, microwave substrates that are going to be used as wearable sensor should be as flexible and as soft as possible. Keeping this factor in mind, the antenna is designed on Roger's RT Duriod 5880 because of its low $\epsilon_r$ and flexibility. Figure \ref{tb1}(a) shows the antenna geometry and schematic dimensions. Cavity backed radiators such as \cite{taguchi2000cavity,bashri2016compact,porter2016wearable,rahman2016electromagnetic,bashri2017flexible,santorelli2015time} have been widely used in the microwave imaging application because of high directivity and stable s-parameters. This is achieved by directing the propagating wave along the \textit{+z}-direction. 

\begin{figure}[htb]
	\centering
	\begin{subfigure}{0.68\textwidth}
		\includegraphics[width=1\linewidth]{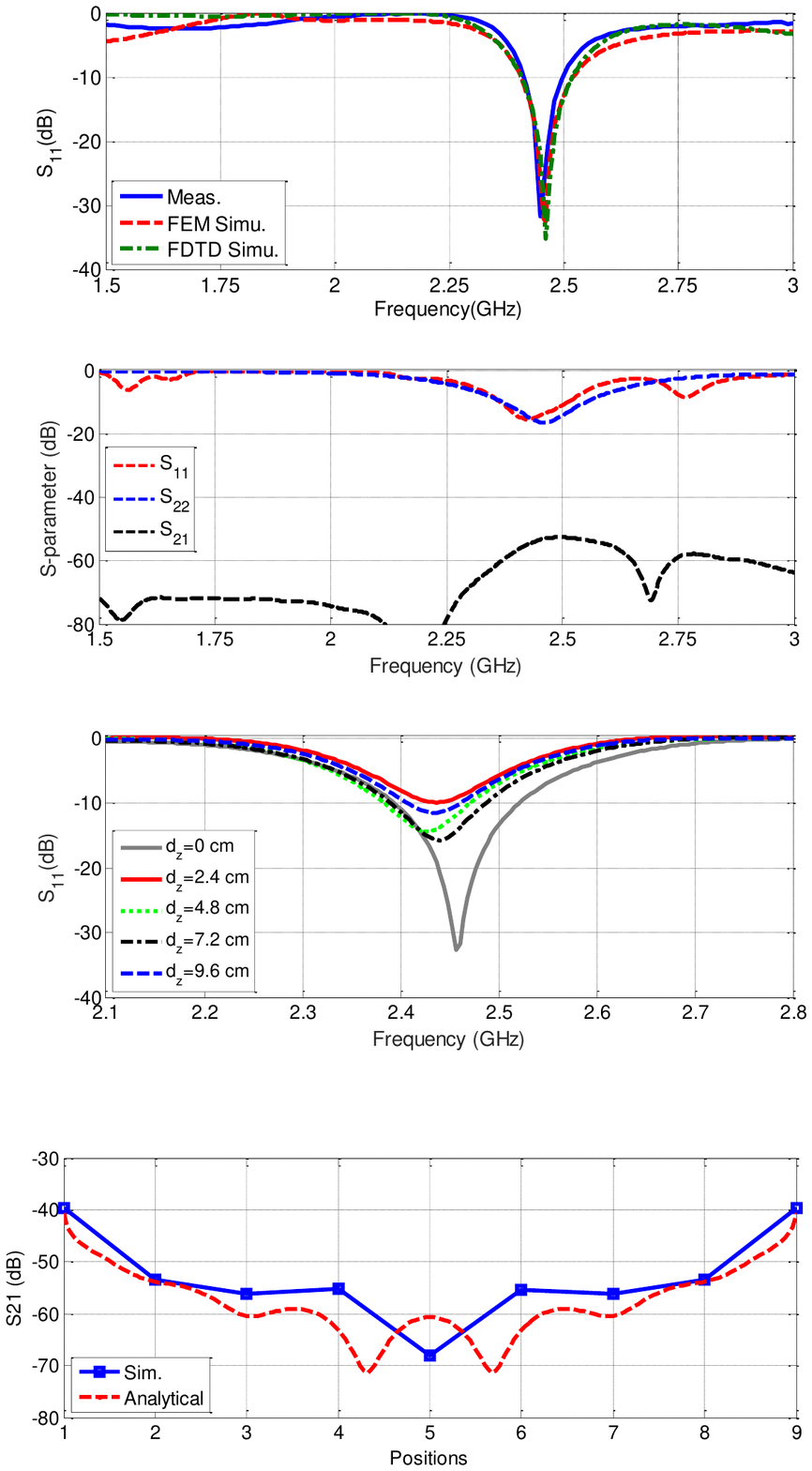} 
		\centering
		\caption{}
	\end{subfigure}
	\begin{subfigure}{0.28\textwidth}
		\includegraphics[width=1\linewidth]{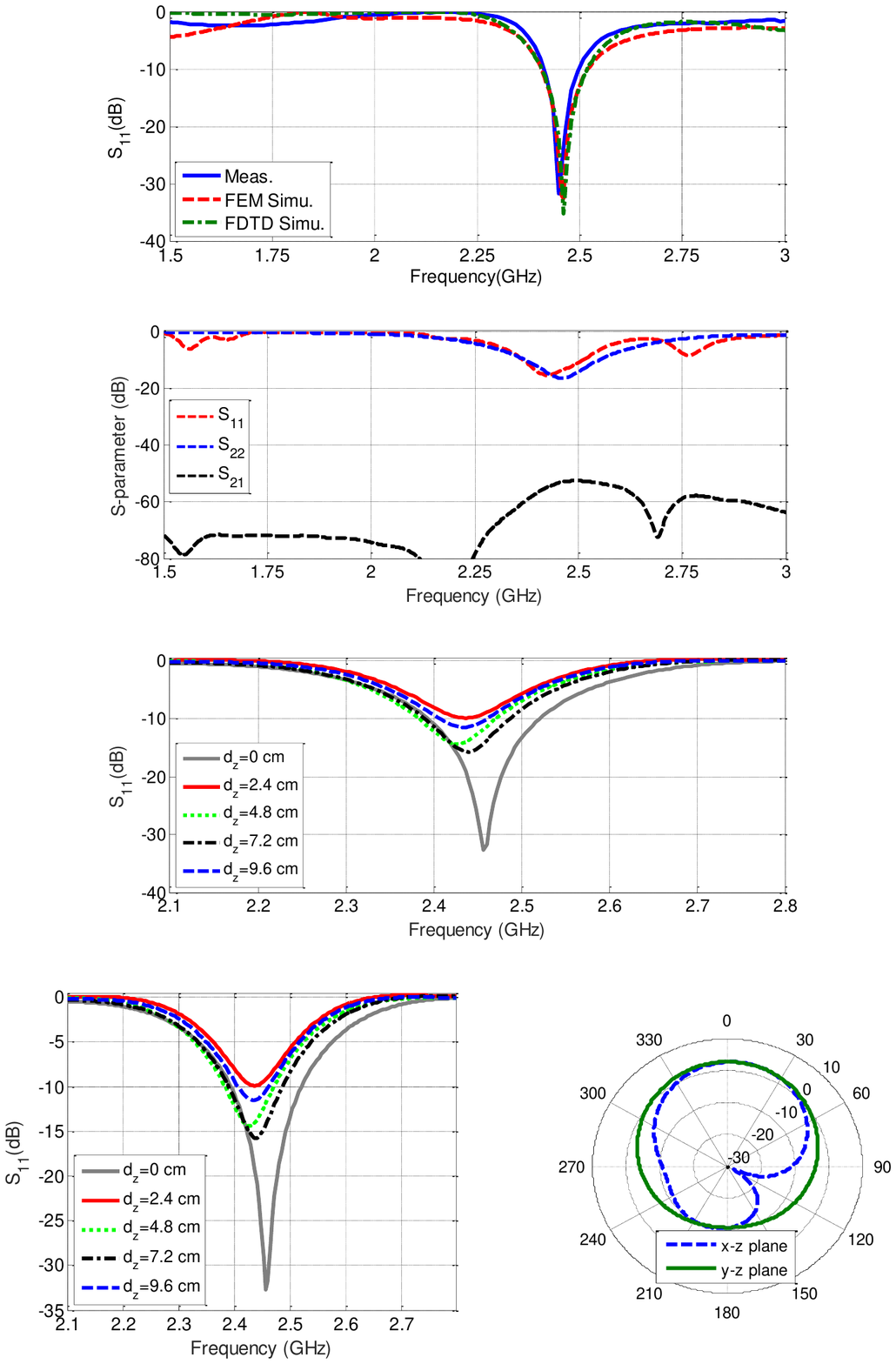} 
		\centering
		\caption{}
	\end{subfigure}
	\caption{(a) Simulated S${}_{21}$ when antenna is placed on chest and moved away from the body. (b) 2D gain plot. }
	\label{tb1_1}
\end{figure}

The initial point of designing was a cavity backed fork shaped antenna. It was observed that the presence of the cavity severely impacted the low-profile feature of an antenna operating at 2.45 GHz ISM band and tend to be wearable. To maintain low-profile with the required directivity, a fully grounded planar antenna working primarily on microstrip patch principle was designed. It is important to mention that the overall board dimensions of the wearable sensor antenna are \textit{ G${}_{l}$}$\times$\textit{G${}_{w}$} from the Figure\ref{ebg1}. The main reason of selecting these dimensions are to be able to eventually integrate the two antennas as a single wearable sensor unit (further explanation is in the proceeding sections). A fork shaped radiator is designed which is comprised of a planar base connected to two legs via curved microstrip section. The transmission line width $\textit{w$_{a1}$}$ and the curvature of $\textit{l$_r$}$ is mainly responsible for an efficient impedance matching between the input port of the antenna and the radiating section. The principle of operation of the antenna lies at the fact that the fork shaped radiator lengths along \textit{y}-axis ($\textit{l$_{a2}$}+\textit{l$_{a4}$})$ is $43.39mm$ which is close to $\lambda_{eff}/2 \approx 44.75mm$. In other words, the radiator hosts a $\lambda_{eff}/2$ wave, efficiently leaving the surface of the antenna. The maximum energy transmission from antenna to the human body resides at the antenna physical center. This enables a realizable alternative of creating a cavity at the back side of a radiator. The principle of operation can further be verified by Figure \ref{tb1}(b) where the EM wave \textit{E}-field null can be seen to be approximately at the physical center of the fork shaped radiator. Figure \ref{tb1_1}(a) presents the antenna impedance matching response when it is placed almost on top of the human body phantom (initial optimization was done at chest region). The simulated results indicate the antenna covers an entire 2.45 GHz ISM band when placed right top of the body. An inevitable mismatch can be observed when the antenna is moved away from the human body, however it is not severe. Most importantly, the antenna detuning factor is negligible when it is moved away from the body ($d_z > 0mm$). The antenna gain patterns simulated by normalizing the input impedance port to 50 $\Omega$ are given in Figure \ref{tb1_1}(b). The reason of normalizing is that when antenna is placed on body phantom and simulated in EM simulator, the background material in the antenna surroundings are considered a part of a radiator, which impacts the accuracy of simulations. A reliable method is to calculate far field patterns in free space without considering the impedance mismatch impact which will occur when the given antenna is moved away from the body. The patterns are very close to standard microstrip patch antenna while the peak gain is 3.41 dBi. . The mentioned properties makes the proposed antenna fit for through body communication with an implant at 2.45 GHz even at a distance of multiple centimeters from the body.

\subsection{Wearable Sensor Node Architecture}

\begin{figure}[htb]
	\centering
	\begin{subfigure}{1\textwidth}
		\includegraphics[width=0.9\linewidth]{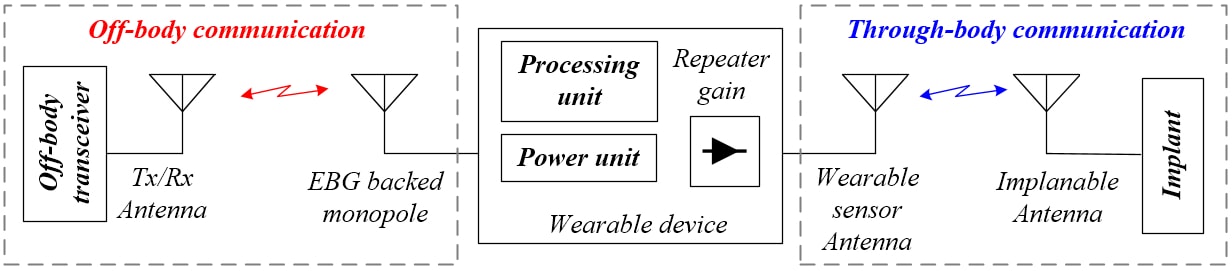} 
		\centering
		\caption{}
	\end{subfigure}
	\begin{subfigure}{0.5\textwidth}
		\includegraphics[width=1\linewidth]{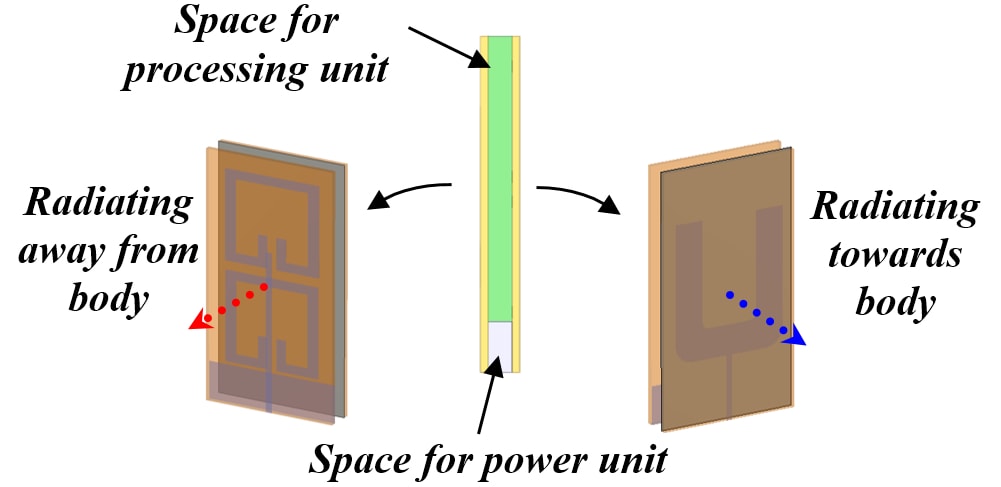} 
		\centering
		\caption{}
	\end{subfigure}
	\caption{(a) System level block diagram of implantable to off-body transceiver communication via wearable node (b) Wearable device architecture representing the operational principle when two antennas radiate in the opposite directions.}%
	\label{tb2}
\end{figure}

Figure \ref{tb2} shows the proposed system model and the device architecture for a reliable communication between an external off-body transceiver and a body implant for an uninterrupted patient's physiological parameters monitoring. The system consists of the EBG backed monopole antenna, transmitting/receiving signal from the off-body transceiver. The signal is to be processed at the wearable sensor node and transmitted via wearable sensor antenna to the implantable antenna after an added (repeater) gain.

As discussed previously, the EBG backed monopole antenna is efficient for off-body communication and it needs a full ground plane to operate effectively. This full ground plane is to host a wearable sensor antenna facing towards the opposite direction as compared to the EBG backed monopole antenna forming the proposed wearable sensor architecture The spacer between EBG surface and the full ground plane can be readily used to host the electronic components of a sensor node (which may include a power management unit, processing unit and circuit board hosting electronic components). It is recommended to place the power unit in-between the two metallic ground plates (\textit{ G${}_{l}$}$\times$\textit{G${}_{w}$} and \textit{ gnd${}_{l}$}$\times$\textit{S${}_{w}$}) so that it has a minimum impact on the radiation performance of both antennas. An important condition that must apply in the implementation of the proposed architecture is that both the antennas should be re-optimized in the presence of the entire system governed by the principles of operation in their respective design guideline sections.

\subsection{Simulation Test-bench}

A detailed numerical body model was used in the simulations to realistically mimic a real life scenario for the investigation of wearable sensor antenna performance. Since the wearable sensor antenna is a suppose to always direct towards the body, and is mainly isolated from the EBG backed monopole antenna, only the operation of wearable sensor antenna is discussed in this study, assuming that it will behave the same way as a part of entire system presented in Figure \ref{tb2}. The performance of wearable sensor antenna was optimized when it was placed on top of the realistic body phantom at the torso region. Note that the antenna metallic layer was not in direct contact with the tissue muscles.

\begin{figure} [htb]
	\centering
	\begin{subfigure}{0.63\textwidth}
		\includegraphics[width=1\linewidth]{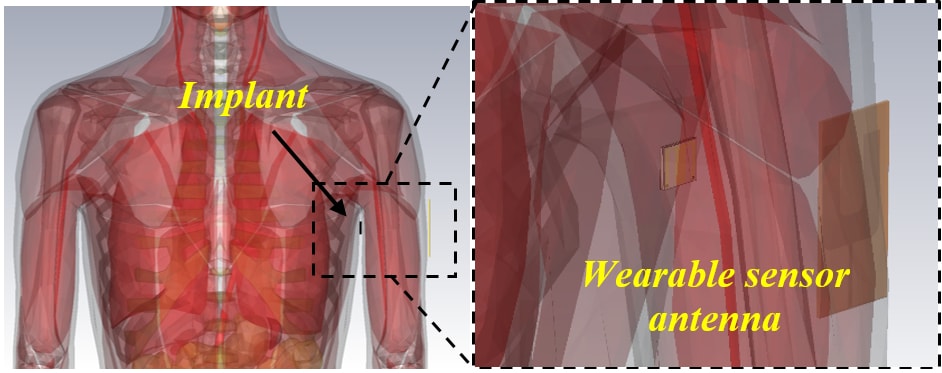} 
		\centering
		\caption{}
	\end{subfigure}
	\begin{subfigure}{0.245\textwidth}
		\includegraphics[width=1\linewidth]{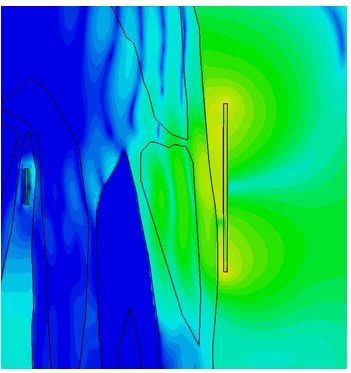} 
		\centering
		\caption{}
	\end{subfigure}
	\begin{subfigure}{0.7\textwidth}
		\includegraphics[width=1\linewidth]{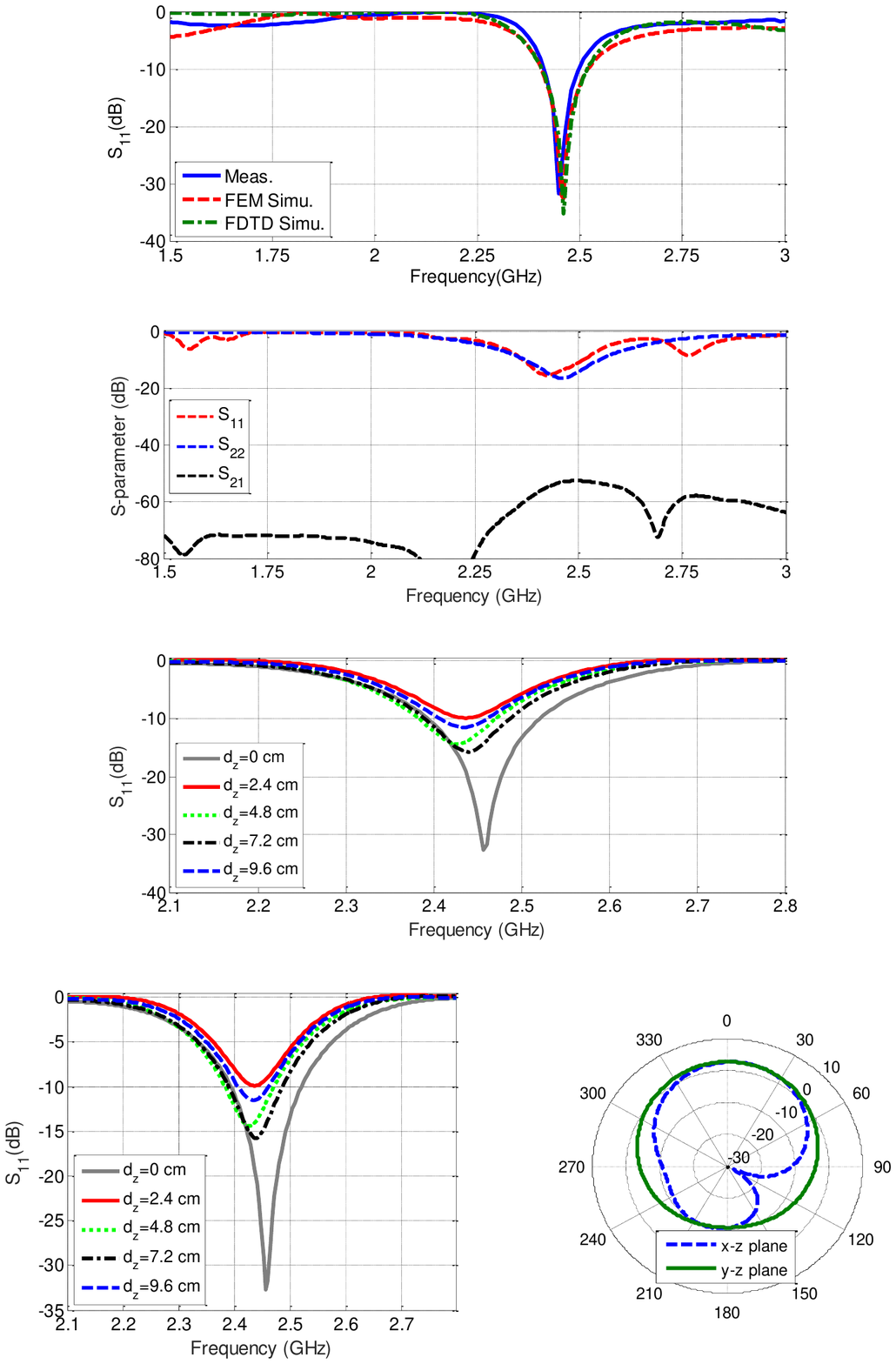} 
		\centering
		\caption{}
	\end{subfigure}
	\caption{(a) Simulation setup for wearable to implantable antenna communication and normalized \textit{E}-field distribution. (b) S-parameter response when port 1 is connected to the implantable antenna and port 2 is connected to the wearable sensor antenna. }%
	\label{tb3}
\end{figure}

This subsection presents a simulation test-bench to investigate the antenna detuning, impedance mismatch and losses in through-body communications. These parameters are relevant to the efficient design of wearable and implantable antennas for biomedical sensors. The study was carried out using finite-difference-time-domain electromagnetic simulations using realistic high-resolution numerical human body phantom models. S-parameter measurement corresponding to the simulation setup was carried out when an implantable antenna connected to port 2 was placed in the armpit of a human subject and communication was established with an external wearable antenna connected to port 1. The first advantage of this setup is the convenience of placing an electrically small antenna in a close-to-implantable scenario. The second advantage is that the armpit area contains a combination of tissues (muscle, fat and skin) with distinct dielectric properties which are randomly distributed, and therefore insightful for the study of through-body wave propagation. The downside is the requirement of high computational resources required for the simulations. 

As discussed in \cite{moradi2017measurement}, the wearable sensor node

\subsection{Results and Discussion}
As discussed in last chapter, initial antenna testing was done in minced pork meat since its electrical properties resemble human muscle and skin tissue between 100 MHz and 3 GHz \cite{kiourti2012review}. In the next step, the implantable antenna was placed in the armpit section of the numerical phantom while the wearable sensor antenna was placed at \textit{$d_z$} = 1 cm from the arm's surface. The results presented in Figure \ref{tb3}(c) indicate a small frequency shift, which disturbthe impedance bandwidth of both the antennas, nonetheless the simulated {\textbar}S${}_{11}${\textbar} and {\textbar}S${}_{22}${\textbar} remained below -10 dB at 2.45 GHz ISM band. The time-varying electric fields in the full-wave simulator depicted valuable information of the wave propagation through the complex heterogeneous tissues. It was observed that the communication between the wearable sensor antenna and the implantable antenna mainly relied on the diffracted and reflected signals propagating through the tissues. Figure \ref{tb3}(b) show that even though the wave propagation was almost blocked by the presence of the bone tissues in the arm, however the wave diffracting and creeping along the bone tissues reached the implantable antenna. The results indicate that the proposed test-bench is found to be insightful tool for antenna optimization and reliable through body communication environment modeling.

\section{Across-Body Communication} \label{ac}
When two wearable nodes are placed at different parts of a human subject, there communication mainly depends upon the multi-path and creeping waves. Since the EM wave propagation in an in-door or dense urban environment is itself a separate research ares, in this study, we only focus on the communication between wearable nodes in close proximity of human body and governed by the creeping wave theory. A number of research efforts has been surfaced to characterize the on-body to on-body wave propagation \cite{conway2009antennas,alves2011analytical} propagation but due to a large number of variables involved even in a single scenario, these studies only focused on a specific sub-area. The most relevant investigation to the current thesis topic is the propagation study of creeping waves. In this work, a novel approach is presented in which NRI-TL metamaterial based electrically small leaky-wave antenna \cite{antoniades2008cps} is used as a wearable antenna. The propagation of the transmitted signal is investigated along the torso region of a human body. For investigation, detailed numerical phantom is used. It was observed that the antenna \cite{antoniades2008cps} do not involve the capability of flexing along the body curvature. In the light of this observation, the same antenna has been modified in such a way that the antenna operation stays the same, however, the ground and the lumped component hosting board is bent along the direction of the body curvature. The bent radius is 50 mm, selected carefully to compensate the antenna placement along the torso region. The simulation setup to investigate the across-body communication is shown in Figure \ref{ab1} (a). 

\subsection{Simulation Setup}

\begin{figure}
	\centering
	\begin{subfigure}{0.35\textwidth}
		\includegraphics[width=0.8\linewidth]{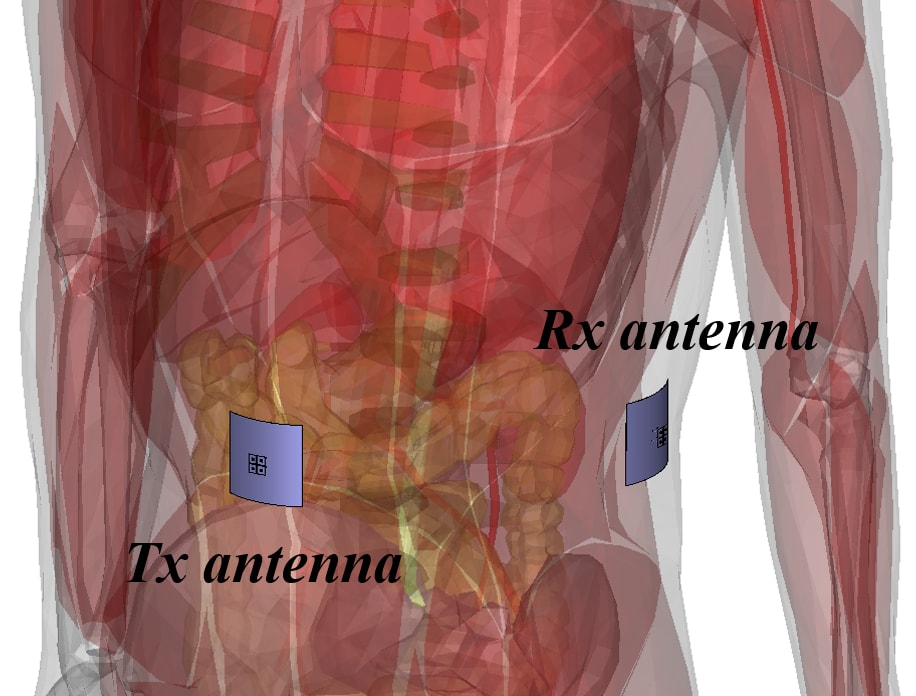} 
		\centering
		\caption{}
	\end{subfigure}
	\begin{subfigure}{0.6\textwidth}
		\includegraphics[width=1\linewidth]{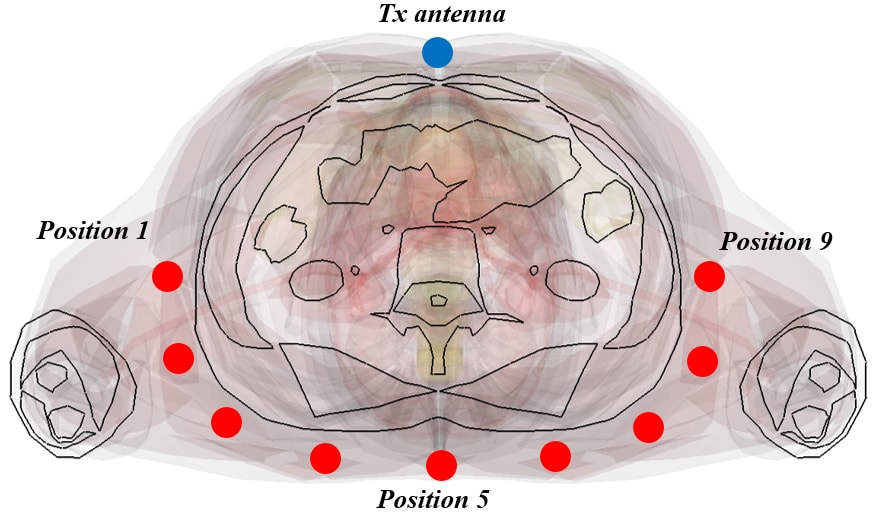} 
		\centering
		\caption{}
	\end{subfigure}
	\caption{(a) Simulation setup for across-body communication (b) Cross-sectional view of the torso region at the antenna level showing the transmitter and receiver element positions.}
	\label{ab1}
\end{figure}

The simulation setup to study the across-body communication involve a pair of antennas described in Section \ref{ac} and detailed numerical phantom. The general principle of the creeping wave theory has thoroughly been investigated in \cite{chandra2013analytical} and an analytical model is presented to quantify the EM wave path loss. However, it has been observed that the body curvature used in the proceeding study is rather ideal. There is a need to investigate the across-body communication in the light of proposed analytical model. First, the most oval-like cross sectional space was selected in the numerical phantom. The transmitter antenna was placed at the stomach region, while the receiver antenna was placed at the positions 1 to 9 along the back side of the body. This setup replicates the most generic scenario presented in \cite{chandra2013analytical}. The radiation patterns of the antenna electrically small leaky-wave antenna with reduced beam squinting in \cite{antoniades2008cps} were found to best suite the across-body communication application. The antenna radiates maximum energy along broadfire direction, so it can be assumed that the maximum EM wave energy will flow along the body torso region when the antenna is placed flat on body. Note that the presented simulation setup is computationally extensive. A total of nine independent simulations were run when the received antenna was placed at each locations indicated in Figure \ref{ab1} (b).

\subsection{Results and Discussion}

\begin{figure}
	\centering
	\includegraphics[width=0.8\textwidth]{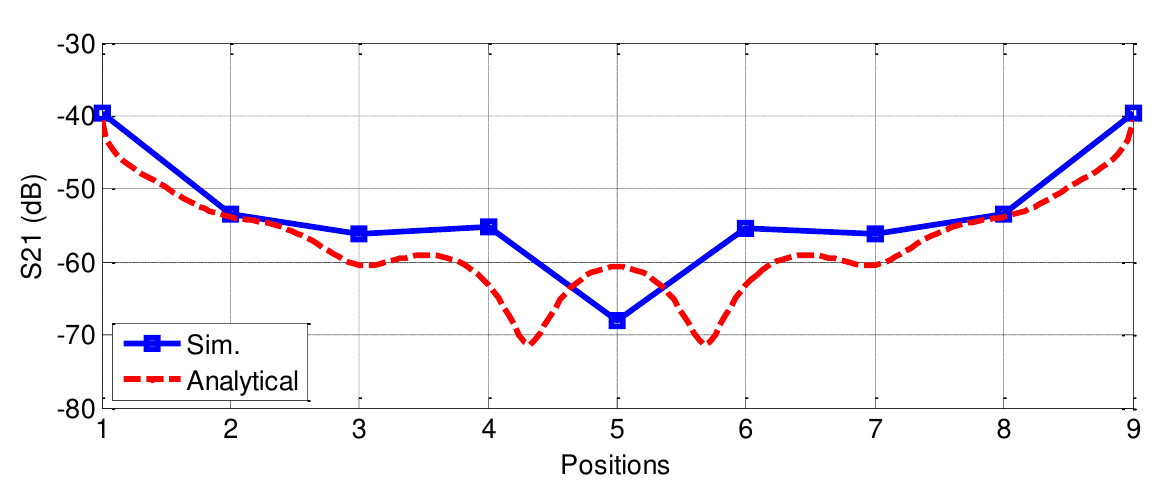}
	\caption{Simulated S${}_{21}$ versus analytical S${}_{21}$ when the transmitted antenna stays on its position while the receiver antenna is moved along the torso from position 1 to 9. }%
	\label{ab3}
\end{figure}

\begin{figure}
	\centering
	\includegraphics[width=0.8\textwidth]{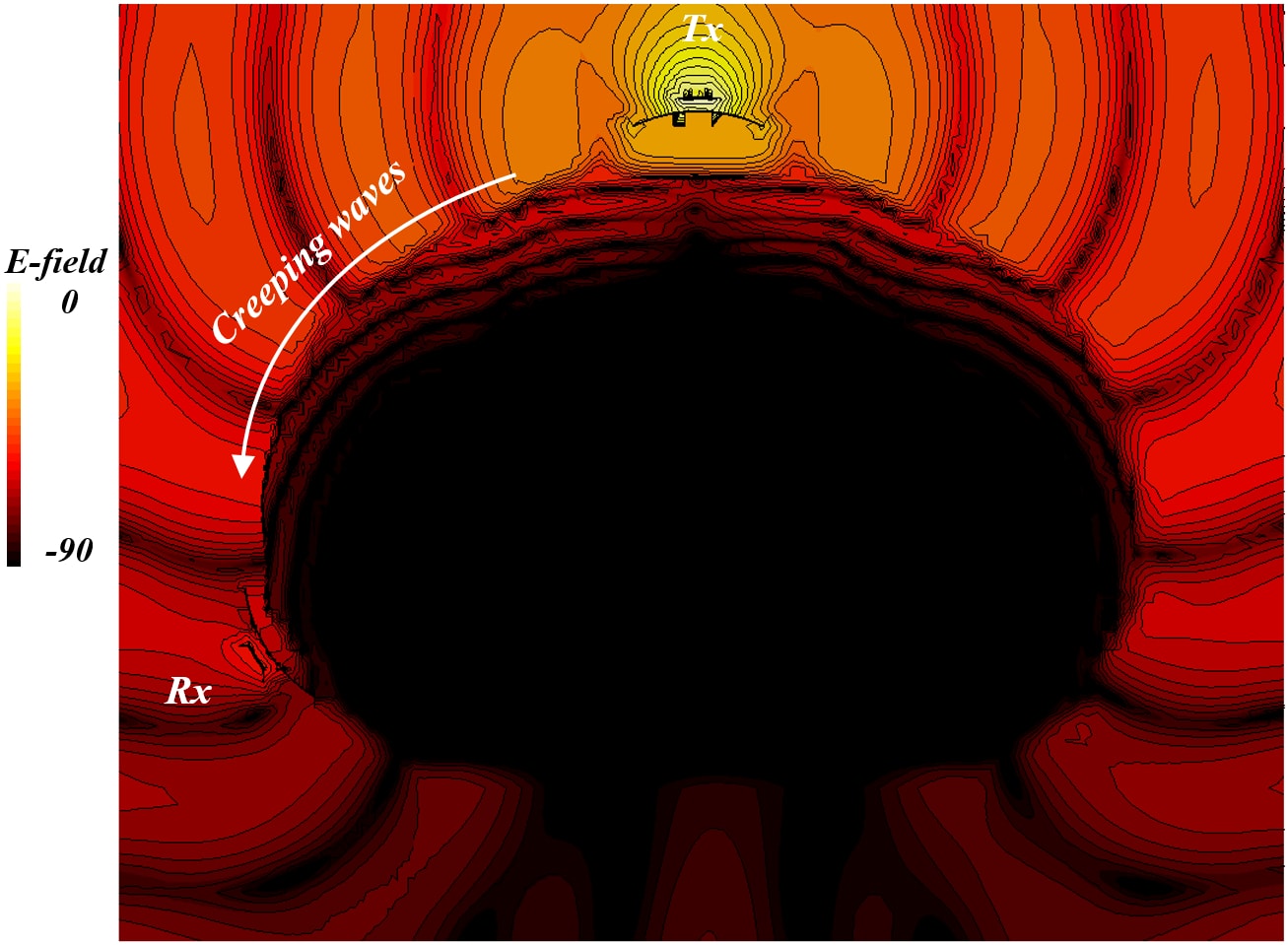}
	\caption{Simulated creepy wave illustration by normalized \textit{E}-field magnitude plot. }%
	\label{ab2}
\end{figure}
The simulated results indicate a number of observations. Figure \ref{ab3} shows the comparison between the analytical and realistically simulated path loss. It is important to mention that the overall EM wave energy loss in the given scenario is only a combination of losses due to impedance mismatch by antenna placement, free space losses in clock-wise and anti-clockwise creeping waves, and a minor addition of through-body wave. No multi-path is considered. It can be seen the the proposed analytical model follows the path loss contour of the realistic scenario. Figure \ref{ab2} demonstrates the physical behavior of creeping waves generating from the transmitter antenna and collecting at the receiver antenna. The given surface plot indicate that the field strength of the wave propagating along the torso region decreases at a very high rate and a reliable communication relies heavily on the receiver sensitivity. Although this propagation model is suitable for wearable sensor application in controlled in-door health care units, but it should be noted that the model can not be considered as power efficient.


\chapter{Miniaturized Ultra Wideband Antenna}

\ifpdf
\graphicspath{{Chapter5/Chapter5Figs/PNG/}{Chapter5/Chapter5Figs/PDF/}{Chapter5/Chapter5Figs/}}
\else
\graphicspath{{Chapter5/Chapter5Figs/EPS/}{Chapter5/Chapter5Figs/}}
\fi

Three planar, CPW-fed, Ultra Wideband (UWB) antennas with increasingly reduced size are presented and the miniaturization method is discussed. The first antenna is a CPW-fed elliptical slot with an uneven U-shaped tuning stub, the second antenna is a cactus shaped monopole and the third one is a miniaturized version of the cactus shaped monopole antenna. All presented antennas have a simulated and measured return loss below -10 dB over the 3.1 to 10.6 GHz UWB frequency range and mostly omni-directional radiation patterns. The proposed antennas are fabricated on liquid crystal polymer (LCP). The CPW-fed slot antenna requires an overall board dimension of 38 mm $\times$ 40 mm, the evolved cactus monopole is confined in a 28 mm $\times$ 32 mm board, while the final miniaturized cactus monopole is printed on 28 mm $\times$ 20 mm board, resulting in a 41 \% and 63\% size reduction respectively. Using both simulations and measurements the paper analyzes the response of all three antennas, and discusses, and demonstrates the effectiveness of the implemented miniaturization method. 
\section{Introduction on UWB Antennass}
The Ultra Wide Band protocol that covers the frequency range from 3.1-10.6 GHz was released from the FCC in 2002 \cite{federal2002first}, partly as an attempt to meet the demand for high data rate communications in short distances for mobile and personal applications. Consequently, there is an increasing need for compact sized, low cost and high efficiency antennas with omni-directional radiation patterns. The combination of these characteristics in a wide frequency range such as the UWB band is a challenging problem and several design concepts and different materials have been used in an attempt to provide a satisfactory solution. The challenge associated with the miniaturization of antennas has been the tradeoff between the reduction of the physical size of the antenna and its operational bandwidth \cite{abbosh2008miniaturization} and radiation efficiency. Some researchers have adopted the use of substrates with relatively high dielectric constant to reduce the antenna's resonant frequency because permittivity is inversely related to the resonant frequency \cite{abbosh2009miniaturized}. However, bandwidth reduction is inevitable in this approach because bandwidth is also inversely proportional to the permittivity. In an attempt to increase the bandwidth of an antenna and cover the whole UWB frequency range, configurations like planar monopoles \cite{ezuma2015design, koohestani2012influence, elsheakh2009ultrawide}, slot fed IF (inverted F) \cite{chung2010compact}, CPW fed monopole\cite{koohestani2012influence, koohestani2014novel, nikolaou2007compact, liang2005study, kim2005design}, U-shaped elliptical slot \cite{li2006study}, patch array with energy band gap (EBG) structures \cite{abbosh2009miniaturized} and CPW fed fractal antenna \cite{sedghi2014fabrication} have been proposed. Also the effects of printed UWB antenna miniaturization on transmitted time domain pulse fidelity and pattern stability have been recently discussed in \cite{liu2014effects}. In several papers \cite{abbosh2008miniaturization, ezuma2015design, chen2011miniaturization, amert2009miniaturization} it has been demonstrated that by following certain miniaturization guidelines, compact planar UWB antennas can be made. Numerous planner UWB antennas and band-notch UWB quasi-monopole antennas have illustrated symmetrical shapes\cite{kim2005design, liu2014effects, sun2010miniaturization}. In these structures, two symmetrical halves exhibits two identical strong current paths. It has been noted that the current distribution patterns in these structures replicate the conditions of magnetic mirror symmetry. Chopping off half of the symmetrical monopole antenna provides a straight forward miniaturization. It has also been explored that by having one strong current path, antenna exhibits an even wider bandwidth\cite{liu2014effects, sun2010miniaturization, mobashsher2015utilizing}, however, not all half cut UWB structures can achieve a suitable impedance matching over a wide bandwidth. A further modification to the feeding structure is required for a better impedance matching \cite{mobashsher2015utilizing}. Increaser in cross-polarization level is reported to be a major limitation in this miniaturization technique \cite{sun2010miniaturization}. Another configuration of self-complementary antenna (SCA) has stimulated a considerable attention for low profile UWB applications. Theoretically, a wideband operation is obtained when an antenna and its complement are identical, providing a constant input impedance independent of the antenna geometry and frequency. Some recent efforts have been devoted to use SCA antenna for UWB application. Antennas formed by a self-complementing semi-circular monopole \cite{guo2008study, guo2012miniature, guo2009small}, quarter-circular monopole\cite{huang2011printed, lin2011ultra}, obtuse pie-shaped monopole \cite{lin2013obtuse}, compact bow-tie \cite{lin2012compact} and $\Gamma$-shaped radiator \cite{guo2010study} has been proposed and studied. In addition to the impressive theoretical prospective of SCAs, it has to be reported that a matching network is always required to match the antenna to a 50 ohm feed line. This has limited the use of SCAs for miniaturized UWB antennas since the matching network is indispensable. In this paper, a simple procedure in aggregation with previously proposed guidelines is used to miniaturize an elliptical slot UWB planner antenna. The initial design is similar to the one proposed in \cite{li2006study} whereas the final novel design is a CPW-fed cactus shaped UWB antenna having dimensions 20 mm $\times$ 28 mm. It is worth mentioning that the final miniaturized UWB cactus antenna is rather compact in size and exhibits wider bandwidth as compared to similar structures presented in \cite{saephan2014tri, mishra2011compact, zachou2006planar, ammann2005dual} while fabricated on lower $\epsilon$${}_{r}$ material (LCP).
\subsection{Electrically Small UWB Antenna Designs}\label{ijap}
\begin{figure}[htp]
	\centering
	\begin{subfigure}{0.34\textwidth}
		\includegraphics[width=1\textwidth]{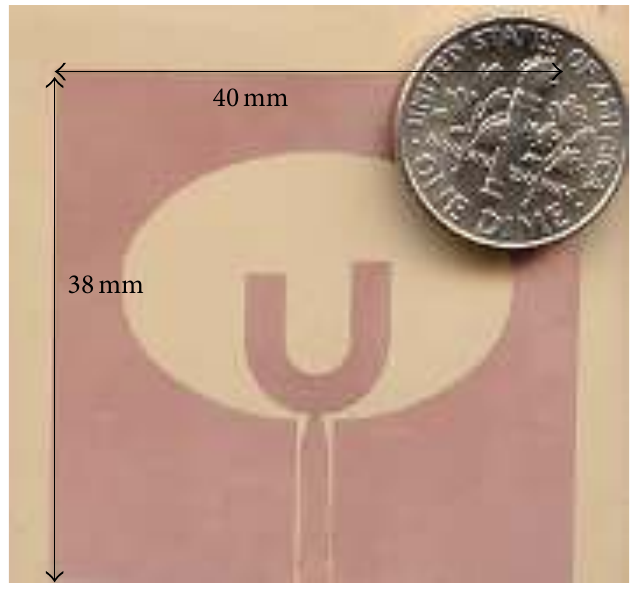}
		\caption{}
	\end{subfigure}
	\begin{subfigure}{0.32\textwidth}
		\includegraphics[width=1\textwidth]{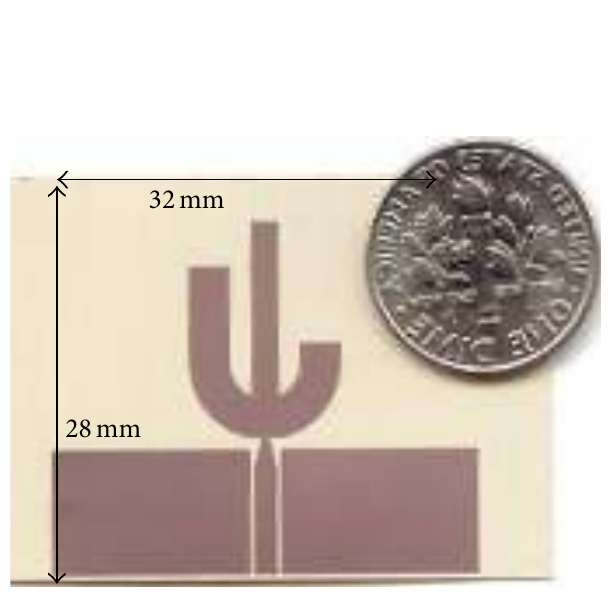}
		\caption{}
	\end{subfigure}
	\begin{subfigure}{0.22\textwidth}
		\includegraphics[width=1\textwidth]{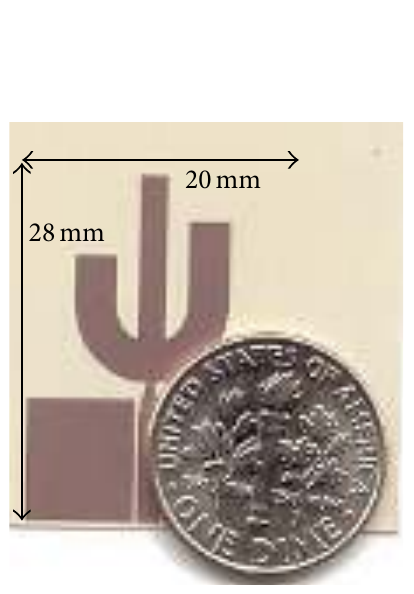}
		\caption{}
	\end{subfigure}
	\caption{Fabricated (a) CPW-fed slot antenna, (b) cactus antenna, and (c) miniaturized cactus antenna. $\textcopyright$ 2016 IJAP.}
	\label{ijap1}
\end{figure}

The UWB antennas are presented in Figure \ref{ijap1}. They are fabricated on low loss (tan$\delta$=0.002), low dielectric constant ($\epsilon$${}_{r}$=3), LCP with a copper layer that is 18 $\mu$m thick. The CPW-fed slot antenna is fabricated on a 350 $\mu$m thick substrate while for the cactus antenna and the miniaturized cactus antenna a thinner 225 $\mu$m thick substrate was used. At the early stages of the design procedure, it was observed that the 350 $\mu$m thick paper substrate exhibit rigidness. To make the cactus antenna conformal as well as miniaturized, substrate thickness was reduced. By specifying the angle and rate of LCP extrusion while manufacturing, the coefficient of thermal expansion (CTE) can be controlled. With this unique characteristic, one can engineer the thermal expansion of LCP to match with many commonly used cladding materials like silver, copper etc. \cite{thompson2004characterization}. Standard photolithography was used for the fabrication. The size reduction of the cactus antenna and the miniaturized cactus is obvious from Figure \ref{ijap1} where the fabricated prototypes are presented and compared in size with a coin.

%
The schematics of the compared antennas and the dimensions are summarized in Tables I, II and III in \cite{nikolaou2016miniaturization}. Three antennas are presented, which are the successive evolutions of the first antenna, namely CPW-fed slot antenna. The second antenna, called cactus antenna is a CPW-fed monopole UWB antenna with 41\% (28x32 mm${}^{2}$ compared to 38x40 mm${}^{2}$) size reduction, and finally the third version called miniaturized cactus is a monopole UWB antenna with even smaller RF ground patches which is 63\% (28x20 mm${}^{2}$ compared to 38x40 mm${}^{2}$) smaller than the original slot antenna. Full wave EM simulators were used for the design of the presented prototypes and for the radiation pattern and the return loss optimization. All three antennas are well matched as can be seen from the S${}_{11}$ plots.


For the CPW-fed slot, the stub dimensions and the linear tapering affect the matching, while the ellipse axes size has a small effect on the radiation patterns. The proposed antenna is fed by a CPW line with an inner conductor width, W, of 2.2 mm and a gap, g, between the ground and the inner conductor of 0.3 mm. At a distance S=9.9 mm from the board edge, the inner conductor is linearly tapered until its width becomes 0.9 mm to improve the matching between the transmission line and the U-shaped stub. The U-shaped stub consists of a semi annular ring and two linear segments. The semi annular has an outer radius R=5.5 mm and inner radius r=2.5 mm. The left linear segment has length S1=5 mm and width d=3 mm while the right linear segment has length 6 mm and width d=3mm. The center C of the semi annular ring is 4.1 mm from the ellipse center O and the ellipse center is 22 mm from the bottom edge. The ellipse has a major axis equal to L1=30 mm and secondary axis equal to L2=20 mm. Overall board dimensions are 40 mm $\times$ 38 mm. 
The evolution of the cactus antenna was based on the observation that most of the radiated energy for the CPW-fed slot antenna was confined on the tuning stub. Therefore a design was attempted without the elliptical slot. Impedance matching over the entire UWB range was not satisfactory with only the two linear segments of the U-shaped tuning stub, and to overcome this problem, a third linear segment was added in the feed line direction. The thinner LCP substrate used for the cactus antenna and the addition of the middle linear segment required the linearly tapered transition and the semi-annular segment re-optimization. Consequently, for the cactus antenna, the CPW center conductor width W is 1.78 mm and length d2 is 7.92 mm. A linear taper is used to reduce the center conductor width to d=0.61 mm and is connected to the cactus shaped stub at distance d1=10.24 mm from the board edge. The two rectangular ground patches have dimensions Gl $\times$ Gw which correspond to 9.44 mm and 14.89 mm respectively. For the primary radiator, a cactus shaped stub is used. It consists of a semi-annular ring with inner radius r=2.60 mm and an outer radius R=5.72 mm and three linear segments of different lengths. The middle linear segment is L2=13.00 mm long and W2=2.08 mm wide while the left and right segments are L1= 7.28 mm and L3=1.56 mm long respectively. Both of them are 3.12 mm wide. From the bottom part of the semi-annular ring a circular sector is detached leaving a chord of length S=2.73 mm. 
The third and final evolution of UWB antenna, the miniaturized cactus is based primarily on the intermediate design, and the objective was to decrease the size of the two ground patches. Careful design allowed the decrease of the ground patches to an overall size of 9 mm $\times$ 8 mm which correspond to Gl and Gw respectively. The description of the miniaturized cactus design is similar to the presented description for the cactus antenna. As a result of the ground size reduction, further tuning was needed for the three stubs that consist the cactus shaped radiating element. 

The overall board dimensions for the cactus antenna is 32 mm $\times$ 28 mm, resulting in a 41\% reduction in area compared to the CPW-fed slot while the miniaturized cactus has overall board dimensions 28 mm $\times$ 20 mm resulting in 63\% size reduction compared to the slot antenna. For the antenna dimensions, the common variables' names are set independently for each antenna schematic and must not be related.

\subsection{Miniaturization Procedure}\label{ijapsection}
The size reduction was envisioned by the investigation of the surface current distribution on the slot antenna. It was observed that the radiation was primarily caused by the current distribution on the U-shaped stub (Figures 4a, 4b), although the elliptical slot also contributes, to a lesser extent. The surface current distributions on the two cactus antennas (Figures 4c-4f) have a similar form with the one on the U-shaped stub, something that explains the similarity in the resulted radiation patterns. Based on the surface current distribution observations, and trying to improve the matching the even U-shaped stub (Ant1 from Figure \ref{ijap5}) was replaced with an uneven U-shaped stub. The added perturbation on the tuning stub added one design degree of freedom that allowed the improvement of the matching as can be seen in Figure \ref{ijap6}. The uneven U-shaped slot, which is presented in Figure \ref{ijap5} under the name Ant2 had improved matching as can be seen in the S${}_{11}$ plots of Figure \ref{ijap5}. In the next iteration (Ant3) the slot was removed, and in order to further improve the matching for the remained U-shaped stub, a third tuning stub was added, along the direction of the feed line, resulting in the cactus shaped radiator (Ant4) that evolved eventually, after some additional tuning, to the miniaturized cactus antenna. This third middle stub allowed for an additional design parameter and as a result of its bigger length the matching in the lower end of the UWB range, in the area around 3.1 GHz could be improved. The matching improvement in the lower frequency end, is evident in Figure \ref{ijap5}, and the presented frequency notch that can be seen in Figure \ref{ijap7} (red dotted line) as a result of the additional third stub, can be easily suppressed with the careful selection of the stub size L2.

\begin{figure}[htb]
	\centering
	\begin{subfigure}{0.37\textwidth}
		\includegraphics[width=1\textwidth]{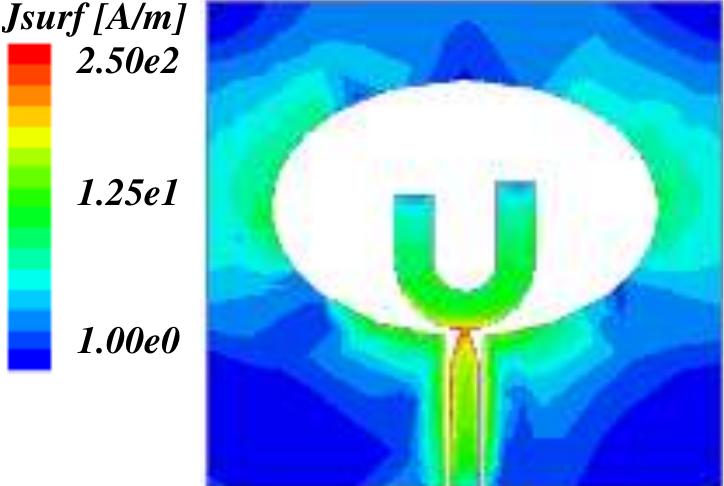}
		\caption{}
	\end{subfigure}
	\begin{subfigure}{0.37\textwidth}
		\includegraphics[width=1\textwidth]{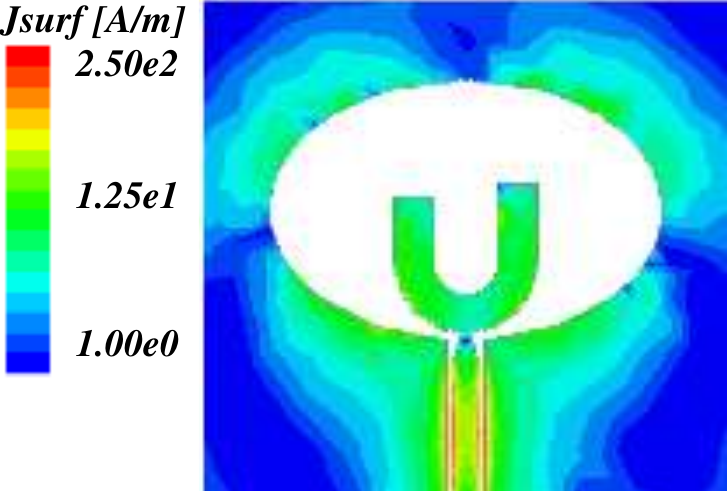}
		\caption{}
	\end{subfigure}
	\begin{subfigure}{0.37\textwidth}
		\includegraphics[width=1\textwidth]{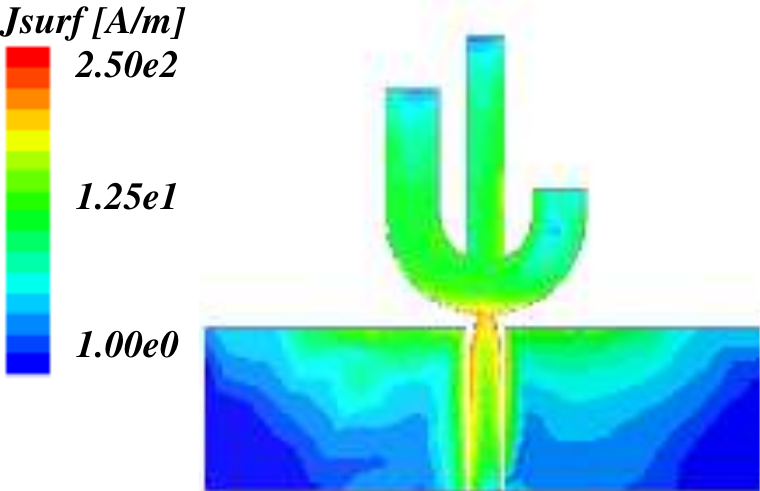}
		\caption{}
	\end{subfigure}
	\begin{subfigure}{0.37\textwidth}
		\includegraphics[width=1\textwidth]{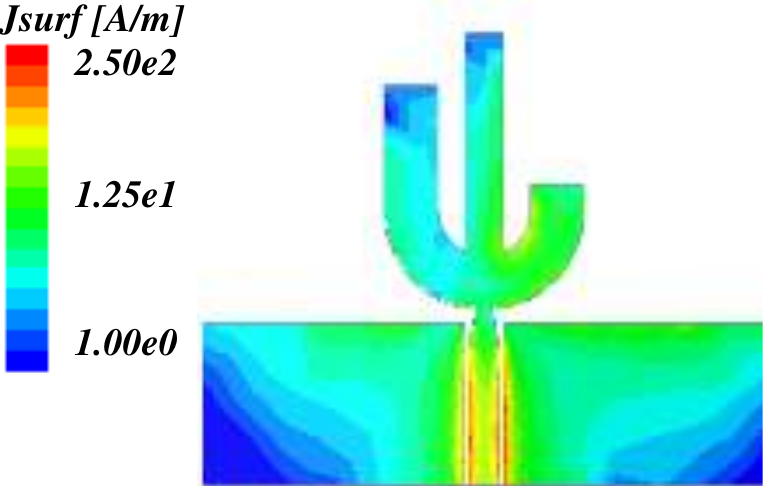}
		\caption{}
	\end{subfigure}
	\begin{subfigure}{0.37\textwidth}
		\includegraphics[width=1\textwidth]{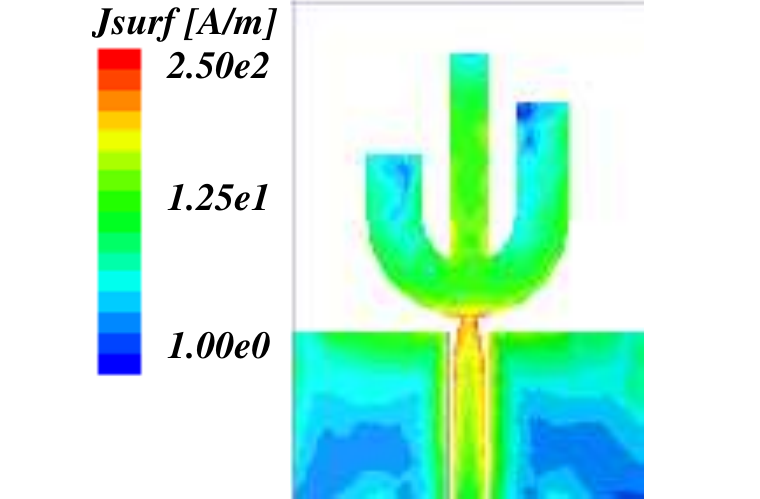}
		\caption{}
	\end{subfigure}
	\begin{subfigure}{0.37\textwidth}
		\includegraphics[width=1\textwidth]{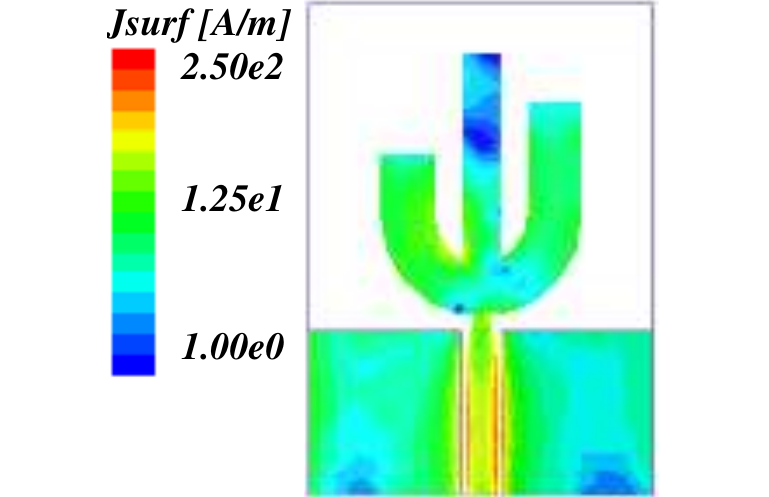}
		\caption{}
	\end{subfigure}
	\caption{Surface current (J) distributions on CPW-fed slot antenna at (a) 5 GHz (b) 9 GHz, cactus antenna at (c) 5 GHz (d) 9 GHz and miniaturized cactus antenna at (e) 5 GHz and (f) 9 GHz. }
	\label{ijap4}
\end{figure}

\begin{figure}[htb]
	\centering
	\includegraphics[width=0.85\textwidth]{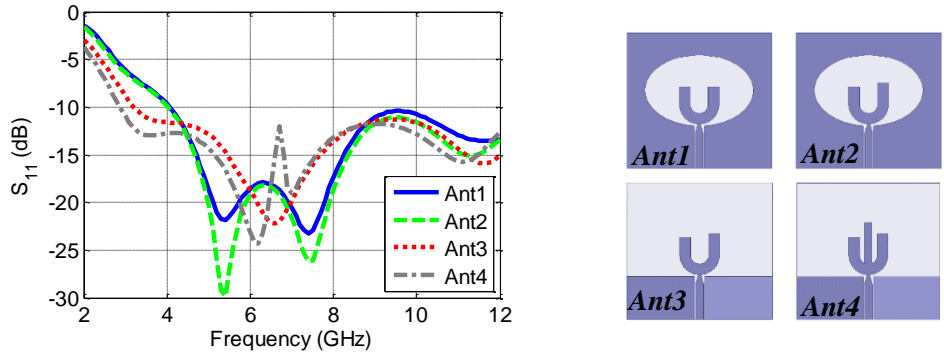}
	\caption{Return loss of Miniaturization process}
	\label{ijap5}
\end{figure}

The miniaturization process so far led to the Ant4 structure shown in Figure \ref{ijap5}. This structure was further optimized and final structure is presented as cactus antenna in Figure \ref{ijap1}(b). However the overall antenna size could be further improved by attempting a ground patch reduction in addition to the removal of the elliptical slot. The idea was also based on the study of the surface current distribution of the cactus antenna (Figures \ref{ijap4}(c) and (d)) where the current intensity along the outer edges of the rectangular ground patches is clearly lower than the current intensity on the edges closer to the signal line. Parametric study of the width of the ground patches (Gw) showed only little effect on the S${}_{11}$ plots (Figure \ref{ijap8}) and the optimization steps resulted in the miniaturized cactus version depicted in Figure \ref{ijap1}(c), with Gw equal to only 8 mm, and overall board dimensions 20 mm $\times$ 28 mm, which is equivalent to 63\% size reduction compared to the original design of the CPW-fed slot antenna. 

\begin{figure}[htb]
	\centering
	\includegraphics[width=0.7\textwidth]{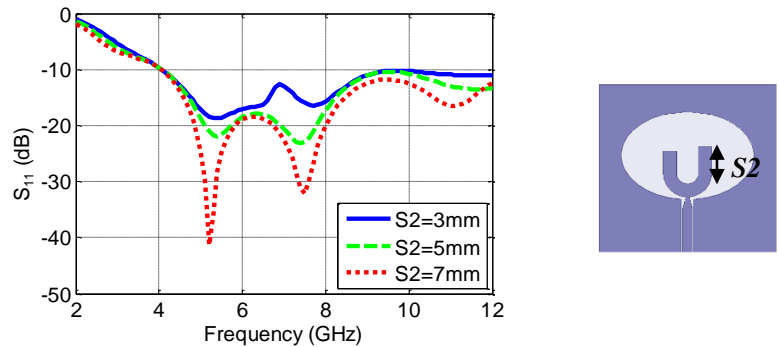}
	\caption{Return loss with \textit{S2} variation}
	\label{ijap6}
\end{figure}

\begin{figure}[htb]
	\centering
	\includegraphics[width=0.7\textwidth]{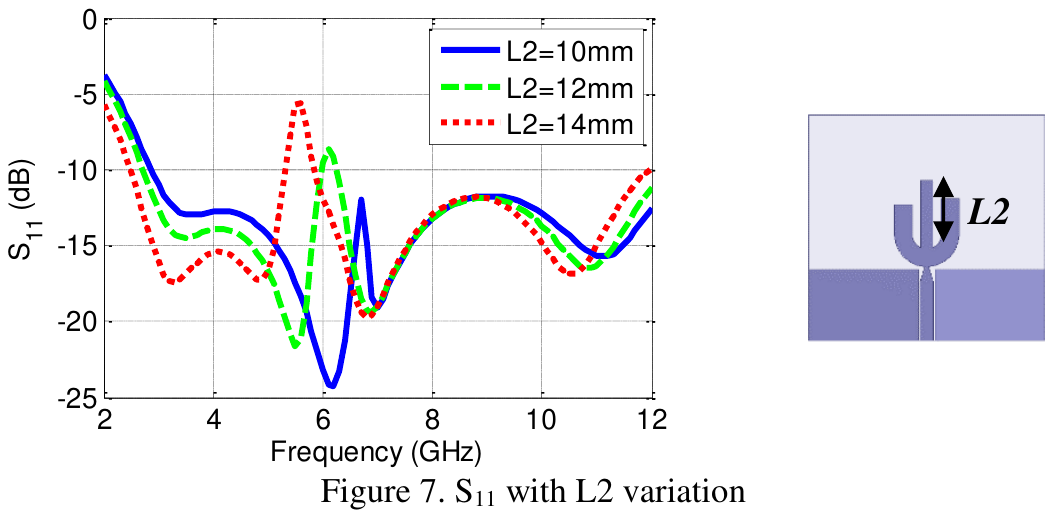}
	\caption{Return loss with \textit{L2} variation}
	\label{ijap7}
\end{figure}

\begin{figure}[htb]
	\centering
	\includegraphics[width=0.7\textwidth]{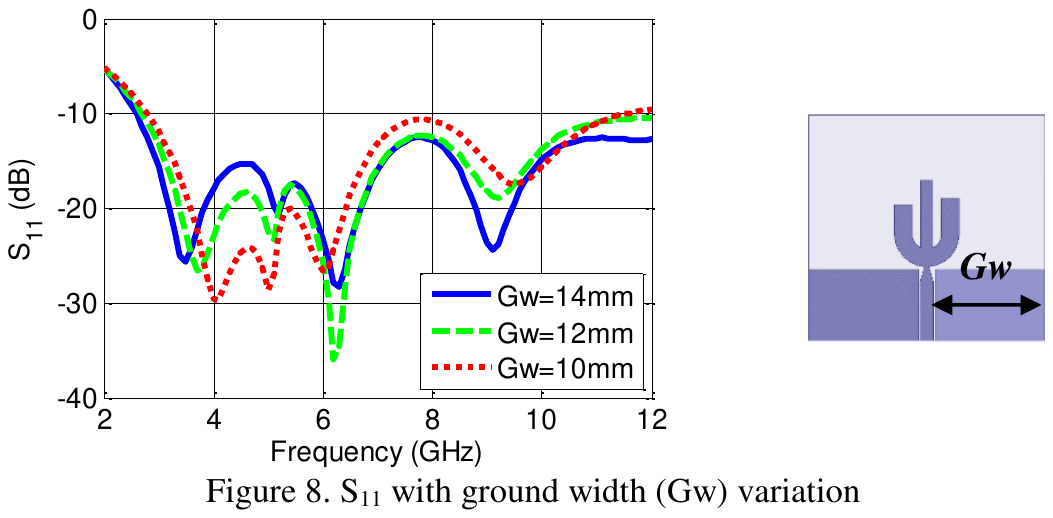}
	\caption{Return loss with \textit{Gw} variation}
	\label{ijap8}
\end{figure}


The simulated and measured return loss for both, cactus antenna (Figure 3b) and miniaturized cactus antenna (Figure 3c) is obviously better, especially at the two ends of the frequency range, with a better than -10 dB return loss from 2.9 GHz to 12 GHz that overlaps the designated UWB range. Three resonances dominate the return loss, for the cactus antenna these appear at 3.7, 5.1 and 6.4 GHz one for each linear segment. Generally the longer the stub is, the lower the corresponding resonance appears. This can be seen in Figure \ref{ijap7} where the simulated S${}_{11}$ is plotted for three different length values (L2) of the longest linear segment. The matching at the higher frequencies is affected by the rectangular ground patches' width Gw as can be seen in Figure \ref{ijap8} where S${}_{11}$ is plotted for three different Gw values for the miniaturized cactus antenna. It was concluded that the width of the ground patch cannot be smaller than 8 mm without compromising matching at higher frequencies and radiation patterns consistency, although it would be highly desired for an even more compact design. 

For the presented S${}_{11}$ plots there is a small discrepancy between the simulated and measured results. This is partly due to the fact that the UWB range is large compared to the central frequency and is difficult for the frequency domain simulation tools to give accurate results over the whole band. Moreover the size of the SMA connector which is significant compared to the size of the antennas causes additional discrepancy between measurements and simulated results, for which a CPW mode excitation port was used.

\subsection{Radiation Performance}

\begin{figure}[htb]
	\centering
	\begin{subfigure}{0.4\textwidth}
		\includegraphics[width=1\textwidth]{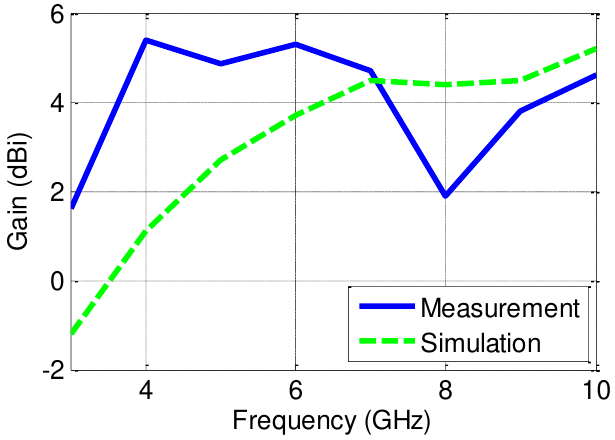}
		\caption{}
	\end{subfigure}
	\begin{subfigure}{0.4\textwidth}
		\includegraphics[width=1\textwidth]{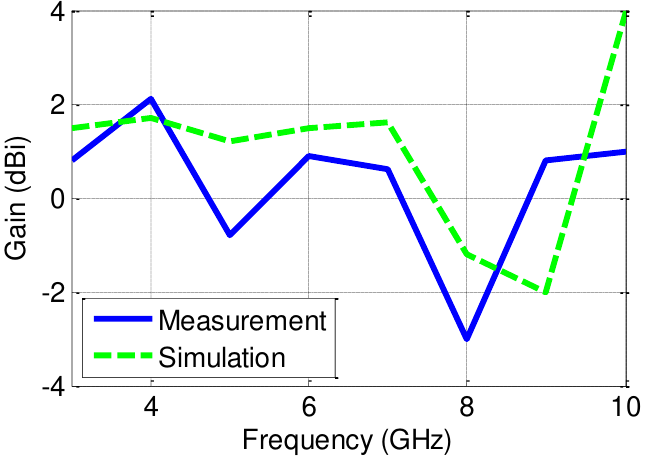}
		\caption{}
	\end{subfigure}
	\begin{subfigure}{0.4\textwidth}
		\includegraphics[width=1\textwidth]{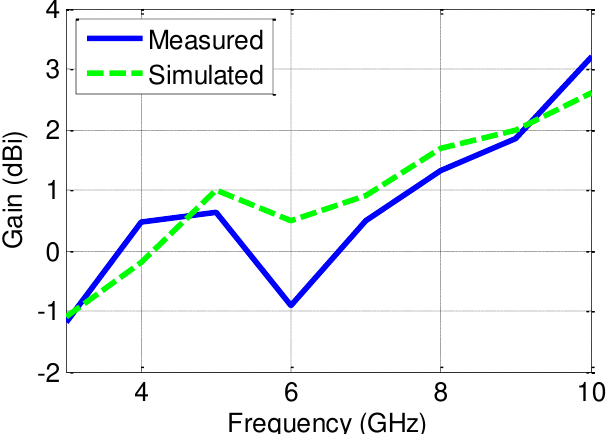}
		\caption{}
	\end{subfigure}
	\begin{subfigure}{0.4\textwidth}
		\includegraphics[width=1\textwidth]{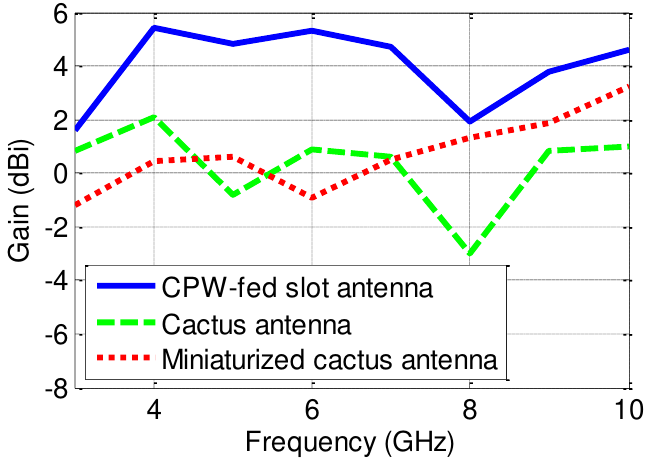}
		\caption{}
	\end{subfigure}
	\caption{Simulated and measured gain comparison for (a) CPW -fed slot antenna (b) Cactus antenna (c) Miniaturization cactus antenna and (d) Comparison of measured gain for all three antennas}
	\label{ijap11}
\end{figure}

Measured and simulated radiation patterns for all three antennas at 5 and 9 GHz, which are representative of the patterns across the frequency range, are presented in \cite{nikolaou2016miniaturization}. The radiation pattern figures presents the \textit{E}-plane (\textit{x-z}) co-polarization, where $\theta$=\textit{0}${}^\circ$ corresponds to \textit{z}-axis and $\theta$=\textit{90}${}^\circ$ corresponds to the \textit{x}-axis. It is seen that for all three antenna designs the \textit{E}-plane has a null along the \textit{x}-axis due to the feed line, and a pattern that is nearly symmetric around the \textit{x}-axis. The \textit{H}-plane (\textit{y-z}) co-polarization plots are presented in \cite{nikolaou2016miniaturization}, where $\theta$=\textit{0}${}^\circ$ is the \textit{z}-axis and $\theta$=\textit{90}${}^\circ$ is the \textit{y}-axis. It is seen that the \textit{H}-plane patterns for both cactus antenna designs are almost perfectly omni-directional at 5 GHz, and mostly omni-directional at 9 GHz, however especially at 9 GHz the slot antenna \textit{H}-plane pattern flattens along horizontal axis. This somewhat directional behavior is verified by the gain measurements which are taken along the \textit{z}-axis direction shown in Figure \ref{ijap11}. As can be deduced from Figure \ref{ijap11}(a) the gain at 5 GHz and 9 GHz for the slot antenna is 5 dBi and 4 dBi respectively. The evident discrepancy between simulated and measured peak gain shown in Figure \ref{ijap11}(a) can be explained by relatively more directive measured \textit{E}-plane pattern when comparing with the simulated \textit{E}-plane pattern. A directive beam with a maxima at \textit{37}${}^\circ$ was observed in measured \textit{E}-plane pattern resulting in a 2.2 dBi higher peak gain value when compared with the simulated predictions. This more directive measured pattern can be directly relate to fabrication anomalies. Both cactus shaped antennas maintain almost perfectly omni-directional radiation patterns which is also verified from the gain plot which is close to 0 dBi. Especially the miniaturized cactus antenna in addition to its compact size, presents rather constant gain which improves the fidelity of transmitted time domain fast pulses \cite{liu2014effects}. The additional size of the slot antenna, as a result of the included elliptical slot, makes the antenna more directive, and for some applications this could be an advantage. However considering that most applications involve mobile hand-held devices omni-directional characteristics can be an overall advantage for a UWB antenna.



\section{Miniaturized Cactus-shaped Antenna}\label{cac}
This section presents an in depth study on novel cactus-shaped Ultra Wideband (UWB) monopole antenna fabricated on Liquid Crystal Polymer (Figure \ref{ijap1}(c)) (LCP) as a next step to the miniaturization of large size UWB antenna. The proposed antenna is a very compact design since it can be fabricated on a board with dimensions only 20$\times$28 mm${}^{2}$, while the three linear segments that comprise the cactus-shaped monopole provide a direct control on antenna matching. The proposed antenna is operating from 2.85 GHz to 11.85 GHz and it presents very consistent omni-directional patterns throughout the UWB frequency range. Return loss and pattern measurements are presented and the operation principles are discussed in detail. The simplicity of this topology, with the easily controllable return loss, allows for its easy implementation for various UWB sub-band designs, just by building suitable monopole versions, for which the only difference is the length of the three linear segments.

Since the FCC \cite{federal2002first} regulated the 3.1--10.6 GHz band for UWB applications, a significant amount of research activity has been recorded in the area of the design and implementation of UWB antennas. Depending on the application, the requirements for antenna designing may vary significantly, however, for wireless applications, especially for mobile handheld devices, the small size and the omni-directional pattern are highly demanded. For the omni-directional pattern requirements the solution of the printed monopole has been proven very popular and adequately efficient. A circular disk monopole proposed in \cite{siddiqui2015compact, siddiqui2014compact} provides consistent omni-directional pattern, however it does not provide any control on the return loss that would potentially allow the operation of the antenna in desired UWB sub-bands (lower and higher). A direct control on the return loss and therefore on the radiated frequencies is achieved by the CPW-fed hexagonal antenna presented in \cite{wu2015compact} and the composite right/left-handed (DCRLH) transmission line loaded antenna \cite{li2015cpw}, which are used for multi-band applications. CPW-fed monopole and electromagnetic band-gap (EBG) combination antenna that features multi-band and wideband behavior is introduced in \cite{de2012dual, de2013polypropylene}, while CPW-fed monopole with parasitic circular-hat patch is used for the broadband operation in \cite{liu2011parasitically} and the L-shaped monopole antenna with hexagonal slot is demonstrated in \cite{zhou2011cpw}. However, none of the aforementioned antennas covers the whole UWB range. Microstrip line fed Fork-shaped \cite{mishra2011compact}, U-shaped \cite{tang2014compact}, pentagon-shaped \cite{moody2011ultrawide} and CPW-fed bowled-shaped \cite{koohestani2012influence} planar monopole antennas that cover the whole UWB range have been presented in recent years, however all of these antennas are almost two times the size of the proposed cactus-shaped monopole.

In next sections, a compact cactus-shaped monopole UWB antenna is proposed. The presented prototype has board dimensions, 20$\times$28 mm${}^{2}$, is fabricated on low $\epsilon$${}_{r}$ flexible organic dielectric material (LCP with $\epsilon$${}_{r}$=3),${}^{ }$presents consistent omni-directional patterns in \textit{H}-plane, and allows direct control on the return loss. The radiation mechanism and the wide band operation are explained in detail, while return loss, pattern and gain measurements are presented to verify its performance.

\subsection{Geometrical Configuration}
\begin{figure}[htb]
	\centering
	\includegraphics[width=0.8\textwidth]{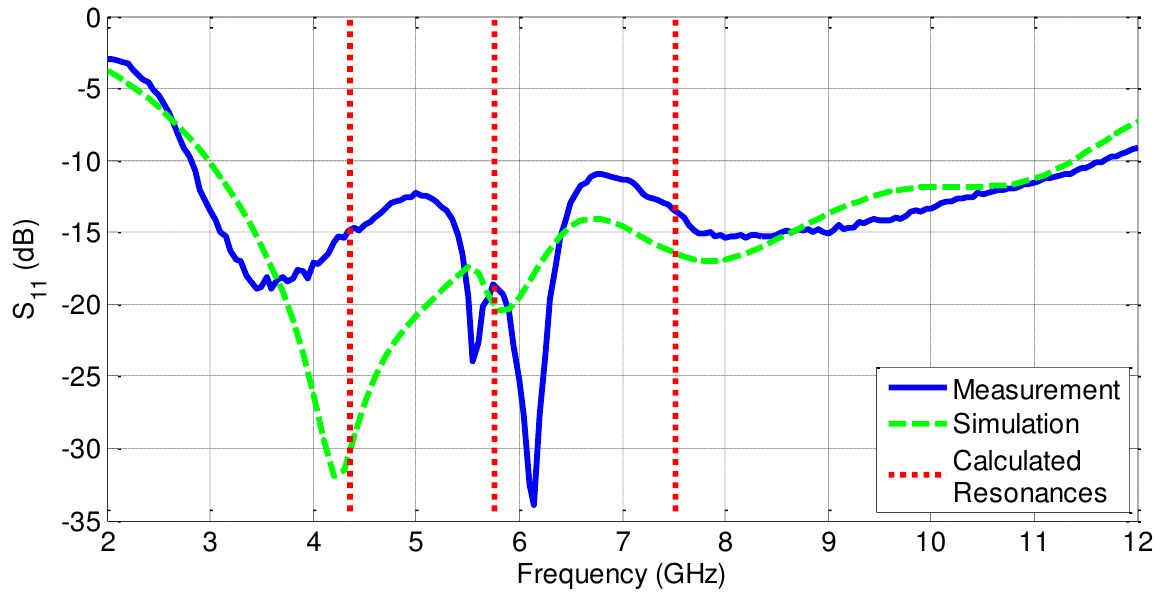}
	\caption{Measured and simulated return loss. Red dotted line shows the calculated resonance frequency using proposed equation. $\textcopyright$ 2016 IEEE.}
	\label{cac2}
\end{figure}

As discussed in subsection \ref{ijapsection}, the antenna consists of a CPW line with a linearly tapered broadband transition terminated with a semi-annular ring resonator. Three uneven linear segments are extended from both ends and the center of the semi-annular ring, along the direction of the feed line. Those three different-size linear segments are approximately $\lambda$/4 long for three intermediate frequencies at which major resonances occur, in the return loss plot of Figure \ref{cac2}. In a transmission line equivalent circuit, the tapered segment with the cactus-shaped radiator can be considered as a broadband load that terminates a typical CPW line with length H1=7.92 mm. If the equivalent load at the end of the CPW line was equal to the characteristic impedance of the line throughout the whole frequency band of operation, the H1 length would make no difference in the resulted return loss. However, since the impedance of the load changes over frequency, the dimension H1 is critical in order to achieve good matching. A linearly tapered segment is used as a transition between the CPW line and the radiator. This size of the tapered segment was chosen after a parametric sweep analysis was conducted, in order to achieve the best matching behavior between the CPW line and the ring. From the bottom part of the semi-annular segment, a circular sector is removed resulting in a chord of length C=2.73 mm. Since the most important parameter is the length of each stub, the width of the central stub was chosen to be narrower than the width of the side stubs, aiming to increase the distance and therefore reduce the coupling between the neighboring stubs. The overall dimensions of the fabricated antenna are only 20$\times$28 mm${}^{2 }$and the dimensions of schematic are summarized in \cite{nikolaou2017design}. 

\section{Principle of Operation} \label{caceq}

The bandwidth of the circular planar monopoles is larger than any other planar monopole configuration \cite{kumar2003broadband}. The main advantage of the circular planar monopole over the rectangular planar monopole can be interpreted in terms of its various resonant modes that are closely spaced, hence supportive for wideband operations like UWB. For a planar monopole, the lower resonant frequency can be approximately calculated by equating its area \textit{L${}_{monopole}$} $\times $ \textit{W${}_{monopole}$} to that of an equivalent cylindrical monopole antenna of same height\textit{ h${}_{eq}$}, and equivalent radius\textit{ r${}_{eq}$}, as described below \cite{kumar2003broadband, garg2001microstrip}:

\begin{equation}\label{cace1}
r_{eq} =\frac{W_{monopole} }{2\pi } _{} 
\end{equation}
\begin{equation}\label{cace2}
h_{eq} =L_{monopole} {} 
\end{equation}

The input impedance of a $\lambda$/4 monopole antenna is half of that of the $\lambda$/2 dipole antenna. Thus, the input impedance of an ideal monopole is 36.5+\textit{j}21.25$\Omega$. To make this impedance real, a slightly smaller length of the monopole is used \cite{balanis2016antenna}. Where,

\begin{equation}\label{cace3}
L_{monopole} =0.24\lambda F
\end{equation}

And,

\begin{equation}\label{cace4}
F=h_{eq} /(h_{eq} +r_{eq} )
\end{equation}

Equations \ref{cace3} and \ref{cace4} yield:

\begin{equation}\label{cace5}
\lambda =(h_{eq} +r_{eq} )/0.24
\end{equation}

So, the resonance frequency for a planar monopole can be approximated by:

\begin{equation}\label{cace6}
f=c/\lambda =c\times 0.24/(h_{eq} +r_{eq} )
\end{equation}

This equation does not include the length of the tapered feed, which affects the length of the antenna and consequently the resonant frequency. Accordingly, this equation is modified for the proposed design by incorporating feed length to form: 

\begin{equation}\label{cace7}
f_{resonance} =\frac{c\times 0.24}{h_{eq} +r_{eq} +(d1-Gl)} Hz
\end{equation}

\begin{figure}[htb]
	\centering
	\includegraphics[width=0.9\textwidth]{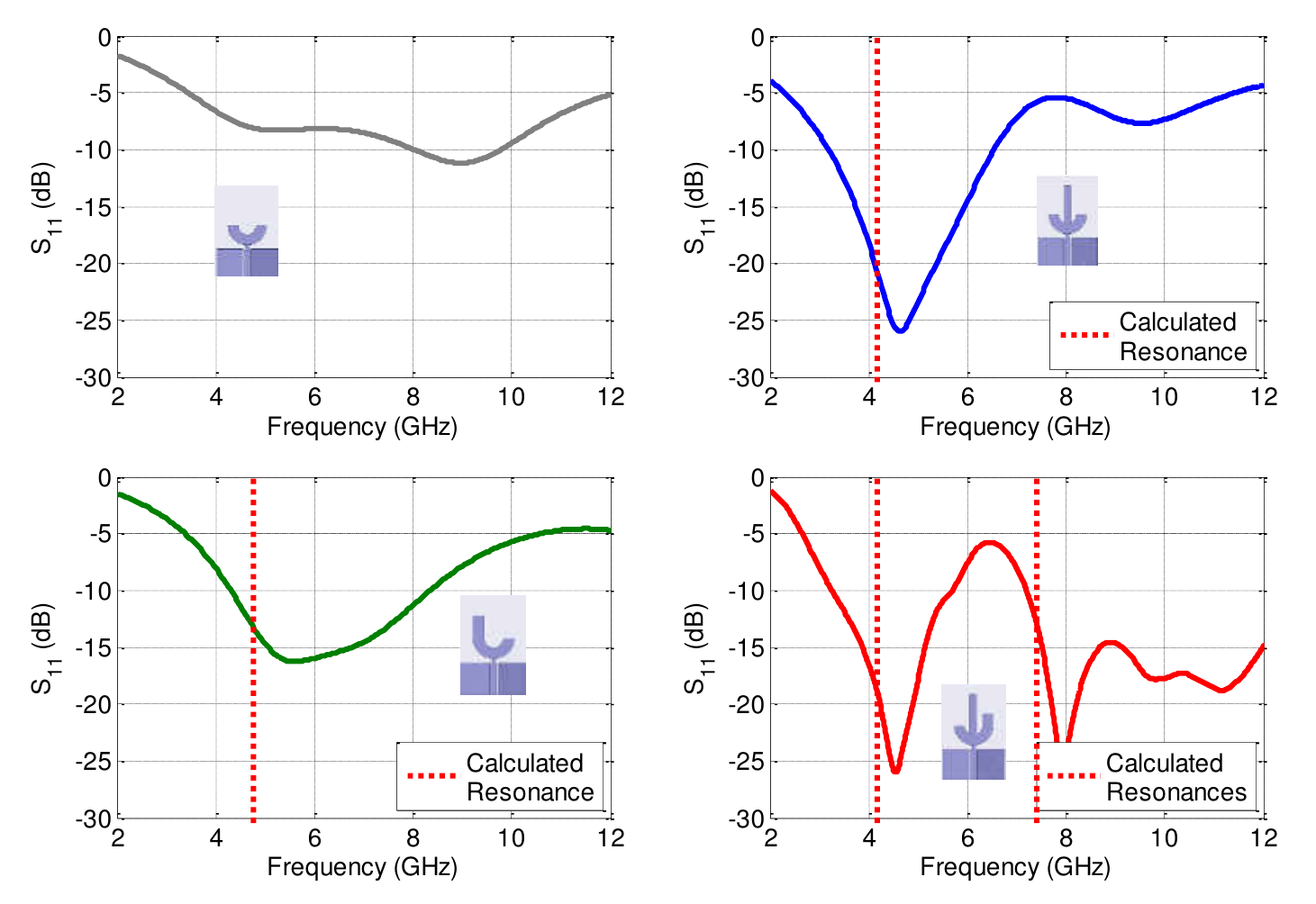}
	\caption{Modified versions of the cactus-shaped monopole operating at selected UWB sub-bands when \textit{L1}=\textit{L3}=0, \textit{L2}=12 (top left), \textit{L1}=6.5, \textit{L2}=\textit{L3}=0 (top right), \textit{L1}=\textit{L2}=\textit{L3}=0 (bottom left) and \textit{L1}=0, \textit{L2}=11.5, \textit{L3}=3.5 (bottom right) when all values in mm. $\textcopyright$ 2016 IEEE.}
	\label{cac3}
\end{figure}

Where \textit{c} is the speed of light in \textit{m/s}. The three rectangular segments have been designed based on equation \ref{cace7}. To effectively combine the three linear segments in a single UWB monopole antenna a transition is needed from the CPW feed line. For that scope, a semi-annular ring is used to connect the two side segments. The radius R of the ring is defined using equation \ref{cace7} presented in \cite{mishra2011compact} where instead of the minimum frequency \textit{f${}_{L}$} a higher, closer to the UWB range center frequency, \textit{f${}_{R}$}=5 GHz was used, that allowed the miniaturization of the semi-annular ring and therefore the miniaturization of the proposed UWB cactus-shaped monopole.

\begin{equation}\label{cace8}
f_{R} =\frac{7.2}{2.25R+(d1-Gl)} GHz
\end{equation}

The design dimensions for the three linear rectangular segments were chosen such as to cover the whole UWB range. The semi-annular ring is used to provide a platform to host these three linear rectangular segments and at the same time provide a smooth and effective transition from the CPW feed line. It has to be noted that the \textit{h${}_{eq}$} value of the left and right radiators comprise of the combination of L1 or L3 respectively, with a portion of the attached semi-annular ring. The effective lengths of left and right segments correspond, and therefore control the highest and intermediate frequencies of UWB spectrum. The calculated values using equation \ref{cace1} and \ref{cace2} for the left segment (\textit{h${}_{eq}$}=9.02 mm, \textit{r${}_{eq}$}=496 µm) and for the right segment\textit{ }(\textit{h${}_{eq}$}=12.02 mm,\textit{ r${}_{eq}$}=496 µm) correspond to 7.56 GHz and 5.75 GHz, respectively. The effective length of the central segment comprises of a rectangular strip with dimensions $L2\times W2$ and the portion of the semi-annular ring connecting it to the antenna feed line. The central segment length affects the lower frequencies of the return loss (Figure \ref{cac2}). The calculated values using equation \ref{cace1} and \ref{cace2} (\textit{h${}_{eq}$}=16.13 mm, \textit{r${}_{eq}$}=331 µm) for the middle segment yields the lowest resonance frequency of the cactus-shaped monopole at 4.37 GHz. 

\begin{figure}[htb]
	\centering
	\includegraphics[width=0.7\textwidth]{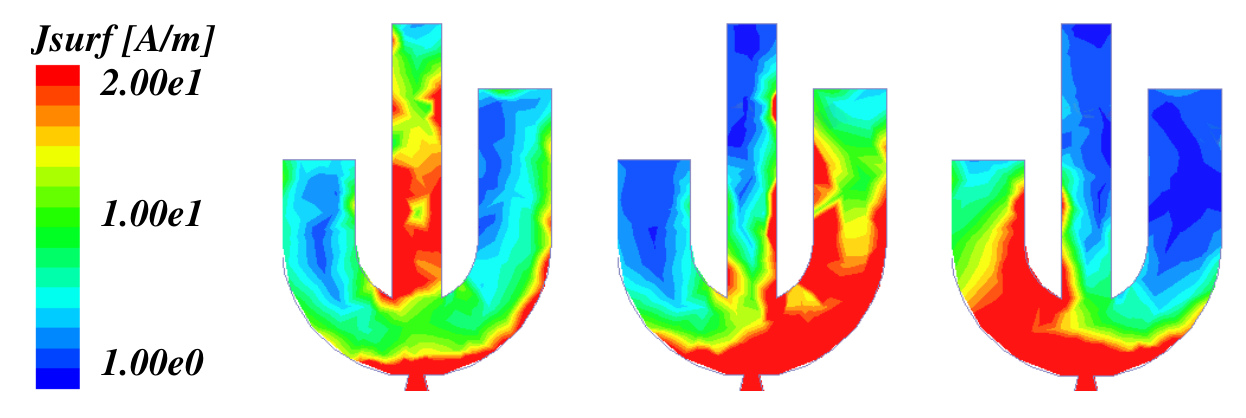}
	\caption{Surface current distribution on cactus-shaped radiator at 4.20 GHz (left), 5.90 GHz (center) and 7.80 GHz (right). }
	\label{cac4}
\end{figure}

\begin{figure}[htb]
	\centering
	\includegraphics[width=0.9\textwidth]{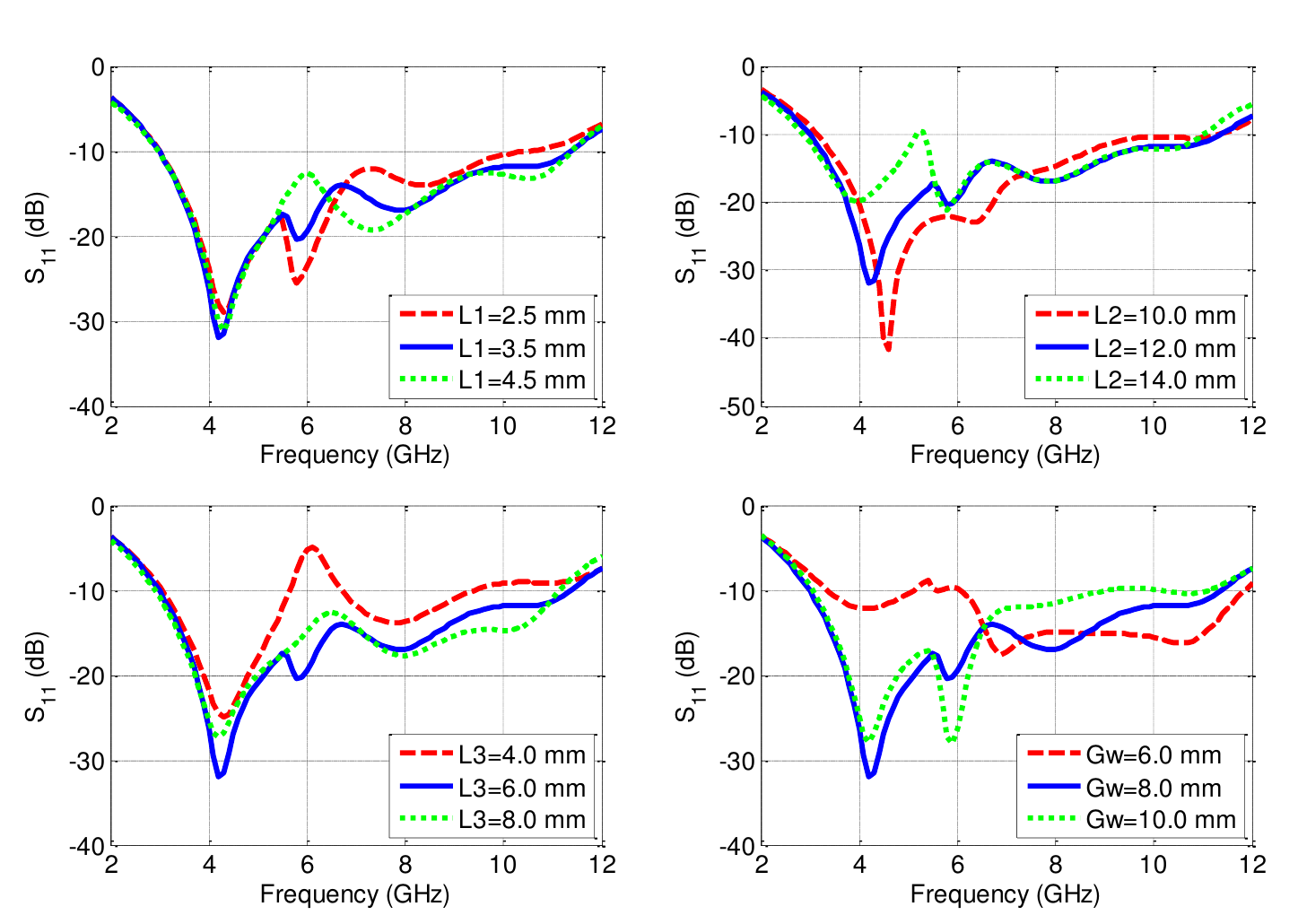}
	\caption{Parametric study for the design parameters effect on the return loss. left monopole effect (top left) middle monopole effect (top right) right monopole effect (bottom left) effect of ground plane (bottom right). In each plot other than the varied parameter the other parameters have same values. $\textcopyright$ 2016 IEEE.}
	\label{cac5}
\end{figure}

\begin{table}[]
	\centering
	\caption{Comparison of simulated and calculated resonance frequencies}
	\begin{tabular}{l|llll}
		\cline{1-4}
		Radiator & Calculated fresonance & Simulated fresonance & Error & \\ \cline{1-4}
		Left monopole & 7.56 GHz & 7.85 GHz & 3.6\% & \\
		Central monopole & 4.37 GHz & 4.20 GHz & 3.8\% & \\
		Right monopole & 5.75 GHz & 5.90 GHz & 0.9\% & \\ \cline{1-4}
	\end{tabular}
	\label{cact1}
\end{table}

All three calculated resonance frequencies are indicated in Figure \ref{cac2} along with simulated return loss in comparison with the measured results. The proposed antenna was designed using full wave EM simulators. Three resonances can be observed from the simulated and measured return loss plot. Each resonance mainly depends on the length of the corresponding linear segment. The comparison of the calculated and simulated resonance frequencies for these segments is summarized in Table \ref{cact1}. The combined dimensions of the tapered line, ground, semi-annular ring, along with the strategically calculated resonances of each radiating stub, ensure good impedance matching for the antenna throughout the UWB frequency range. 

The imminent effect of the lengths L1, L2 and L3 -as implied by equation \ref{cace6}, on the resonances of the, thereby, easily controllable return loss, allows the modification of the presented antenna by only changing these three lengths, in order to design antennas that may be used for selected UWB sub-bands. The common platform and three distinct cases, are presented in Figure \ref{cac3} where lengths L1, L2 and L3 were chosen such as to match the antennas in different sub-bands. Figure \ref{cac3}(top left) shows the return loss, when all the lengths are set to zero, creating a semi-annular ring monopole which in clearly not well matched. Using this common platform, a lower sub-band (Figure \ref{cac3}(top right)), a middle sub-band (Figure \ref{cac3}(bottom left)), and a UWB antenna with a band-stop frequency notch, (Figure \ref{cac3}(bottom right)) can be designed. In all three cases only lengths L1, L2 and L3 are changed and every other design dimension remains the same. As can be observed in Figure \ref{cac3}, the predicted resonances from equation \ref{cace6} (dotted lines) that fall within the UWB range are in good agreement with the simulated resonant frequencies.

The surface current distribution on the antenna is presented in Figure \ref{cac4} where the higher tone (red in online colored version) represents the strongest surface current. It is obvious from the surface current plot at 7.80 GHz, that most of the power is radiated from the left stub of the antenna, while at 5.90 GHz the right stub radiates the maximum power. The longest linear segment in the middle of the antenna is most critical for the radiation at the lowest resonance frequency centered at 4.20 GHz.

The effect of the ground patch length and the effect of each of the three linear segments' length, were parametrically studied, and the results are compiled in Figure \ref{cac5}. The contour with the solid line in all four plots in Figure \ref{cac5}. presents the optimized design's simulation results. The width of the ground plane, S2 directly affects the impedance matching at the lower frequencies, when S2 = 6 mm. Similar mismatch is observed at higher frequencies when S2 = 10 mm. The optimized value of S2 is 8 mm, for which the antenna is matched everywhere in the UWB band, as can be seen in Figure \ref{cac5}(top left). Figures \ref{cac5}(top right, bottom right and bottom left) reveal the impact of each stub's length on the position of the respective resonance. In general, the increase in the length of a stub causes the related resonance to shift towards lower frequencies. This tendency, remains consistent as can be seen in every one of the three plots mentioned above. The same behavior has been theoretically predicted from equation \ref{cace6}. One of the most common issues in the resonances distribution approach, for wide band antenna designs, is the `collapse of resonance'. This collapse often reveals a peak in the return loss plot at the frequency band where the resonance previously existed. This phenomenon can be observed in Figure \ref{cac5}(top right) for L1 = 4.5 mm. A similar mismatch problem is often observed when the two resonances occur too far apart from each other. Such case can be witnessed in Figure \ref{cac5}(top right) for L2=14 mm in which the first resonance shifts to lower frequencies, resulting in a peak of return loss between the first two resonances. The most intense mismatch peak can be observed from Figure 5(bottom left) when L3 = 4 mm. The reason behind this peak is the position of the second resonance far away from the first resonance and its merge with the third resonance, eventually resulting in significant mismatching of the {\textbar}S${}_{11}${\textbar} around 6 GHz. The size of the ground plane (S2) is an important parameter for the overall matching of the antenna as it affects the position as well as the depth of each resonance. On the other hand, the length of each stub does not significantly affect the depth or the position of the resonances created by the other two stubs. It is evident that while a certain stub only affects its corresponding resonance, the ground patches and the cactus-shaped radiator should be considered as a single object, in order to achieve good matching results throughout the complete UWB frequency range.

The design allows the interested designer to achieve good radiation performance in different UWB sub-bands by only varying the lengths of the three consisting monopoles while keeping every other design feature (feed line, ground sizes, semi-annular ring transition, substrate) the same, and in addition the resonances that dictate the return loss are analytically predicted. This is how the antenna design allows the claimed "easily controllable return loss".

\begin{figure}[htb]
	\centering
	\includegraphics[width=0.9\textwidth]{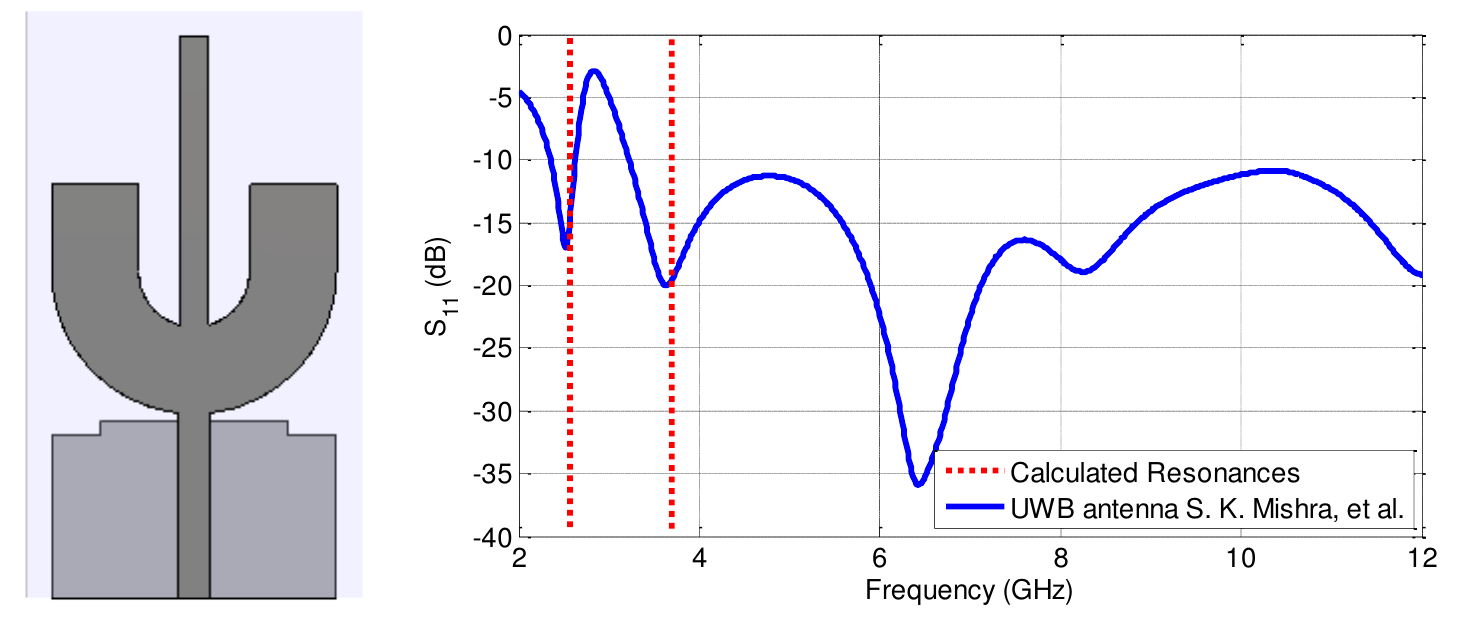}
	\caption{Verification of equation \ref{cace7}.}
	\label{cacmishra}
\end{figure}

Equation \ref{cace7} independently can not be directly used as a method to calculate the resonant frequencies. The equation \ref{cace7} was used in \cite{mishra2011compact} only in order to calculate the radius R of the semi-annular ring starting for the minimum frequency \textit{f${}_{L}$} for which ({\textbar}S${}_{11}${\textbar}$<$-10 dB). Actually the used \textit{f${}_{L}$} was the Bluetooth frequency that lies outside the UWB range. The parametric analysis presented in Figure \ref{cac5} shows that the resonant frequencies are controlled by the length of the linear segments and this parameter is not included in equation \ref{cace7} that uses R instead. It is equation \ref{cace6} that predicts the resonances, equation \ref{cace6} is not used in \cite{mishra2011compact} and it does not claim to predict any resonances. By a careful inspection or Figure \ref{cacmishra} two additional resonances can be observed that cannot be predicted from the formula because the antenna in \cite{mishra2011compact} was not built based on the formula. To further verify the validity of our equation \ref{cace6}, it was used to verify the resonances of the antenna in \cite{mishra2011compact} and the accuracy is evident in Figure \ref{cacmishra}. 

\noindent In the design equation \ref{cace7} was used in order to calculate the smallest radius R, by using the highest possible frequency \textit{f${}_{R}$} = 5 GHz in order to combine effectively the three linear segments, and ensure the good matching ({\textbar}S${}_{11}${\textbar}$<$-10 dB) between the predicted resonances. As a result the antenna synthesized (as in \cite{mishra2011compact}) is considerably larger. Actually it is 180\% the size of the proposed antenna as can be verified and as is included in Table \ref{cact3}.

\begin{figure}[t]
	\centering
	\includegraphics[width=1\textwidth]{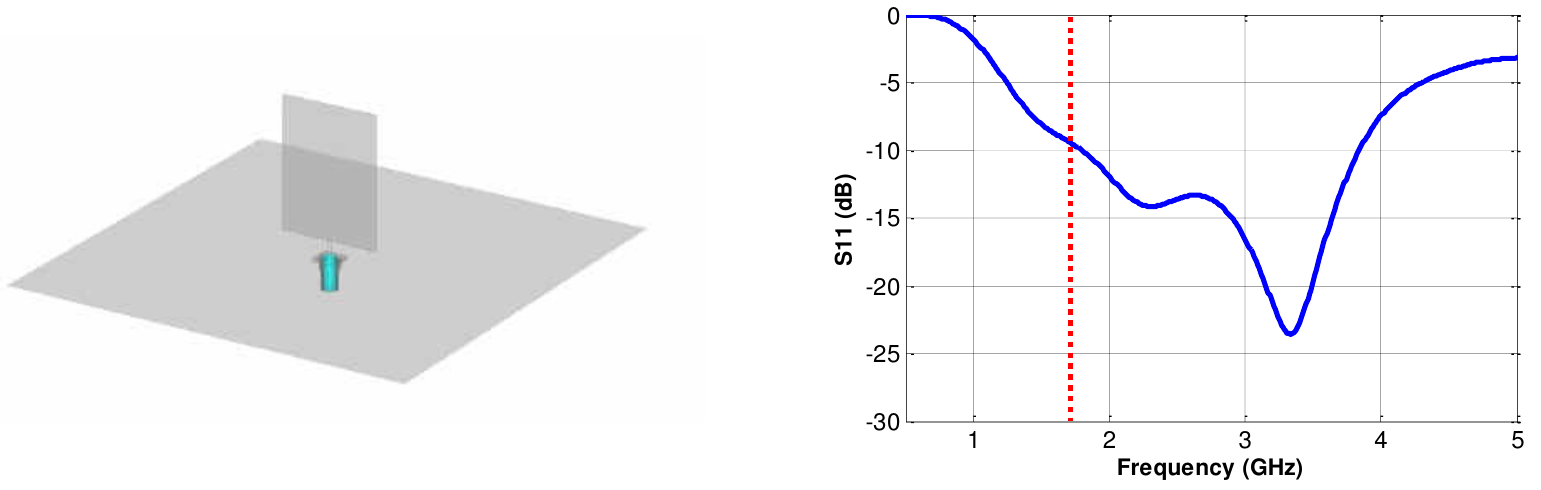}
	\caption{Predicted \textit{f${}_{L}$} from literature = 1.56 GHz and the simulated \textit{f${}_{L}$} = 1.79 GHz.}
	\label{cacrect}
\end{figure}

\begin{figure}[htb]
	\centering
	\includegraphics[width=0.9\textwidth]{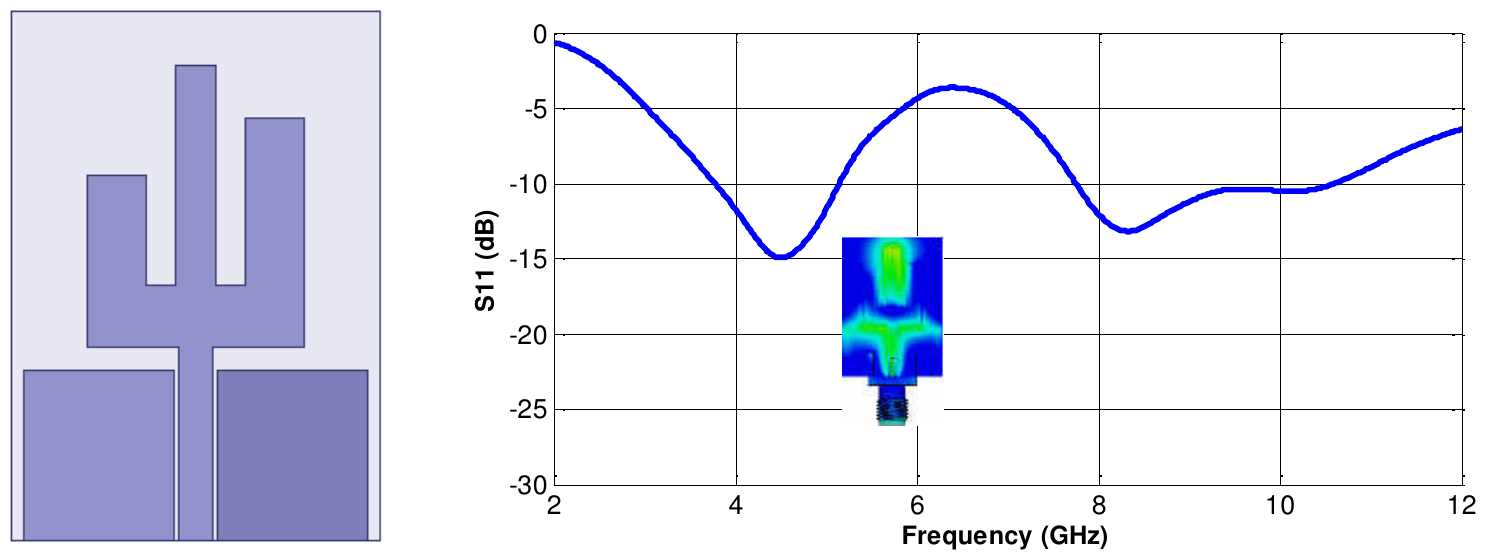}
	\caption{The first resonance occurs because the central planar radiating monopole at ~4.2 GHz (shown in \textit{E}-field plot) with inherently wideband characteristics along with the harmonics at the higher frequencies. }
	\label{cacnocurve}
\end{figure}

The authors based their analysis on the equation in section \ref{caceq} from \cite{kumar2003broadband} that was used for a wideband (69.1\% fractional bandwidth -much narrower than UWB) 3D planar monopole, fed by a probe, through a large ground plane. The predicted frequency refers to the lowest frequency for which ({\textbar}S${}_{11}${\textbar}$<$-10 dB). The predicted frequency was 1.56 GHz 12.85\% off compared to the simulated 1.79 GHz frequency. The discussed antenna in \cite{kumar2003broadband} is presented in Figure \ref{cacrect}. The authors implemented the formula cited in literature (originally used for a 3D monopole on top of a large ground plane) for a printed CPW fed antenna with very compact ground patches, where three such monopoles were combined effectively to exceed the designated UWB bandwidth (3.1-10.6 GHz) achieving (-10dB) bandwidth from 2.85 - 11.85 GHz. The direct use of the formulas used in \cite{garg2001microstrip} with a simplistic combination of three monopoles would result in the antenna presented in Figure \ref{cacnocurve} which is clearly not a UWB antenna. 

The use of equation \ref{cace7} from reference \cite{mishra2011compact} was used to calculate R of the semi-annular ring that allows the smooth transition between the CPW line and the cactus-shaped radiator. It was used for higher frequency \textit{f${}_{R}$} that allowed the successful implementation of the transition with significantly reduced R that resulted in the design of a UWB antenna which is 44.4\% smaller than the one from \cite{mishra2011compact}.

Consequently, the design approach used in this work referred to the aforementioned references \cite{kumar2003broadband} and \cite{mishra2011compact} to build an antenna with different characteristics (CPW-fed printed) and improved features. Improved bandwidth (compared to \cite{kumar2003broadband}, 3D antenna) and improved size that additionally presents easily controllable return loss.

\subsection{Measurement Accuracy Impediment}
\begin{figure}[htb]
	\centering
	\begin{subfigure}{0.8\textwidth}
		\includegraphics[width=1\textwidth]{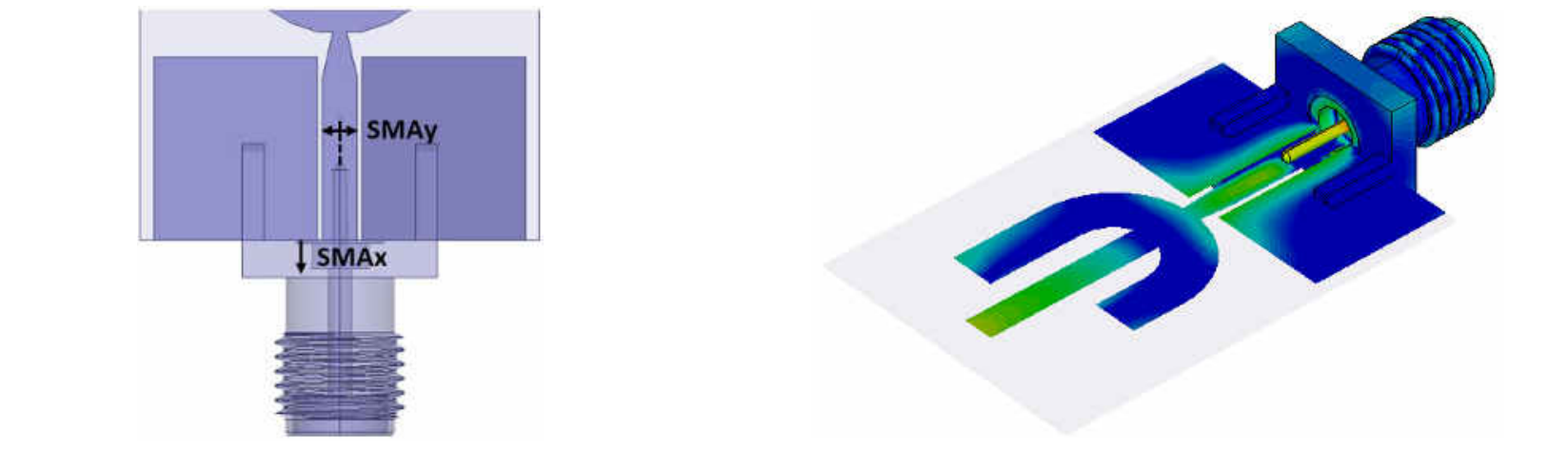}
		\caption{}
	\end{subfigure}
	\begin{subfigure}{0.6\textwidth}
		\includegraphics[width=1\textwidth]{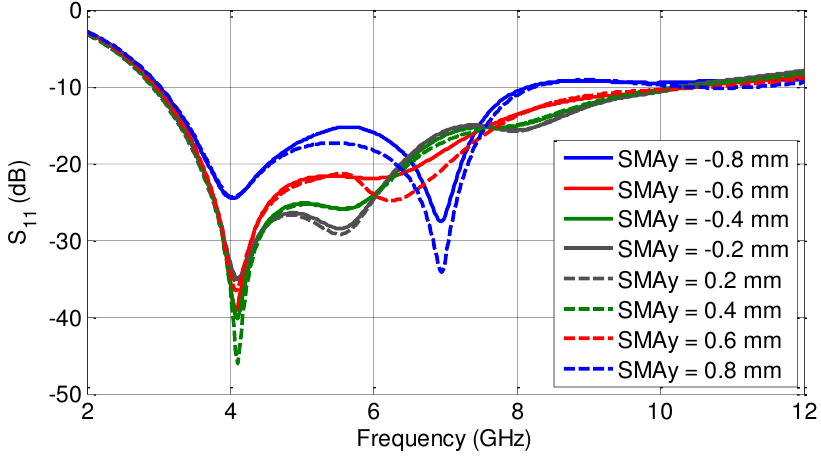}
		\caption{}
	\end{subfigure}
	\begin{subfigure}{0.6\textwidth}
		\includegraphics[width=1\textwidth]{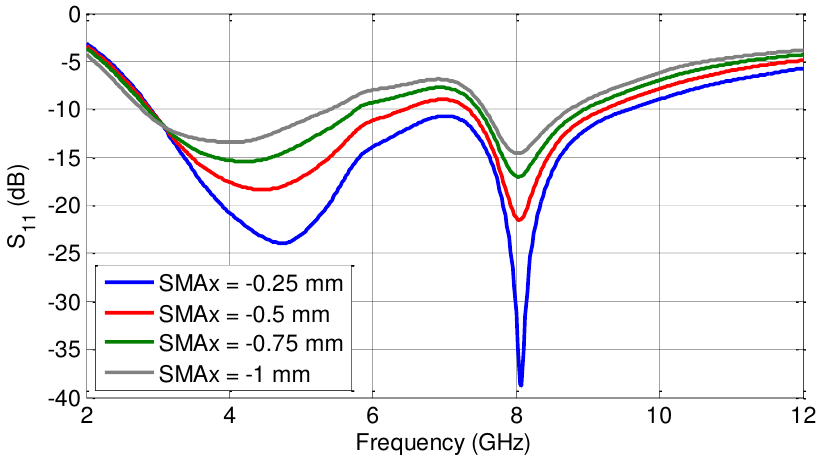}
		\caption{}
	\end{subfigure}
	\caption{Sensitivity of antenna’s return loss when SMA connector slides along the transmission line (a) The antenna size compared with the size of attached SMA connector. Return loss when (b) SMA connector slides along \textit{y}-axis (c) SMA connector slide along \textit{x}-axis. }
	\label{cacsma}
\end{figure}

The discrepancy between simulations and measurements in UWB antennas is rather common as can be supported by several reported results in literature. The accuracy is also affected by the addition of the necessary SMA connector. Slight variations in the soldering position have significant effect on the return loss as can be concluded from Figure \ref{cacsma}. Moreover, the compact size of the proposed antenna makes it comparable in size with the SMA connector and the connector's effect while evident in the measurements cannot be predicted analytically.

\section{Radiation Performance}

Last column of the radiation pattern figure in \cite{nikolaou2016miniaturization} presents the comparison of simulated and measured normalized radiation patterns in both \textit{E} and \textit{H-}planes along the bandwidth of the antenna. Based on the antenna orientation of Figure \ref{ijap1} \textit{x-z} and \textit{y-z} planes represent \textit{E} and \textit{H}-planes respectively when antenna is placed in \textit{x-y} plane while feed line is along the \textit{x}-axis. It is evident from the radiation patterns that the antenna is exhibiting, close to typical monopole antenna radiation patterns, in both frequencies, and in both planes, as well. The plots also confirm the sustainability in radiation performance of the cactus-shaped antenna throughout the UWB frequency range. Measured radiation patterns are in good agreement with the simulated predictions, validating the performance of the fabricated prototype. Although the antenna is not symmetrical, the presented radiation patterns are almost symmetrical. The expected asymmetry in the radiation patterns should be reflected on \textit{x-y} plane cuts, however the measured patterns are presented only in \textit{x-z} and \textit{y-z} planes. The primary parameters that control the return loss (i,e, L1, L2 and L3) have minor effect on the antenna's radiation patterns which remain consistently omnidirectional along the \textit{H}-plane.

The gain of the antenna is measured using the substitution method, in which a 2 to 18 GHz Q-par Angus Ltd rectangular horn antenna of known gain was used. The gain accuracy of the horn antenna is within $\pm$0.8 dB range, thus this can be considered as the accuracy of the measurement setup for the cactus-shaped antenna. A direct comparison of simulated and measured gain in complete UWB frequency range is presented in Figure \ref{ijap11}(c) Maximum gain is measured in $\phi$ = $\theta$ = 0\textit{${}^\circ$}${}^{ }$i.e. in the direction of \textit{z}-axis and the agreement between measurements and simulated results is evident. The gain of the antenna grows linearly with respect to frequency and the overall gain remains within 4 dB range. The very consistent omnidirectional radiation patterns of the antenna, in \textit{H}-plane, is the reason that maximum gain is kept low as can be deduced from Table \ref{cact3} that provides a comparison between this work and the similar antennas reported in the references. The consistent omni-directional patters in combination with the compact size of the cactus-shaped monopole make the proposed antenna a good candidate for most UWB applications.

\section{Chapter Conclusions}
Three proposed antennas in Section \ref{ijap} are fabricated on flexible, low loss and low cost LCP organic material and a miniaturization method is discussed. All three antennas have a return loss better than -10 dB in the whole ultra-wideband range and have close to omni-directional radiation patterns. The evolved cactus antenna and miniaturized cactus antenna are developed based on the fact that the original slot antenna's operation depends primarily upon the current distribution on the U-shaped tuning stub. Based on this observation, regions with relatively lower surface current amplitude were removed to achieve more compact, size reduced device. During this process, the U-shaped stub elliptical slot antenna was modified to form a cactus shaped radiator. The radiation characteristics of cactus were thoroughly investigated and the new antenna was optimized to be well-matched in the whole UWB range. The bigger cactus antenna covers only 59\% of the area of the original CPW-fed slot antenna whereas the miniaturized cactus antenna covers only 37\% of the initial area. As a consequence of the removal of the elliptical slot, the monopole cactus antennas became more omni-directional. The good agreement between simulated and measured results verify the good performance of the proposed antennas and validate the success of the proposed miniaturization method.

A novel CPW fed, compact, cactus-shaped monopole antenna fabricated on low loss, flexible, 225 $\mu$m thick LCP organic material is presented in Section \ref{cac}. The radiation mechanism of the antenna, depends on three radiating stubs, connected to a semi-annular ring creating a cactus shape radiator. The resonances of the antenna were initially predicted with an analytical formula using classical theory of CPW-line-fed, semi-annular, and rectangular, radiators. Full wave EM simulator was further used to optimize the performance of the antenna to exhibit good impedance matching throughout the whole UWB frequency range. The lengths of the three radiating stubs control the position and depth of {\textbar}S${}_{11}${\textbar} resonances, hence allow a direct control on the return loss and therefore on the antenna matching. This feature allows a readjustment of the antenna characteristics to focus on different UWB sub-bands, making the presented antenna a potential candidate for next generation UWB transceivers, which may operate in designated sub-bands, depending upon the specifications of the desired application.

The proposed equation 6 predicts with high accuracy the occurrence of the resonances throughout the UWB range for both standard whole UWB fabricated antenna and the UWB sub-bands versions described in Figure \ref{cac3}. To the best of our knowledge the analytical prediction of the radiating resonances throughout the UWB range has not been reported before. The design is one of the smaller UWB antennas reported in literature (Table \ref{cact3}). The design presents nearly perfect omnidirectional patterns with average gain 0.84 dBi and gain span from -1.2 to 3.2. (Table \ref{cact3}).

\begin{table}
	\centering
	\begin{tabular}{cccccc} \hline 
		Ref. & Size (mm) & Freq. \newline (GHz) & {Gain range} & {Gain avg.} & Size comp. \\ \hline 
		\cite{siddiqui2015compact, siddiqui2014compact} & 50$\times$50 & 2.60 -- 10.20 & -2.6 -- 6.0 & 2.27 & 446.4\% \\ 
		\cite{koohestani2012influence} & 44$\times$38 & 3.05 -- 12+ & - & - & 298.6\% \\ 
		\cite{tang2014compact} & 40$\times$30 & 2.26 -- 22.18 & 3.3 -- 6.8 & 4.53 & 214.3\% \\ 
		\cite{mishra2011compact} & 42$\times$24 & \begin{tabular}[c]{@{}l@{}}2.30 -- 2.50\\ 3.10 -- 12.00\end{tabular} \newline & 1.0 -- 5.4 & 3.68 & 180.0\% \\ 
		\cite{liu2011parasitically} & 30$\times$30 & 3.40 -- 7.62 & 2.7 -- 5.3 & 4.12 & 160.7\% \\ 
		\cite{wu2015compact} & 25$\times$26 & \begin{tabular}[c]{@{}l@{}}2.40 -- 2.55\\ 3.30 -- 3.66\end{tabular} & 1.6 -- 2.2 & 1.97 & 116.1\% \\ 
		\cite{li2015cpw} & 26.2$\times$18 & \begin{tabular}[c]{@{}l@{}}3.06 -- 3.43\\ 4.55 -- 5.86 \end{tabular} & 2.2 -- 3.6 & 2.67 & 84.1\% \\
		This work & 28$\times$20 & 2.85 -- 11.85 & -1.2 -- 3.2 & 0.84 & 100.0\% \\ \hline 
	\end{tabular}
	\caption{Comparison between reported planar monopole type antennas}
	\label{cact3}
\end{table}


\appendix

\chapter{Efficient Rectifiers for Wireless Power Transfer}

\ifpdf
\graphicspath{{Appendix1/Appendix1Figs/PNG/}{Appendix1/Appendix1Figs/PDF/}{Appendix1/Appendix1Figs/}}
\else
\graphicspath{{Appendix1/Appendix1Figs/EPS/}{Appendix1/Appendix1Figs/}}
\fi

The rectifier design approach for the implementation of a System on Package (SOP) Wireless Power Transfer (WPT) receiver is discussed. Real Wireless Power Transfer testing conditions revolved around the rectifier component. For the WPT receiver module, the 1.6 GHz rectifier operated with a 49\% RF-to-DC efficiency for an input RF power of -10 dBm. Three efficient rectifiers, first at 2.45 GHz ISM band, second at higher frequency (5.6 GHz), and third at lower frequency (1.6 GHz) is presented.

\section{Introduction and Background}

Wireless energy transfer has been studied for decades now since it dates back to the experiments of Heinrich Hertz. Shifting to a much more recent decade, this research topic gained a lot of attention indirectly when radio frequency identification (RFID) applications surfaced in day-to-day applications. RFID systems were initially introduced in the late 1990's for very short range systems ($<$ 1.5 meters) operating at either 0.9 GHz or at 2.4 GHz, ISM band, and were used for industrial applications. However, it is with the rapid development of the Internet of Things (IoT) devices,- that took place mostly in the last five years,- that increased attention was paid in wireless charging in an attempt to spare the battery and thus reduce the overall cost and the maintenance cost of such wireless nodes. The rectenna design, and more specifically the RF-to-DC efficiency of the rectifier has been the key focus of the related research with ultimate scope the implementation of efficient Wireless Power Transmission (WPT) systems. It has been noted that for low RF input power levels, the RF-to-DC efficiency of most rectifiers is normally very low. Therefore, most of the research effort, straightforwardly, aimed the intensification of the rectifier's efficiency. Some very recent approaches are presented in literature, like balanced rectifiers \cite{wei2017balanced,erkmen2017electromagnetic} that compensate the microwave circuits terminations to recycle the resistive dissipated power, based on single \cite{liu2017novel}, dual \cite{mitani2017analysis}, multiple diodes \cite{erkmen2017electromagnetic}, and even system in package (SiP) \cite{li2017compact} low-loss, integrated, passive devices. In this approach, the impedance matching at the rectifier's input and output terminals with the antenna and the power management unit (PMU) respectively has always been the major bottleneck in the overall performance. The obvious choice of wide impedance matching network has been known to severely deteriorate the RF-to-DC efficiency for a wide range of input power levels. Some recent studies, \cite{huang2017impedance,sun2017new} have focused mostly on possible solutions to this very specific and seemingly simple problem. The high importance of the matching problem at the input and output terminals of a rectifier is the reason why in this paper the key system elements of rectifiers are not studied only as standalone devices but they are also studied as part of a rectenna (antenna + rectifier module) and as part of a complete system with a DC-to-DC booster connected on the output of the rectifier. In addition to the matching problem, most rectifier topologies have evidently different behavior for different frequencies. In general, higher frequency rectifiers normally incur lower efficiency, and consequently charge pump topologies with the class-F loads \cite{wang2017study}, boost converter \cite{dolgov2010power}, buck-boost \cite{huang2014constant} etc. are used to enhance the harvested energy at DC level. In all the aforementioned topologies, rectifiers' operation and their RF-to-DC efficiency has been demonstrated to be the key feature for the WPT receiver. Even mathematical models like \cite{chen2017maximum} have been proposed to pinpoint the highest possible RF-to-DC efficiency at a given power level, frequency and output load. It has been reported in literature that there is always a trade-off between complexity, cost, operational power level and the frequency of a rectifier. 

In this chapter the voltage double topology \cite{salmon1993circuit} has been preferred for the rectifier designs and fabricated rectifiers at three different frequencies are presented.

\section{Rectifier Design}

\begin{figure}[t]
	\centering
	\includegraphics[width=1\textwidth]{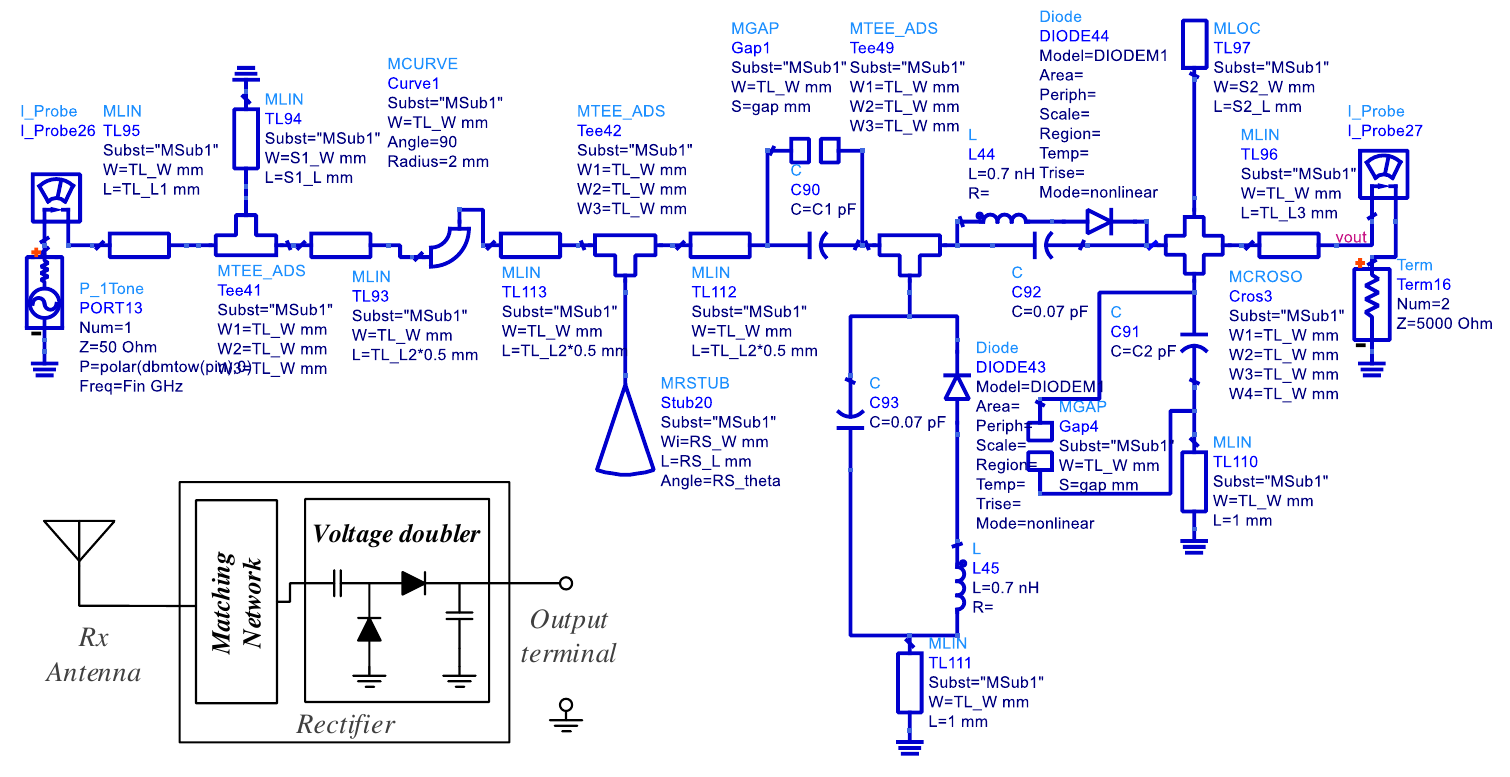}
	\caption{Rectifier and voltage doubler architecture (bottom left) and ADS circuit schematic of the rectifier.}
	\label{vd}
\end{figure}

There are different RF-to-DC rectifier topologies to choose from depending on the application's specifications. The most popular topology is the voltage doubler that consists of two Schottky diodes and two capacitors therefore this is the only type of rectifier considered in this work as shown in Figure \ref{vd}. Voltage doubler has a fairly simple topology, can have a relatively compact size, and is preferred when high efficiency can be traded. It has been demonstrated that the number of rectification ladders have a direct relation with rectifier efficiency, however losses increase too \cite{chen2017maximum}. When a rectifier is intended for energy harvesting, as a component of a WPT system receiver, the most important feature is the RF-to-DC efficiency which needs to be as high as possible. However, rectifiers are non-linear circuits and practically this means that their efficiency depends non-linearly on a) the input RF power, and b) the termination load. Generally, a rectifier should have high RF-to-DC efficiency for a wide range of input power levels and similarly high efficiency, for a wide range of termination loads, since the subsequent to the rectifier component, which is usually a power management unit, (PMU), has a varying input impedance that depends on its unforeseen operation conditions at the time. 

\subsection{2.45 GHz Rectifier}

The rectifier was designed at 2.45 GHz keeping in mind its potential use for wireless power transfer in Wireless Body Area Network (WBAN) applications. Keysight – Advanced Design System (ADS) was used to design the hosting traces of the rectifier’s commercially available lumped components. Lumped components were simulated using the vendor’s ADS library models, and 0.787 mm thick Rogers RT/duroid 5880 substrate was used with $\epsilon_r$ = 2.2, tan$\delta$ = 0.0009. In order to analyze the behavior of the output rectified voltage, large scale signal analysis was used. The nonlinear behavior of the rectifier was taken into account by using harmonic balance simulation with the fundamental frequency at 2.45 GHz and order five. A linear power sweep was applied from -30 dBm to 10 dBm considering the possible range of the RF input power levels. The 2.45 GHz rectifier was optimized to be matched to a 50 $\Omega$ input impedance of a receiver antenna, according to equation \ref{a1}, and at the same time, to perform with high RF-to-DC efficiency (equation \ref{a2}) at the available power range, for a potentially changing load \textit{{R$_L$}}.

\begin{equation}\label{a1}
\Gamma =\frac{Z_{in} - 50}{Z_{in} + 50} 
\end{equation}

\begin{equation}\label{a2}
\eta =\frac{P_{out}}{P_{in}} = \frac{\frac{V_{dc}^{2}}{R_{L}}}{P_{in}} 
\end{equation}

\begin{figure}[htb]
	\centering
	\includegraphics[width=0.55\textwidth]{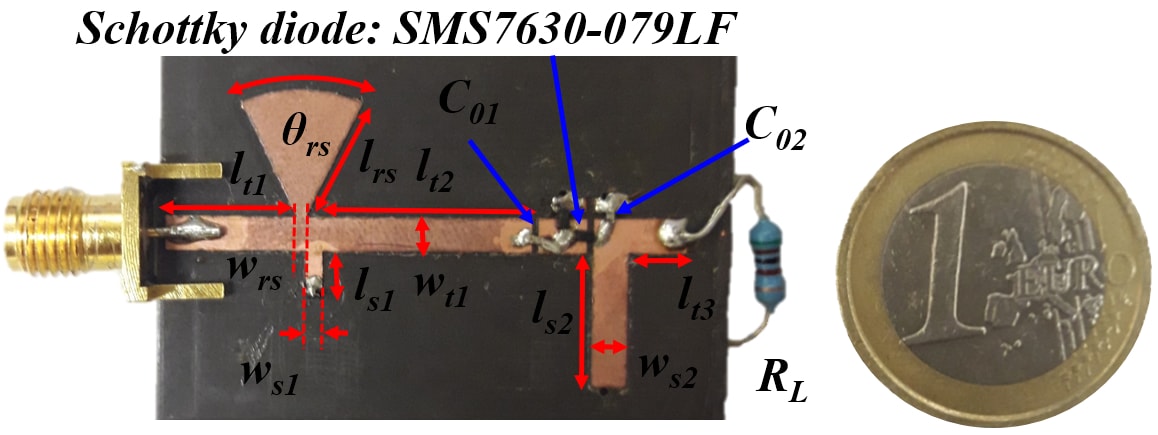}
	\caption{Rectifier at 2.45 GHz}
	\label{rct1}
\end{figure}

\textit{Z$_{in}$} is the impedance at the input terminal of the rectifier, \textit{P$_{in}$} and \textit{P$_{out}$} are the input RF power, and the output DC power values in (W), respectively, and Vdc is the voltage across the load resistance \textit{R$_L$} which is the input impedance load of the cascaded PMU and it depends on the biasing conditions. For the presented measurements \textit{R$_L$}=5.1 K$\Omega$. Initially, a conventional voltage doubler at 2.45 GHz was designed using microstrip technology with 50 $\Omega$ characteristic impedance traces. The main encountered problems were a) the narrow bandwidth of the rectifier that made the good matching within the wide RF input power range difficult to maintain, and b) the relatively large physical size of the rectifier. (Especially \textit{l$_{t}$}, \textit{l$_{s1}$} and \textit{l$_{s2}$}). Linear shorted or open stubs matching network designed to match a transmission line to a standard 50$\Omega$ port is normally narrowband. This eventually fails to match a rectifier circuit for a wide range of input power, so a radial stub was introduced along\textit{ l$_{t}$} to broaden the matching bandwidth of the rectifier. It can be seen from Figure \ref{rct1} that a combination of a radial stub and a linear shorted stub match the transmission line (\textit{l$_{t}$}) to the input port. The overall horizontal size of the rectifier primarily depends upon the length of these two stubs, therefore design consideration must be made to ensure that these two stubs have the smallest possible footprint.

The rectifier in Figure \ref{rct1} is a result of a multivariable optimization in ADS circuit/momentum solver. The dimensions of the rectifier enlisted in the caption of Figure \ref{rct1} were dictated based on Quasi-Newton optimization algorithm when all the variables were bounded by arbitrarily set realistic dimension parameters in the perspective or a feasible miniaturization goal. The first optimization goal was set “$|S_{11}|$ < -15 dB at 2.45 GHz”, while the second goal was the “highest possible RF-to-DC efficiency”. Upon achieving a feasible microwave trace for the rectifier, the capacitors \textit{C$_{01}$} and \textit{C$_{02}$} were replaced by the closest SMD capacitor values that could be found in the Murata component library and then Quasi-Newton optimization algorithm was used for the second time. This process converged after three iterations and the final values of \textit{C$_{01}$} and \textit{C$_{02}$} were 47 pF and 80 pF.

\begin{figure}[t]
	\centering
	\begin{subfigure}{0.45\textwidth}
		\includegraphics[width=1\textwidth]{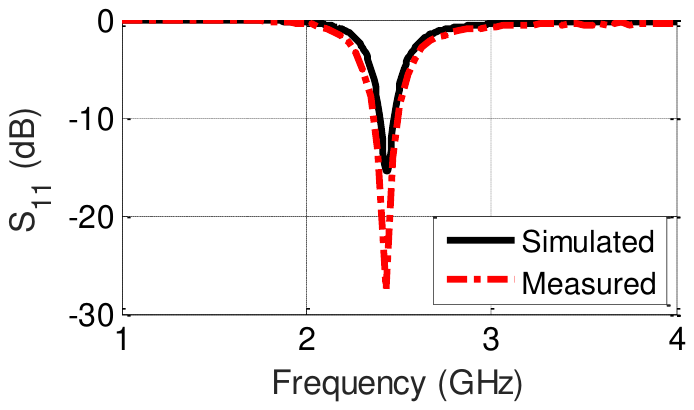}
		\caption{}
	\end{subfigure}
	\begin{subfigure}{0.45\textwidth}
		\includegraphics[width=1\textwidth]{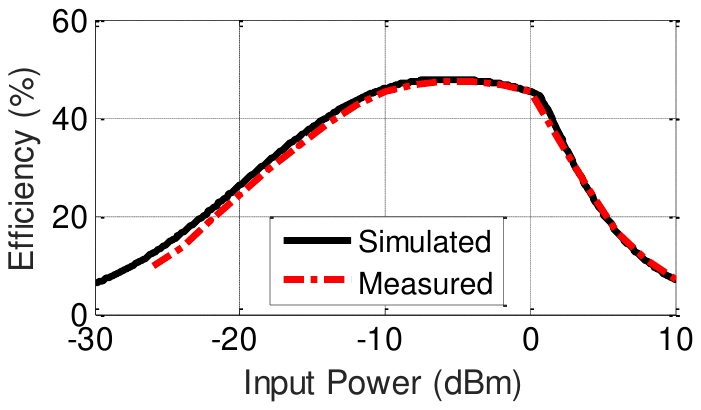}
		\caption{}
	\end{subfigure}
	\caption{Comparison between simulated and measured (a)  $|S_{11}|$ vs frequency and (b) Efficiency vs \textit{P$_{in}$} at 2.45 GHz.}
	\label{rct11}
\end{figure}

A sma connector was soldered on the fabricated board for connecting the rectifier under test with the vector network analyzer (VNA) for reflection coefficient and RF-to-DC efficiency measurements. Figure \ref{rct11}(a) shows the comparison of the simulated and measured S11 of the tested 2.45 GHz rectifier where the $|S_{11}|$ in dB is below -10 dB from 2.37 to 2.49 GHz (measured bandwidth of 120 MHz). The measured efficiency of the rectifier was calculated by measuring the DC voltage across the \textit{R$_L$} while changing the input power level \textit{P$_{in}$} from an Agilent E8363B vector network analyzer (VNA) when one of the ports was set to generate a single power tone (unmodulated sine wave) at 2.45 GHz. Due to the power limitation of the VNA, the power range from -26 dBm to 10 dBm was varied using a 2 dBm discretization step. Output voltage across the \textit{R$_L$} = 5.1 k$\Omega$ simulated and measured. Equation \ref{a2} was used to calculate the efficiency using the measured voltages across \textit{R$_L$} and the RF-to-DC efficiency vs RF input power plot is presented in Figure \ref{rct11} (b). Measured and simulated results are in excellent agreement verifying the simulation setup and modelling and at the same time the fabrication and testing accuracy. From Figure \ref{rct11} (b), it can be seen that the measured RF-to-DC efficiency of the rectifier ranges from $\sim$11.6\% for -25 dBm, to the peak value of $\sim$47.4\% when the input power is around -5 dBm and then, it goes rapidly down to below $\sim$10\% for input power somewhat less than +10 dBm. The rectifier has been tested using a single tone signal at 2.45 GHz to ensure fair comparison with the referenced rectifier designs summarized in Table I and the other two fabricated rectifiers (5.6 GHz part of the rectenna rectifier, and 1.6 GHz part of the complete SOP energy harvester module rectifier) that will be presented in the two subsequent sections.

\subsection{5.6 GHz Rectifier}

\begin{figure}[t]
	\centering
	\includegraphics[width=0.45\textwidth]{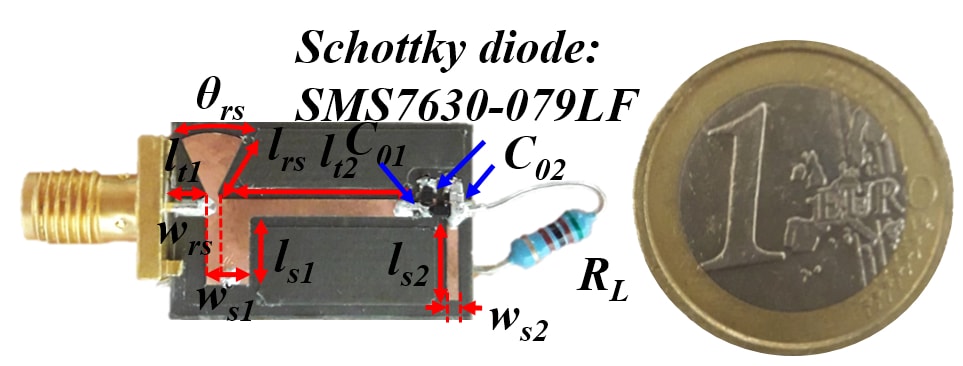}
	\caption{Rectifier at 5.6 GHz}
	\label{rct2}
\end{figure}

Most reported rectifiers are designed for UHF and ISM bands\cite{song2017recent} consequently rectifiers operating at 5.6 GHz are not very common in the literature. However, several reconfigurable band-notch UWB antennas \cite{nikolaou2009uwb} have been reported and a rectifier at 5.6 GHz could be used to make the UWB antenna dynamically reconfigurable exploiting the harvested energy of the interfering 5.6 GHz WLAN signal. Since the voltage doubler topology was chosen for the proposed 5.6 GHz rectifier the design limitations and the general objectives - high efficiency, compact size, good matching for a wide range of input power levels, peak efficiency for 5.1 K$\Omega$ termination load - that were discussed in section I for the 2.45 GHz rectifier still apply. Therefore the same design method was used with the additional consideration of the cascaded patch antenna’s input impedance. After the previously described large scale ADS simulations and the harmonic balance analysis where the fundamental frequency was set to 5.6 GHz, a Quasi-Newton algorithm-based multi-variable optimization was run. The multi-variable optimization was performed with two equal-weight objectives: a) 50 $\Omega$ matching at the input terminal of the rectifier (equation \ref{a1}) and, b) highest possible RF-to-DC efficiency (equation \ref{a2}) at the desired wide power range for a termination load of 5.1 K$\Omega$. For the simulations a power sweep from -30 dBm to 10 dBm was applied. The geometric parameters of the rectifier’s schematic along with the capacitor values (Cs) were defined as variables for the optimization algorithm. Their range of possible values, were constrained by size limitations and by the available values of the commercially available capacitors. During the optimization process the peak efficiency was attempted to be pushed towards lower input power levels. The peak efficiency was calculated from the measured rectified DC voltage across a termination load \textit{R$_L$}=5.1 K$\Omega$ using equation \ref{a2}. The rectifier (Figure \ref{rct2})was directly connected to a signal generator (R$\&$S SMF100A) while the DC voltage across \textit{R$_L$} was measured for different input power levels (\textit{P$_{in}$}), at 5.6 GHz.

\begin{figure}[t]
	\centering
	\begin{subfigure}{0.45\textwidth}
		\includegraphics[width=1\textwidth]{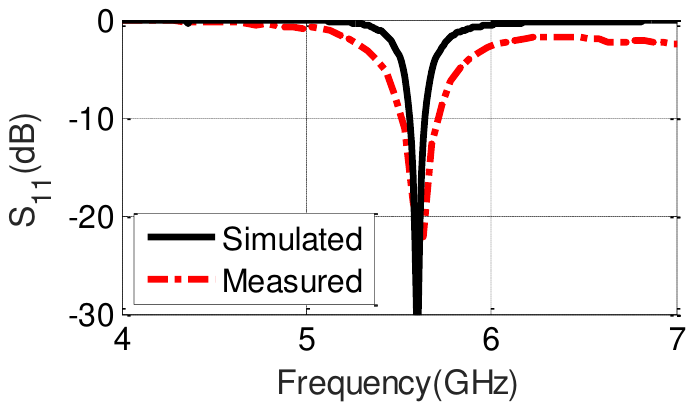}
		\caption{}
	\end{subfigure}
	\begin{subfigure}{0.45\textwidth}
		\includegraphics[width=1\textwidth]{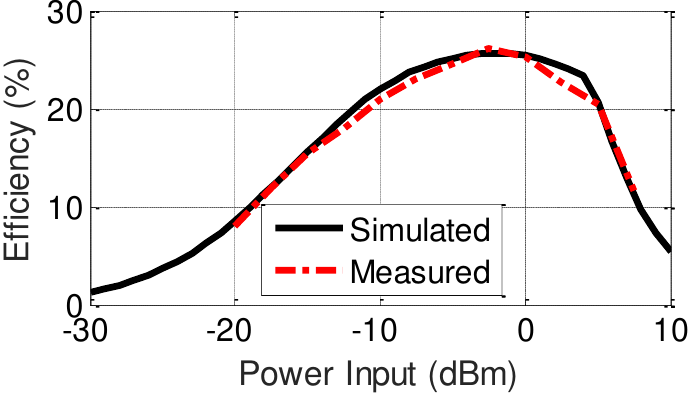}
		\caption{}
	\end{subfigure}
	\caption{Comparison between simulated and measured (a)  $|S_{11}|$ vs frequency and (b) Efficiency vs \textit{P$_{in}$} at 5.6 GHz.}
	\label{rct21}
\end{figure}

An unmodulated sine wave was used to test the efficiency of the rectifier. Then, the efficiency was calculated by changing the power level of the signal generator from, -30 dBm to +10 dBm. The calculated RF-to-DC efficiency vs input power is presented in Figure \ref{rct21}(b) and the peak efficiency is observed for -5 dBm input power. The efficiency of the rectifier is approximately 15\% at -15 dBm and increases to 27\% at -2.5 dBm. The attempt to shift the peak efficiency for lower RF input power level namely for -20 dBm as the goal of another optimization run wad terminated without convergence. High efficiency and peak efficiency for low RF input power levels (<-20 dBm), remain the holy grail of the research on RF-to-DC rectifiers. The major concern for the combination of the two consisting components was the successful matching. In order to meet this need the narrowband, -10 dB band, of the rectifier’s reflection coefficient, had to be increased. Generally, Schottky diodes are wideband devices. What actually limits the -10dB bandwidth of the rectifier is the bandwidth of the matching network, therefore in order to increase the bandwidth, the linear open stubs were replaced with radial stubs. Using the radial stubs and after the numerical optimization the measured $|S_{11}|$ depicted in Figure \ref{rct21}(a) had 190 MHz bandwidth – since it remained below -10 dB from 5.51 to 5.70 GHz. A Rogers RT/duroid 5880 substrate was used with $\epsilon_r$ = 2.2, tan$\delta$ = 0.0009 and substrate thickness 0.787 mm. The fabricated stand-alone rectifier is presented in Fig 4(a) and the optimized set of parameters is listed in the caption. The optimized values for both C01 and C02 were 150 pF. The rectenna microstrip layout was fabricated using an LPKF ProtoMat H100 milling machine, in single step fabrication, and at a later stage the Schottky diodes (Skyworks SMS7630-079LF) and the SMD capacitors (Murata series GJM03 - 150 pF) were mounted by-hand on the rectenna sole module. 

\subsection{1.6 GHz Rectifier}

\begin{figure}[t]
	\centering
	\includegraphics[width=0.38\textwidth]{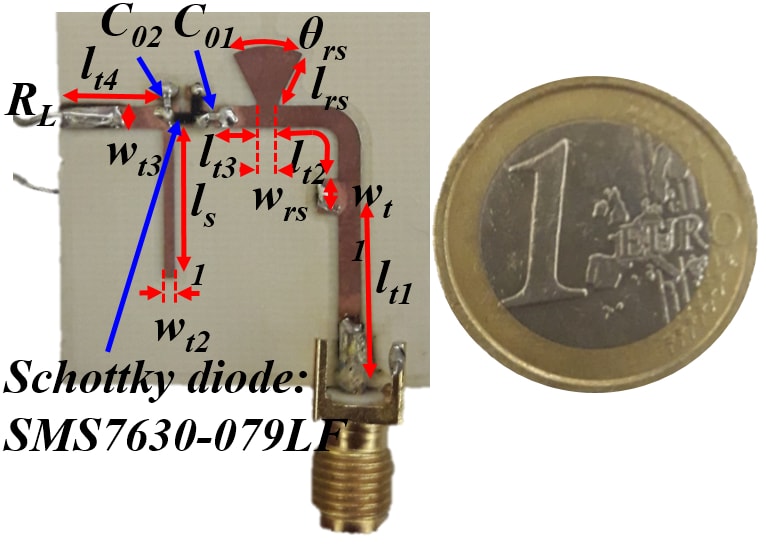}
	\caption{Rectifier at 1.6 GHz}
	\label{rct3}
\end{figure}

\begin{figure}[t]
	\centering
	\begin{subfigure}{0.45\textwidth}
		\includegraphics[width=1\textwidth]{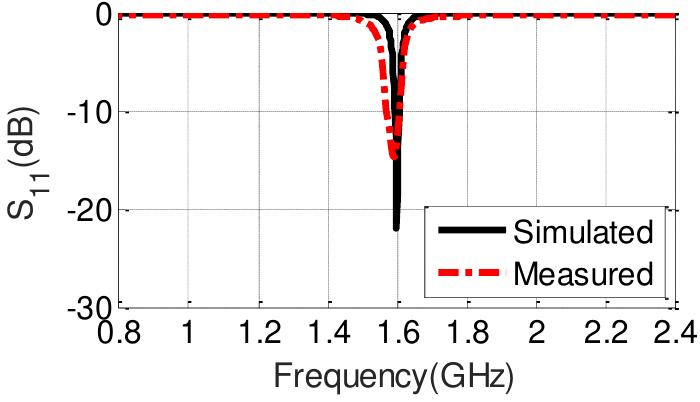}
		\caption{}
	\end{subfigure}
	\begin{subfigure}{0.45\textwidth}
		\includegraphics[width=1\textwidth]{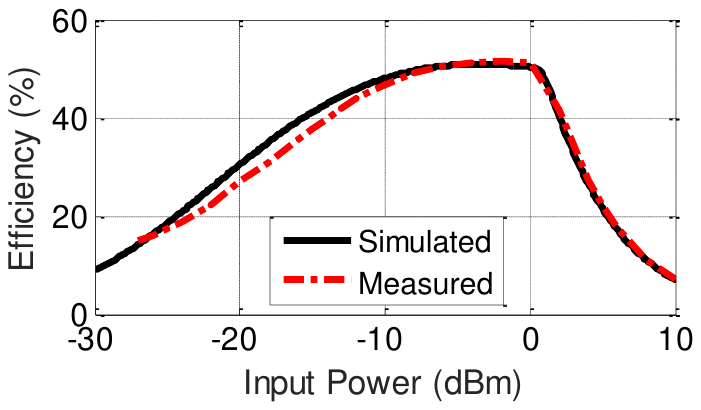}
		\caption{}
	\end{subfigure}
	\caption{Comparison between simulated and measured (a)  $|S_{11}|$ vs frequency and (b) Efficiency vs \textit{P$_{in}$} at 1.6 GHz.}
	\label{rct31}
\end{figure}

The design method that was used for the two rectifiers and was discussed in the two previous sub-sections, was also used for the 1.6 GHz voltage double rectifier. The additional requirement had to do with the addition of the intended DC-to-DC power booster, which has an input impedance (termination load for the rectifier) that depends on the biasing and operating conditions. The RF-to-DC efficiency was optimized for a fixed load of \textit{R$_L$} = 5.1 K$\Omega$ which can be considered the average load for the varying load (4.5 – 6.0 K$\Omega$) corresponding to the different biasing conditions (refer to min and max voltage for \textit{V$_{CC}$} of the DC-to-DC power booster. The schematic of the rectifier is presented in Figure \ref{rct3} and the set of the consisting feature values is listed in the caption. Its reflection coefficient is presented in Figure \ref{rct31}(a) proving the good matching in a narrowband range around 1.6 GHz. It can be seen that $|S_{11}|$< -10 dB bandwidth ranges from 1.57 to 1.61 GHz resulting in a measured bandwidth of 40 MHz.

RO4003C substrate was used as substrate board, the same as for the PIFA. To overcome the limited bandwidth of the rectifier and to decrease its overall size, a radial stub Figure \ref{rct3} \textit{l$_{rs}$} ,\textit{$\theta$$_{rs}$} ) was introduced along with the curved segment with length \textit{l$_{t}$}. After introducing the radial stub, the overall size was further increased. The rectifier circuit begins from the end of the receiver antenna, and a curved transmission line section hosts the shorted stub and a radial stub placed at the point where sufficient area was available. This rectifier shape and dimensions were dictated based on the results of the previously introduced multi-variable ADS optimization. For this case the two output goals were: a) $|S_{11}|$ $\<$ -15 dB at 1.6 GHz and b) maximum possible RF-to-DC efficiency with varying parameters the dimensions of the consisting elements and the values of the lumped capacitors. The optimized capacitors were: \textit{C$_{01}$} = 139 pF and C02 = 150 pF. The measured DC voltage and the related efficiency of the rectifier were measured using on board measurements. A signal generator (R$\&$S SMF100A) with varying power unmodulated sine signals at 1.6 GHz was used to provide the input RF power at the rectifier and the rectified DC voltage was measured across the termination load (\textit{R$_L$} = 5.1 k$\Omega$) using a DMM. Simulated and measured output voltage across the \textit{R$_L$} = 5.1 k$\Omega$ is used to calculate RF-to-DC efficiency of the rectifier, and is presented in Figure \ref{rct31}(b). It can be seen that the efficiency of the rectifier is $\sim$20\% and goes up to $\sim$49\% when the input power level increases from -25 dBm to -10 dBm.

\chapter{List of Publications}


A portion of this thesis is published in parts. It is to declare that the author already completed the copy rights transfer for the papers already in press. An accurate acknowledgment of the papers should be made with given Digital Object Identifiers (DOI) while citing the relevant portion of this thesis. Personal use of this material is permitted. Permission from publisher must be obtained for all other uses, in any current or future media, including reprinting/republishing this material for advertising or promotional purposes, creating new collective works, for resale or redistribution to servers or lists, or reuse of any copyrighted component of this work in other works.

\textbf{Journal papers in press:}
\begin{enumerate}
	
	\item \textbf{M. Ali Babar Abbasi}, S. Nikolaou, Marco Antoniades, M. N. Stevanovic, P. Vryonides, "Compact EBG-Backed Planar Monopole for BAN Wearable Applications," \textit{IEEE Transactions on Antennas and Propagation}, Vol.65, Iss. 2, pp.1 - 11, February 2017.  DOI: 10.1109/TAP.2016.2635588 
	
	\item\textbf{ M. Ali Babar Abbasi},   Marco A. Antoniades and S. Nikolaou, , "A Compact Reconfigurable NRI- TL Metamaterial PhaseShifter  for  Antenna  Applications"  \textit{IEEE  Transactions  on  Antennas  and  Propagation},  2017 (accepted for publication).  DOI: 10.1109/TAP.2017.2777520 
	
	\item S. Nikolaou, \textbf{M. Ali Babar Abbasi}, "Design and Development of a Compact UWB Monopole Antenna with Easily-Controllable Return Loss" \textit{IEEE Transactions on Antennas and Propagation}, Vol.65, Iss. 4, pp.1 - 5, April 2017. DOI: 10.1109/TAP.2017.2670322 
	
	\item Symeon Nikolaou, \textbf{Muhammad Ali Babar Abbasi}, "Miniaturization of UWB Antennas on Organic Material," \textit{International Journal of Antennas and Propagation (IJAP)}, vol. 16, 2016. DOI: 10.1155/2016/5949254
	
	\item S. Arain, Photos Vryonides, \textbf{M. Ali Babar Abbasi}, A. Quddious, Marco A. Antoniades and S. Nikolaou, "Reconfigurable Bandwidth Bandpass Filter with Enhanced out-of-band Rejection Using $\pi$-Section-Loaded Ring Resonator", \textit{IEEE Microwave and Wireless Components Letters}, vol. 28, no. 1, pp. 28-30, Jan. 2018.  DOI: 10.1109/LMWC.2017.2776212 
	
	\item S. Arain, Photos Vryonides, \textbf{M. Ali Babar Abbasi}, A. Quddious, Marco A. Antoniades and S. Nikolaou, “Wideband BPF Using Quadruple-Mode Ring Resonator Loaded with Short-Circuited Stubs and $\Gamma$-Shaped Band-Stop Sections”, \textit{Wiley Microwave and Optical Technology Letters (MOTL)}, , 59(9), pp. 2316-2320, 2017. DOI: 10.1002/mop.30737
	
	\item S. Arain, Photos Vryonides, \textbf{M. Ali Babar Abbasi}, A. Quddious, Marco A. Antoniades and S. Nikolaou, “On the Use of Tunable Power Splitter for Simultaneous Wireless Information and Power Transfer Receivers”, \textit{International Journal of Antennas and Propagation}, vol. 2018, Article ID 6183412, 12 pages, 2018. DOI: 10.1155/2018/6183412
	
\end{enumerate}	

\textbf{Conference papers in press:}
\begin{enumerate}
	\setcounter{enumi}{6}
	\item \textbf{Muhammad Ali Babar Abbasi}, Abdul Quddious, Marco Antoniades, Symeon Nikolaou, "Wearable sensor for communication with implantable devices," submitted in \textit{12th European Conference on Antennas and Propagation (EuCAP)}, London, 2018.	
	
	\item M. A. Antoniades, \textbf{M. A. B. Abbasi}, M. Nikolic, P. Vryonides and S. Nikolaou, "Conformal wearable monopole antenna backed by a compact EBG structure for body area networks," 2017 \textit{11th European Conference on Antennas and Propagation (EUCAP)}, Paris, 2017, pp. 164-166.
	
	\item\textbf{ M. A. B. Abbasi}, S. Nikolaou and M. A. Antoniades, "A high gain EBG backed monopole for MBAN off-body communication,"\textit{ IEEE International Symposium on Antennas and Propagation (APSURSI)}, Fajardo, PR, USA, pp. 1907-1908, 2016.  
	
	\item S. Arain, \textbf{M. A. B. Abassi}, S. Nikolaou and P. Vryonides, "A square ring resonator bandpass filter with asymmetrically loaded open circuited stubs," \textit{5th International Conference on Modern Circuits and Systems Technologies (MOCAST)}, Thessaloniki, Greece, pp. 1-4, 2016.
	
	\item \textbf{M. A. B. Abbasi}, P. Vryonides and S. Nikolaou, "Humidity sensor devices using PEDOT:PSS," \textit{IEEE International Symposium on Antennas and Propagation $\&$ USNC/URSI National Radio Science Meeting,} Vancouver, BC, 2015, pp. 1366-1367.
	
	\item \textbf{M. Ali Babar Abbasi}, D. Philippou, and S. Nikolaou, "Comparison Study of Layered Homogeneous Models with Detailed Human Tissue Models for Through-body Communications",\textit{ Progress in Electromagnetic Research Symposium (PIERS)}, Prague, Czech Republic, 2015. 
	
	\item \textbf{M. Ali Babar Abbasi}, S. Arain, P. Vryonides, and S. Nikolaou, "Design and optimization of miniaturized dual-band implantable antenna for MICS and ISM bands" \textit{Progress in Electromagnetic Research Symposium (PIERS)}, Prague, Czech Republic, 2015. 
	
	\item S. Arain, \textbf{M. Ali Babar Abbasi}, S. Nikolaou, and P. Vryonides, "A Reconfigurable Bandpass to Bandstop Filter Using PIN Diodes Based on the Square Ring Resonator", \textit{Progress in Electromagnetic Research Symposium, Prague (PIERS)}, Prague, Czech Republic, 2015. 
	
\end{enumerate}	

\textbf{Forthcoming:}
\begin{enumerate}
	\setcounter{enumi}{14}
	\item \textbf{Muhammad Ali Babar Abbasi}, Abdul Quddious, Aqeela Saghir, Salman Arain, Marco A. Antoniades, Photos Vryonides, Symeon Nikolaou, “Dynamically Reconfigurable Single- to Dual-Band SIR Filter,” awaited to be submitted in \textit{IEEE Transactions on Microwave Theory and Techniques}, 2017.
	
	\item \textbf{Muhammad Ali Babar Abbasi}, Marco A. Antoniades, Nikolaou “Compact NRI-TL Metamaterial Microstrip Crossover,” submitted in \textit{IET Electronics Letters}, 2017.
	
	\item Abdul Quddious, \textbf{M. Ali Babar Abbasi}, Farooq A. Tahir, Marco A. Antoniades, Photos Vryonides, Symeon Nikolaou, "UWB Antenna with Dynamically Reconfigurable Notch-Band," submitted in \textit{IEEE Transactions on Antennas and Propagation}, 2017.
	
	\item H. Nawaz, S. Sadiq, \textbf{M. A. B. Abbasi} and S. Nikolaou, “Mutual Coupling Reduction between Finite Spaced Planar Antenna Elements using Modified Ground Structure”, submitted in\textit{ Wiley Microwave and Optical Technology Letters (MOTL)}, 2017.
	\item Abdul Quddious, \textbf{Muhammad Ali Babar Abbasi}, Haroon Tariq Awan, Photos Vryonides, Alexis Polycarpou, M.A. Antoniades, Symeon Nikolaou1 "On the Design and Implementation of a System on Package (SOP) Receiver for Wireless Power Transmission Applications", to be submitted in \textit{IET Microwaves, Antennas $\&$ Propagation}, 2017.
	
	\item Muhammad Sufian Anwar, Symeon Nikolaou, Hamza Nawaz and \textbf{M. Ali Babar Abbasi}, “Synthesis of Cross-Coupled Trisection and Quadruplet Bandpass Filters Using Hybrid Structures of $\lambda$/4 and $\lambda$/2 Resonators,” submitted in \textit{IEEE Transactions on Microwave Theory and Techniques}, 2017.
	
	\item S. Arain, Photos Vryonides, A. Quddious, \textbf{M. Ali Babar Abbasi}, Marco A. Antoniades and S. Nikolaou, "Square Ring Resonator Based Switchable Bandpass-to-Bandstop Filters Using PIN Diodes", to be submitted in IET Transactions on Microwave Theory and Techniques MTT, 2017

\end{enumerate}

\bibliographystyle{ieeetr}
\renewcommand{\bibname}{References} 
\bibliography{references} 

\end{document}